\documentclass[proof]{pasj01}

\Received{$\langle$reception date$\rangle$}
\Accepted{$\langle$acception date$\rangle$}
\Published{$\langle$publication date$\rangle$}
\usepackage{graphicx}
\usepackage{xcolor}
\usepackage{txfonts}

\begin{document}


\title{Closure relations during the plateau emission of Swift GRBs and the fundamental plane}
\author{M. G. Dainotti\altaffilmark{1} \altaffilmark{3},
   A. Ł. Lenart\altaffilmark{2},
   N. Fraija\altaffilmark{4},
   S. Nagataki\altaffilmark{6} \altaffilmark{7},
   D. C. Warren\altaffilmark{7},\\
   B. De Simone\altaffilmark{8},
   G. Srinivasaragavan\altaffilmark{9}, \and
   A. Mata\altaffilmark{4}}

\altaffiltext{1}{National Astronomical Observatory of Japan, 2 Chome-21-1 Osawa, Mitaka, Tokyo 181-8588, Japan}
\altaffiltext{2}{Astronomical Observatory, Jagiellonian University, ul. Orla 171, 31-501 Krak{\'o}w, Poland}
\altaffiltext{3}{The Graduate University for Advanced Studies, SOKENDAI, Shonankokusaimura, Hayama, Miura District, Kanagawa 240-0193, Japan}
\altaffiltext{4}{Space Science Institute, 4765 Walnut St STE B, Boulder, CO 80301, United States}
\altaffiltext{5}{Instituto de Astronomia, Universidad Nacional Autonoma de Mexico, Apartado Postal 70264, C.P. 04510, Mexico D.F., Mexico}
\altaffiltext{6}{RIKEN Cluster for Pioneering Research, Astrophysical Big Bang Laboratory (ABBL), 2-1 Hirosawa, Wak\={o}, Saitama, Japan 351-0198}
\altaffiltext{7}{RIKEN Interdisciplinary Theoretical and Mathematical Sciences (iTHEMS) Program, Wak\={o}, Saitama, Japan 351-0198}
\altaffiltext{8}{Università degli Studi di Salerno, Dipartimento di Fisica "E. R. Caianiello", Fisciano, Salerno, Via Giovanni Paolo II, 132, 84084, Italy}
\altaffiltext{9}{Cahill Center for Astrophysics, California Institute of Technology, 1200 E. California Blvd. Pasadena, CA 91125, USA}

\email{maria.dainotti@nao.ac.jp; first and second author share the same contribution}


\maketitle

\begin{abstract}
The Neil Gehrels \emph{Swift} observatory observe Gamma-Ray bursts (GRBs) plateaus in X-rays.
We test the reliability of the closure relations through the fireball model when dealing with the GRB plateau emission. We analyze 455 X-ray lightcurves (LCs) collected by \emph{Swift} from 2005 (January) until 2019 (August) for which the redshift is both known and unknown using the phenomenological Willingale 2007 model. Using these fits, we analyze the emission mechanisms and astrophysical environments of these GRBs through the closure relations within the time interval of the plateau emission. Finally, we test the 3D fundamental plane relation (Dainotti relation) which connects the prompt peak luminosity, the time at the end of the plateau (rest-frame), and the luminosity at that time, on the GRBs with redshift, concerning groups determined by the closure relations. This allows us to check if the intrinsic scatter $\sigma_{int}$ of any of these groups is reduced compared to previous literature. The most fulfilled environments for the electron spectral distribution, $p>2$, are Wind Slow Cooling (SC) and ISM Slow Cooling for cases in which the parameter $q$, which indicates the flatness of the plateau emission and accounts for the energy injection, is $=0$ and $=0.5$, respectively, both in the cases with known and unknown redshifts. 
{\bf We also find that for the sGRBs All ISM Environments with $q=0$ have the smallest $\sigma_{int}=0.04 \pm 0.15$ in terms of the fundamental plane relation holding a probability of occurring by chance of $p=0.005$}. 
We have shown that the majority of GRBs presenting the plateau emission fulfill the closure relations, including the energy injection, {\bf with a particular preference for the Wind SC environment.} The subsample of GRBs that fulfill given relations can be used as possible standard candles and can suggest a way to reduce the intrinsic scatter of these studied relationships.
\end{abstract}

\section{Introduction}
\label{sec:intro}
Gamma-ray bursts (GRBs) are the furthest and most explosive objects witnessed by humanity so far. GRBs are observed from $\gamma$--rays to sometimes radio wavelengths \citep[for review]{kumar15}. Their energy emission mechanisms have been widely studied since their discovery. One satellite that has been a hallmark of studying GRBs is the Neil Gehrels \emph{Swift} Observatory \citep{Gehrels2004} that has discovered many GRBs from low to high redshifts, with almost of all of them possessing an afterglow emission. The afterglow is a long-lasting (from minutes, hours, days to even months and years) emission observed in several wavelengths from $\gamma$-ray to radio \citep[e.g.,see]{1999A&AS..138..537S, KPi00a, 2001ApJ...561L.171P, soderberg06, Izzo2012,163, 2015ApJ...804..105F, 2020ApJ...905..112F}. It most probably comes from an external forward shock (ES), where the GRB ejecta that move at a relativistic speed are made up of electrons and positrons ($e^-$, $e^+$) and heavier nuclei, interact with the interstellar medium \citep{Paczynski1993,1994MNRAS.269L..41M,1995ApJ...455L.143S, meszaros97}. The ES model can be verified through relationships, based on theory, called closure relations (hereafter CRs): those relate the afterglow's temporal index ($\alpha$) considering the power-law (PL) and the spectral index ($\beta$) due to synchrotron emission \citep{Sari+98, panaitescu02}. Furthermore, each of these relations indicates different possible environments from which the GRBs stem. The \emph{Swift} GRB lightcurves (LCs) in many cases also include a plateau emission (from now on, PE), a flat portion of the LC, which follows the prompt emission decays \citep{Nousek2006,OBrien2006,Zhang2006,sakamoto07,Zhang2019}. The interpretations related to PE vary from an energy injection model due to the central engine \citep{dai98,rees98,Panaitescu+98, sari2000,zhang2001,Zhang2006,liang2007}, to a magnetar \citep[e.g.,]{zhang2001,Toma2007,troja07,dallosso2011,rowlinson2013,rowlinson14,rea15,BeniaminiandMochkovitch2017,Stratta2018,Metzger2018,Fraija2020} or a mass fall-back accretion onto a black hole \citep{Kumar2008,Cannizzo2009,cannizzo2011,Beniamini2017,Metzger2018}. We here stress that evolving microphysical parameters \citep{fan06,panaitescu06} and the off-axis jet scenario \citep{2019ApJ...871..200F, 2020ApJ...896...25F} have been employed to interpret the PE phase. Additional models regarding a structured jet have been presented \citep{Ito2014,BeniaminiandMochkovitch2017}.
\indent

Within the cosmological context, GRBs can be potentially used as standardizable candles, because they are detected up to redshift, $z$, such that $z\sim9$. 
The most popular standard candles are Supernovae Type Ia (SNe Ia), but the problem is that they can be observed only for $z\lesssim3$ \citep{Rodney2015}. Therefore, using GRBs for the same purpose would allow astronomers to cover orders of magnitude on the distance ladder problem. However, the prompt luminosities of GRBs range over eight orders of magnitude, which is why trying to analyze the entire set of GRBs as a whole has proven to be difficult, especially because many of them have fundamentally different intrinsic physics.  \citet{Dainotti2010} classified GRBs that share similarities considering their phenomenological properties and such a morphological classification can lead to more reliable correlations \citep{cardone09, cardone10, dainotti16c, Dainotti2017c, Dainotti2017b, Dainotti2017a}.
An attempt to check if the 2D Dainotti relations is independent on the $\alpha$ parameter after the PE by using the Willingale et al. 2007 is provided by \cite{delvecchio16}.
\citet{dainotti16c} built a tight 3-D fundamental plane relation and pinpointed a subsample of GRBs called the ``Gold" class which forms a tight fundamental plane between the luminosity at the end of the PE in the afterglow, the time at the end of the PE in the afterglow and the peak luminosity in 1 second of the prompt emission -- ($\log(T_{\rm a})-\log(L_{{\rm peak}})-\log(L_{\rm a})$). This relationship is called the Dainotti 3--D relation (where for simplicity of notation we denote $T_a = T_a^*$, the rest-frame time at the end of the PE), and is an extension of the Dainotti 2--D relations -- $\log(T_{\rm a})-\log(L_{\rm a})$ \hspace{1ex} \citep{Dainotti2008, dainotti11a, Dainotti2013a, Dainotti15a, Dainotti2017a},  and $\log (L_\mathrm{{peak}})- \log(L_{\rm a})$ \hspace{1ex} \citep{Dainotti11b,Dainotti2015b}. {\bf There are several interpretations to the Dainotti 3--D relation starting from the 2D relation. It can be explained within several scenarios: the accretion model on the black hole \citep{cannizzo2011}, the supercritical pile model \citep{kazanas15}, and the magnetar model \citep{rowlinson14,rea15,Stratta2018}, where the luminosity of an ultra magnetized millisecond pulsar is supposed to generate the plateau emission.}
\noindent Our main goals are:

\begin{enumerate}

\item To investigate the PE phase of \emph{Swift} GRBs considering their emission mechanisms and progenitor environments from an astrophysical point of view \bf{both for sGRBs and LGRBs}; 

\item To pinpoint, if exists, a subclass of GRBs with peculiar features whose intrinsic scatter from the fundamental plane correlations is reduced. The aim is to help further the process of understanding GRBs to create a new type of standard candles from a cosmological point of view. \bf{In this case as well we divide GRBs in sGRBs and LGRBs}.

\end{enumerate}

\noindent To achieve goal (1), we try out several CRs described in \citet{Racusin+09} for 455 GRB LCs observed by Swift (for which the redshift is both known or unknown) from the beginning of 2005 until the end of 2019 August.  This work, thus, covers 15 years of detections, using the spectral indices $\beta$ corresponding to the PE duration, and brings new insight into the emission processes of GRBs. 

\noindent To achieve goal (2), we divide GRBs into groups according to the astrophysical environments they belong to. These groups are built depending on whether GRBs fulfill the CRs. Additionally, we check the 3--D Dainotti relation for these environments to see if the intrinsic scatter decreases comparing to past literature. Furthermore, we make a comparison between our results and the ones reported in \citet{Srinivasaragavan2020}, who did a similar analysis but using a different time, namely the time after the end of the PE.

\noindent The energy injection \citep{Dai1998a,Dai1998b,Panaitescu+98, sari2000,zhang2001, Zhang2006,lu2014,lu2015,Chen2017,Zhao2020,Ma2021}, the emission at high latitude \citep{Kumar2000,OBrien2006,Genet2009,Willingale2010,Zhang2011,Ascenzi2020} are the most acknowledged theoretical frameworks for the PE. We here test the CRs within the energy injection scenario. 
During the deceleration phase, a continuous energy injection may contribute to the forward shock, preventing it from decelerating as quickly as it would in the framework of impulsive energy injection. When the engine is a long-lasting one, its luminosity can be described with the following:

\begin{equation}
    \hspace{20ex}
    \mathcal{L}(t)=\mathcal{L}_0\biggr(\frac{t}{t_B}\biggr)^{-q},
    \label{eq_L(t)}
\end{equation}
whereas $\mathcal{L}_0$ is the luminosity at the beginning of the PE phase, $t_B$ is the transitional time between the different phases of a LC and $q$ indicates the flatness of the PE phase. In general, for the injection mechanism to change the blast-wave dynamics the condition $q<1$ is required \citep{Zhang2006}. Thus, in this work, we will use this particular assumption and we will not discuss the cases for $q\geq1$.
Here, we consider the cases of $q=0$ and $q=0.5$. We here stress that for $q=1$, the standard synchrotron forward-shock model is recovered \citep{Sari+98}, but $q=1$ means that we have instantaneous energy injection. 

\section{Data Set And Methodology}
\label{Swift Sample}
The data set and methodology used follow closely the procedures detailed in \citet{Srinivasaragavan2020}, \citet{Dainotti2020a}.
Regarding the spectral analysis, we computed a time-window spectrum for every particular GRB present in our sample during the PE. From the beginning until the end of the PE, denoted with $T_t$ and $T_a$ respectively, we use the {\it Swift} BAT+XRT online repository \footnote{https://www.swift.ac.uk/xrt\_spectra/}. We calculate the photon index by averaging values from the two modes: the windowed and the photon-counting modes of the XRT. We discard GRBs that have $\delta_{\beta}/\beta>1/2$ and $\delta_\alpha/\alpha>1/2$ where $\delta_{\alpha}$ and $\delta_{\beta}$ are the error bars respectively of $\alpha$ and $\beta$. The discarded GRBs correspond to GRBs whose error bars are larger than 50\% of the measurement itself. We additionally discard also GRBs which have spectral indexes with an extremely high value ($> 6$). The spectral index can be written as $\beta=\Gamma-1$, where $\Gamma$ is the photon index.

\noindent Originally, GRBs were classified solely using their duration as either Short (sGRBs, $T_{90} \leq 2s$, \cite{mazets81,kouveliotou93}) or long (lGRBs, $T_{90} > 2s$), {\bf thus performing a classification that is independent of energy ranges and instruments \citep{qin13,Zhang2014}}. Afterwards, as more GRBs began to be discovered, the number of classification groups grew up. Some sGRBs display an extended emission, and are classified as Short with Extended Emission \citep{Norris:06, Norris2010, Levan2007}. It is worth noticing that some lGRBs show an X-ray fluence greater than the $\gamma$-ray fluence, and are thus classified as X-Ray flashes (XRFs). Furthermore, some GRBs show a clear association to Supernovae (SNe) and Kilonovae (KNe) and are classified as GRB-SNe and GRB-KNe. Studying the GRB-KNe class, in particular, is important due to the correlation between sGRBs and gravitational waves. Finally, ultra Long (UL) GRBs display an atypically long duration of $T_{90} \geq 1000s$ \citep{Nakauchi2013,Stratta2013,levan14,Zhang2014}. We subdivide the 222 GRBs with known redshift according to these classes and indicate them with different symbols on the fundamental plane (see \S \ref{Results from SWIFT}).
{\bf In addition to these classes, we also consider the cases of internal and external plateaus. While the external plateaus are engined from the deceleration of the external shock \citep{Tang19}, the internal plateaus are likely powered from the GRB magnetar central engine and are characterized by values of $\alpha >(3,4)$ according to the literature reference we consider such as \citet{Liang18}.}

\section{Closure Relations}
\label{closure}
\subsection{Theoretical Description in Context of Astrophysical Environments}
Following a previous approach by \citet{lu2014} and \citet{lu2015} for the lGRBs and sGRBs, respectively and by \citet{wang15} for the PE (Phase II), we here consider the PE region, instead of phase III as done in \cite{Srinivasaragavan2020} and we compare the results with phase III done by our previous analysis. In this way, we have completed the analysis comprehensively during and soon after the PE region and we can draw conclusions and comparisons about regions II and III. The aforementioned CRs consider the synchrotron radiation as the main underlying mechanism of the afterglow and are computed assuming that $F_{\nu} \propto t^{-\alpha}\nu^{-\beta}$ \citep{Sari+98}. The indexes $\alpha$ and $\beta$ can be associated with the spectral index of the electron distribution, $p$, that is represented as a simple PL: $dn_e/d\gamma_e \propto \gamma_e^{-p}$.

\begin{figure}

    \centering

    \includegraphics[width=1.0\columnwidth]{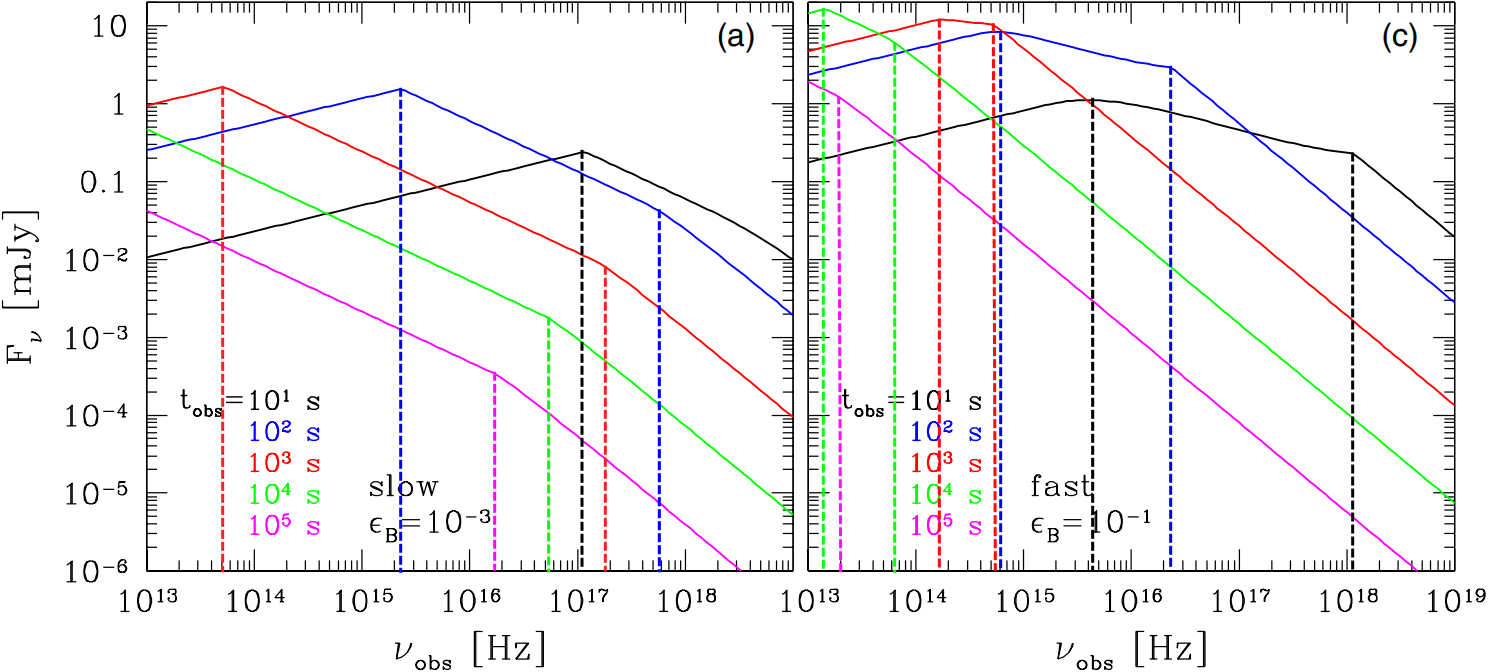}

    \caption{Modified Figure 3 from \citet{Uhm13} in mJy for the flux, representing afterglow spectra in the ES corresponding to a range of times, in both the SC and FC cases in a constant ISM medium and showing with dashed lines the breaks.}

    \label{mJytoergs3}

\end{figure}

\begin{figure}

    \centering

    \includegraphics[width=1.0\columnwidth]{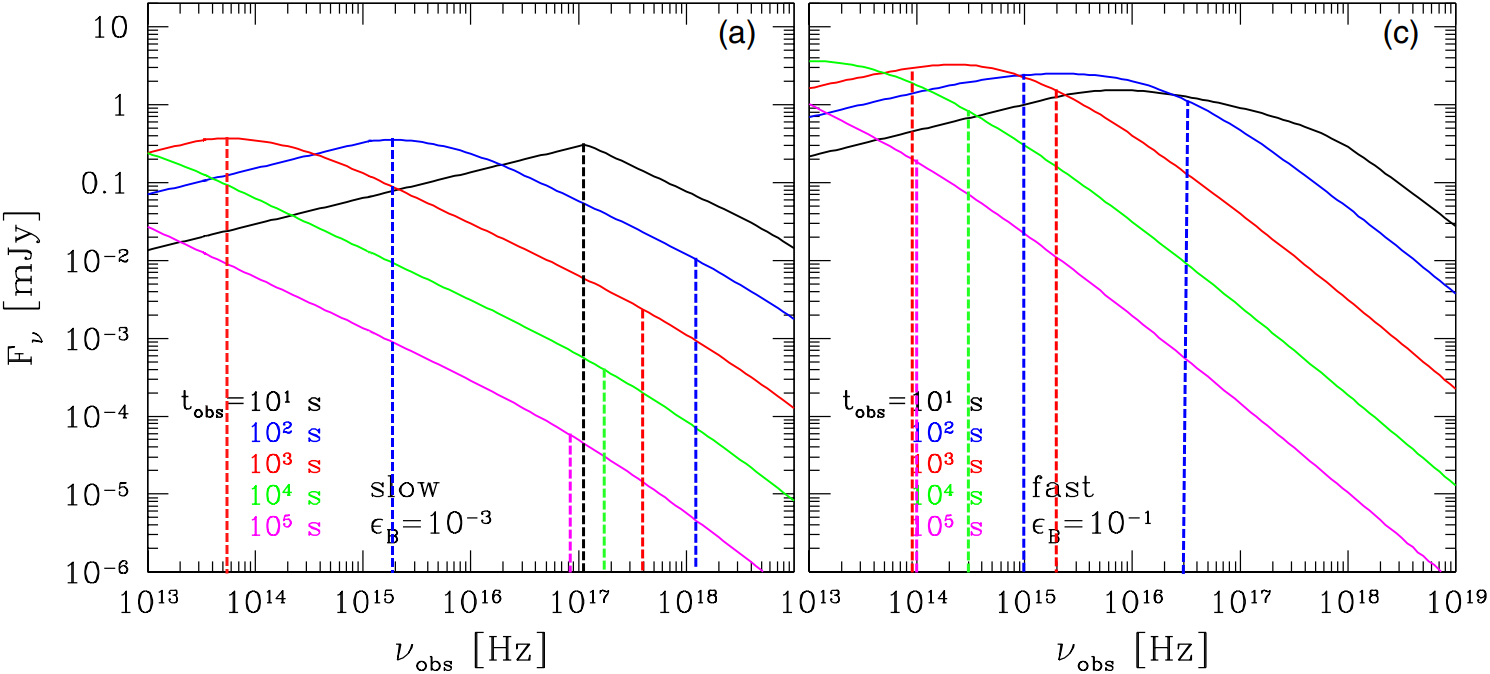}

    \caption{Modified Figure 4 from \citet{Uhm13} for the flux, representing afterglow spectra in the ES corresponding to a range of times, in both the SC and FC cases for Wind medium.}

    \label{mJytoergs4}

\end{figure}

\noindent We here study the CRs presented in \citet{Racusin+09}, where the cooling regime is the time that electrons need to cool down to the critical Lorentz factor $\gamma_c$, where cooling due to synchrotron radiation is significant. 
To test the CRs for the frequency ($\nu$) and $p$ range according to Table \ref{CR2}, we recreate the top panels of Figures 3 and 4 in the theoretical models of \citet{Uhm13} and add the breaks as dashed lines corresponding to the several frequencies. To match the analysis in our paper, we convert the fluxes of our fitting from $\mathrm{ergs}\, \mathrm{cm}^{-2} \, \mathrm{s}^{-1}$ to mJy to check at which flux and time the spectral break corresponds (see Figures \ref{mJytoergs3} and \ref{mJytoergs4}). Then, we count how many GRBs fall within a given time and frequency range and then we compute the p values associated with these regimes and times. More specifically, we check the regime where the flux value at the end of the PE emission, $F_a$, compared to the flux values of Figures 3 and Figure 4 can be found. Then, we subdivide every GRB into either $\nu_m < \nu < \nu_{\rm c}$ and $\nu > \nu_{\rm c}$ for Slow Cooling (hereafter SC), or 
$\nu > \nu_{\rm m}$ for Fast Cooling (hereafter FC). We choose $\nu>\nu_m$ since most of the synchrotron energy is emitted in this range due to the hardness of the spectra \citep{Nakar2009}. Finally, we impose $p>2$ for all GRBs through the relationship between $\beta$ and $p$ given in Table \ref{CR2}: this choice is dictated by the fact that for $1<p<2$ the energy injection scenario is highly improbable and cumbersome for its treatment \citep{Racusin+09}.

\noindent We here summarize for clarity in bullet points all the steps of our analysis:

\begin{enumerate}

\item Obtain $\alpha$ and $T_{\rm a}$ by fitting the PE with the Willingale 2007 function.

\item Calculate the temporal interval region (time slice) for the PE phase taking it by the fitting which determines $T_t$ and $T_a$ and we calculate $\beta$ in this time slice directly from the observations.

\item Calculate the CRs.

\item For all four combinations of FC vs. SC, and ISM vs. wind-like circumburst medium we take the following steps:

\begin{enumerate}

        \item Determine which curve in Figures \ref{mJytoergs3} and Figure \ref{mJytoergs4} is the closest to $T_{\rm a}$ as found previously in step 1.

        \item For the particular curve identified in Figure \ref{mJytoergs3} and Figure \ref{mJytoergs4}, determine whether $\nu_{\rm m}$ ($\nu_{\rm c}$) is above or below the \textit{Swift} window of 0.3-10 keV in the FC or SC scenario, thus corresponding to X-ray.

        \item Once we know whether we are above or below the relevant spectral breaks, we can identify from Table~\ref{CR2} and Table~\ref{CR2.1} which set of CRs we will test. Then, we compute the electron spectral index $p$ using the second column of Table~\ref{CR2} and Table~\ref{CR2.1} together with $\beta$ as computed previously in step 2.
        
        
        \item 
        Given the value of $p>2$ and once the CRs have been identified, we can compare the values of $\alpha$ and $\beta$ with those.

        \item Alternately, if the X-ray band falls between $\nu_{\rm c}$ and $\nu_{\rm m}$ (regardless of FC vs SC) we then use the GRBs for which $\beta=1/2$.

    \end{enumerate}

\end{enumerate} 

\noindent After all these steps, we verify all of the CRs which correspond to the appropriate $p$ and $\nu$ range for all GRBs here investigated. 
{\bf All these steps are performed by dividing the sample in sGRBs and lGRBs and this division is kept throughout all the analysis regarding the CRs, the values of the distributions of the q parameters, and the fundamental planes correspondent to particular environments.}

\begin{figure}
    \centering
    \includegraphics[scale=0.17]{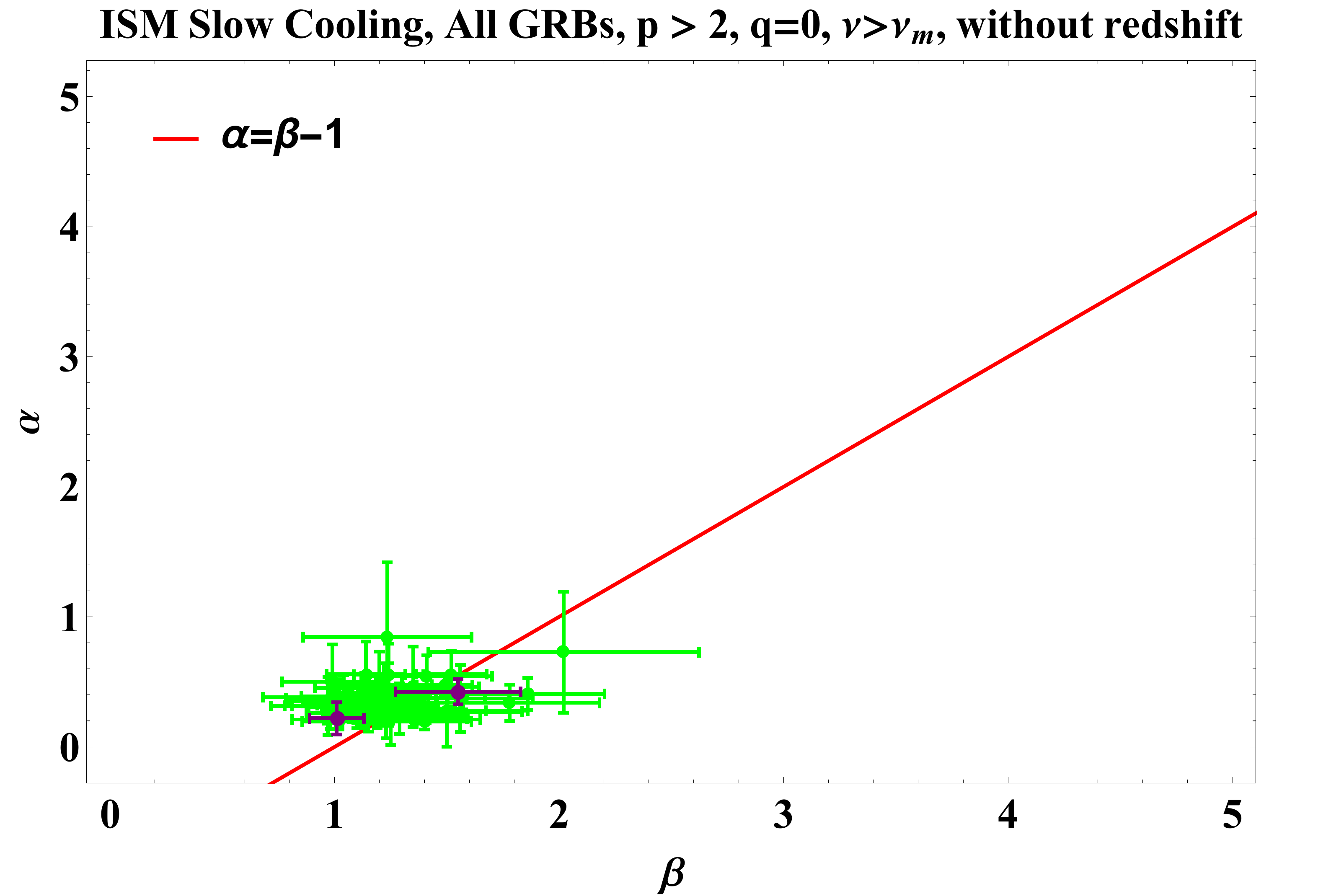}
    \includegraphics[scale=0.17]{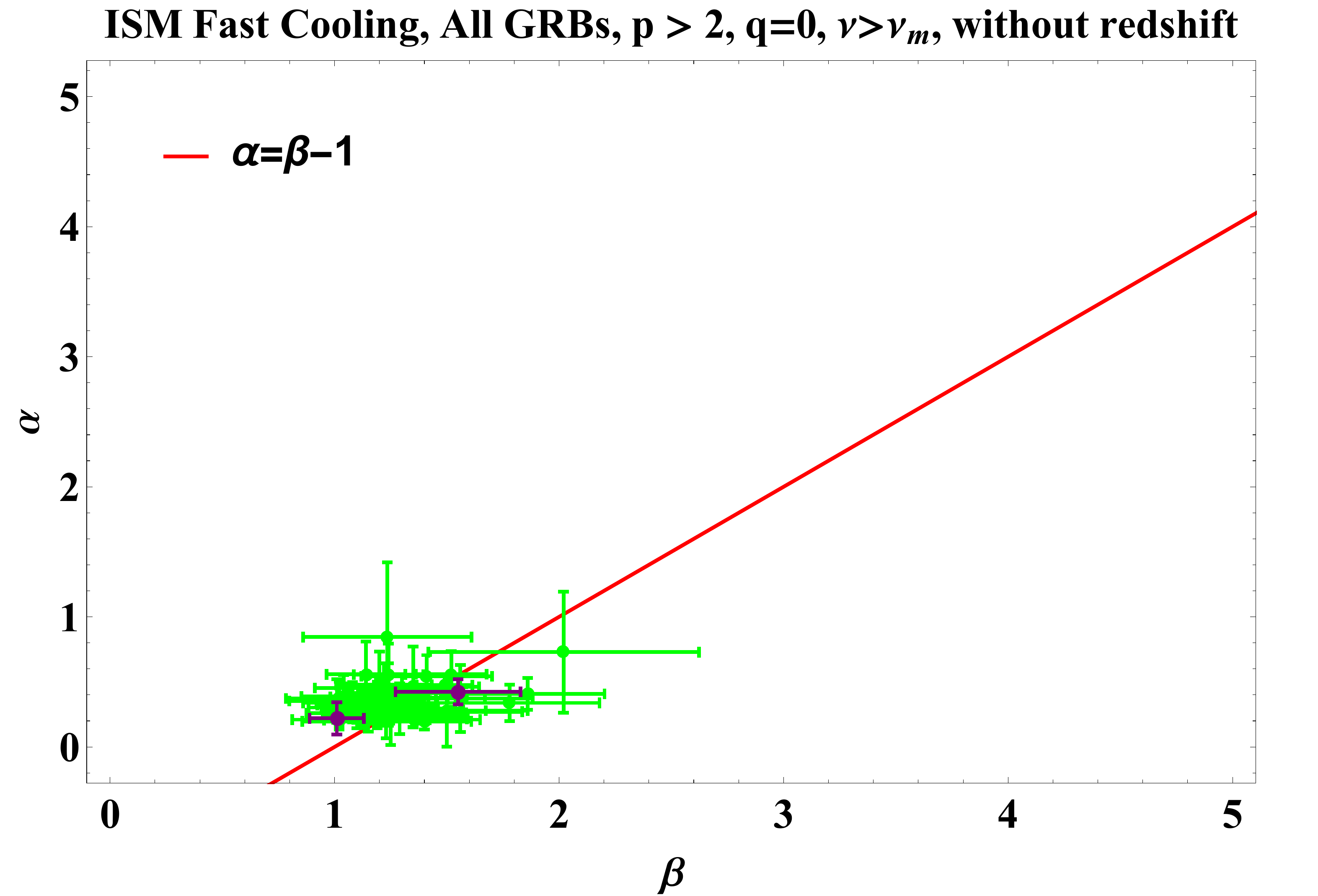}
    \includegraphics[scale=0.17]{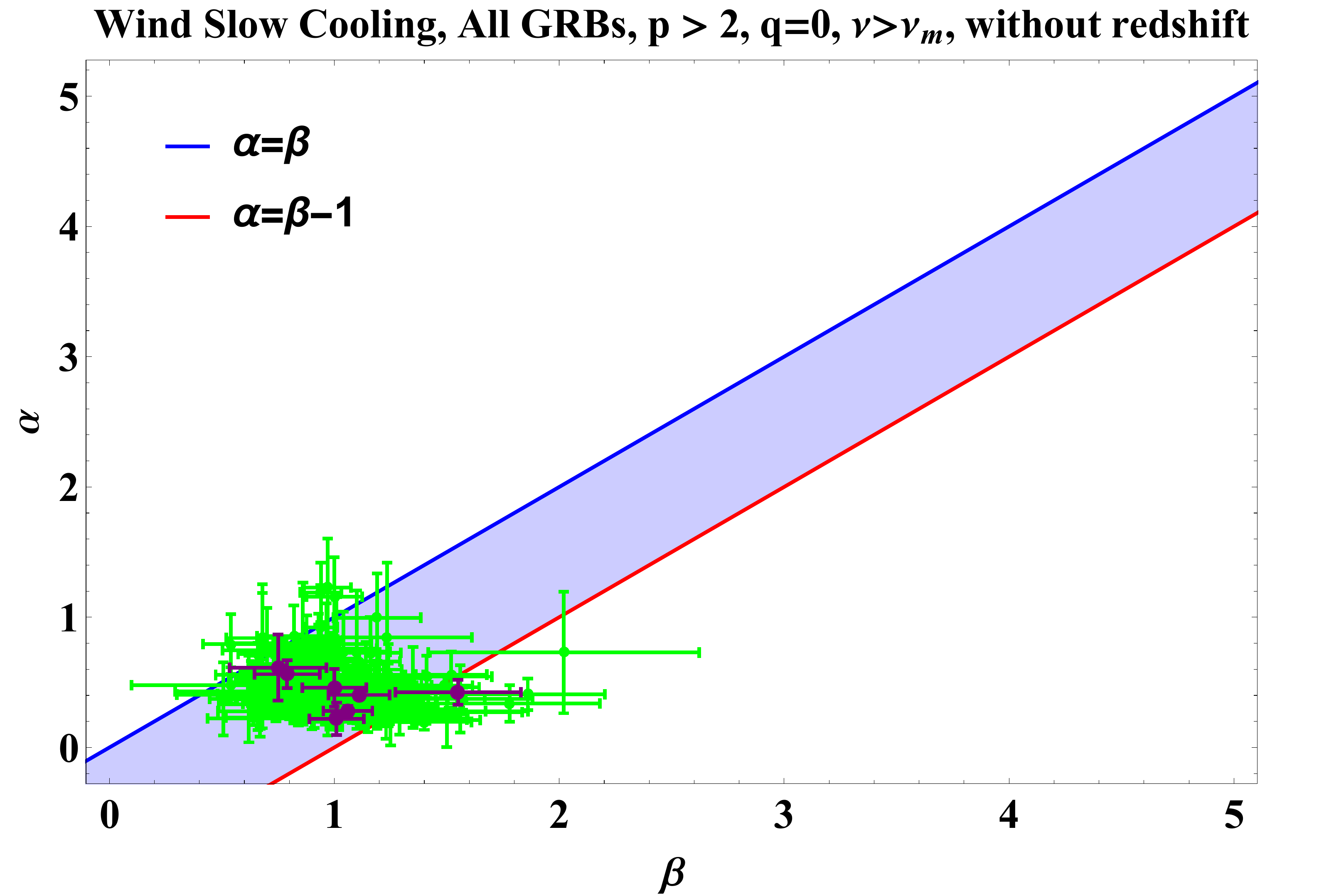}
    \includegraphics[scale=0.17]{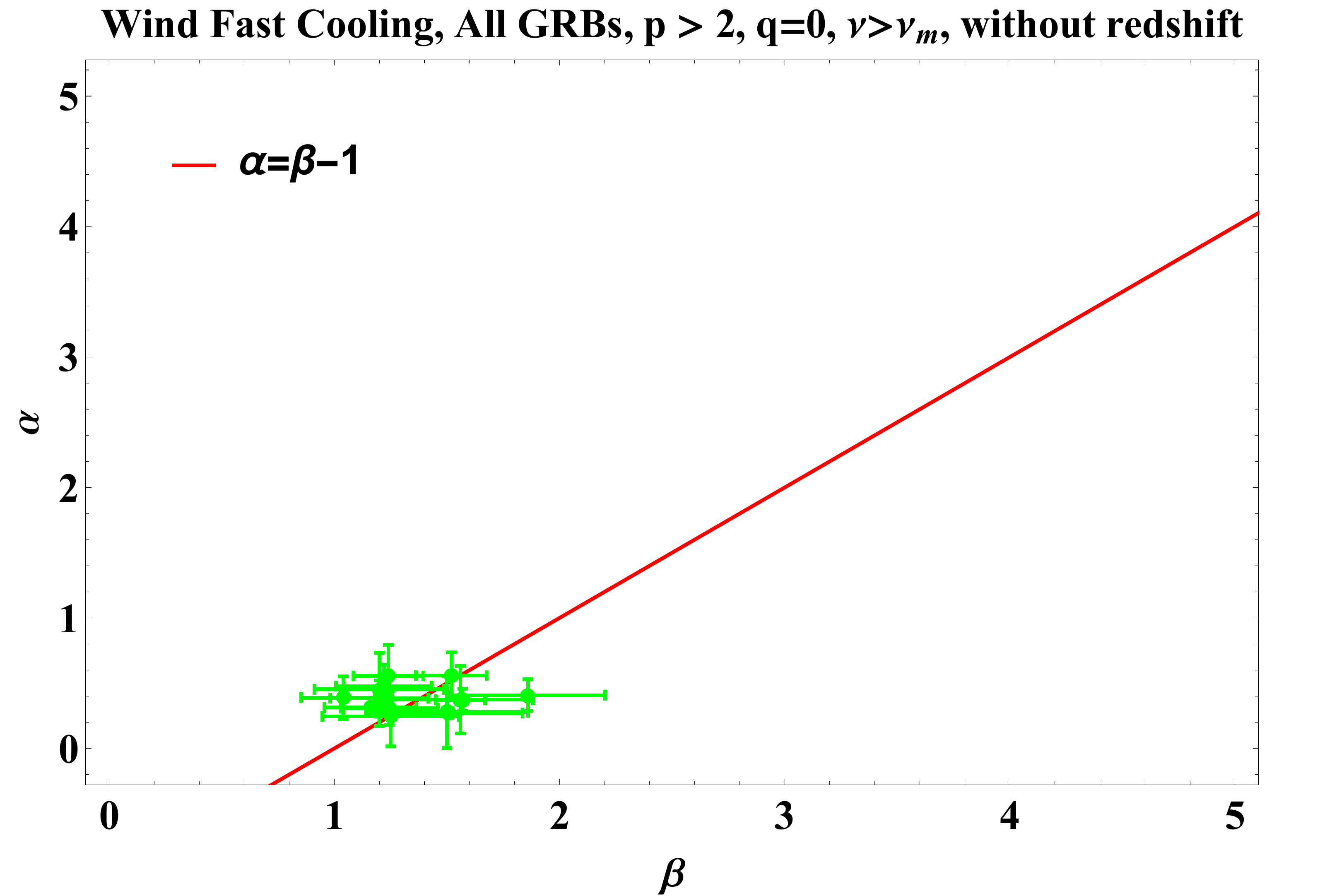}
    \includegraphics[scale=0.24]{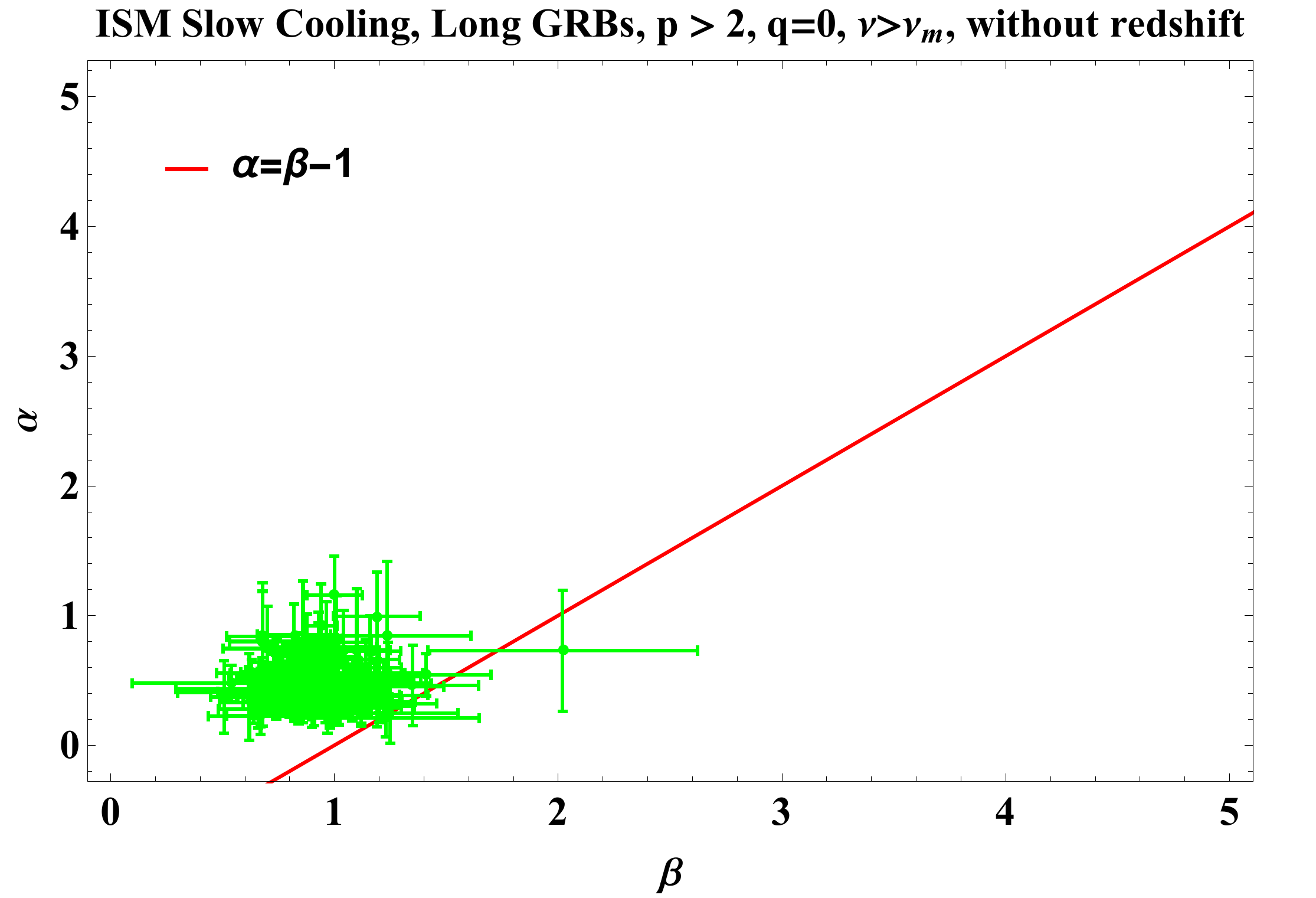}
    \includegraphics[scale=0.23]{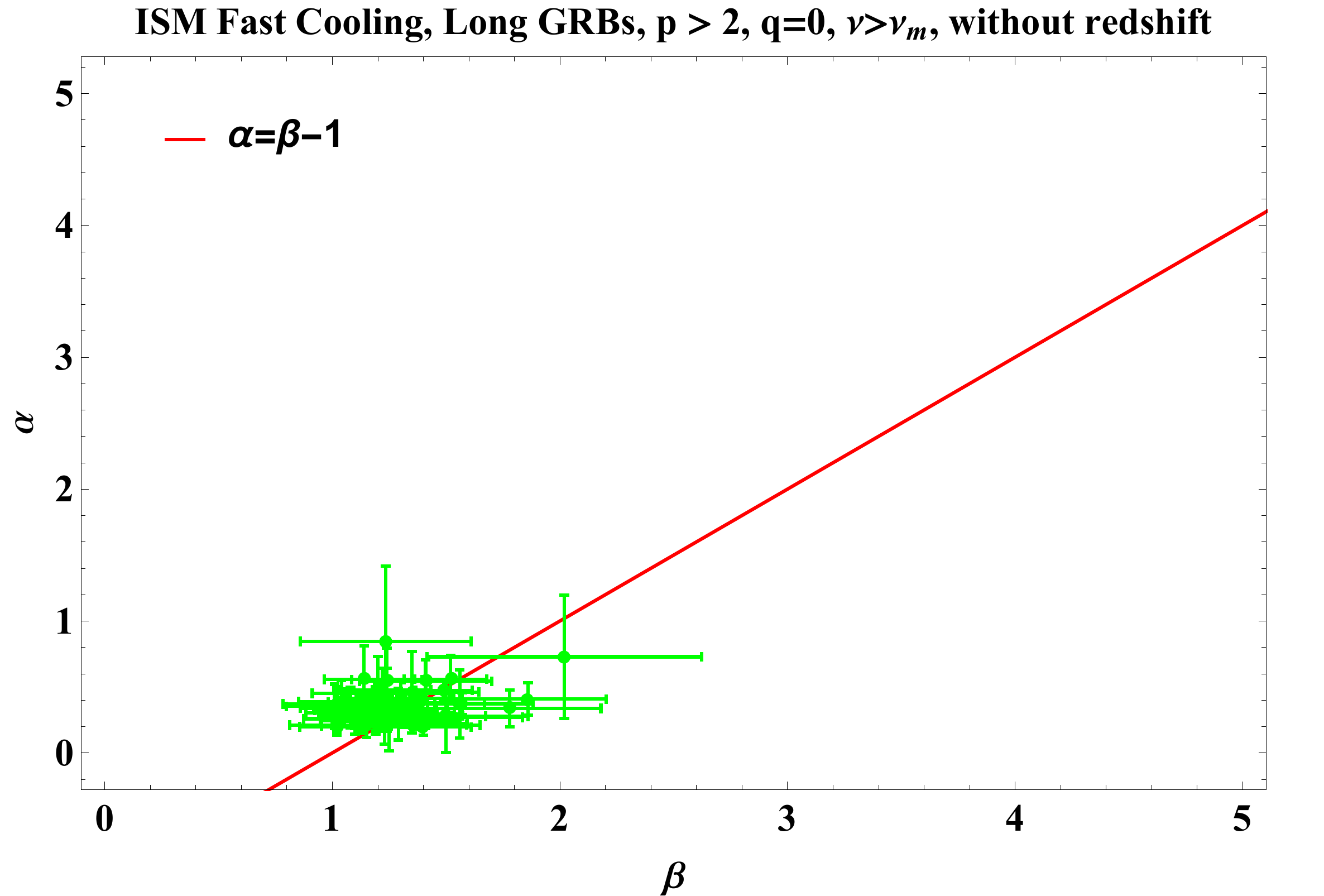}
    \includegraphics[scale=0.23]{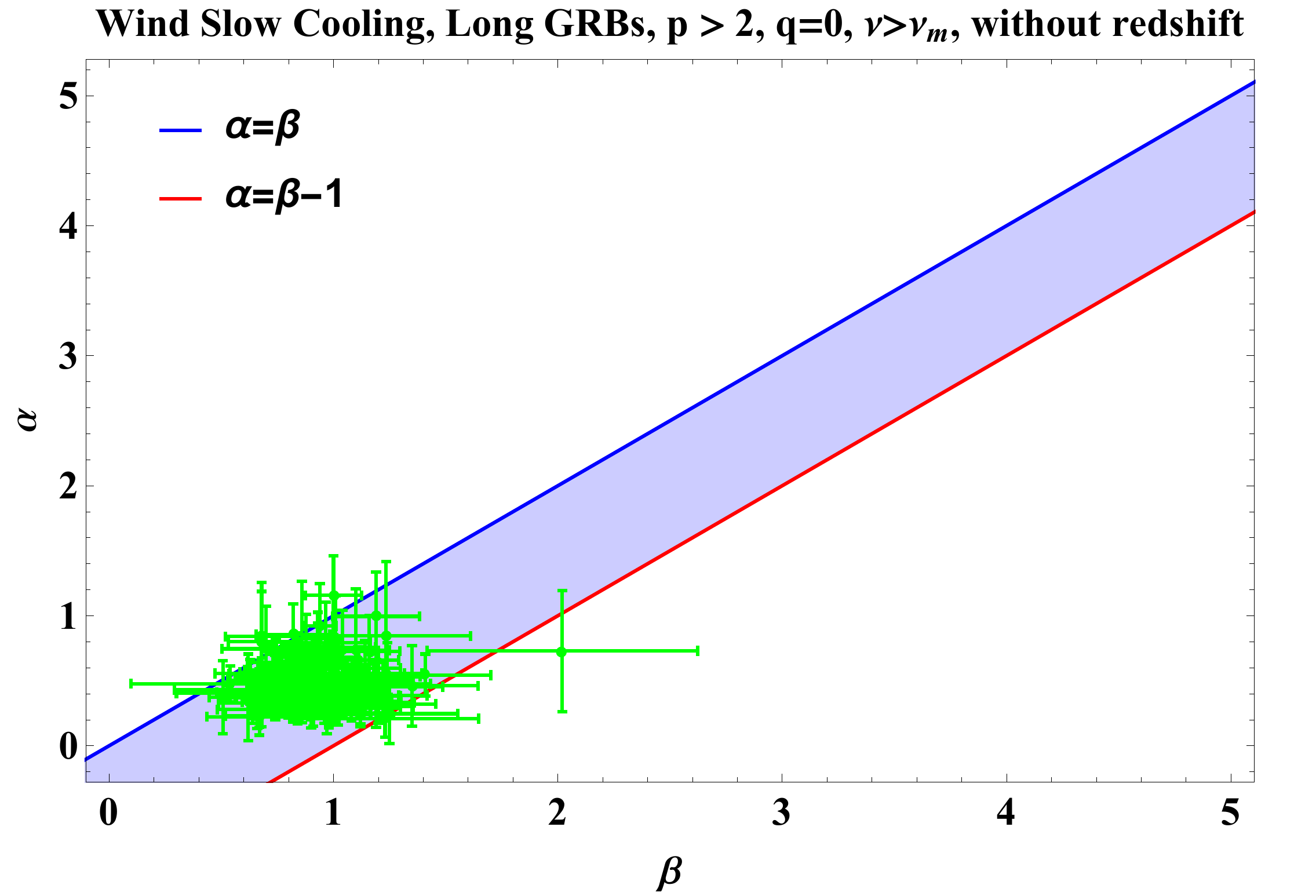}
    \includegraphics[scale=0.24]{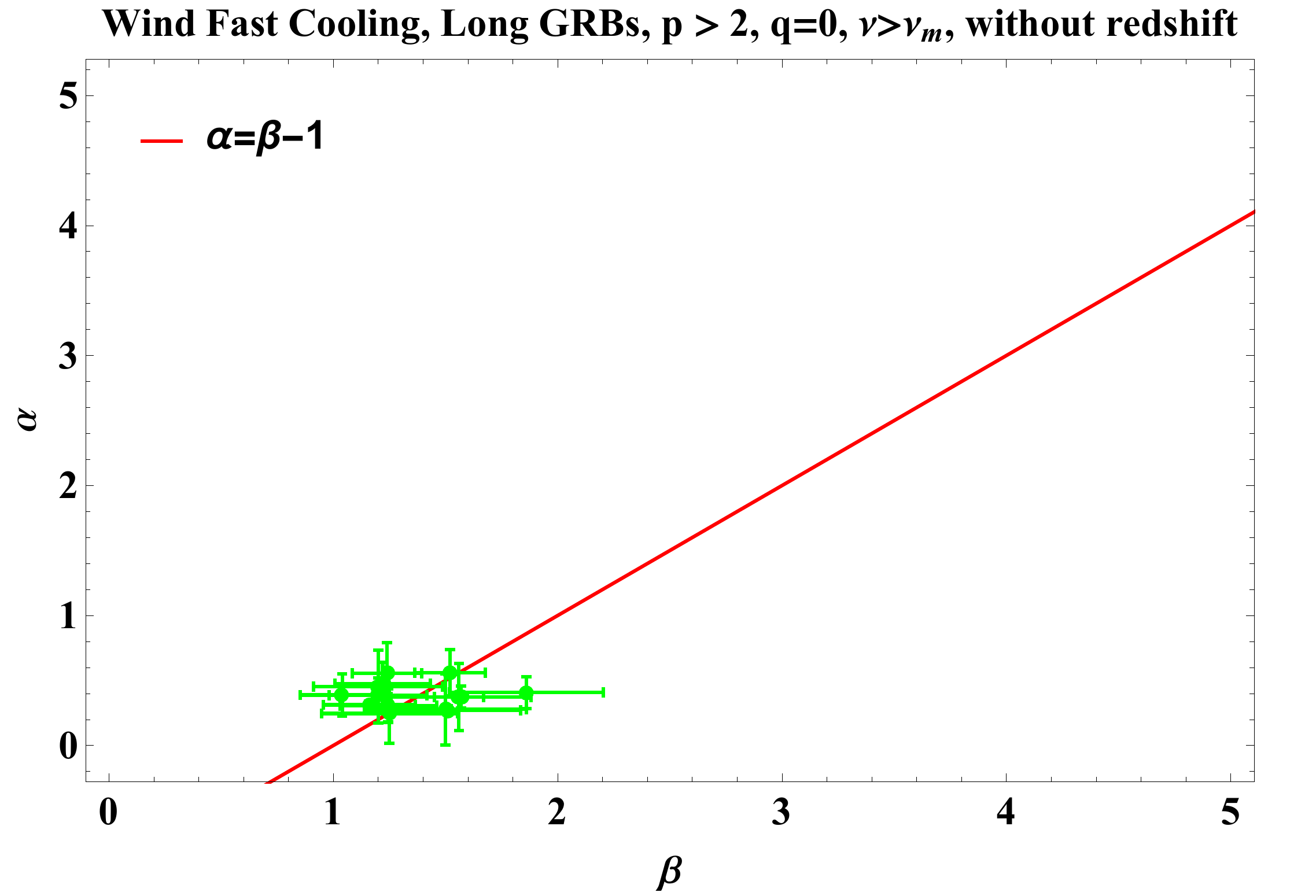}
    \includegraphics[scale=0.17]{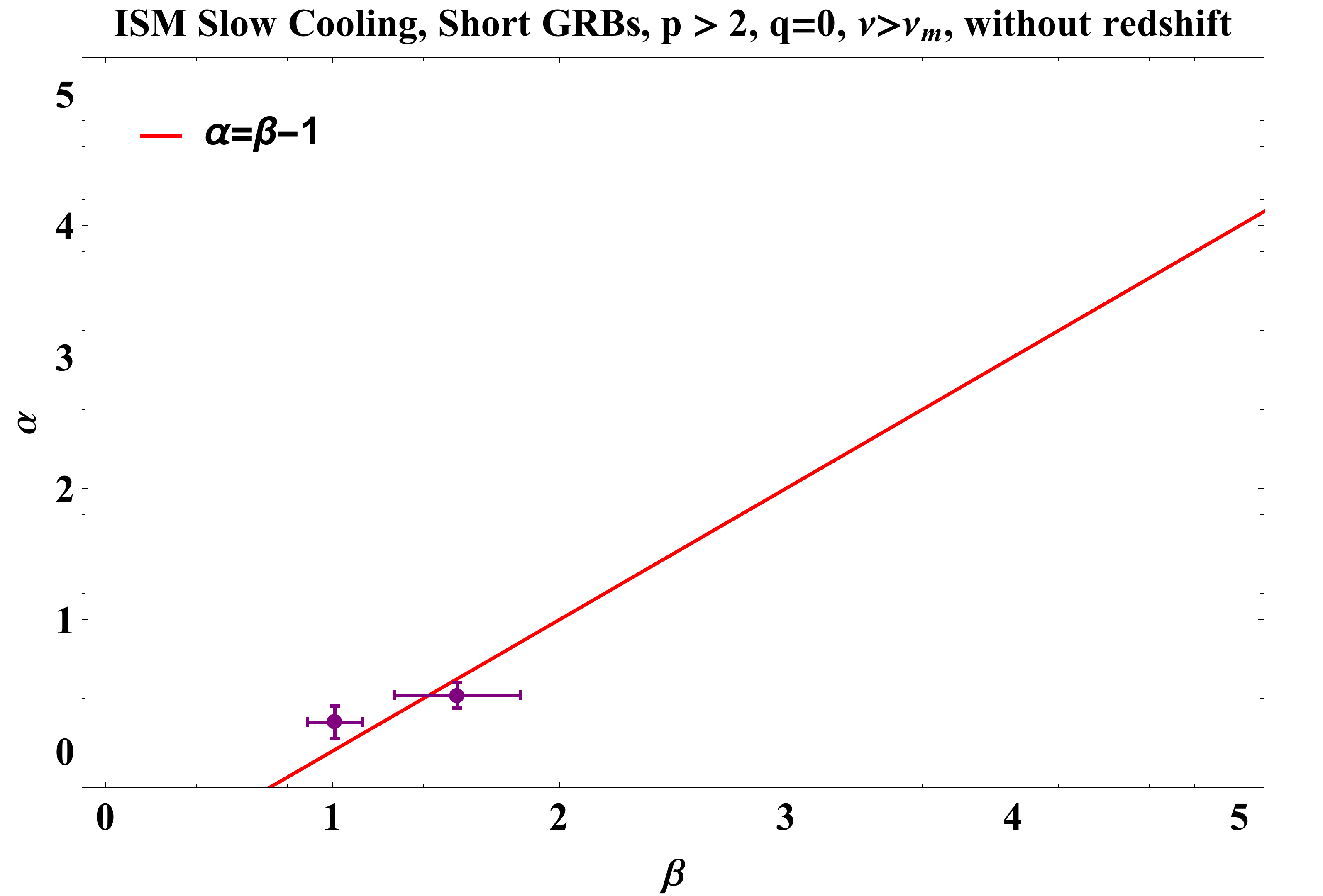}
    \includegraphics[scale=0.17]{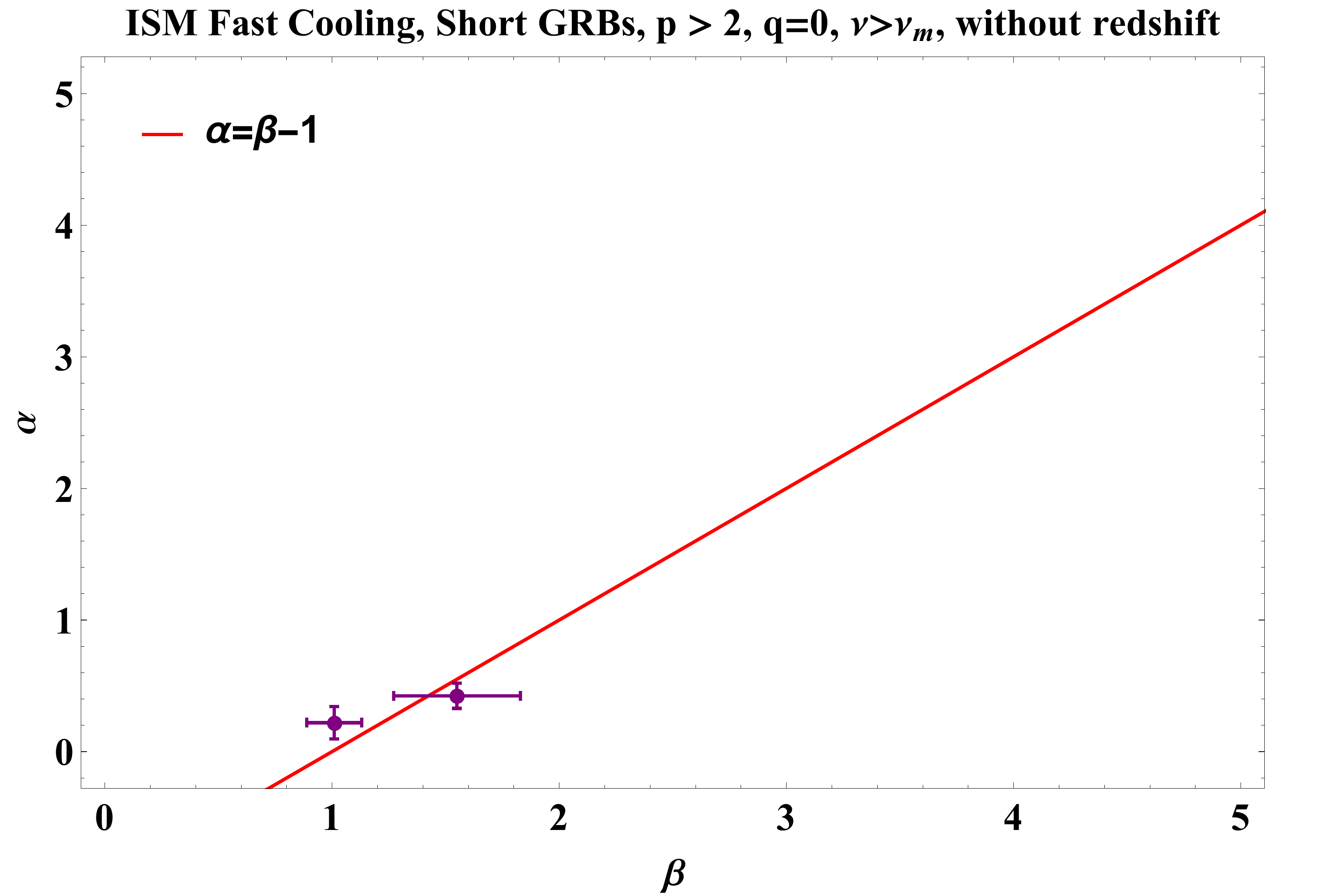}
    \includegraphics[scale=0.17]{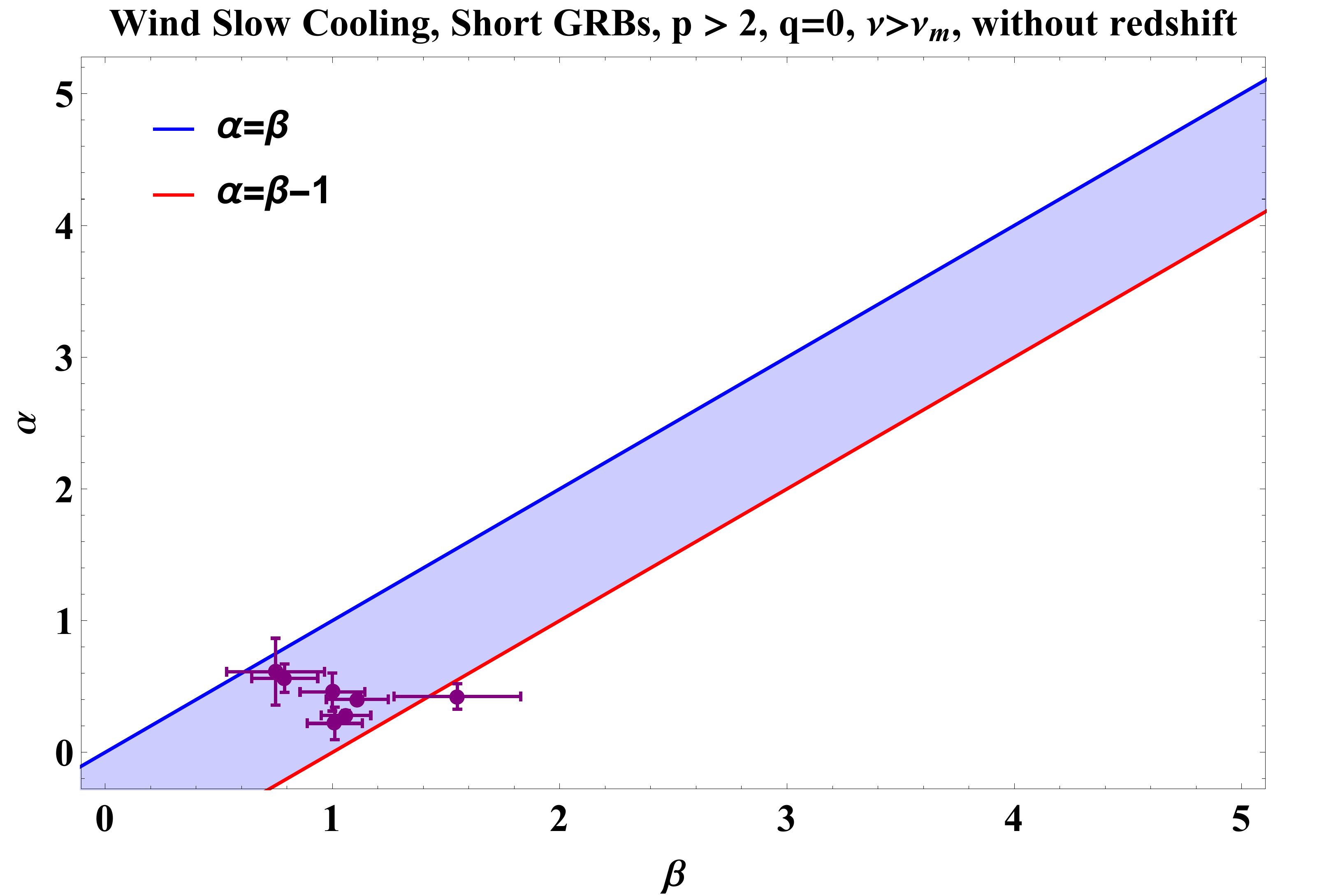}
    \includegraphics[scale=0.17]{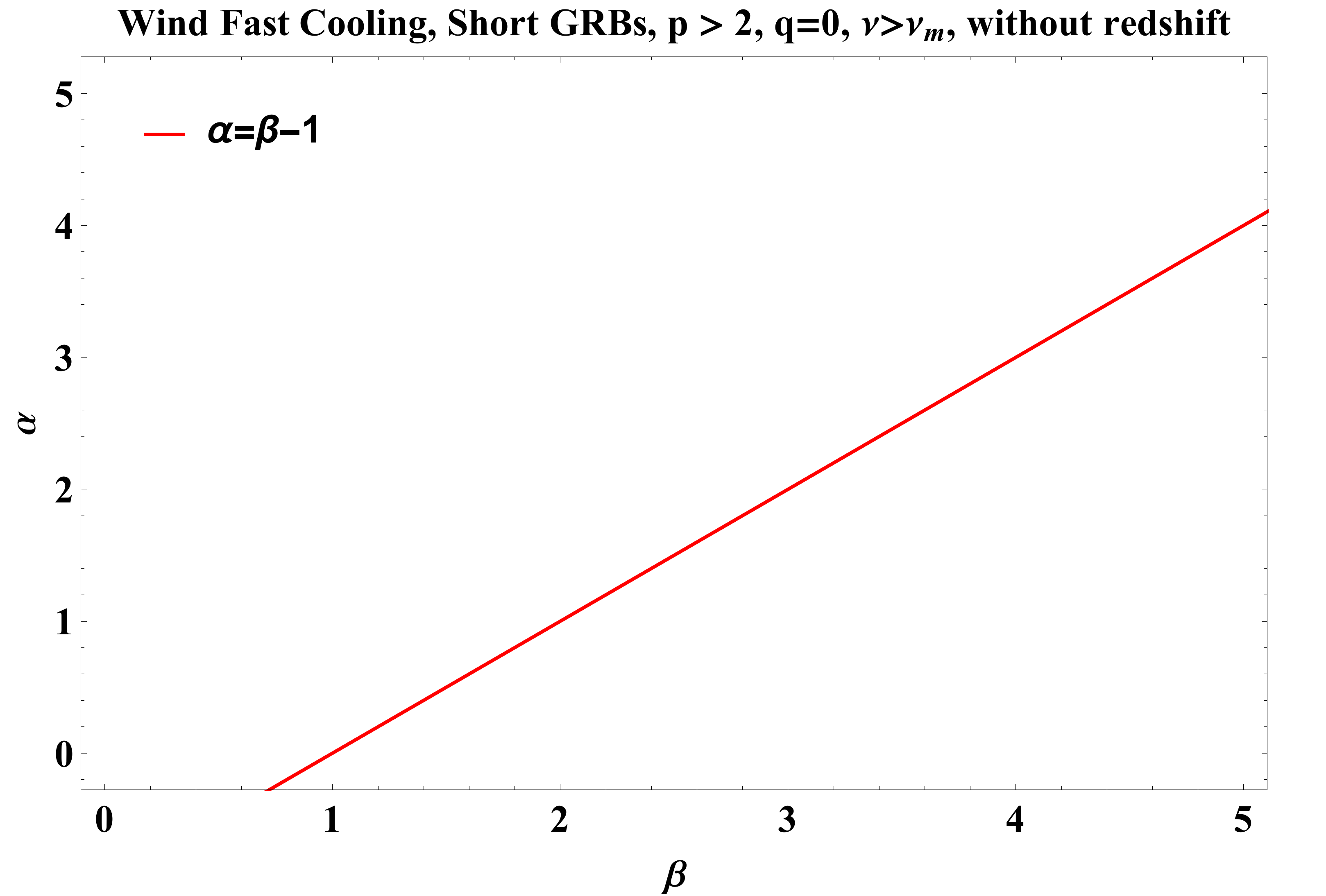}
    \caption{Compatible GRBs for CRs in the case $q=0,\nu>\nu_m, p>2$ without redshift. {\bf In the first 4 panels, the classes lGRBs and sGRBs are gathered together, while in the latter 8 panels these classes are separated (however the fourth panel presents no sGRBs). From this Figure on, green markers will represent the lGRBs while purple markers the sGRBs. We here stress that the colored lines indicate the CRs and the darker sectors mark the compatibility region of GRBs with CRs.}}
    \label{only_nu>num_p>2_NR_q0}
\end{figure}

\begin{figure}
    \includegraphics[scale=0.17]{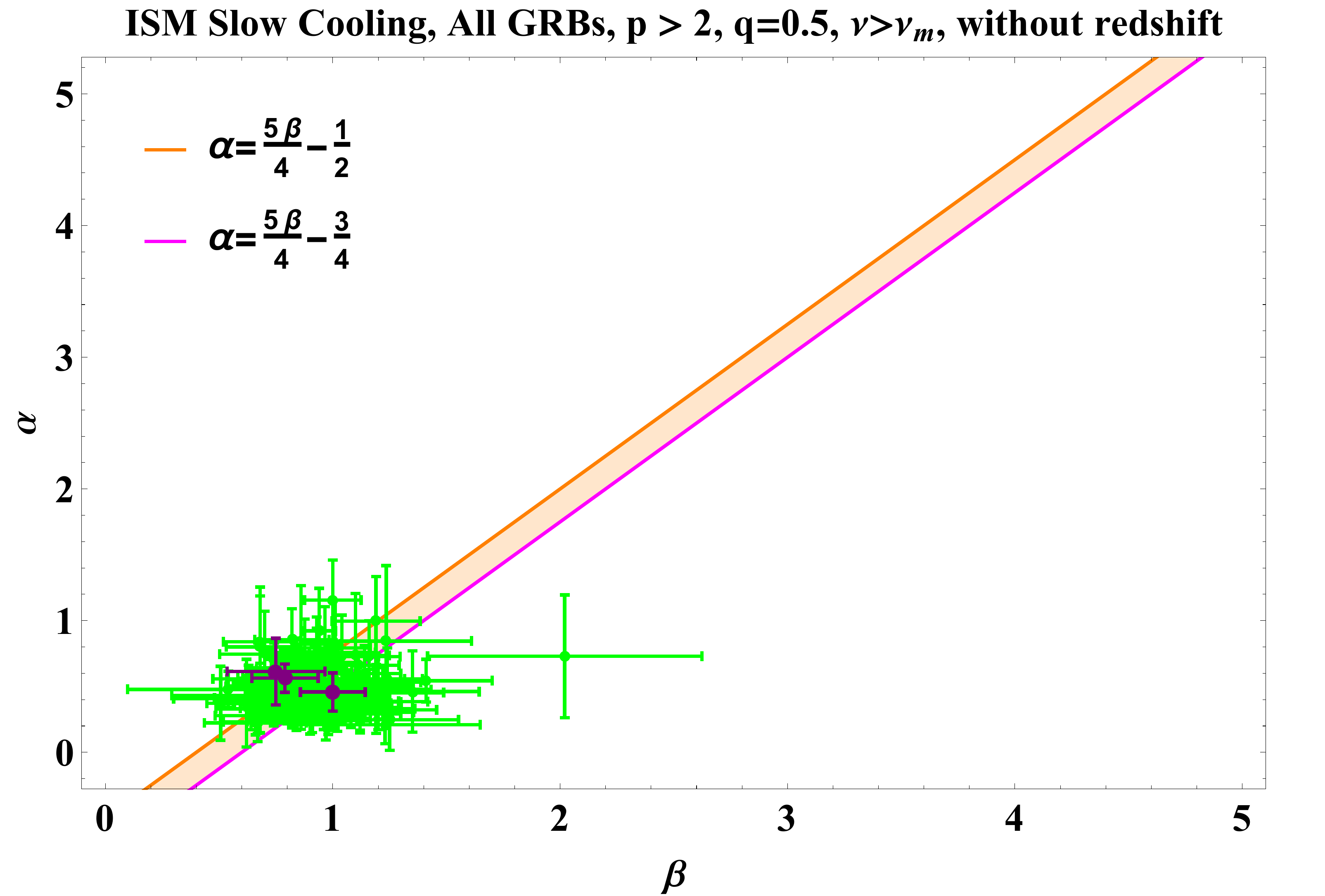}
    \includegraphics[scale=0.17]{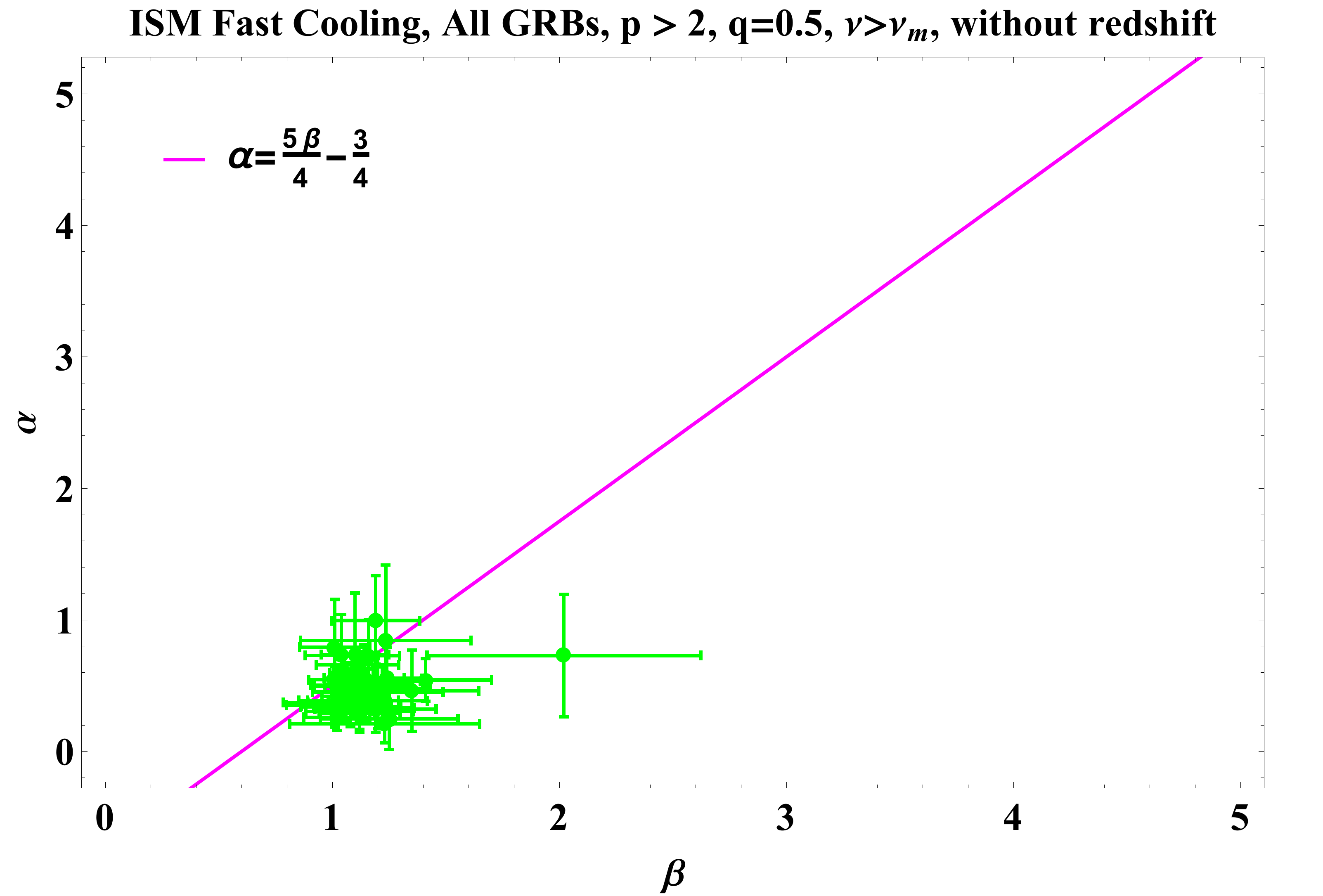}
    \includegraphics[scale=0.17]{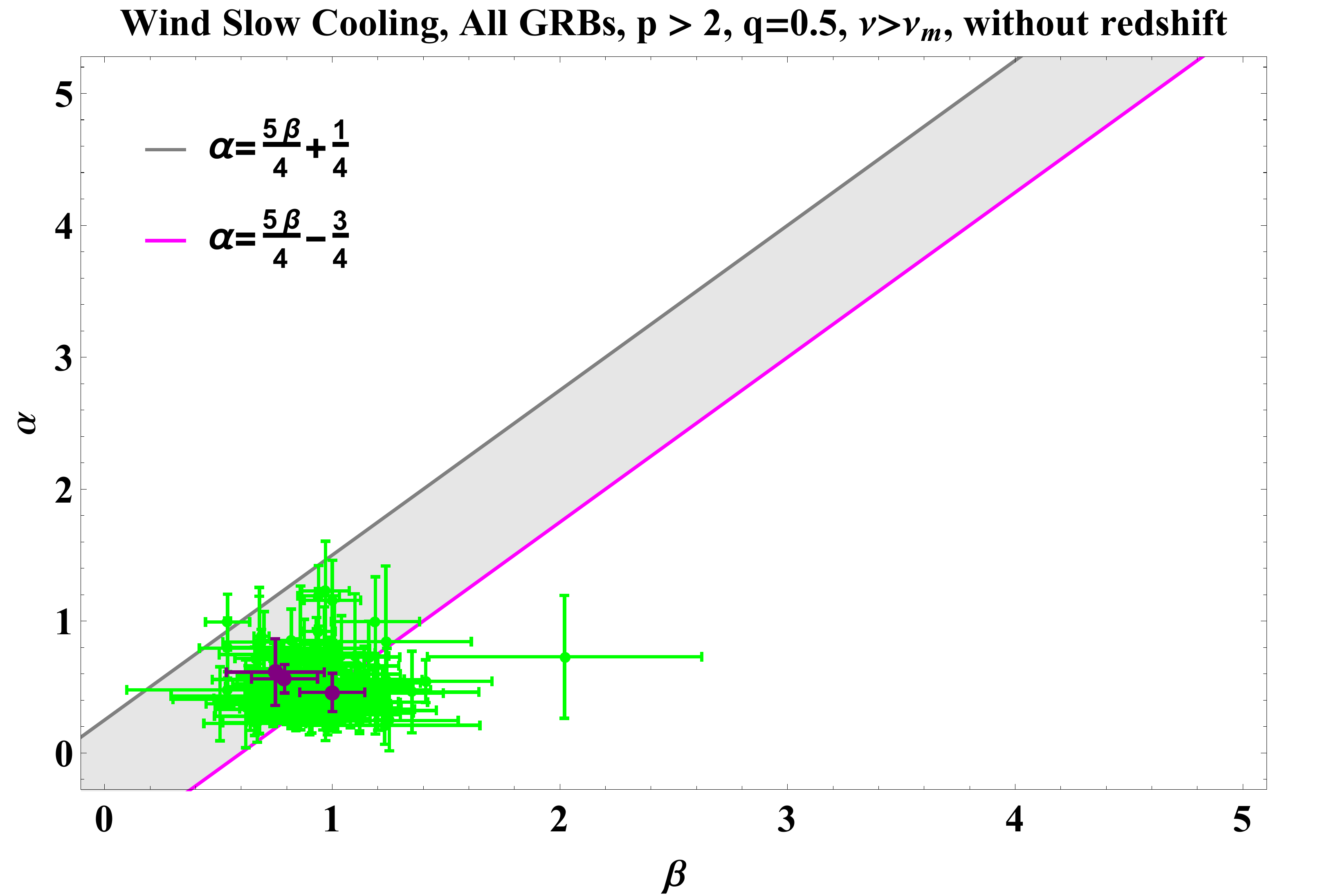}
    \includegraphics[scale=0.17]{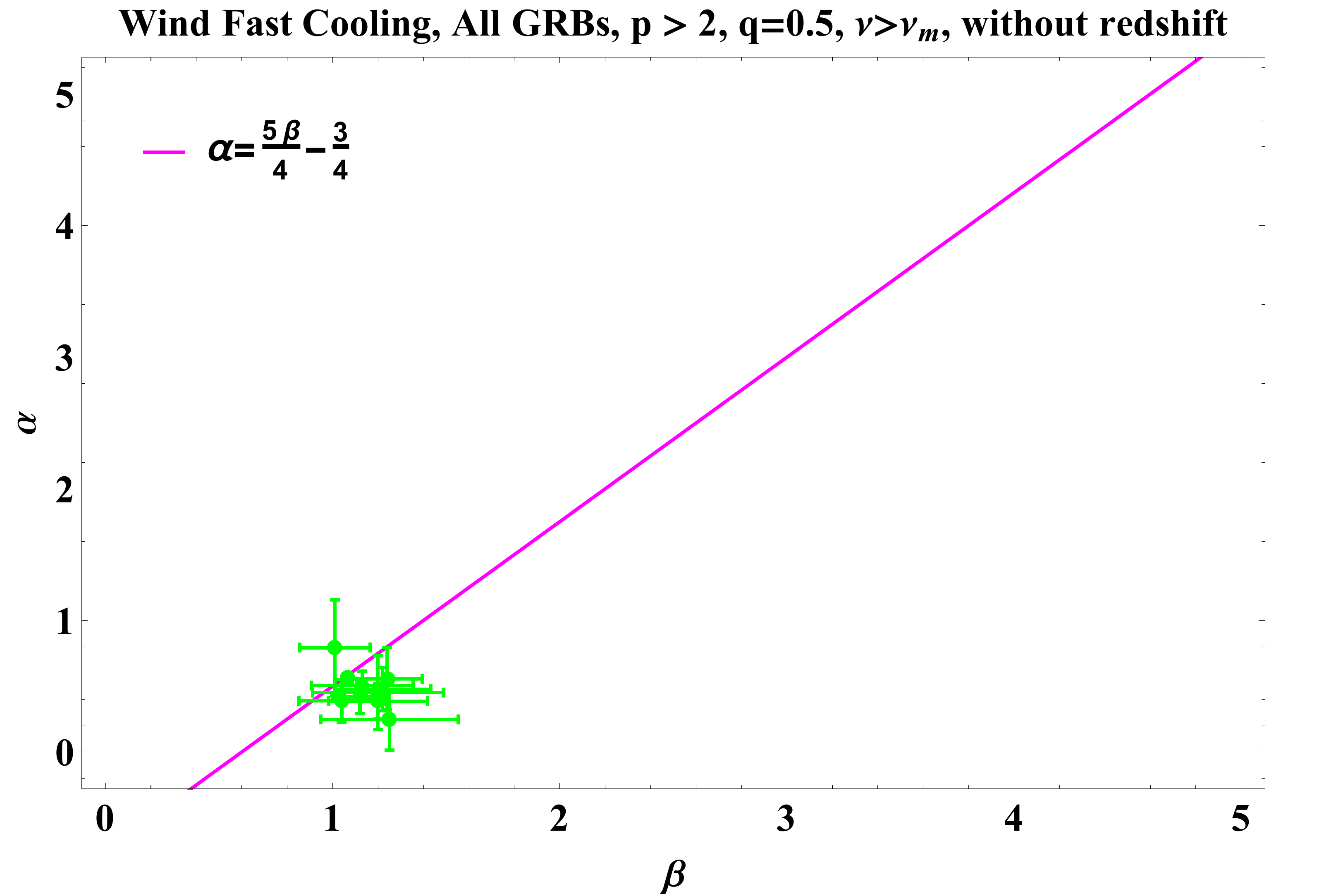}
    \includegraphics[scale=0.17]{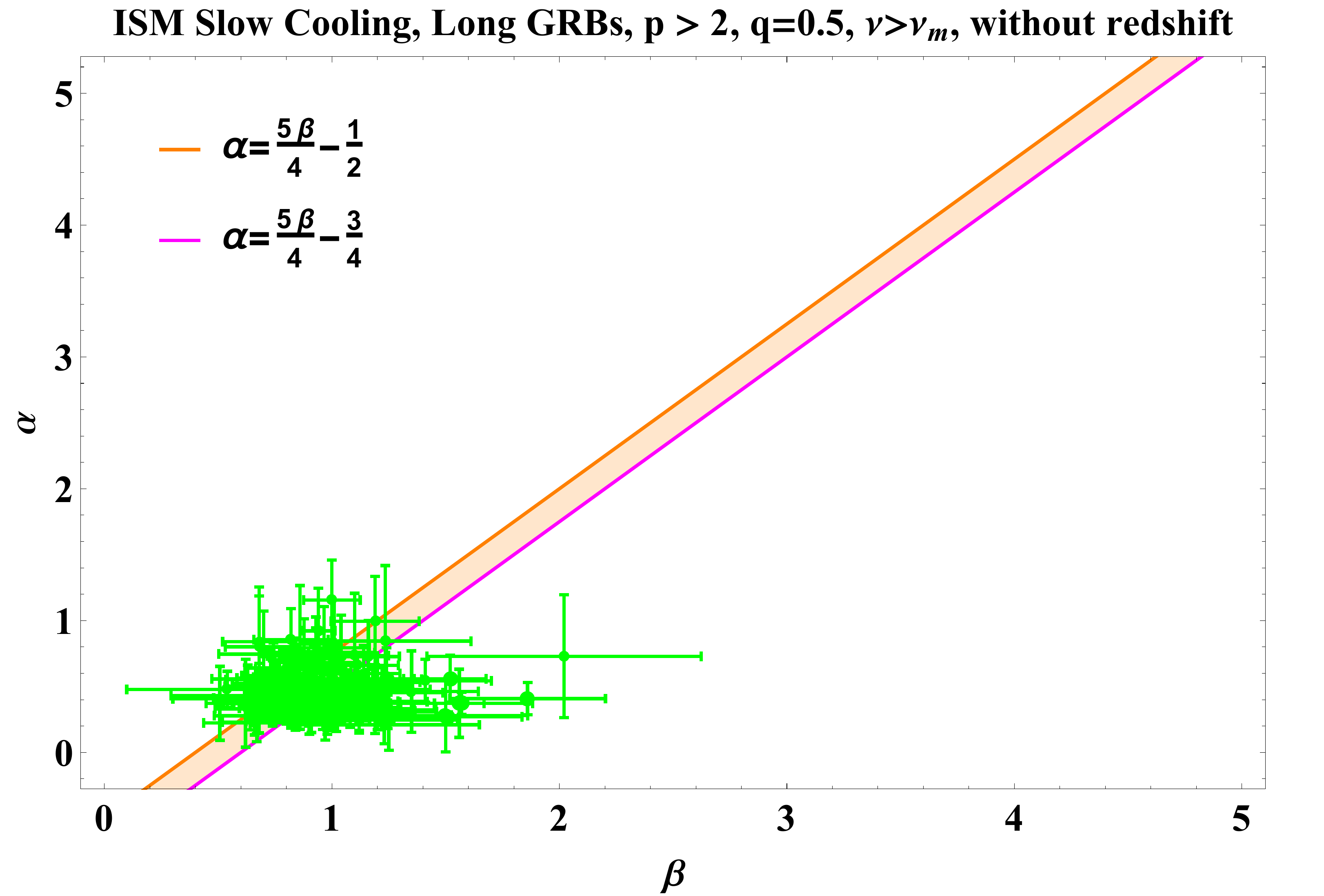}
    \includegraphics[scale=0.17]{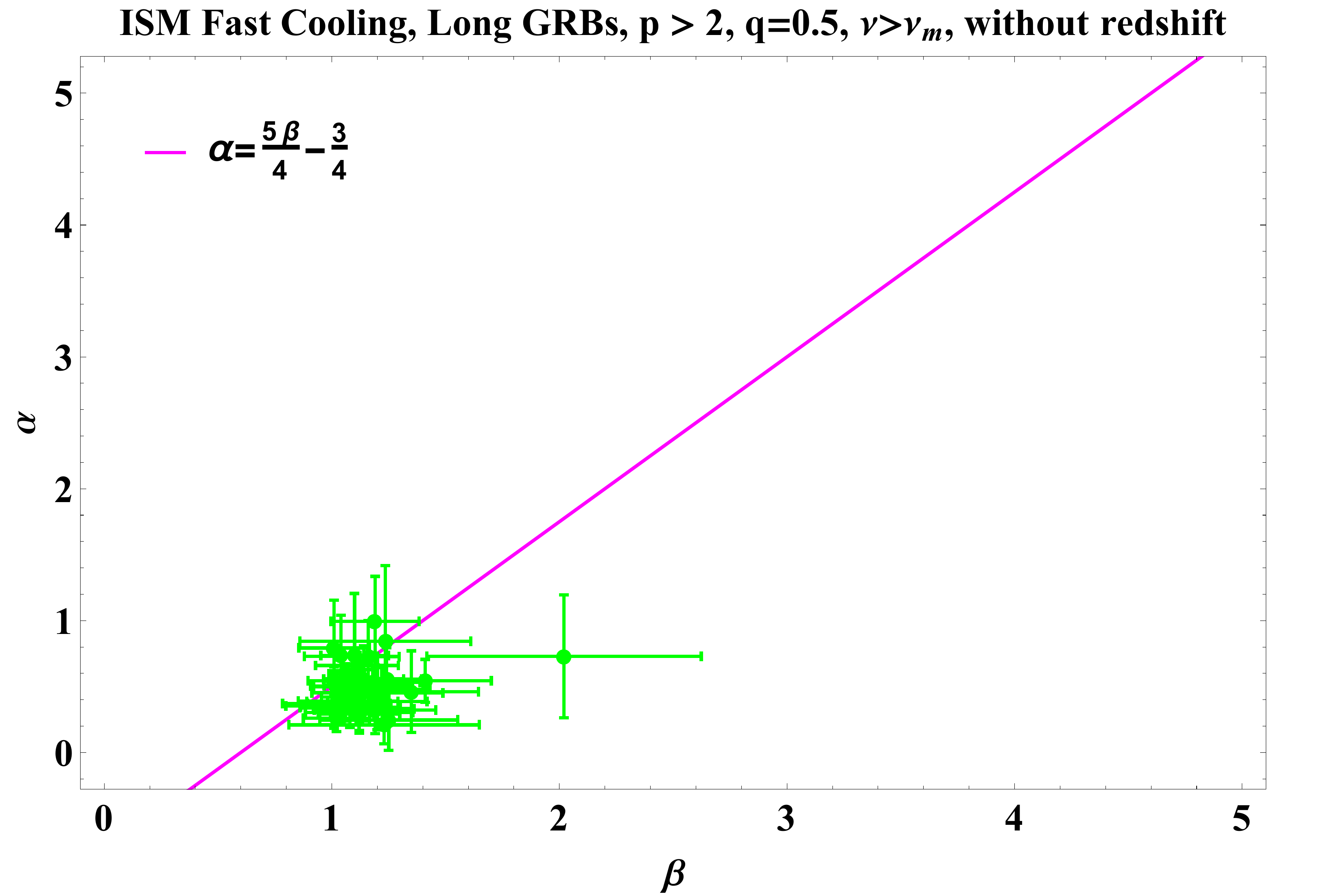}
    \includegraphics[scale=0.17]{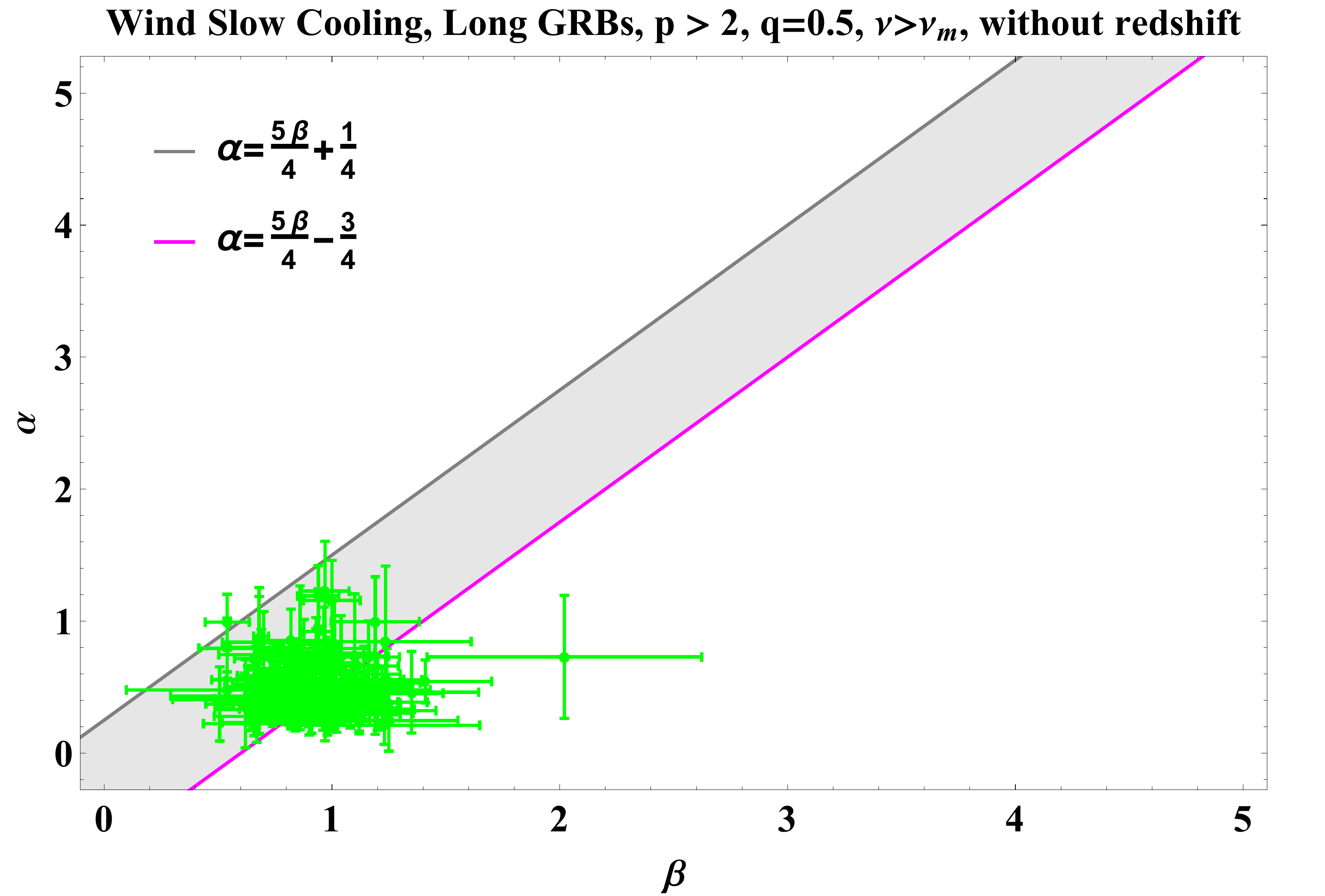}
    \includegraphics[scale=0.17]{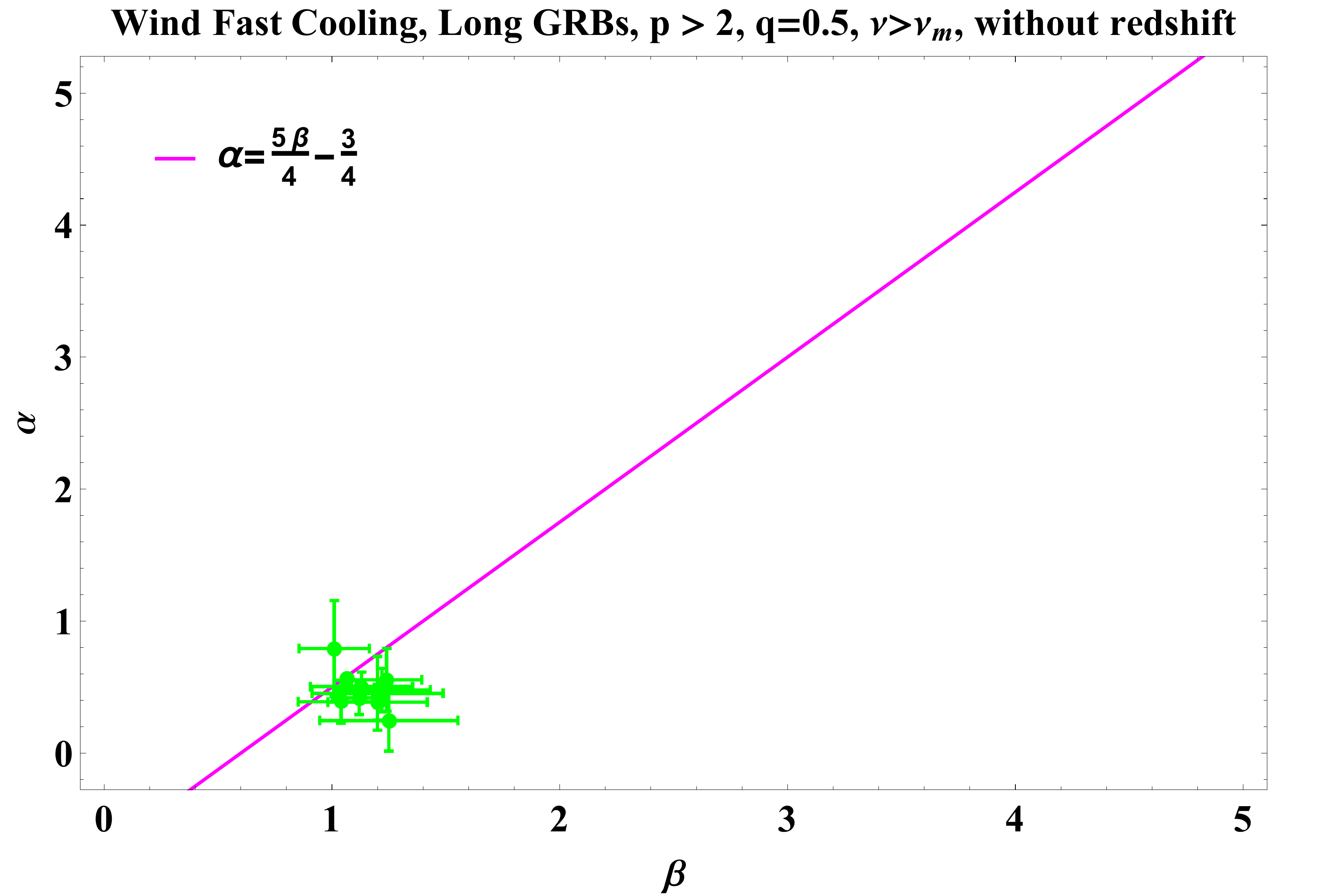}
    \includegraphics[scale=0.16]{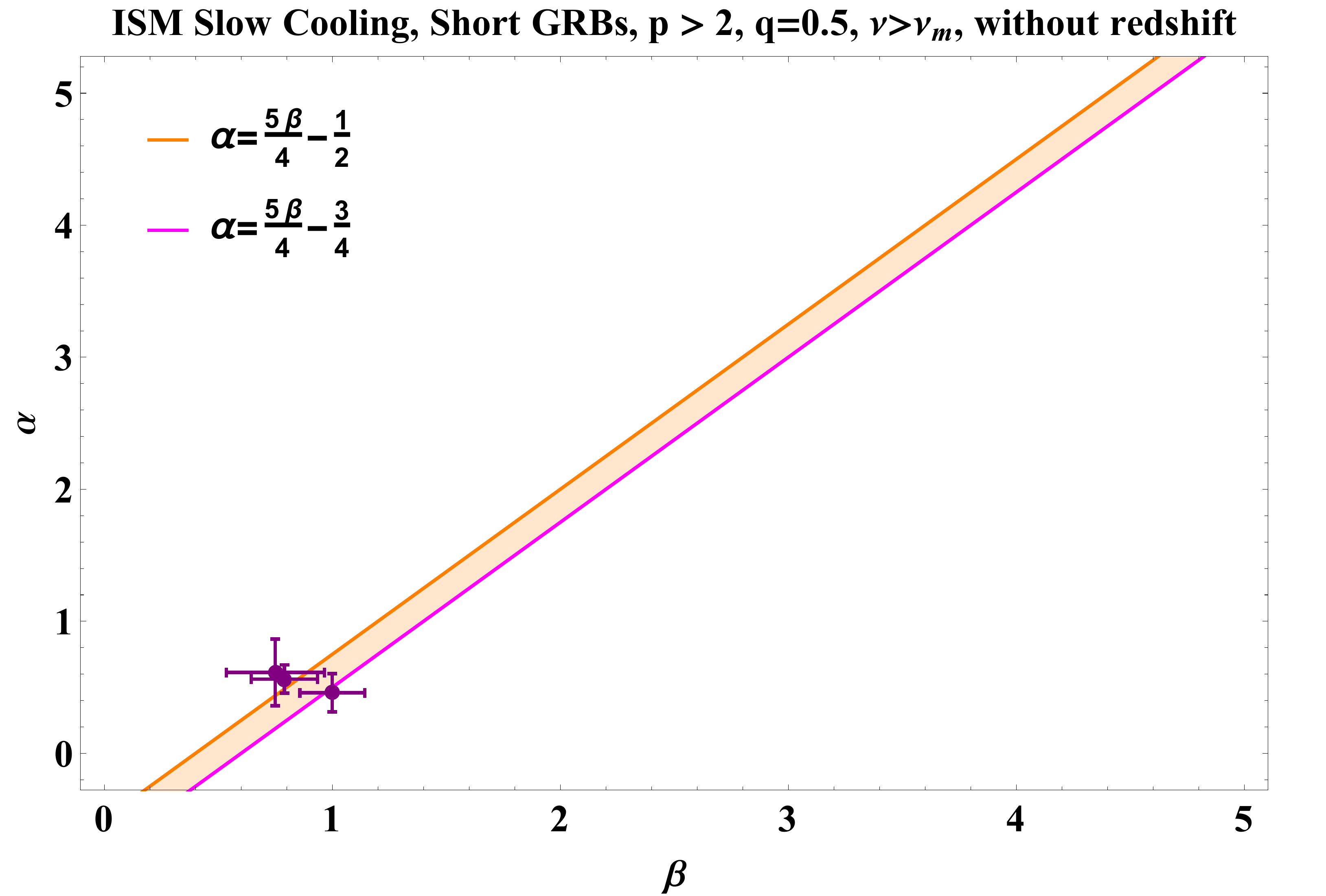}
    \includegraphics[scale=0.17]{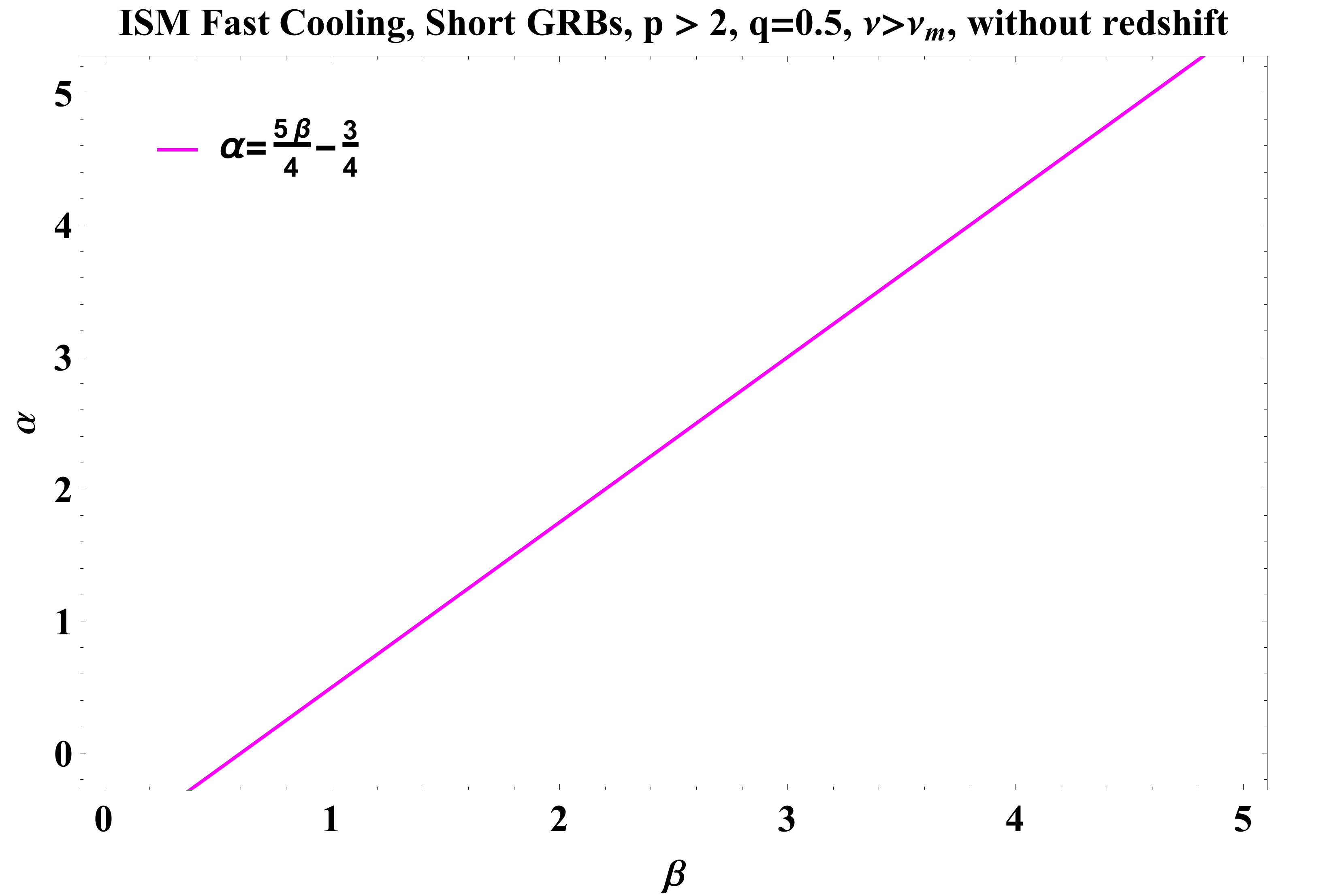}
    \includegraphics[scale=0.17]{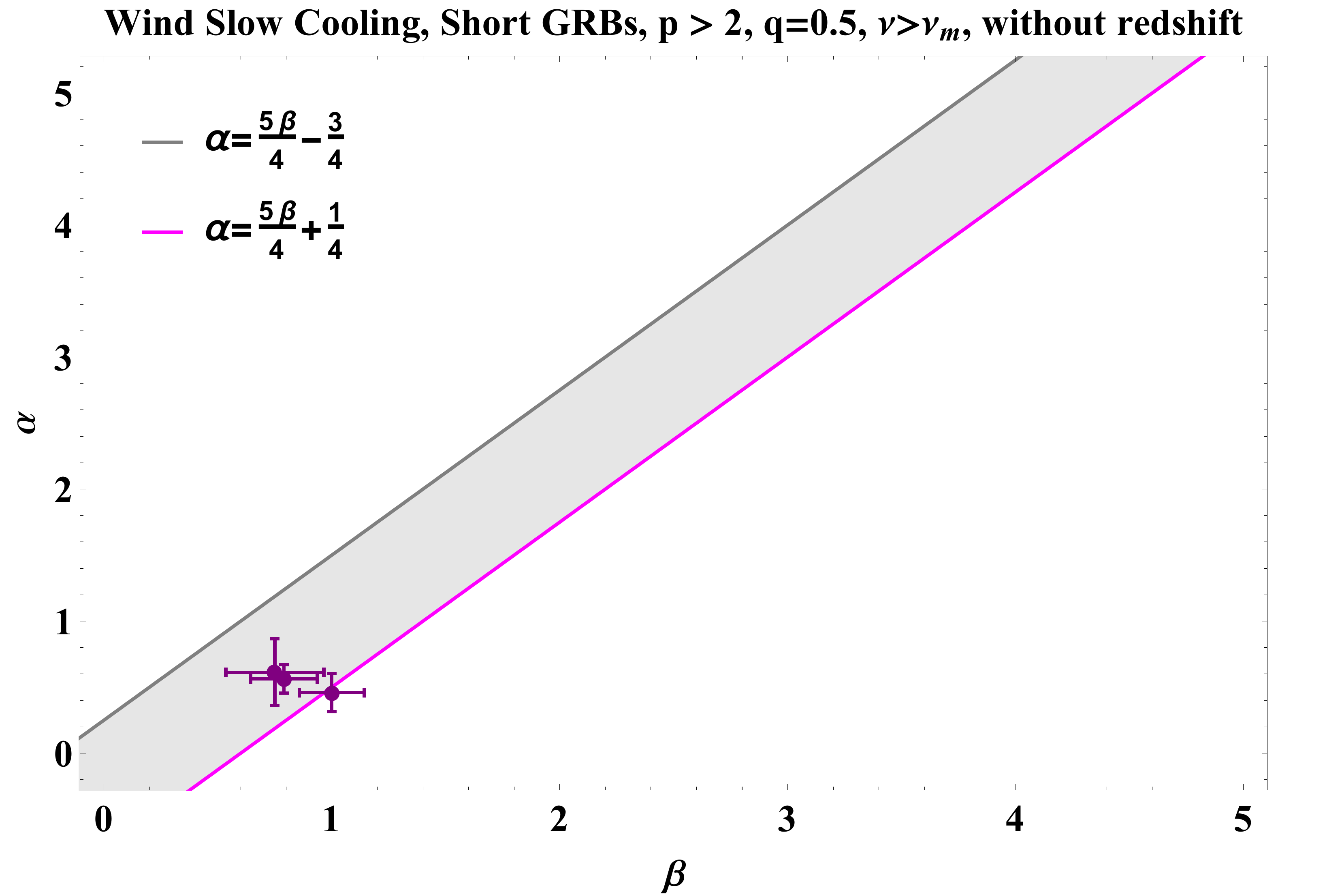}
    \includegraphics[scale=0.16]{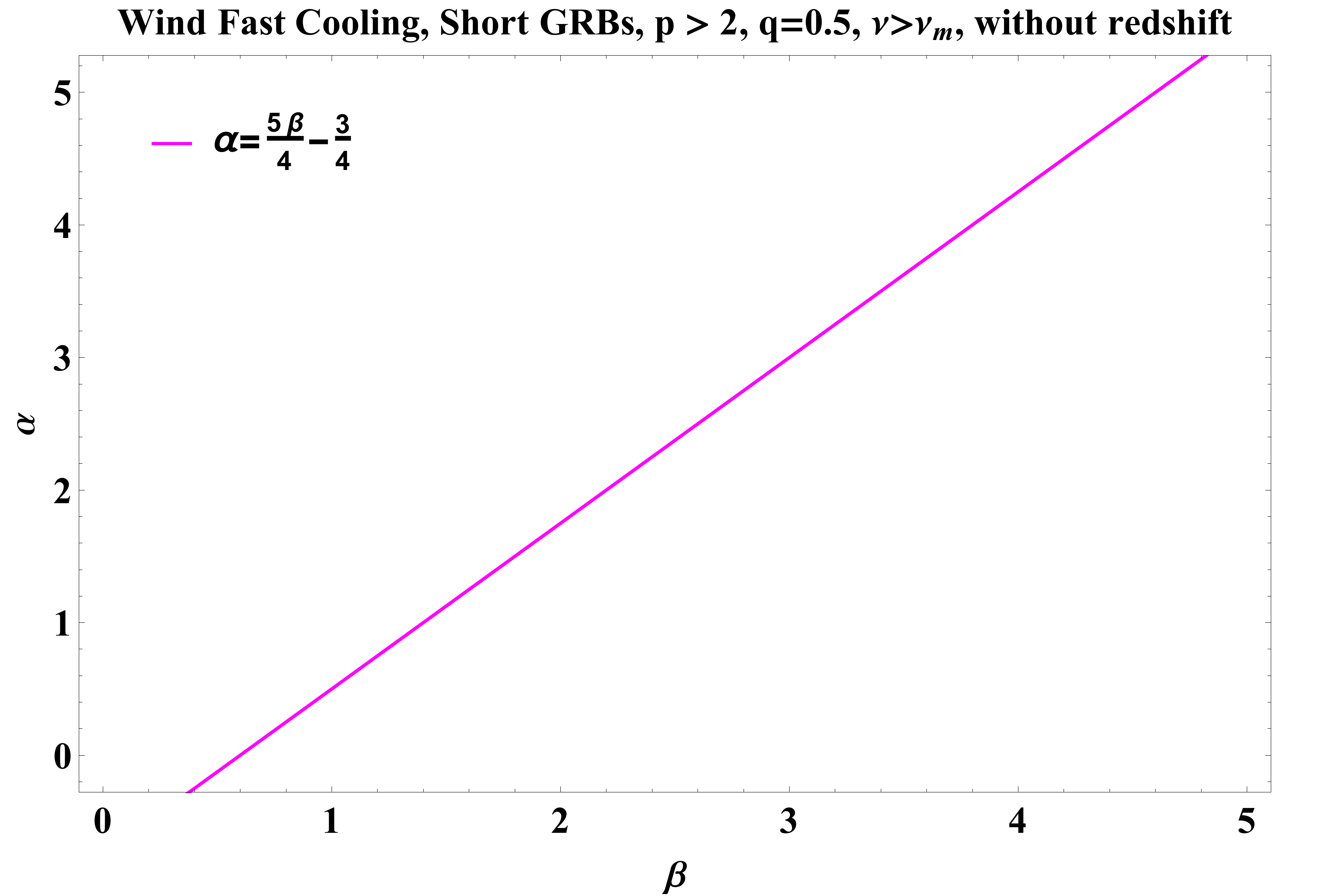}
    \caption{Compatible GRBs for CRs in the case $q=0.5, \nu>\nu_m, p>2$ without redshift. {\bf In the first 4 panels, the classes lGRBs and sGRBs are gathered together, while in the latter 8 panels these classes are separated.}}
    \label{only_nu>num_p>2_NR_q0.5}
\end{figure}

\begin{figure}
    \centering
    \includegraphics[scale=0.23]{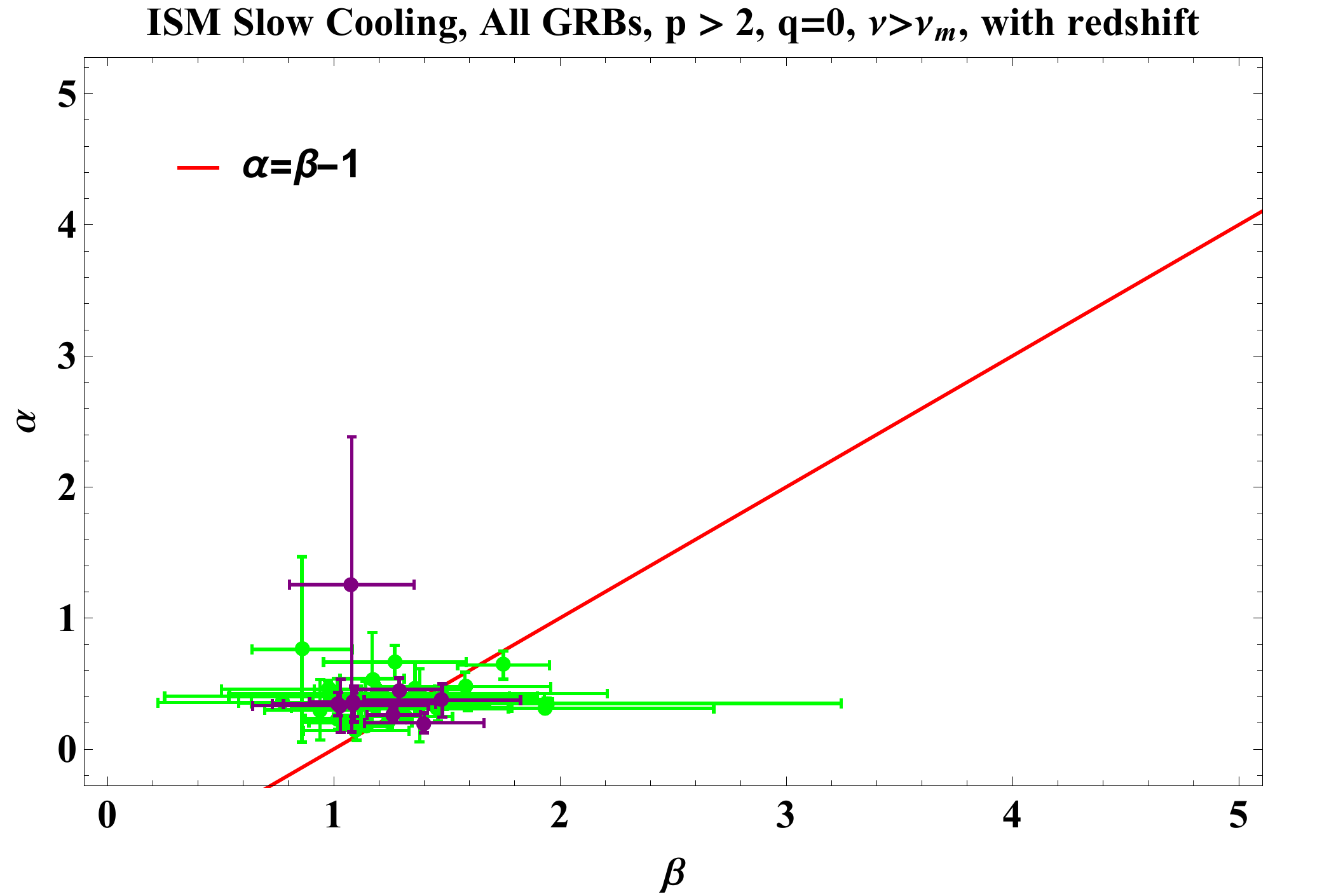}
    \includegraphics[scale=0.23]{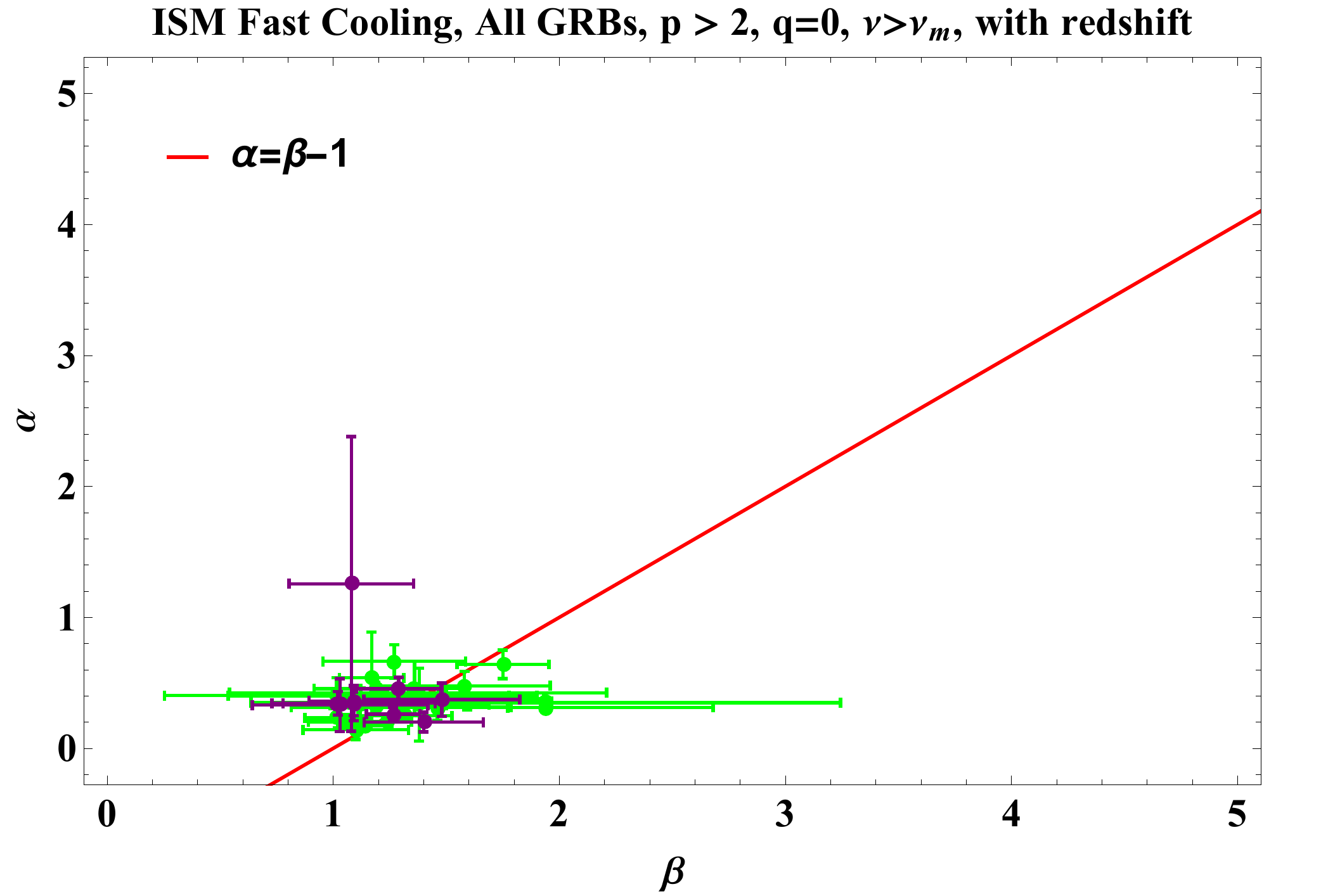}
    \includegraphics[scale=0.23]{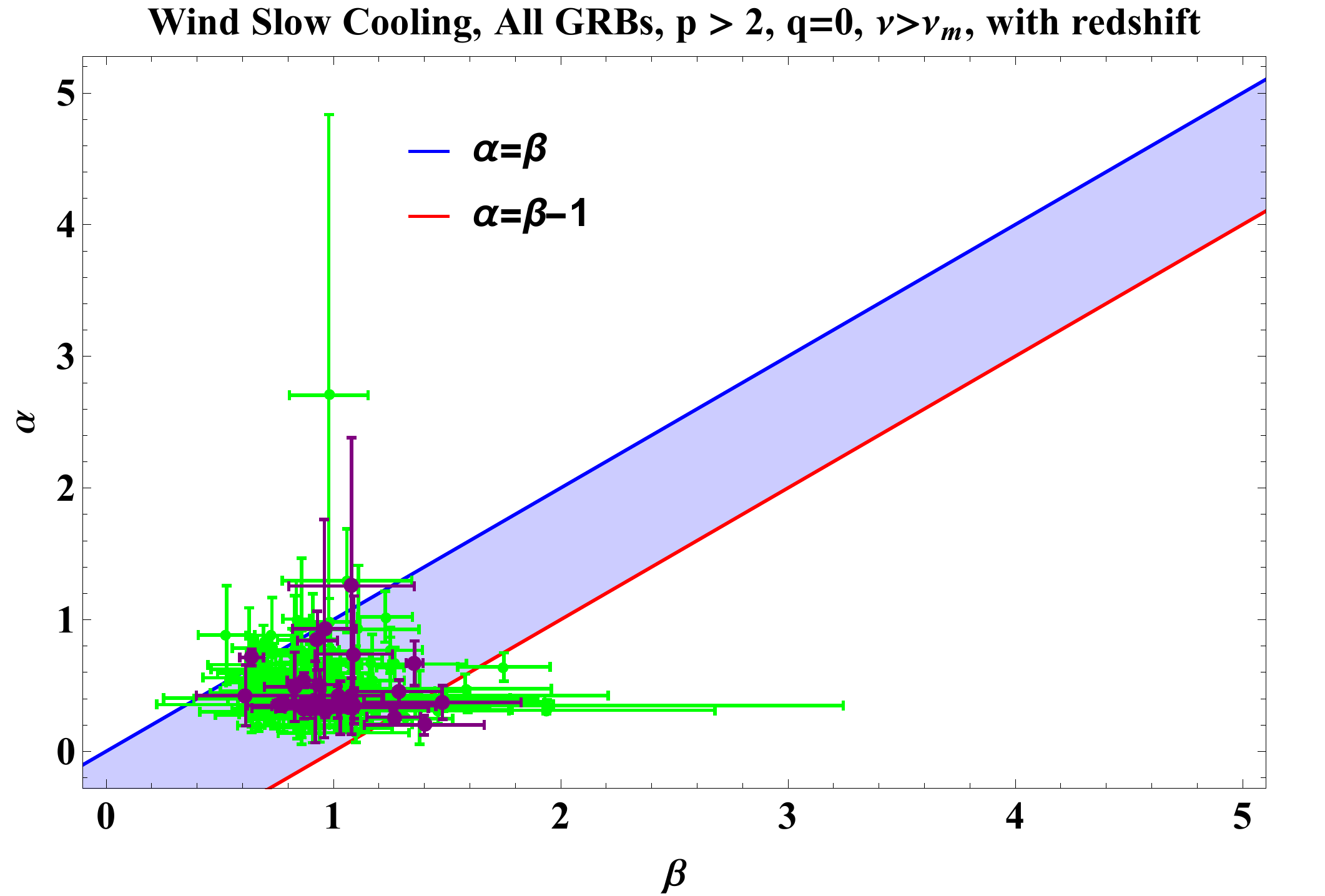}
    \includegraphics[scale=0.23]{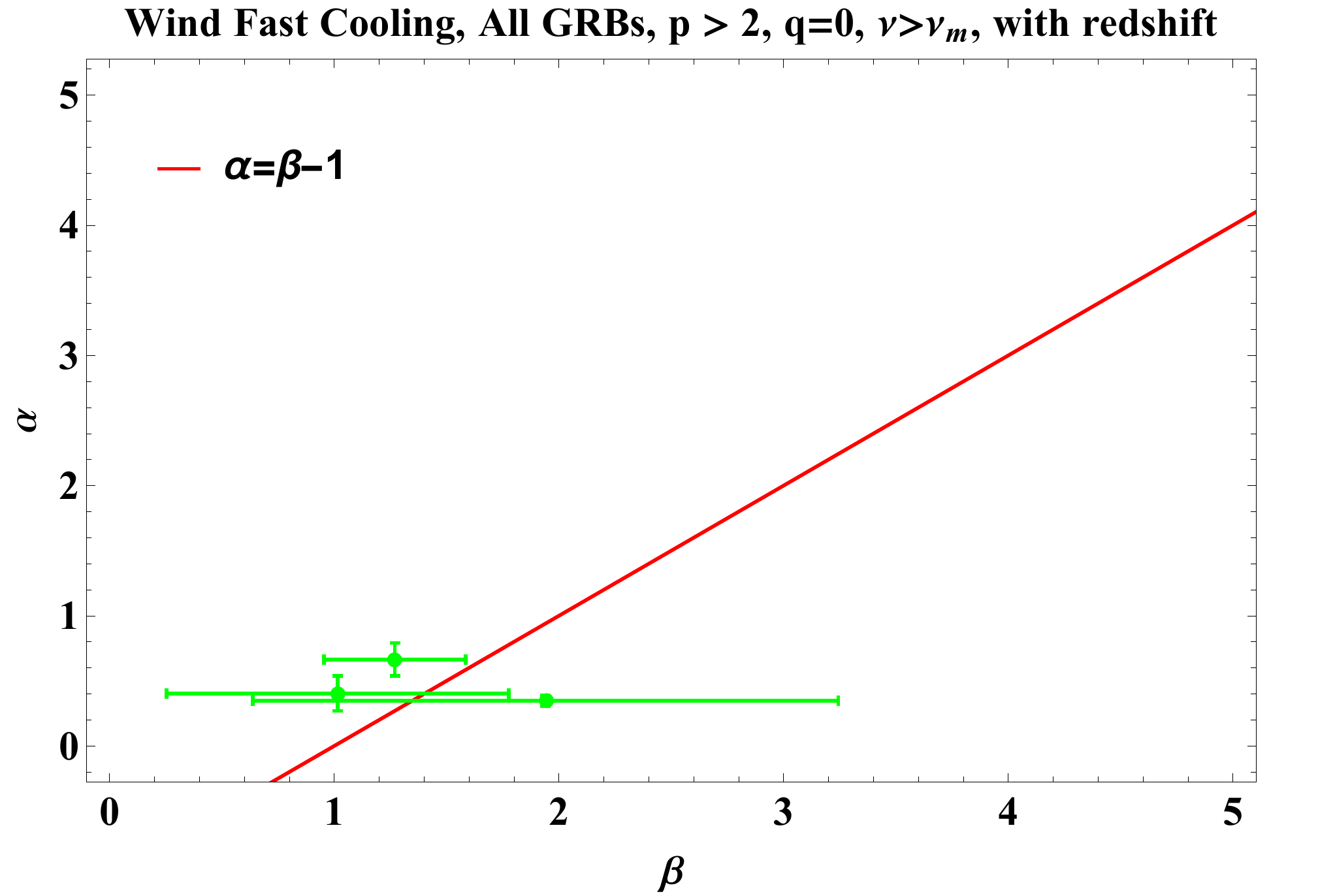}
    \includegraphics[scale=0.23]{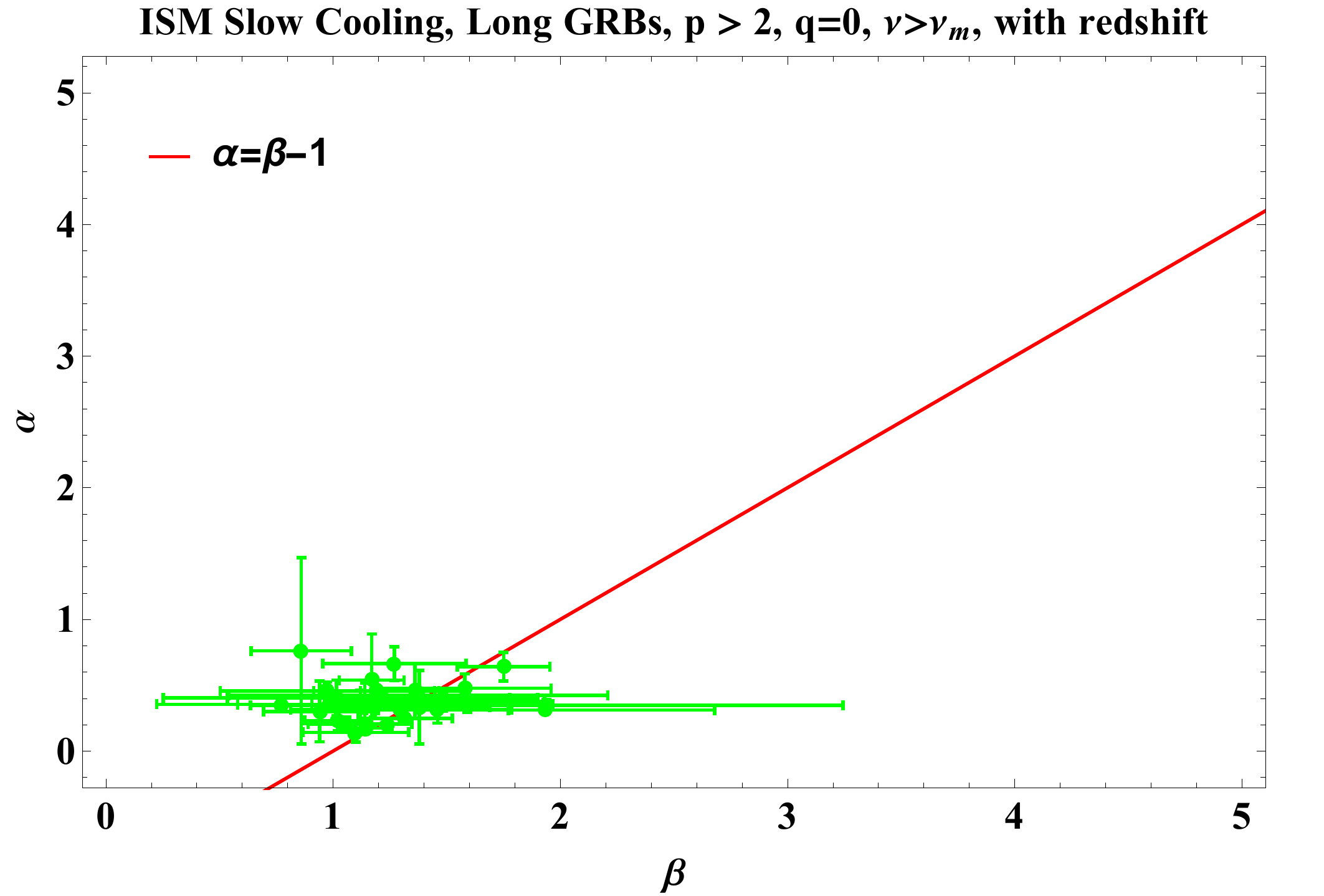}
    \includegraphics[scale=0.23]{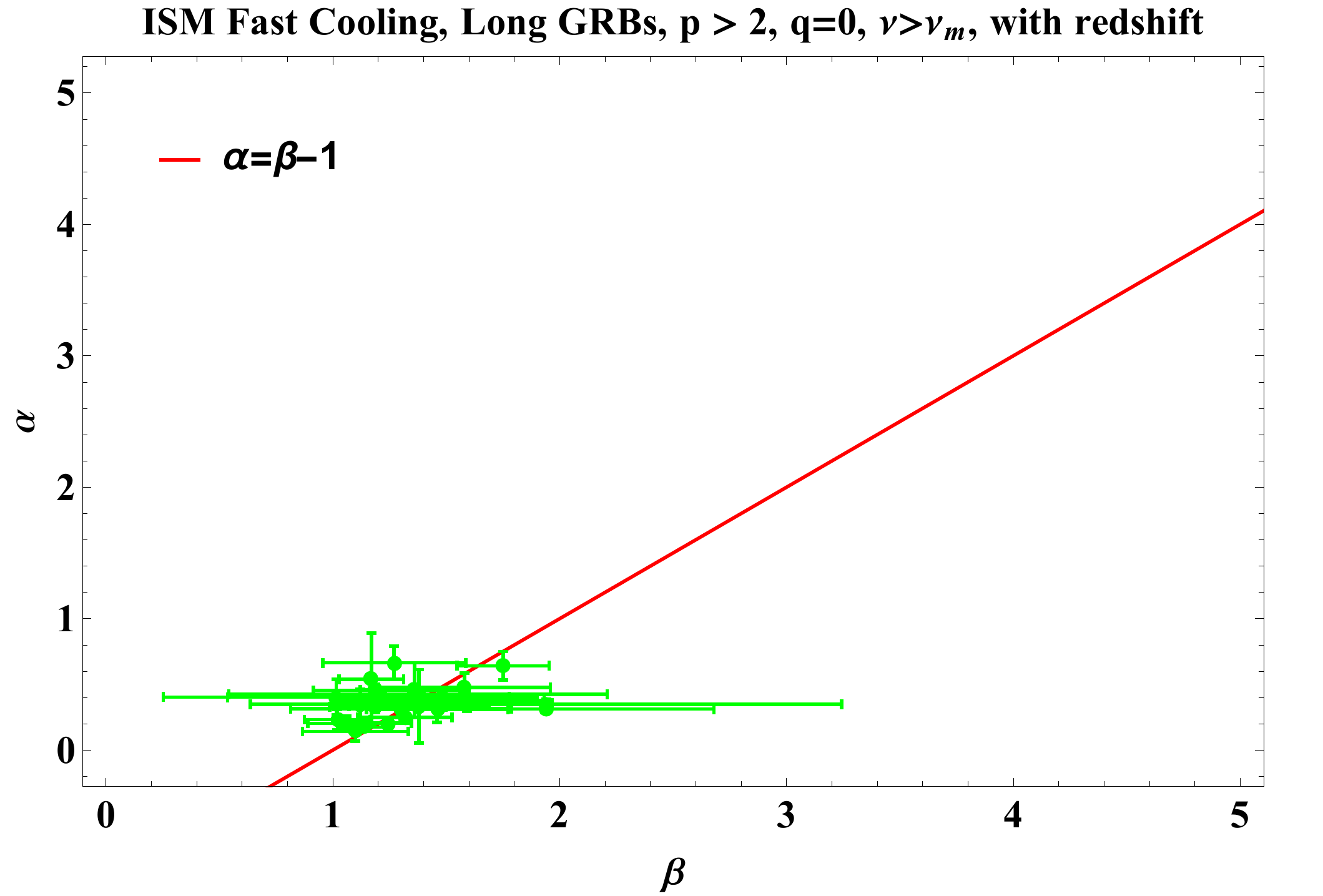}
    \includegraphics[scale=0.23]{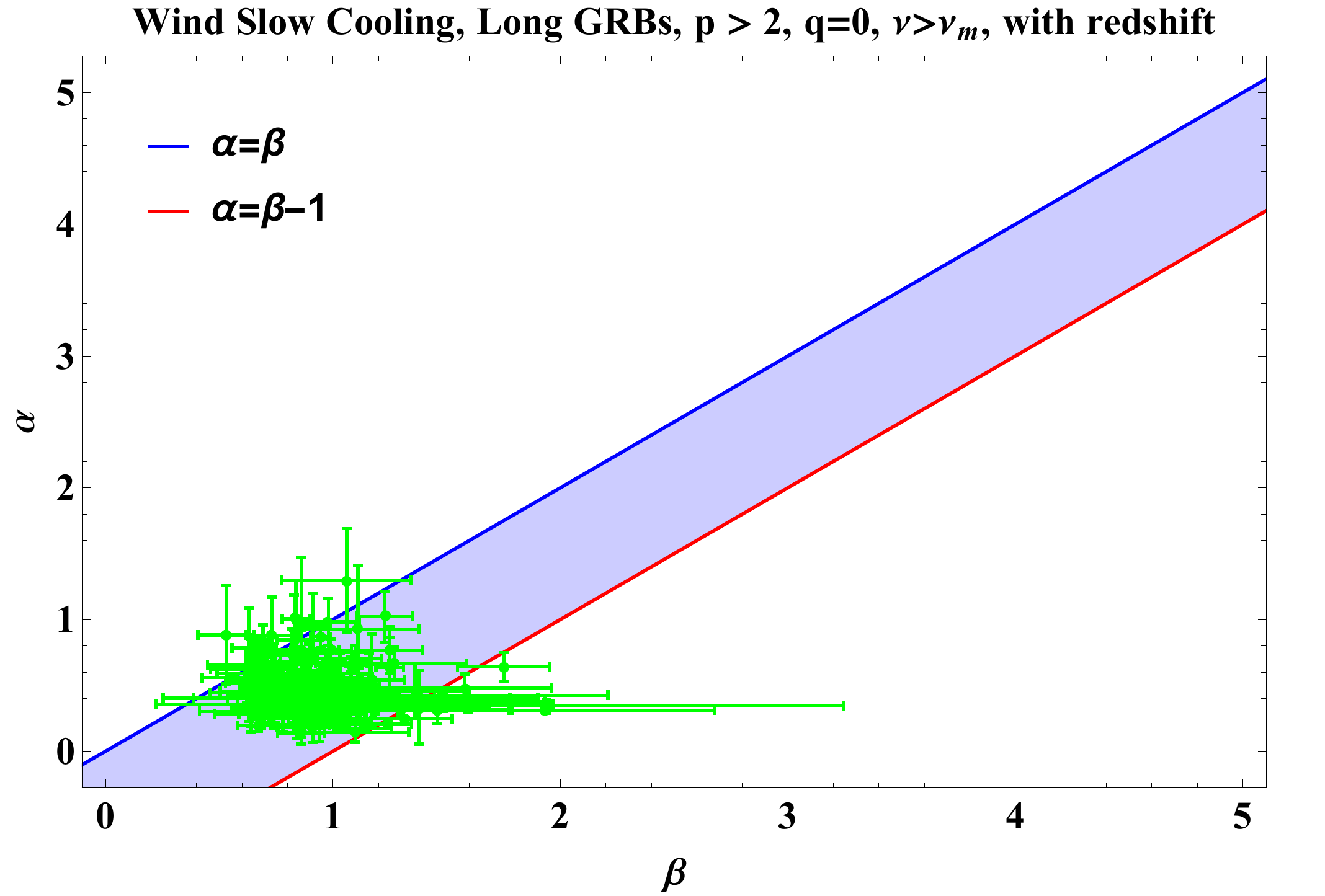}
    \includegraphics[scale=0.23]{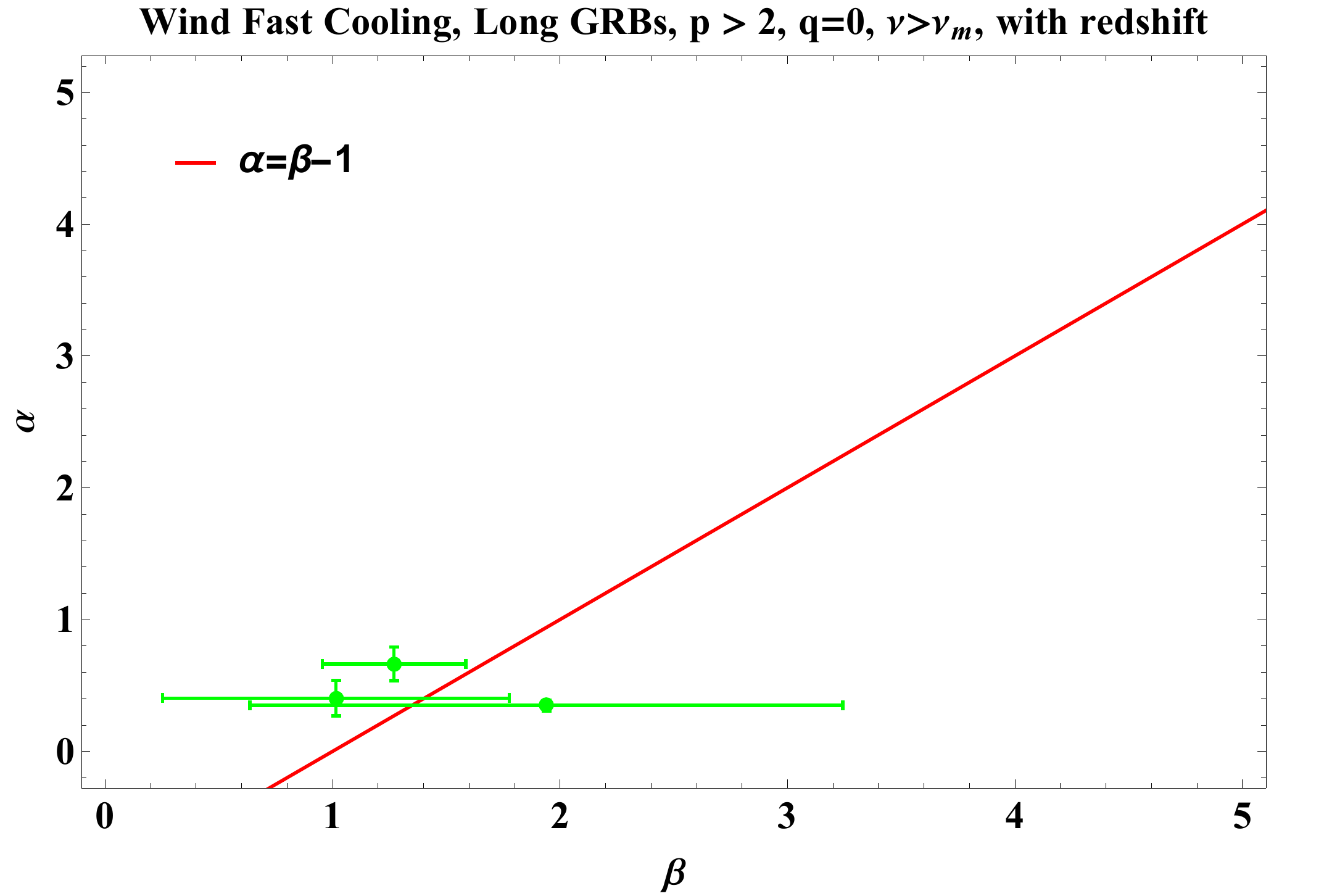}
    \includegraphics[scale=0.23]{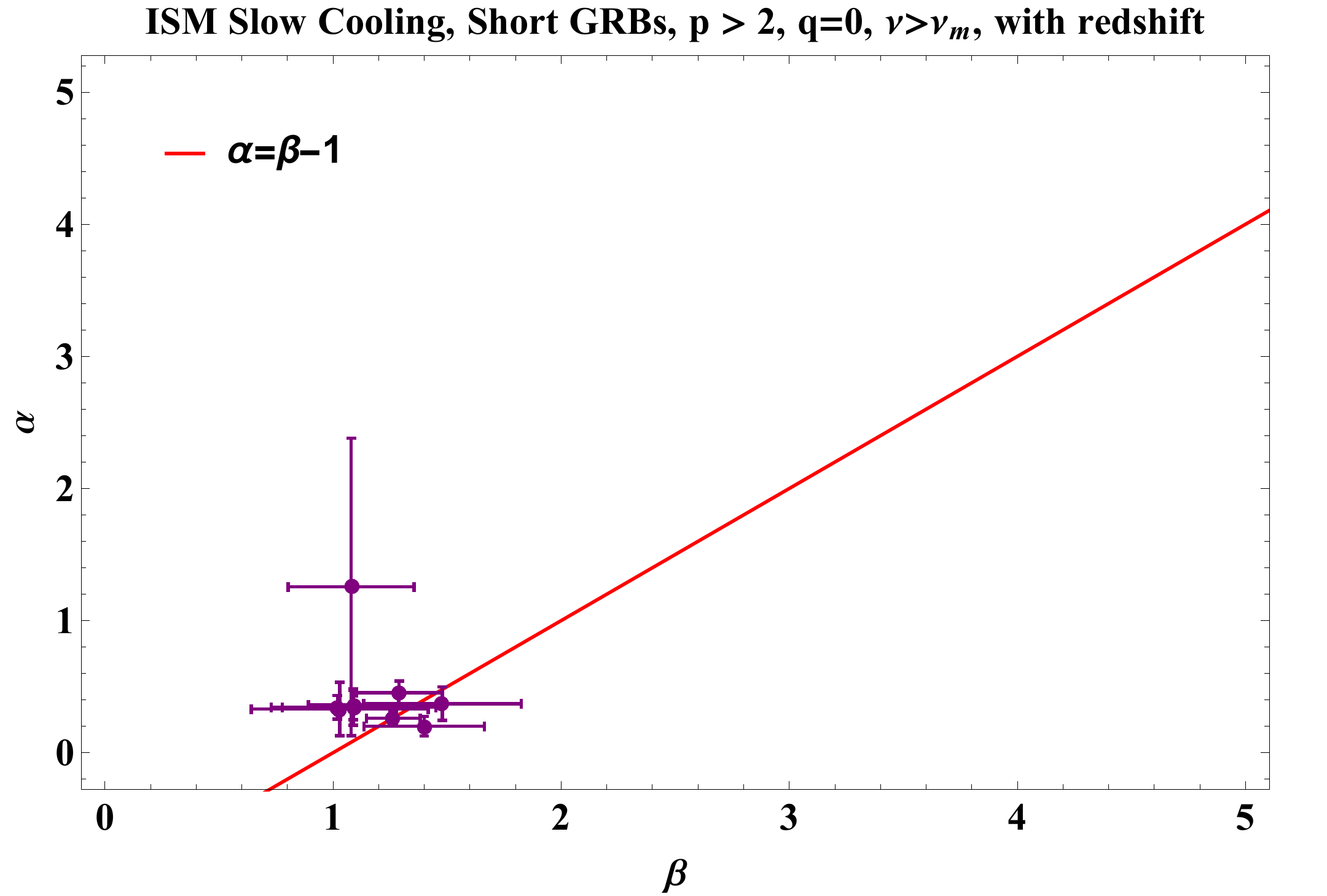}
    \includegraphics[scale=0.23]{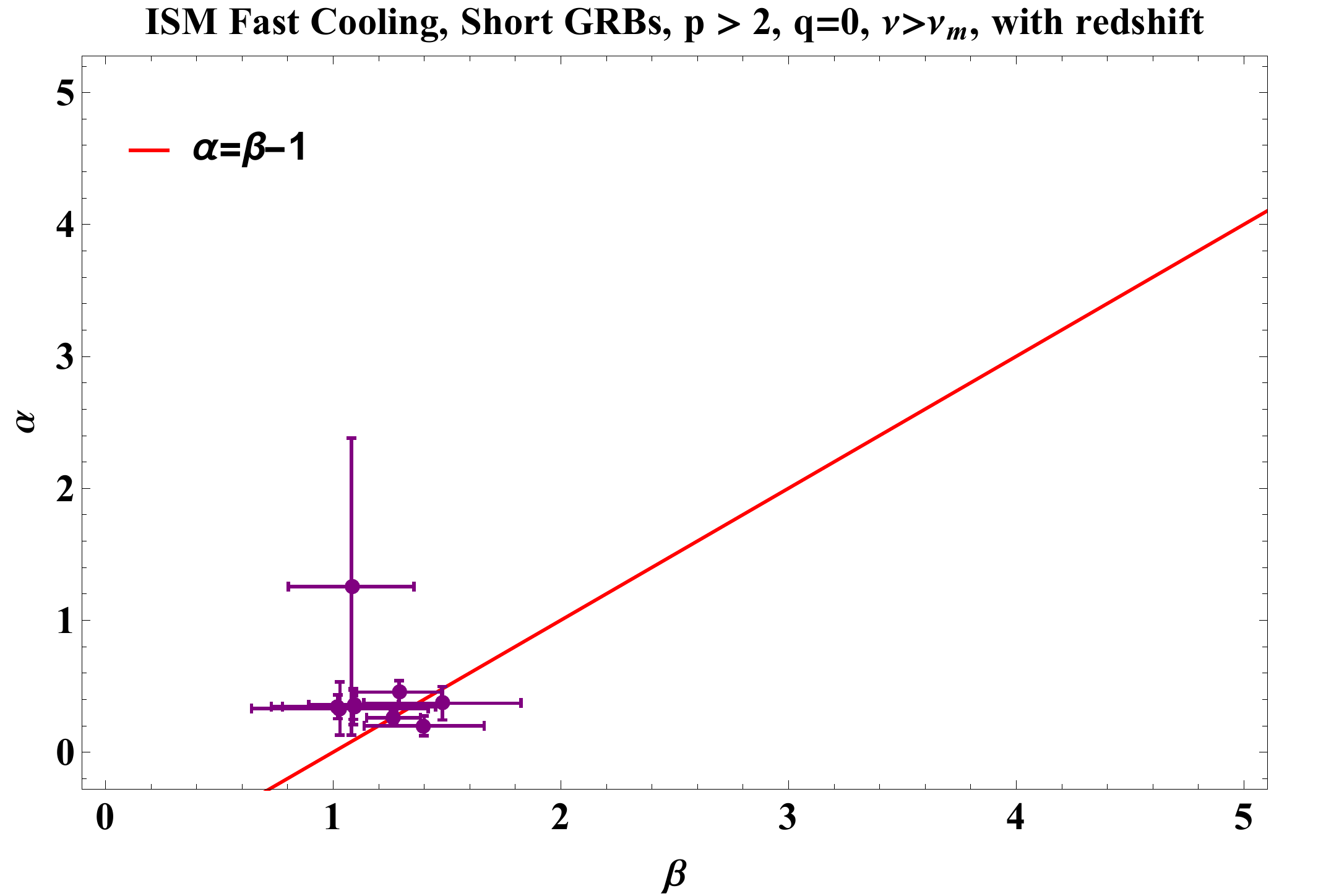}
    \includegraphics[scale=0.23]{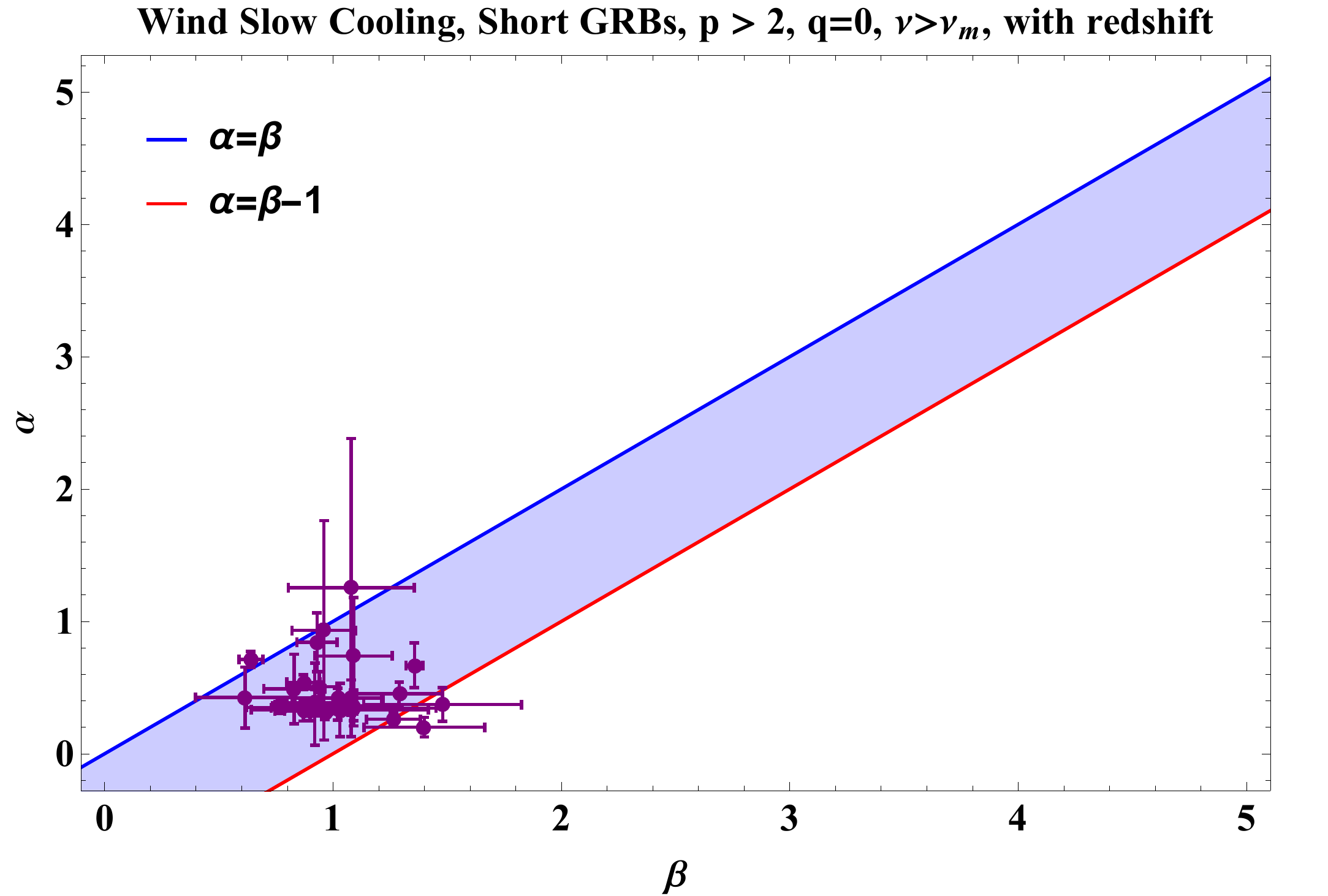}
    \includegraphics[scale=0.23]{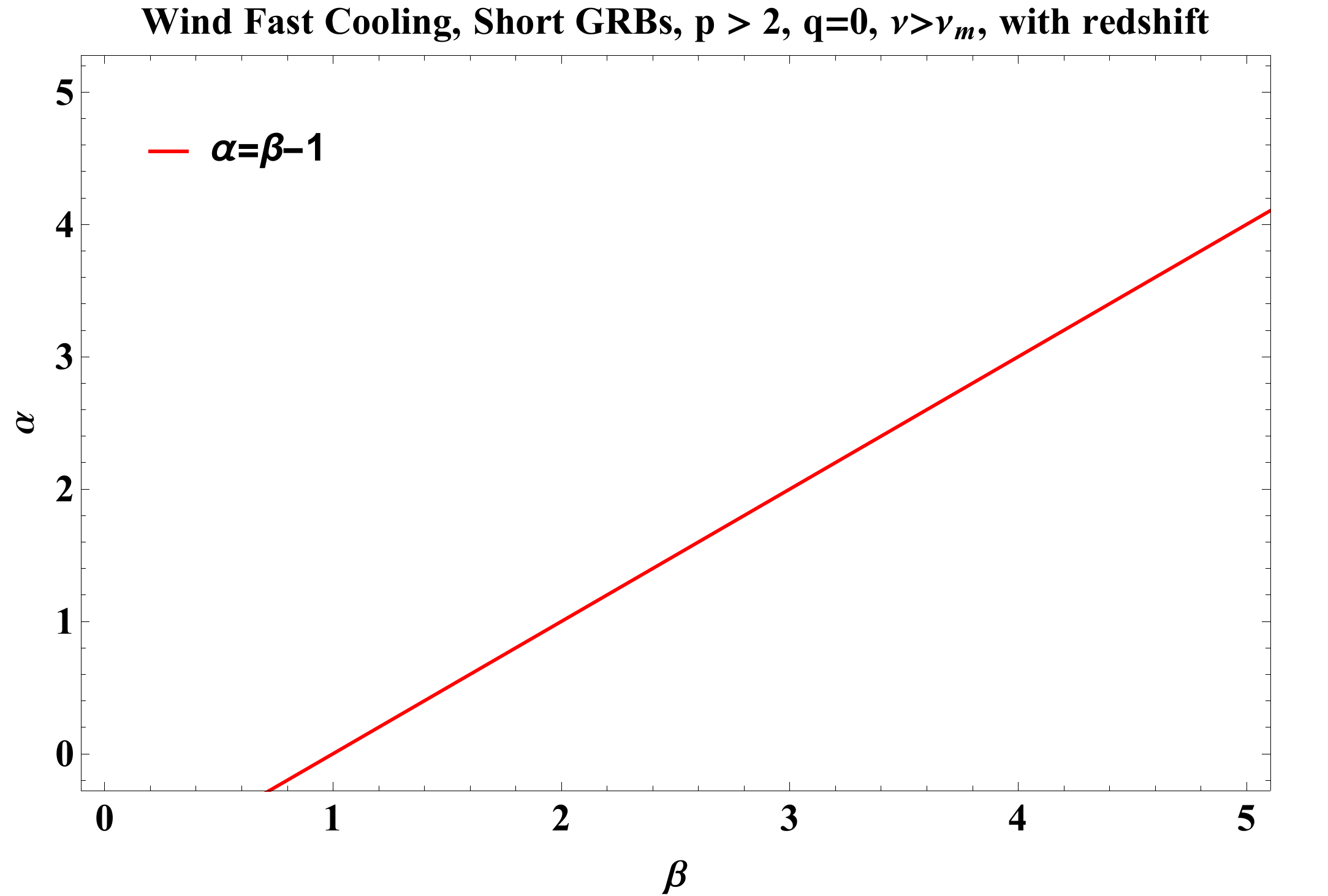}
    \caption{Compatible GRBs for CRs in the case $q=0,\nu>\nu_m, p>2$ where the redshift is known. \textbf{In the first 4 panels, the classes lGRBs and sGRBs are gathered together (although in the fourth panel the sGRB class is empty), while in the latter 8 panels these classes are separated.}}
    \label{only_nu>num_p>2_R_q0}
\end{figure}

\begin{figure}
    \centering
    \includegraphics[scale=0.22]{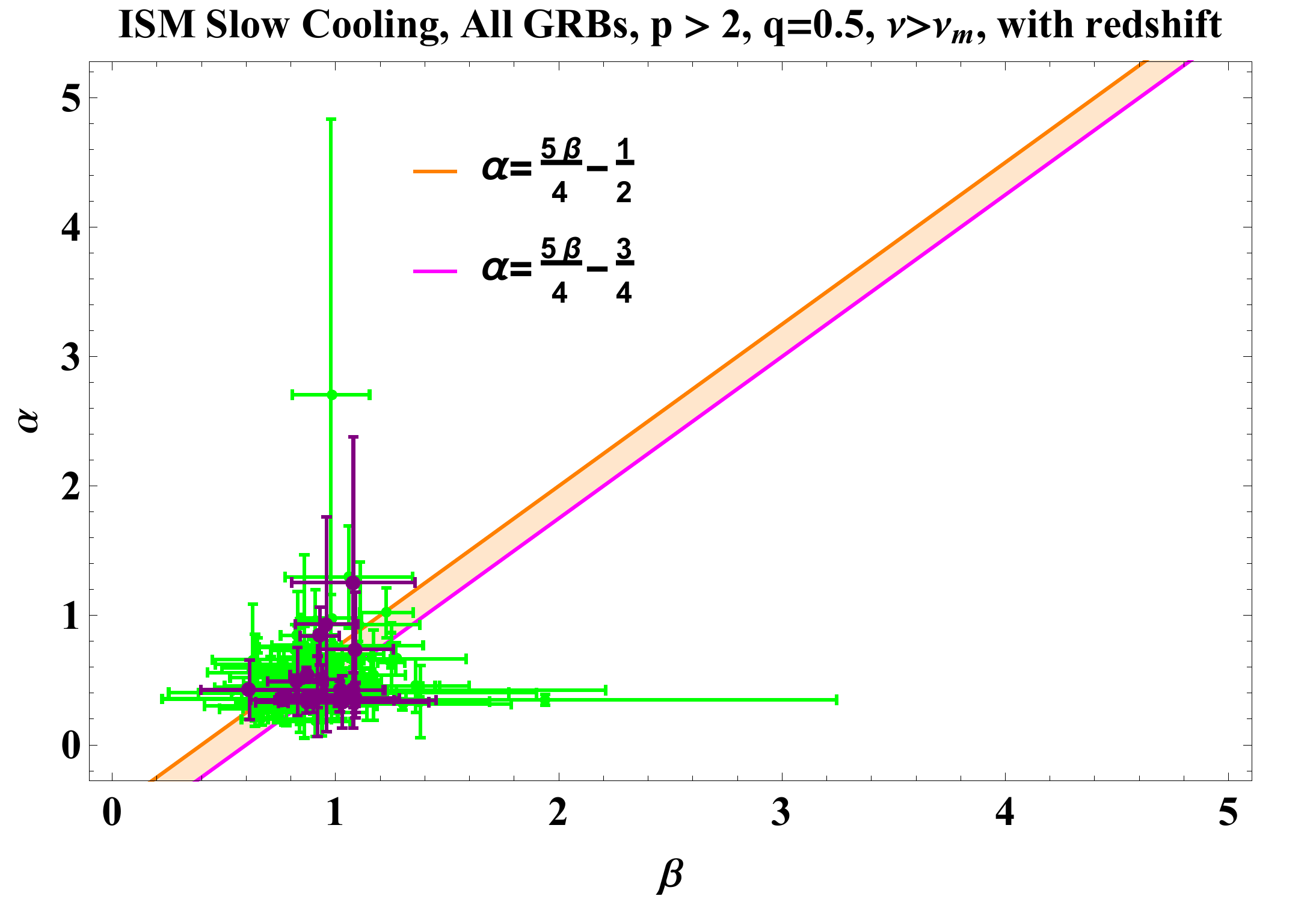}
    \includegraphics[scale=0.22]{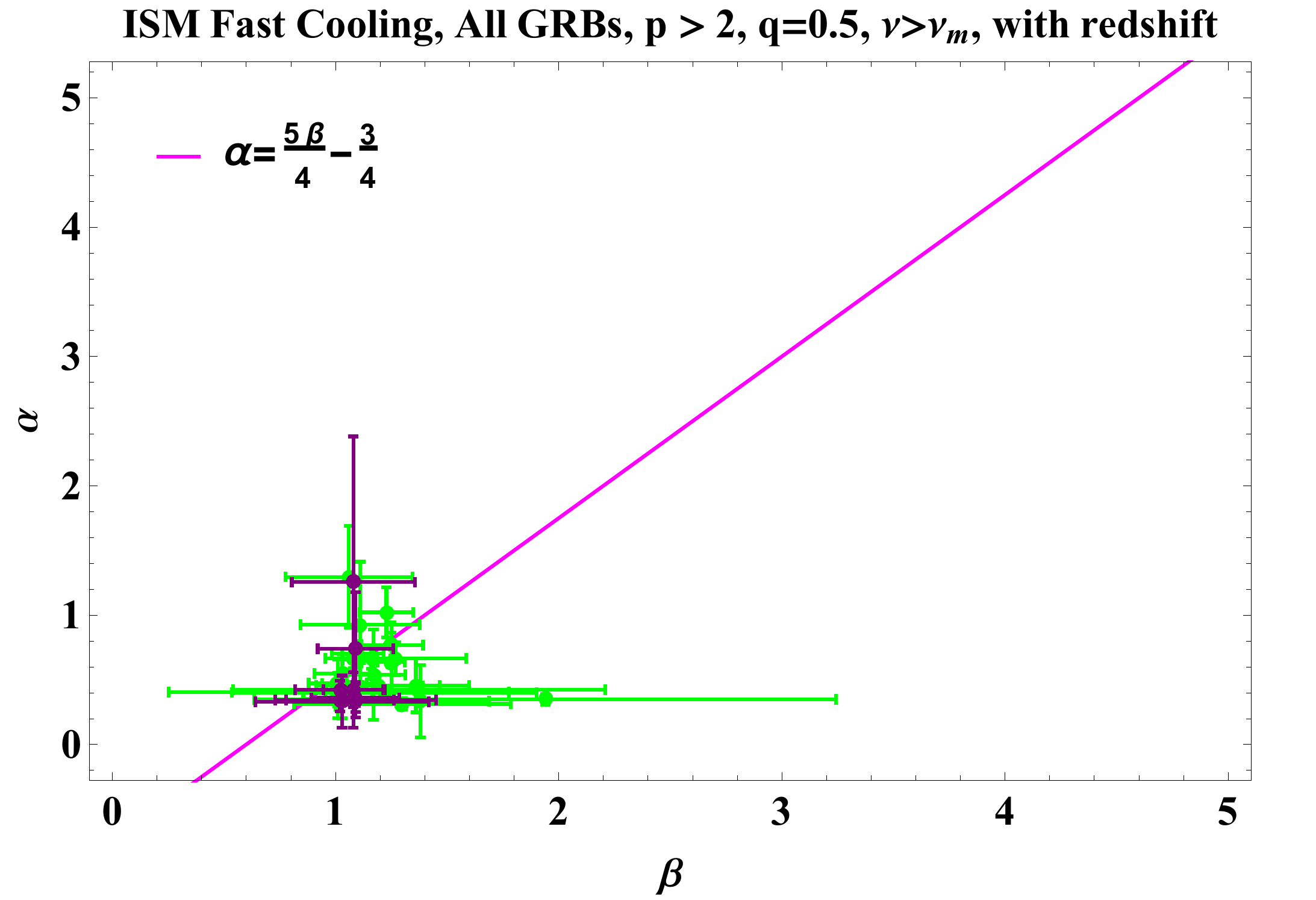}
    \includegraphics[scale=0.22]{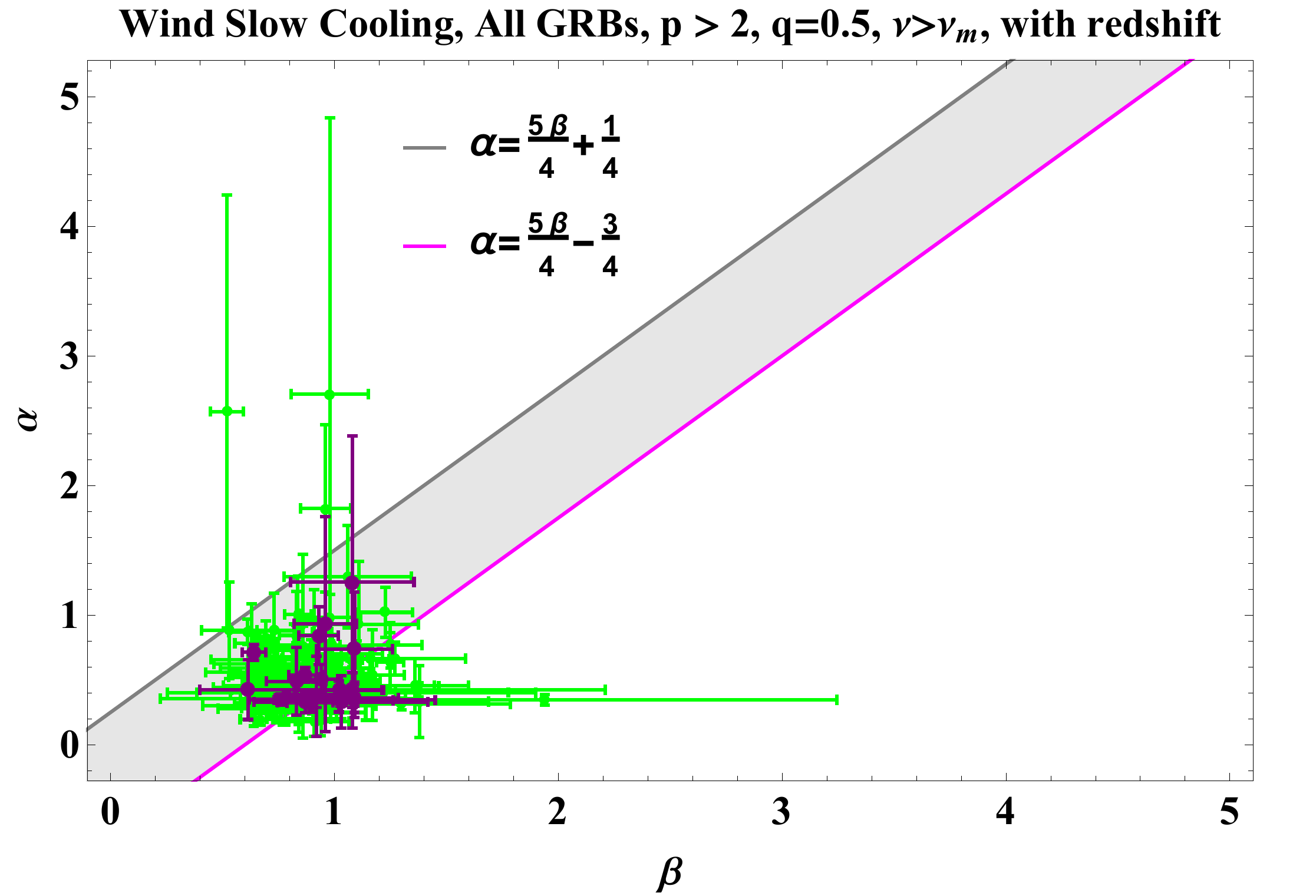}
    \includegraphics[scale=0.22]{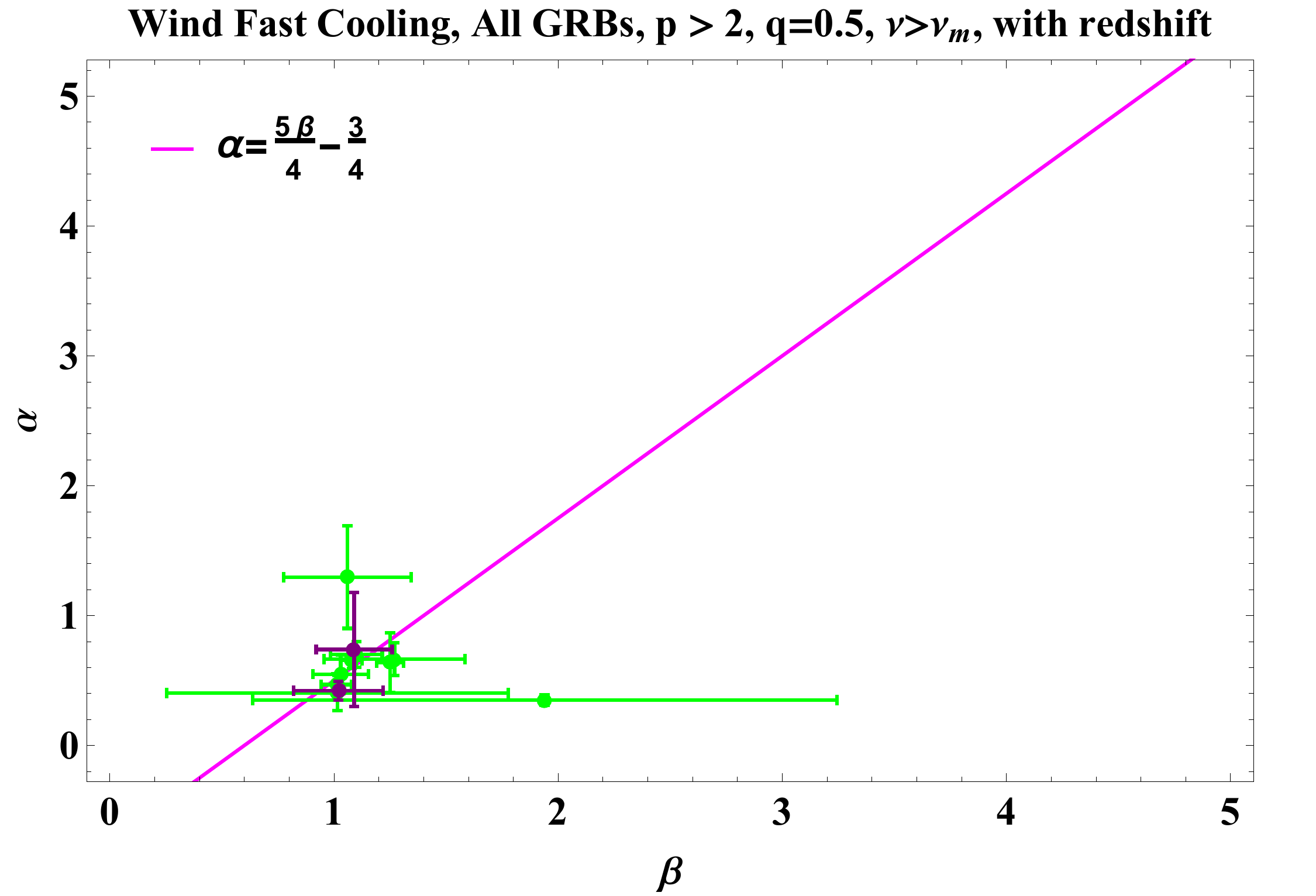}
    \includegraphics[scale=0.22]{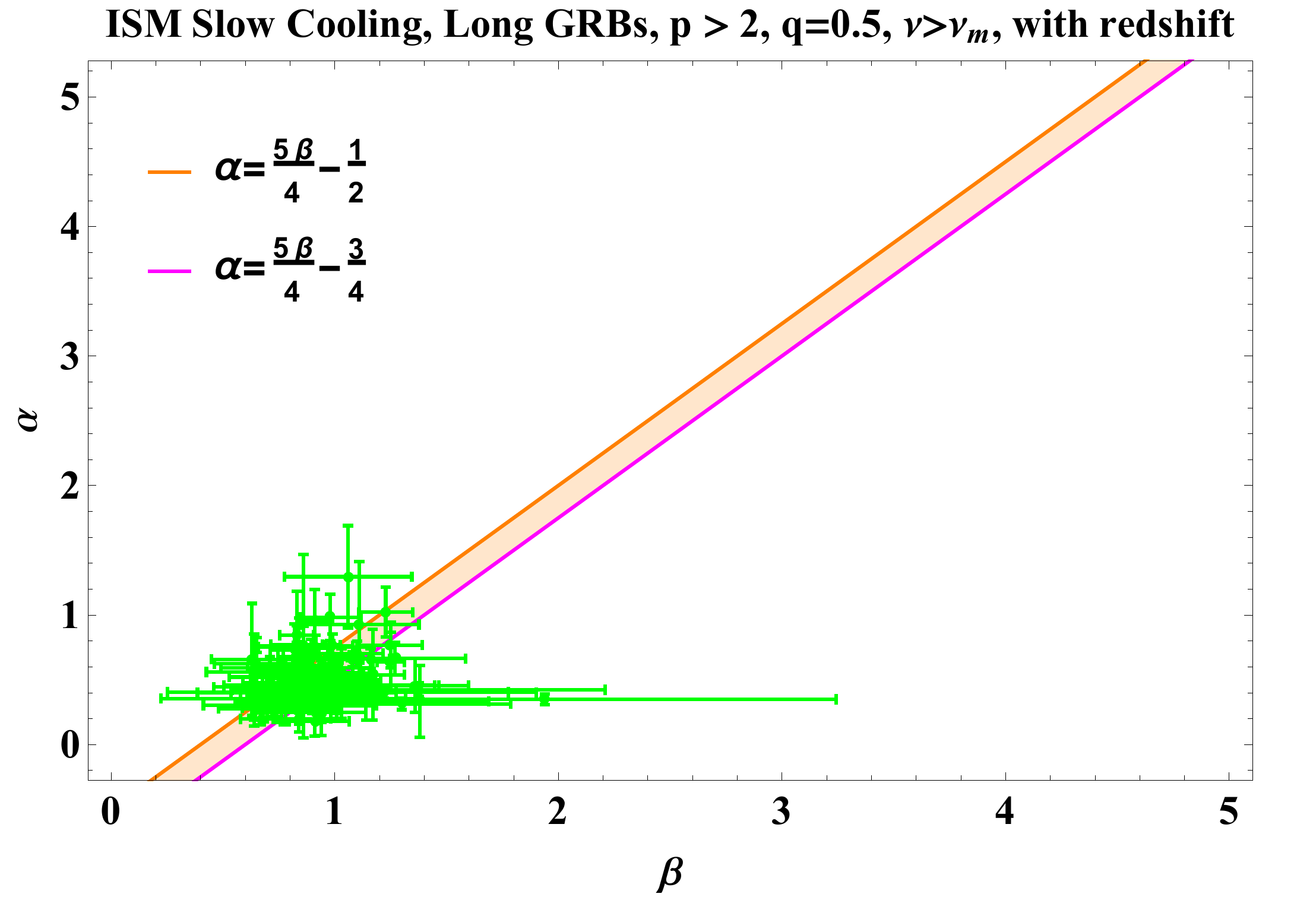}
    \includegraphics[scale=0.22]{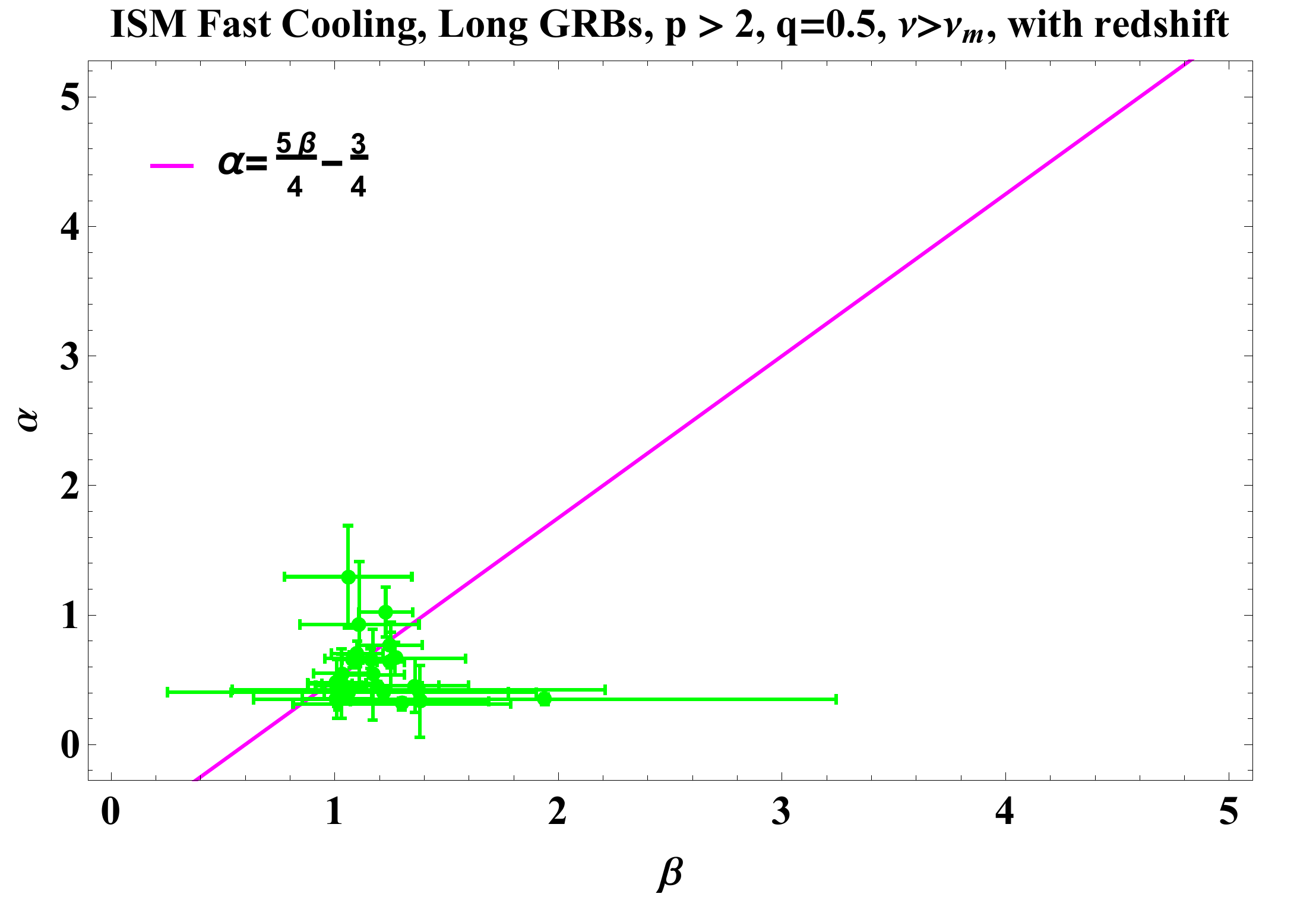}
    \includegraphics[scale=0.22]{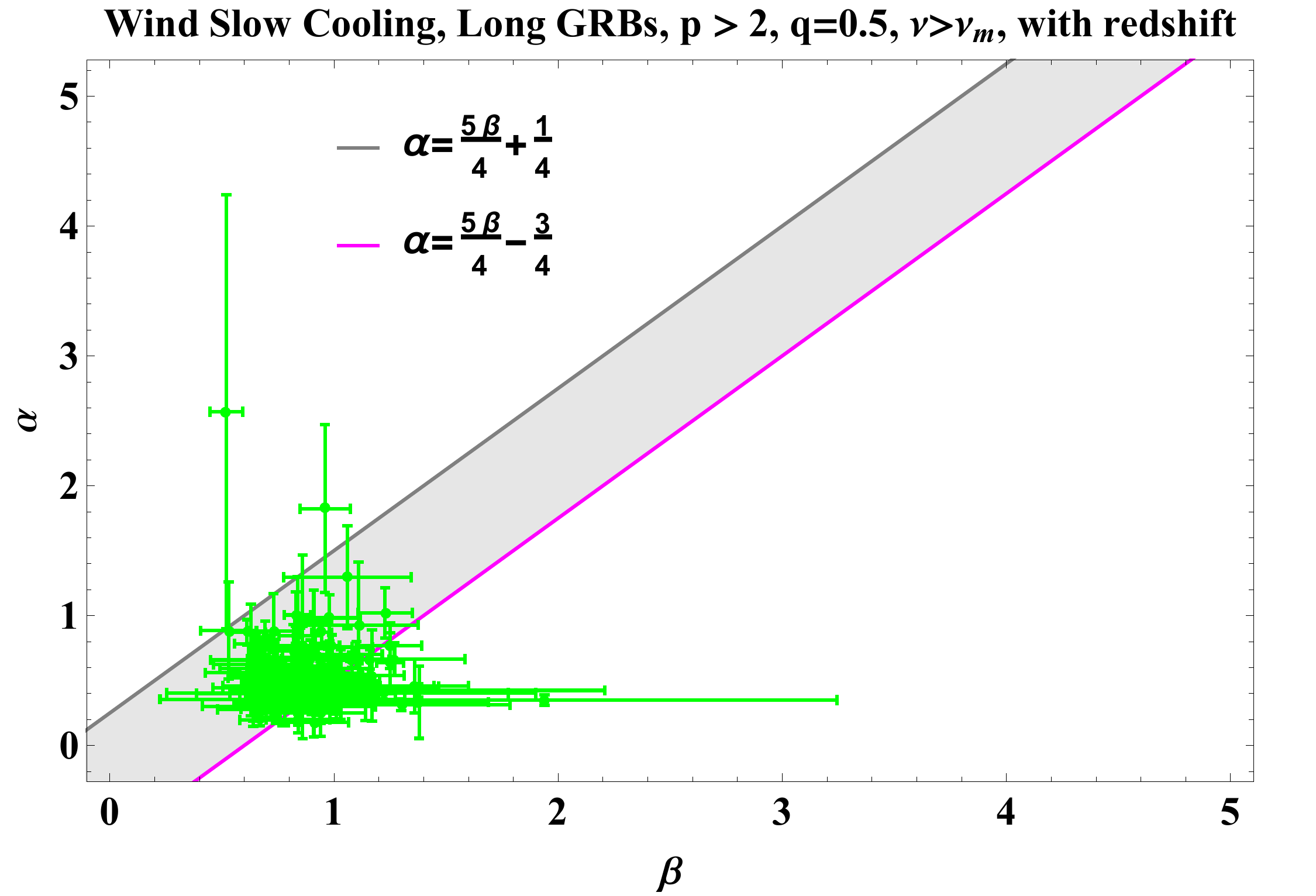}
    \includegraphics[scale=0.22]{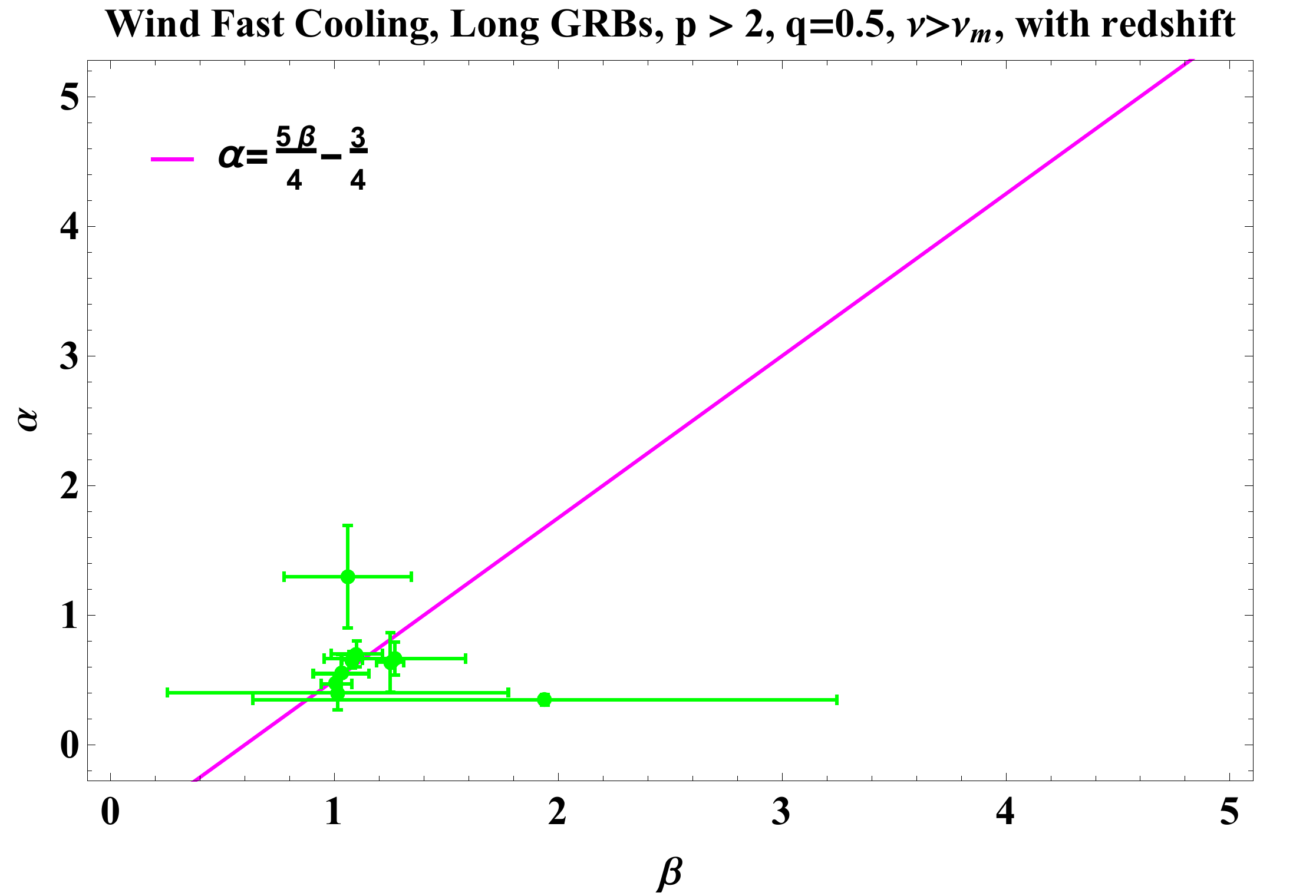}
    \includegraphics[scale=0.22]{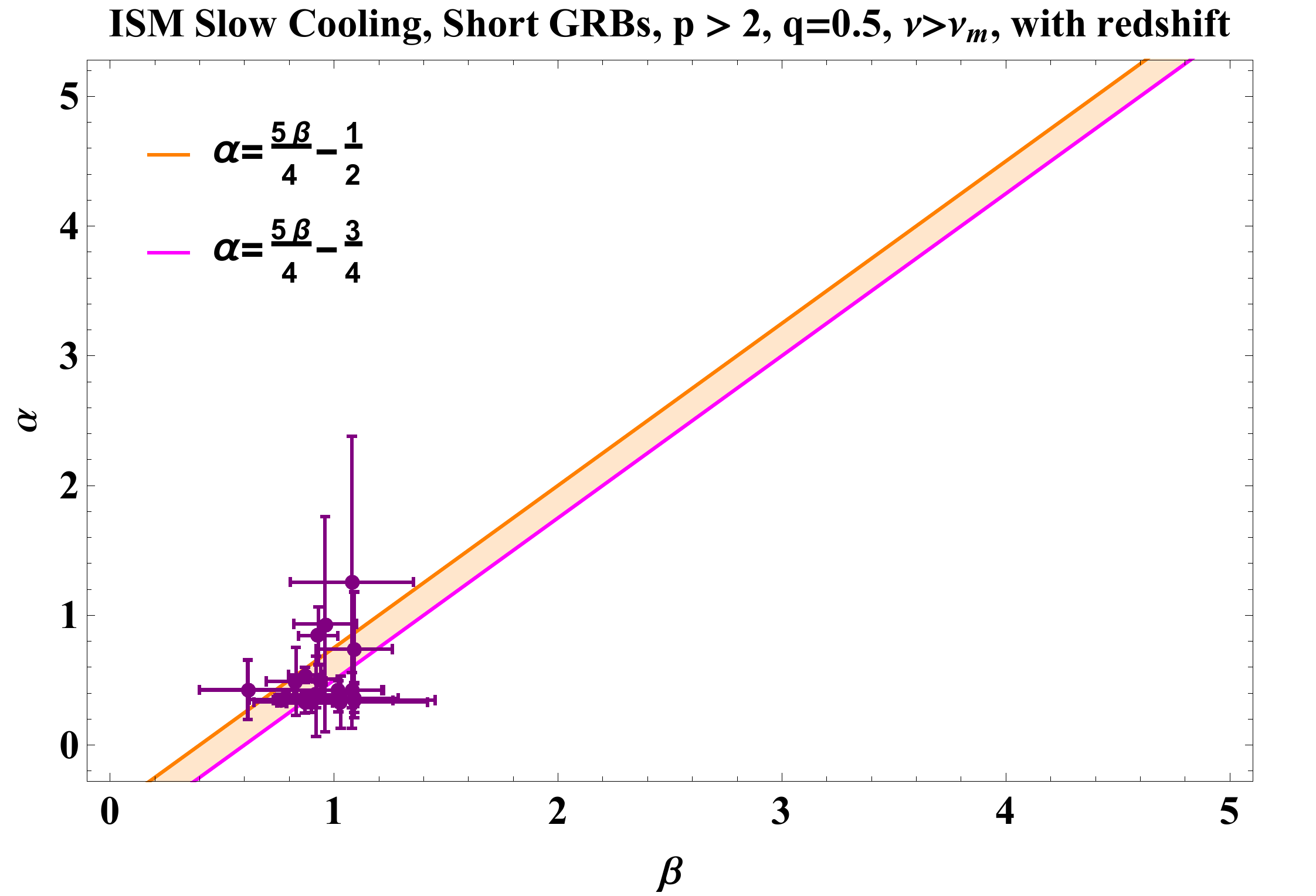}
    \includegraphics[scale=0.22]{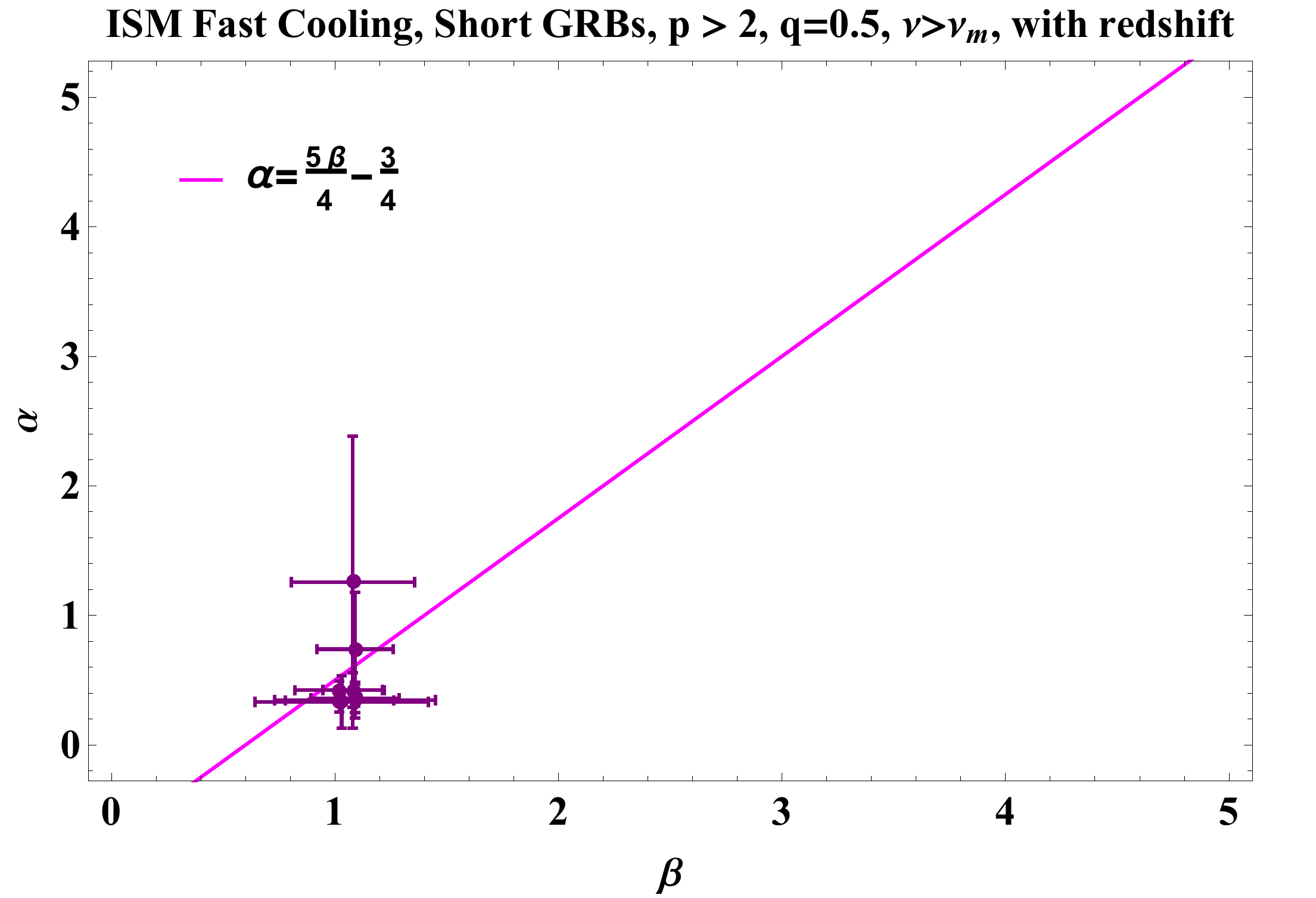}
    \includegraphics[scale=0.22]{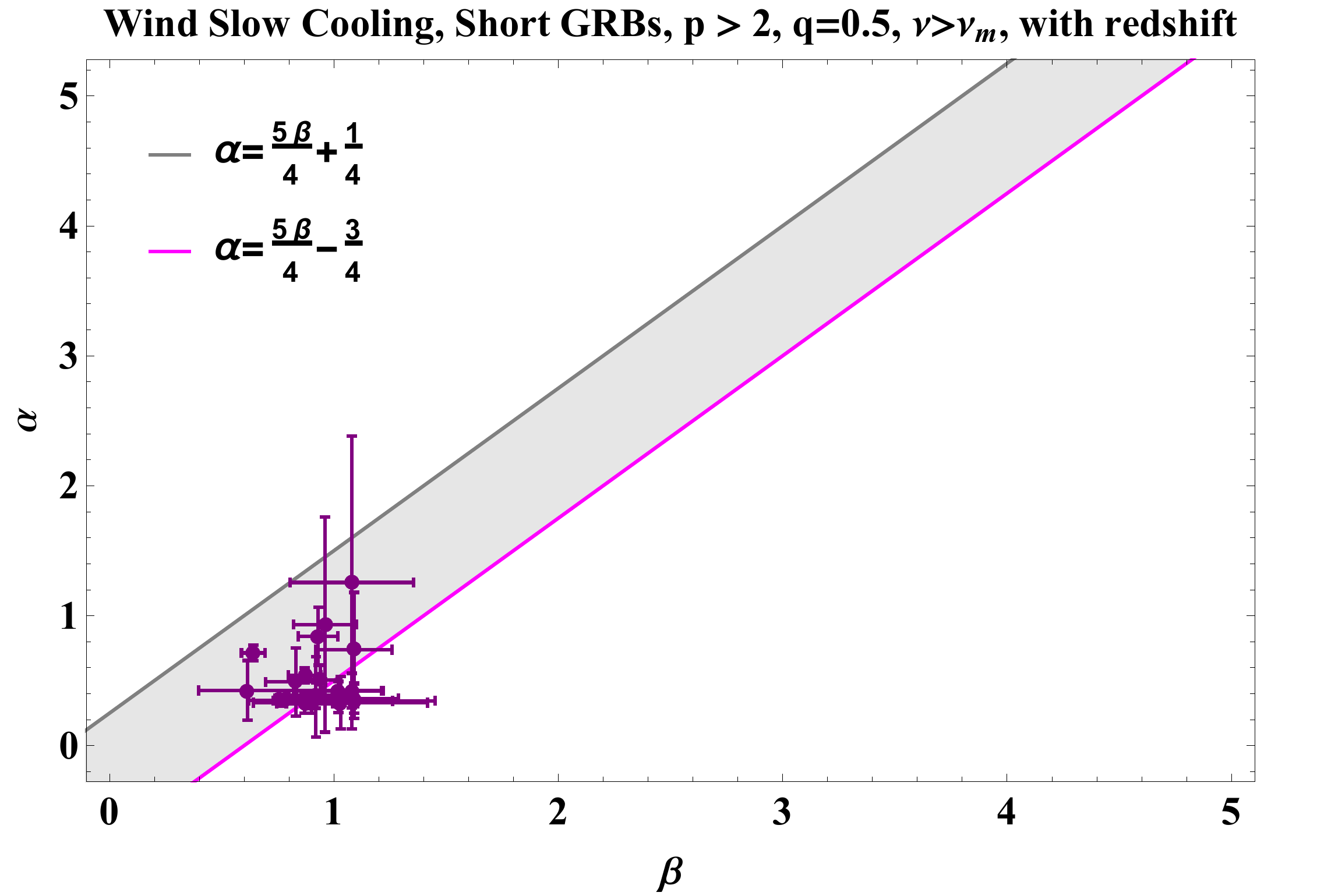}
    \includegraphics[scale=0.22]{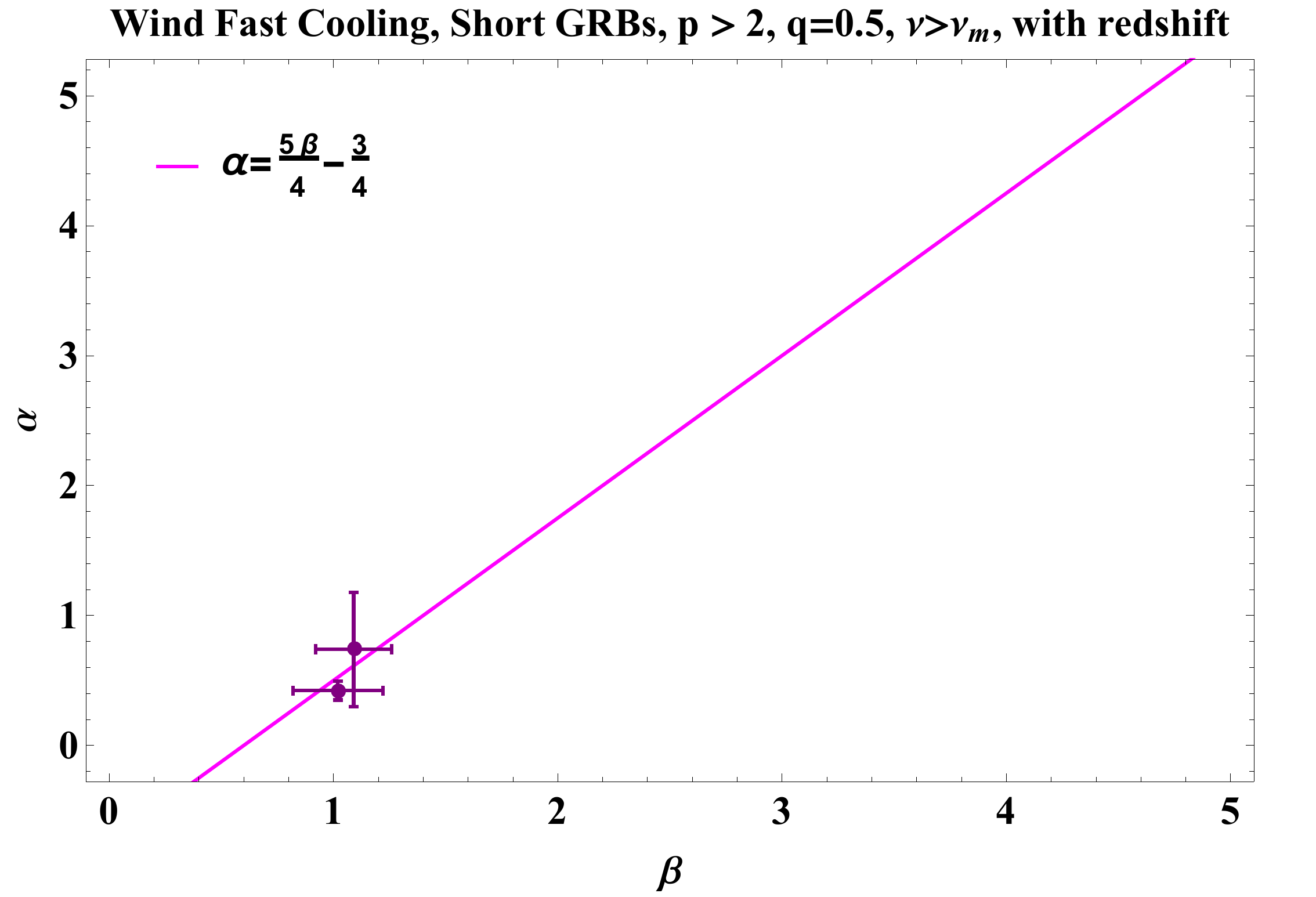}
    \caption{Compatible GRBs for CRs in the case $q=0.5, \nu>\nu_{\rm m}, p>2$ where the redshift is known. {\bf In the first 4 panels, the classes lGRBs and sGRBs are gathered together, while in the latter 8 panels these classes are separated.}}
    \label{only_nu>num_p>2_R_q0.5}
\end{figure}

\noindent We then plot the CRs by showing the $\alpha$ and $\beta$ parameters with the 1 $\sigma_{\rm int}$ error bars, as well as solid lines that represent the equations of the CRs, (see Figure \ref{only_nu>num_p>2_NR_q0}, \ref{only_nu>num_p>2_NR_q0.5}, \ref{only_nu>num_p>2_R_q0}, and \ref{only_nu>num_p>2_R_q0.5}), along with Tables \ref{NoRedshifttable_p>2} and \ref{Redshifttable_p>2}.

\noindent The GRB afterglow LCs are usually segregated into 4 segments: I, the initial steep decay; II, the PE phase; III, the adiabatic fireball deceleration with the observed decay phase; IV, the post-jet break phase \citep{Rhoads99,sari99,Kumar+00,m02,piran04,liang06,Zhang2006,zhang07b,Racusin+09}. We consider Phase II of the LCs in our analysis.

\noindent The Figures from \ref{only_nu>num_p>2_NR_q0} to \ref{only_nu>num_p>2_R_q0.5} show the CRs computed within the time range $T_t$-$T_a$ in the afterglow, together with the lines representing the CR equations and their error bars. Relations that are within the same $p$ range, cooling regime, and astrophysical environments are grouped on the same plot, drawing a ``gray-region" where the lying GRBs are regarded as consistent cases. 
{\bf We here stress that the lines representing the closure relationships are not the best fit lines, but are lines representing the relations themselves. We colour the CRs according to the following color coding: the red=$\alpha=\beta-1$, the blue=$\alpha=\beta$, the magenta=$\alpha=(5/4)\beta-(3/4)$, the gray=$(5/4)\beta+(1/4)$, and the orange=$(5/4)\beta-(1/2)$.} 
\noindent Let's take, as an example, the case of ISM SC with $p>2$: not only one should consider $\alpha = (q-1)+((2+q)\beta$)/2 ($\nu_m < \nu < \nu_c$) and $\alpha = (q-2)/2$ + ((2+q)$\beta$)/2 ($\nu > \nu_c$), but the region between these two lines, in the $\alpha-\beta$ plane, should be considered as ``consistent" region since these in-between regimes are allowed by the model and they are shown in grey in our plots. It may happen that a given couple of CRs for the considered environment degenerates in only one line. Nevertheless, this condition does not change the compatibility.

\noindent In Table \ref{CR3} we perform a classification that allows us to categorize GRBs in terms of the $q$ value which can be determined naturally from the relationships among $\alpha$ and $\beta$ from Table \ref{CR2}. Here, we derive for each GRB the value of $q$ value and its error $\delta q$: according to those values, we check the compatibility of each GRB's $q$ value with the reference values $q=0$ and $q=0.5$, see Table \ref{CR2.1}. Interestingly, we note that the cases for $q \geq 1$ are few and, thus, are not here considered. We here present also the distribution of the $q$ parameters showing the applicability of the method, see Figure \ref{fig_histogram_q0}. In this Figure, we highlight the boundary condition $q=1$ through a red vertical line, thus showing all the $q$ values of our categories despite considering only the ones with $q<1$ for our purposes. {\bf The results of the histograms in Figure \ref{fig_histogram_q0} are shown in the Table \ref{gaussian_parameters}}. From this Table, it is clear that the subdivision of all GRBs in lGRBs and sGRBs for each environment generates values of $q$ compatible in 1 $\sigma$ with the ones corresponding to sGRBs and lGRBs gathered together. The cases with $q \ge 1$ are roughly equivalent to the impulsive case as far as the energy injection is concerned. They add all their energy very quickly, and after that may be treated as an impulsive case with a different $E_\mathrm{iso}$ \citep[for review]{Barkov2011}.

\begin{table}
\begin{center}
\begin{tabular}{ccccc} 
$\nu$ & $\beta(p)$ & $\alpha(p)$ & $\alpha(\beta)$ & \\
& & & $(p > 2)$
\\[0.1cm]
\cline{1-4}
\multicolumn{4}{c}{ISM, SC} \\
\cline{1-4}
$\nu_m < \nu < \nu_c$ & $\frac{p-1}{2}$ &
$\alpha=\frac{(2p-6)+(p+3)q}{4}$  & $\alpha=(q-1)+\frac{(2+q)\beta}{2}$\\ [0.1cm]
$\nu > \nu_c$& $\frac{p}{2}$ & $\alpha=\frac{(2p-4)+(p+2)q}{4}$  & $\alpha=\frac{(q-2)}{2}+\frac{(2+q)\beta}{2}$
\\[0.1cm]
\cline{1-4}
\multicolumn{4}{c}{ISM, FC} \\
\cline{1-4}
$\nu > \nu_m$ & $\frac{p}{2}$&
$\alpha=\frac{(2p-4)+(p+2)q}{4}$  & $\alpha=\frac{q-2}{2}+\frac{(2+q)\beta}{2}$  \\[0.1cm]
\cline{1-4}
\multicolumn{4}{c}{Wind, SC} \\
\cline{1-4}
$\nu_m < \nu < \nu_c$ & $\frac{p-1}{2}$ &
$\alpha=\frac{(2p-2)+(p+1)q}{4}$  & $\alpha=\frac{q}{2}+\frac{(2+q)\beta}{2}$\\[0.1cm]
$\nu > \nu_c$ & $\frac{p}{2}$ &
$\alpha=\frac{(2p-4)+(p+2)q}{4}$  & $\alpha=\frac{q-2}{2}+\frac{(2+q)\beta}{2}$  
\\[0.1cm]
\cline{1-4}
\multicolumn{4}{c}{Wind, FC} \\
\cline{1-4}
$\nu > \nu_m$ & $\frac{p}{2}$ &
$\alpha=\frac{(2p-4)+(p+2)q}{4}$  & $\alpha=\frac{q-2}{2}+\frac{(2+q)\beta}{2}$ \\[0.1cm]
\cline{1-4}
\end{tabular}
\end{center}
\caption{CRs studied in this paper similarly to the one investigated by \citet{Zhang2006}. The first column contains the ranges of frequency, the second the spectral index, the third, and fourth the temporal index for $p>2$.} \label{CR2}
\end{table}

\begin{table}
\begin{center}
\hspace{5ex}
\begin{tabular}{cccccc}
$\nu$ & $\beta(p)$ & $\alpha(p)$ & $\alpha(\beta)$ & $\alpha(\beta)$ \\
& & & $(q=0)$ & $(q=0.5)$
\\
\cline{1-5}
\multicolumn{5}{c}{ISM, SC} \\[0cm]
\cline{1-5}
$\nu_m < \nu < \nu_c$ & $\frac{p-1}{2}$ & $\alpha=\frac{p-3}{2}$  & $\alpha =\beta -1 $  & $\alpha=\frac{5\beta}{4} -\frac{1}{2}$\\ 
$\nu > \nu_c$& $\frac{p}{2}$ & $\alpha=\frac{p-2}{2}$  & $\alpha=\beta -1$  & $\alpha=\frac{5\beta}{4} -\frac{3}{4}$
\\
\cline{1-5}
\multicolumn{5}{c}{ISM, FC} \\[0cm]
\cline{1-5}
$\nu > \nu_m$ & $\frac{p}{2}$ & $\alpha=\frac{p-2}{2}$  & $\alpha=\beta -1$  & $\alpha=\frac{5\beta}{4} -\frac{3}{4}$  \\
\cline{1-5}
\multicolumn{5}{c}{Wind, SC} \\[0cm]
\cline{1-5}
$\nu_m < \nu < \nu_c$ & $\frac{p-1}{2}$ & $\alpha=\frac{p-1}{2}$  & $\alpha=\beta$  & $\alpha=\frac{5\beta}{4} +\frac{1}{4}$ \\
$\nu > \nu_c$ & $\frac{p}{2}$ & $\alpha=\frac{p-2}{2}$  & $\alpha=\beta -1$  & $\alpha=\frac{5\beta}{4} -\frac{3}{4}$  \\
\cline{1-5}
\multicolumn{5}{c}{Wind, FC} \\[0cm]
\cline{1-5}
$\nu > \nu_m$ & $\frac{p}{2}$ & $\alpha=\frac{p-2}{2}$  & $\alpha=\beta -1$  & $\alpha=\frac{5\beta}{4} -\frac{3}{4}$ \\
\cline{1-5}
\end{tabular}
\end{center}

\caption{CRs as in Table \ref{CR2}, but with specific values of $q=0$ and $q=0.5$.} \label{CR2.1}
\end{table}

\begin{table*}
\begin{center}
\resizebox{14cm}{!}{
 \begin{tabular}{||c c c c c c c||} 
 \hline
 All GRBs Environment without z & $\nu\hspace{1ex}(\nu>\nu_m)$ & $q\hspace{1ex}(p>2)$ & CR &  $\#$ of GRBs & GRBs fulfilling CRs & $\%$ fulfilling CRs \\ 
 \hline\hline
 ISM, SC &$\nu_{\rm m} < \nu < \nu_{\rm c}$ & $0$ & $\alpha=\beta-1$ & 229 & 65 & 28.4\% \\ 
  &$\nu > \nu_{\rm c}$ & $0$ & $\alpha=\beta-1$ &  &  & \\
 \hline
 ISM, FC & $\nu > \nu_{\rm m}$ & $0$ & $\alpha=\beta-1$ & 226 & 60 & 26.6\% \\
 \hline
 Wind, SC &$\nu_{\rm m} < \nu < \nu_{\rm c}$ & $0$ & $\alpha=\beta$ & 229 & 204 & 89.1\% \\
 &$\nu > \nu_{\rm c}$ & $0$ & $\alpha=\beta-1$ &  &  &  \\
 \hline
 Wind, FC &$\nu > \nu_{\rm m}$ & $0$ & $\alpha=\beta-1$ & 76 & 14 & 18.4\% \\
 \hline
 \hline\hline
 ISM, SC &$\nu_{\rm m} < \nu < \nu_{\rm c}$ & $0.5$ & $\alpha=\frac{5\beta}{4}-\frac{1}{2}$ & 229 & 143 & 62.4\% \\ 
  &$\nu > \nu_{\rm c}$ & $0.5$ & $\alpha=\frac{5\beta}{4}-\frac{3}{4}$ &  &  & \\
 \hline
 ISM, FC & $\nu > \nu_{\rm m}$ & $0.5$ & $\alpha=\frac{5\beta}{4}-\frac{3}{4}$ & 226 & 42 & 18.6\% \\
 \hline
 Wind, SC &$\nu_{\rm m} < \nu < \nu_{\rm c}$ & $0.5$ & $\alpha=\frac{5\beta}{4}+\frac{1}{4}$ & 229 & 154 & 67.2\% \\
 &$\nu > \nu_{\rm c}$ & $0.5$ & $\alpha=\frac{5\beta}{4}-\frac{3}{4}$ &  &  &  \\
 \hline
 Wind, FC &$\nu > \nu_{\rm m}$ & $0.5$ & $\alpha=\frac{5\beta}{4}-\frac{3}{4}$ & 76 & 10 & 13.2\% \\
 \hline
 \hline
 lGRBs Environment without z &$\nu\hspace{1ex}(\nu>\nu_m)$ & $q\hspace{1ex}(p>2)$ & CR &  $\#$ of GRBs & GRBs fulfilling CRs & $\%$ fulfilling CRs \\ 
 \hline\hline
 ISM, SC &$\nu_{\rm m} < \nu < \nu_{\rm c}$ & $0$ & $\alpha=\beta-1$ & 220 & 63 & 28.6\% \\ 
  &$\nu > \nu_{\rm c}$ & $0$ & $\alpha=\beta-1$ &  &  & \\
 \hline
 ISM, FC & $\nu > \nu_{\rm m}$ & $0$ & $\alpha=\beta-1$ & 218 & 58 & 26.6\% \\
 \hline
 Wind, SC &$\nu_{\rm m} < \nu < \nu_{\rm c}$ & $0$ & $\alpha=\beta$ & 22  & 196 & 89.1\% \\
 &$\nu > \nu_{\rm c}$ & $0$ & $\alpha=\beta-1$ & & & \\
 \hline
 Wind, FC &$\nu > \nu_{\rm m}$ & $0$ & $\alpha=\beta-1$ & 74 & 14 & 18.9\% \\
 \hline
 \hline\hline
 ISM, SC &$\nu_{\rm m} < \nu < \nu_{\rm c}$ & $0.5$ & $\alpha=\frac{5\beta}{4}-\frac{1}{2}$ & 220 & 139 & 63.2\% \\ 
  &$\nu > \nu_{\rm c}$ & $0.5$ & $\alpha=\frac{5\beta}{4}-\frac{3}{4}$ &  &  & \\
 \hline
 ISM, FC & $\nu > \nu_{\rm m}$ & $0.5$ & $\alpha=\frac{5\beta}{4}-\frac{3}{4}$ & 218 & 42 & 19.3\% \\
 \hline
 Wind, SC &$\nu_{\rm m} < \nu < \nu_{\rm c}$ & $0.5$ & $\alpha=\frac{5\beta}{4}+\frac{1}{4}$ & 220 & 150 & 68.2\% \\
 &$\nu > \nu_{\rm c}$ & $0.5$ & $\alpha=\frac{5\beta}{4}-\frac{3}{4}$ & & &  \\
 \hline
 Wind, FC &$\nu > \nu_{\rm m}$ & $0.5$ & $\alpha=\frac{5\beta}{4}-\frac{3}{4}$ & 74 &  10 & 13.5\% \\
 \hline
 \hline
 sGRBs Environment without z &$\nu\hspace{1ex}(\nu>\nu_m)$ & $q\hspace{1ex}(p>2)$ & CR &  $\#$ of GRBs & GRBs fulfilling CRs & $\%$ fulfilling CRs \\ 
 \hline\hline
 ISM, SC &$\nu_{\rm m} < \nu < \nu_{\rm c}$ & $0$ & $\alpha=\beta-1$ & 8 & 2 & 25\% \\ 
  &$\nu > \nu_{\rm c}$ & $0$ & $\alpha=\beta-1$ &  &  & \\
 \hline
 ISM, FC & $\nu > \nu_{\rm m}$ & $0$ & $\alpha=\beta-1$ & 7 & 2 & 28.6\% \\
 \hline
 Wind, SC &$\nu_{\rm m} < \nu < \nu_{\rm c}$ & $0$ & $\alpha=\beta$ & 8 & 7 & 87.5\% \\
 &$\nu > \nu_{\rm c}$ & $0$ & $\alpha=\beta-1$ &  &  &  \\
 \hline
 Wind, FC &$\nu > \nu_{\rm m}$ & $0$ & $\alpha=\beta-1$ & 1 & 0 & - \\
 \hline
 \hline\hline
 ISM, SC &$\nu_{\rm m} < \nu < \nu_{\rm c}$ & $0.5$ & $\alpha=\frac{5\beta}{4}-\frac{1}{2}$ & 8 & 3 & 37.5\% \\ 
  &$\nu > \nu_{\rm c}$ & $0.5$ & $\alpha=\frac{5\beta}{4}-\frac{3}{4}$ &  &  & \\
 \hline
 ISM, FC & $\nu > \nu_{\rm m}$ & $0.5$ & $\alpha=\frac{5\beta}{4}-\frac{3}{4}$ & 7 & 0 & - \\
 \hline
 Wind, SC &$\nu_{\rm m} < \nu < \nu_{\rm c}$ & $0.5$ & $\alpha=\frac{5\beta}{4}+\frac{1}{4}$ & 8 & 3 & 37.5\% \\
 &$\nu > \nu_{\rm c}$ & $0.5$ & $\alpha=\frac{5\beta}{4}-\frac{3}{4}$ &  &  &  \\
 \hline
 Wind, FC &$\nu > \nu_{\rm m}$ & $0.5$ & $\alpha=\frac{5\beta}{4}-\frac{3}{4}$ & 1 & 0 & - \\
 \hline
\end{tabular}
}
\end{center}
\caption{CRs characteristics for the sample without redshift (in the case of $\nu>\nu_{\rm m}$ and $p>2$). \textbf{The upper part of the table refers to All with $q=0$ and $q=0.5$, the middle part refers to lGRBs for $q=0$ and $q=0.5$, and the lower part considers the sGRBs for again $q=0$ and $q=0.5$ .}}\label{NoRedshifttable_p>2}
\end{table*}

\begin{table*}
\begin{center}
\resizebox{13cm}{!}{
 \begin{tabular}{||c c c c c c c||} 
 \hline
 All GRBs Environment with z  & $\nu\hspace{1ex}(\nu>\nu_m)$ & $q\hspace{1ex}(p>2)$ & CR &  $\#$ of GRBs & GRBs fulfilling CR & $\%$ fulfilling CRs \\ 
  \hline\hline
 ISM, SC &$\nu_{\rm m} < \nu < \nu_{\rm c}$ & $0$ & $\alpha=\beta-1$ & 209 & 39 & 18.7\% \\ 
  &$\nu > \nu_{\rm c}$ & $0$ & $\alpha=\beta-1$ &  &  & \\
 \hline
 ISM, FC & $\nu > \nu_{\rm m}$ & $0$ & $\alpha=\beta-1$ & 204 & 33  & 16.2\% \\
 \hline
 Wind, SC &$\nu_{\rm m} < \nu < \nu_{\rm c}$ & $0$ & $\alpha=\beta$ & 209 & 135 & 64.6\% \\
 &$\nu > \nu_{\rm c}$ & $0$ & $\alpha=\beta-1$ &  &  &  \\
 \hline
 Wind, FC &$\nu > \nu_{\rm m}$ & $0$ & $\alpha=\beta-1$ & 53 & 3 & 5.7\% \\
 \hline
 \hline\hline
 ISM, SC &$\nu_{\rm m} < \nu < \nu_{\rm c}$ & $0.5$ & $\alpha=\frac{5\beta}{4}-\frac{1}{2}$ & 209 & 135 & 64.6\% \\ 
  &$\nu > \nu_{\rm c}$ & $0.5$ & $\alpha=\frac{5\beta}{4}-\frac{3}{4}$ &  &  & \\
 \hline
 ISM, FC & $\nu > \nu_{\rm m}$ & $0.5$ & $\alpha=\frac{5\beta}{4}-\frac{3}{4}$ & 204 & 32 & 15.7\% \\
 \hline
 Wind, SC &$\nu_{\rm m} < \nu < \nu_{\rm c}$ & $0.5$ & $\alpha=\frac{5\beta}{4}+\frac{1}{4}$ & 209 & 159 & 76.1\% \\
 &$\nu > \nu_{\rm c}$ & $0.5$ & $\alpha=\frac{5\beta}{4}-\frac{3}{4}$ &  &  &  \\
 \hline
 Wind, FC &$\nu > \nu_{\rm m}$ & $0.5$ & $\alpha=\frac{5\beta}{4}-\frac{3}{4}$ & 53 & 11 & 20.8\% \\
 \hline \hline
 lGRBs Environment with z &$\nu\hspace{1ex}(\nu>\nu_m)$ & $q\hspace{1ex}(p>2)$ & CR &  $\#$ of GRBs & GRBs fulfilling CRs & $\%$ fulfilling CRs \\ 
 \hline\hline
 ISM, SC &$\nu_{\rm m} < \nu < \nu_{\rm c}$ & $0$ & $\alpha=\beta-1$ & 177 & 30 & 16.9\% \\ 
  &$\nu > \nu_{\rm c}$ & $0$ & $\alpha=\beta-1$ &  &  & \\
 \hline
 ISM, FC & $\nu > \nu_{\rm m}$ & $0$ & $\alpha=\beta-1$ & 173 & 24 & 13.9\% \\
 \hline
 Wind, SC &$\nu_{\rm m} < \nu < \nu_{\rm c}$ & $0$ & $\alpha=\beta$ & 177 & 159 & 89.8\% \\
 &$\nu > \nu_{\rm c}$ & $0$ & $\alpha=\beta-1$ &  &  &  \\
 \hline
 Wind, FC &$\nu > \nu_{\rm m}$ & $0$ & $\alpha=\beta-1$ & 44 & 3 & 6.8\% \\
 \hline
 \hline\hline
 ISM, SC &$\nu_{\rm m} < \nu < \nu_{\rm c}$ & $0.5$ & $\alpha=\frac{5\beta}{4}-\frac{1}{2}$ & 177 & 114 & 64.4\% \\ 
  &$\nu > \nu_{\rm c}$ & $0.5$ & $\alpha=\frac{5\beta}{4}-\frac{3}{4}$ &  &  & \\
 \hline
 ISM, FC & $\nu > \nu_{\rm m}$ & $0.5$ & $\alpha=\frac{5\beta}{4}-\frac{3}{4}$ & 173 & 24 & 13.9\% \\
 \hline
 Wind, SC &$\nu_{\rm m} < \nu < \nu_{\rm c}$ & $0.5$ & $\alpha=\frac{5\beta}{4}+\frac{1}{4}$ & 177 & 137 & 77.4\% \\
 &$\nu > \nu_{\rm c}$ & $0.5$ & $\alpha=\frac{5\beta}{4}-\frac{3}{4}$ &  &  &  \\
 \hline
 Wind, FC &$\nu > \nu_{\rm m}$ & $0.5$ & $\alpha=\frac{5\beta}{4}-\frac{3}{4}$ & 44 & 9 & 20.4\% \\
 \hline
 \hline
 sGRBs Environment with z & $\nu\hspace{1ex}(\nu>\nu_m)$ & $q\hspace{1ex}(p>2)$ & CR &  $\#$ of GRBs & GRBs fulfilling CRs & $\%$ fulfilling CRs \\ 
 \hline\hline
 ISM, SC &$\nu_{\rm m} < \nu < \nu_{\rm c}$ & $0$ & $\alpha=\beta-1$ & 31 & 9 & 29\% \\ 
  &$\nu > \nu_{\rm c}$ & $0$ & $\alpha=\beta-1$ &  &  & \\
 \hline
 ISM, FC & $\nu > \nu_{\rm m}$ & $0$ & $\alpha=\beta-1$ & 30 & 9 & 30\% \\
 \hline
 Wind, SC &$\nu_{\rm m} < \nu < \nu_{\rm c}$ & $0$ & $\alpha=\beta$ & 31 & 29 & 93.5\% \\
 &$\nu > \nu_{\rm c}$ & $0$ & $\alpha=\beta-1$ &  &  &  \\
 \hline
 Wind, FC &$\nu > \nu_{\rm m}$ & $0$ & $\alpha=\beta-1$ & 8 & 0 & - \\
 \hline
 \hline\hline
 ISM, SC &$\nu_{\rm m} < \nu < \nu_{\rm c}$ & $0.5$ & $\alpha=\frac{5\beta}{4}-\frac{1}{2}$ & 31 & 20 & 64.5\% \\ 
  &$\nu > \nu_{\rm c}$ & $0.5$ & $\alpha=\frac{5\beta}{4}-\frac{3}{4}$ &  &  & \\
 \hline
 ISM, FC & $\nu > \nu_{\rm m}$ & $0.5$ & $\alpha=\frac{5\beta}{4}-\frac{3}{4}$ & 30 & 8 & 26.7\% \\
 \hline
 Wind, SC &$\nu_{\rm m} < \nu < \nu_{\rm c}$ & $0.5$ & $\alpha=\frac{5\beta}{4}+\frac{1}{4}$ & 31 & 21 & 67.7\% \\
 &$\nu > \nu_{\rm c}$ & $0.5$ & $\alpha=\frac{5\beta}{4}-\frac{3}{4}$ &  &  &  \\
 \hline
 Wind, FC &$\nu > \nu_{\rm m}$ & $0.5$ & $\alpha=\frac{5\beta}{4}-\frac{3}{4}$ & 8 & 2 & 25\% \\
 \hline
\end{tabular}%
}
\end{center}
\caption{CRs characteristics for the sample with known redshift (in the case of $\nu>\nu_{\rm m}$ and $p>2$). {\bf The upper part of the table refers to All GRBs with $q=0$ and $q=0.5$, the middle part refers to lGRBs with $q=0$ and $q=0.5$ and the lower part considers the sGRBs both with $q=0$ and $q=0.5$.}}\label{Redshifttable_p>2}
\end{table*}

\begin{table}
\begin{center}
\hspace{7ex}
\begin{tabular}{ccccccccccccc} \\
$\nu$ & All & $q=0$ & $q=0.5$ & $\overline{q}$ & lGRBs & $q_{\rm L}=0$ & $q_L=0.5$ & $\overline{q}_L$ & sGRBs & $q_{\rm S}=0$ & $q_S=0.5$ & $\overline{q}_S$\\
\cline{1-13}
\multicolumn{1}{c}{ISM, SC} \\[0cm]
\cline{1-13}
$\nu_{\rm m} < \nu < \nu_{\rm c}$ & 167 & 22 & 45 & 0.37 & 145 & 17 & 42 & 0.39 & 22 & 5 & 3 & 0.39 \\ 
\cline{1-13}
\multicolumn{1}{c}{ISM, FC} \\[0cm]
\cline{1-13}
$\nu > \nu_m$ & 165 & 20 & 61 & 0.58 & 142 & 15 & 52 & 0.60 & 22 & 5 & 9 & 0.41 \\ 
\cline{1-13}
\multicolumn{1}{c}{Wind, SC} \\[0cm]
\cline{1-13}
$\nu_m < \nu < \nu_c$ & 167 & 26 & 5 & -0.45 & 145 & 24 & 5 & -0.43 & 22 & 2 & 0 & -0.59 \\ 
\cline{1-13}
\multicolumn{1}{c}{Wind, FC} \\[0cm]
\cline{1-13}
$\nu > \nu_m$ & 44 & 0 & 12 & 0.73 & 145 & 24 & 5 & -0.43 & 6 & 0 & 2 & 0.64 \\ 
\cline{1-13}
\end{tabular}
\end{center}
\caption{Energy injection parameter $q$ computed for particular $\nu$ ranges with equations taken from Table \ref{CR2}. The first column contains the ranges of frequency, the second shows the number of all GRBs in the particular $\nu$ range, the third and fourth columns correspond to the number of GRBs for which computed $q$ is compatible with values $0$ and $0.5$ respectively and the fifth column reports the averaged value of $q$ in the given environment. {\bf The columns from 6th to 9th contain the same quantities of columns from 2nd to 5th, but in the case of lGRBs only. The columns from 10th to 13th contain the same quantities of columns from 2nd to 5th in the case of sGRBs. The whole computation was performed only in the case of temporal index $p>2$.}} \label{CR3}
\end{table}

\subsection{Results and Interpretations}
We test the CRs given in Table \ref{CR2} for the 455 GRB LCs in our sample from 2005 January until 2019 August, detailing the relations calculated for the time range of the PE from $T_t$ to $T_a$. Figures \ref{only_nu>num_p>2_NR_q0}, \ref{only_nu>num_p>2_NR_q0.5}, \ref{only_nu>num_p>2_R_q0}, and \ref{only_nu>num_p>2_R_q0.5} detail the specific CRs, including the error bars of $\alpha$ and $\beta$ in 1 $\sigma_{int}$ along with the equality lines corresponding to each CR, and Table \ref{Redshifttable_p>2} and Table \ref{NoRedshifttable_p>2} convey the number of GRBs satisfying such relations. We group in the same plot the GRBs that correspond to the same $p$ range and astrophysical environment, to check if GRBs lie within the so-called ``gray-region" (called gray-region in \citet{evans09}) marked in darker colors in our plots.

\noindent As seen in Table \ref{NoRedshifttable_p>2} and Table \ref{Redshifttable_p>2}, the majority of GRBs do satisfy the CRs, when looking at the entire sample \textbf{and when considering the lGRBs and sGRBs separately. The most fulfilled set of relations for {\bf both lGRBs and sGRBs} are the ones of Wind SC environment for $q=0$, and $q=0.5$ for redshift (with percentages of 64.6\% and 76.1\%, respectively), while in the case of unknown redshift for $q=0$ and $q=0.5$ again the Wind Slow Cooling is the most populated with percentages of 89.1\% and 67.2\%, respectively.}
We also note that the lowest percentage from $5.7 \%$ to $20.8 \%$ corresponds to Wind FC in all cases for $q=0$ and $q=0.5$ and for GRBs with unknown and known redshift. 
{\bf In the lower part of the Table \ref{NoRedshifttable_p>2} we consider the division in lGRBs and sGRBs for GRBs without redshift. From the percentage computed we can state that the Wind Slow Cooling regime is the most favored ($89.1 \%$ for $q=0$ and $68.2 \%$ for $q=0.5$) for lGRBs, while the least favored is the Wind Fast cooling ($18.9 \%$ for $q=0$ and 13.5 $\%$ for $q=0.5$). For sGRBs again the Wind Slow Cooling is the most favored for $q=0$ (87.5\%), while for $q=0.5$ both ISM Slow Cooling and Wind Slow Cooling are favored (37.5\%). Regarding the cases of redshift, we can state looking at Table \ref{Redshifttable_p>2} that both lGRBs and sGRBs favor Wind Slow cooling for both $q=0$ and $q=0.5$ with the percentages of $\% 64.6$ and $\% 76.1$, respectively. In the lower part of the Table, we investigate the cases of lGRBs, and in the bottom part of the Table sGRBs. It is clear from the table that for lGRBs Wind Slow Cooling is favored for both $q=0$, $q=0.5$ with the percentages of $89.8\%$ and $77.4\%$ respectively. In the case of sGRBs the most favored scenario is again the Wind Slow cooling both for $q=0$ and $q=0.5$ with a percentage of $93.5\%$, and $67.7\%$, respectively. As a general trend in the majority of cases, Wind Slow cooling is favored.}
We here note that since some of the $p$ and $\nu$ ranges are repeated, the total number of GRBs is repeated both for the known redshift and unknown redshift GRBs in the following circumstances: ISM SC has the same number of GRBs as the Wind SC both for $q=0$ and $q=0.5$ in the sets of GRBs with known and unknown redshift. The percentages are computed under the assumption that all scenarios are equally likely to occur for each given GRB. 

\noindent When grouping CRs and different GRBs, we conclude that for our sample, the ISM FC and Wind FC are the least favored in the case $p>2$ for both values of $q$ for the majority of cases. These environments are equally disfavoured also in the case without redshift.

\noindent However, the CRs should be considered a preliminary check on the robustness of the ES model without assuming any more complex physical processes. Therefore, it may still be possible that the ES model is accurate if taking into account such processes, as the effects of particle acceleration in non-linear regime \citep{2017ApJ...835..248W} or an energy injection: \citet{Zhang2006} state that the presence of phase II of the LC may be linked to three possible scenarios of continuous energy injection due to the engine at the center of GRBs. In this scenario, the fireball will take longer to decelerate due to the following:

\begin{itemize}
    \item the long-lasting central engine expressed in Equation \ref{eq_L(t)}; 
    \item the different velocities of ejecta, when the slower materials accumulate and re-ignite the emission;
    \item the relevant fraction of Poynting flux of the outflow.
\end{itemize}

\noindent With regards to particle acceleration in non-linear regime, \citet{2017ApJ...835..248W} tested an evolutionary model of afterglows assuming a more complicated particle distribution than the standard power-law by including a thermal population of electrons. If only a PL distribution of electrons is present, $\beta$ varies monotonically between the $\nu < \max(\nu_{\rm m},\nu_{c})$ value and the $\nu > \max(\nu_{\rm m},\nu_{\rm c})$ value. The inclusion of a thermal population of electrons increases $\beta$ when the characteristic synchrotron frequency of the thermal population’s exponential tail falls in a particular waveband; when the emission is once again due to the accelerated, non-thermal, population of electrons, the value of $\beta$ returns to its usual value. Such non-monotonic, soft-hard-soft behavior of $\beta$ was observed in several GRB afterglows, particularly at the beginning of the afterglow emission \citep{Giannios2009}. 

\indent

\noindent However, it may also be possible that the ES model may not be the ideal scenario in some cases, and it is not external, but internal shocks that produce the afterglow. \citet{lyons2010} discuss this case as driven by a millisecond magnetar. In the ES model, the relativistic jet of the GRB is powered by a black hole formed immediately after the progenitor event. They state that in some cases, there may be a period after the progenitor event during which a millisecond magnetar forms before collapsing into a black hole. As the velocity of the magnetar diminishes over time, the engine will continually inject energy into the LC of the GRB, until its collapse which will be seen as an extremely steep decay in the LC. This feature of emission that is relatively constant followed by an abrupt decay ($\alpha \geq 4$) characterizes the internal PE. In our sample, we have one such GRB that has an $\alpha \geq 4$, GRB 061110A, {\bf thus we have considered using the condition set by \citet{Liang18} in which they define the internal plateaus for cases with $\alpha>3$. This new definition allows us to have in our sample '6' GRBs with internal plateaus.} We now compare the results of phase II and phase III. In \cite{Srinivasaragavan2020}, a similar analysis was performed on the CRs concerning phase III of the LCs, taken between time $T_a$ and $T_{end}$. From the comparison, we see that in the current work the environments favored for $p>2$ are Wind SC and ISM SC for cases $q=0$ and $q=0.5$, respectively (considering both GRBs with known and unknown redshifts). Differently from those results, in \cite{Srinivasaragavan2020} the ISM and Wind FC are favored in the case of unknown redshift, for $p>2$.

\begin{figure}
    \centering
    
    \includegraphics[scale=0.15]{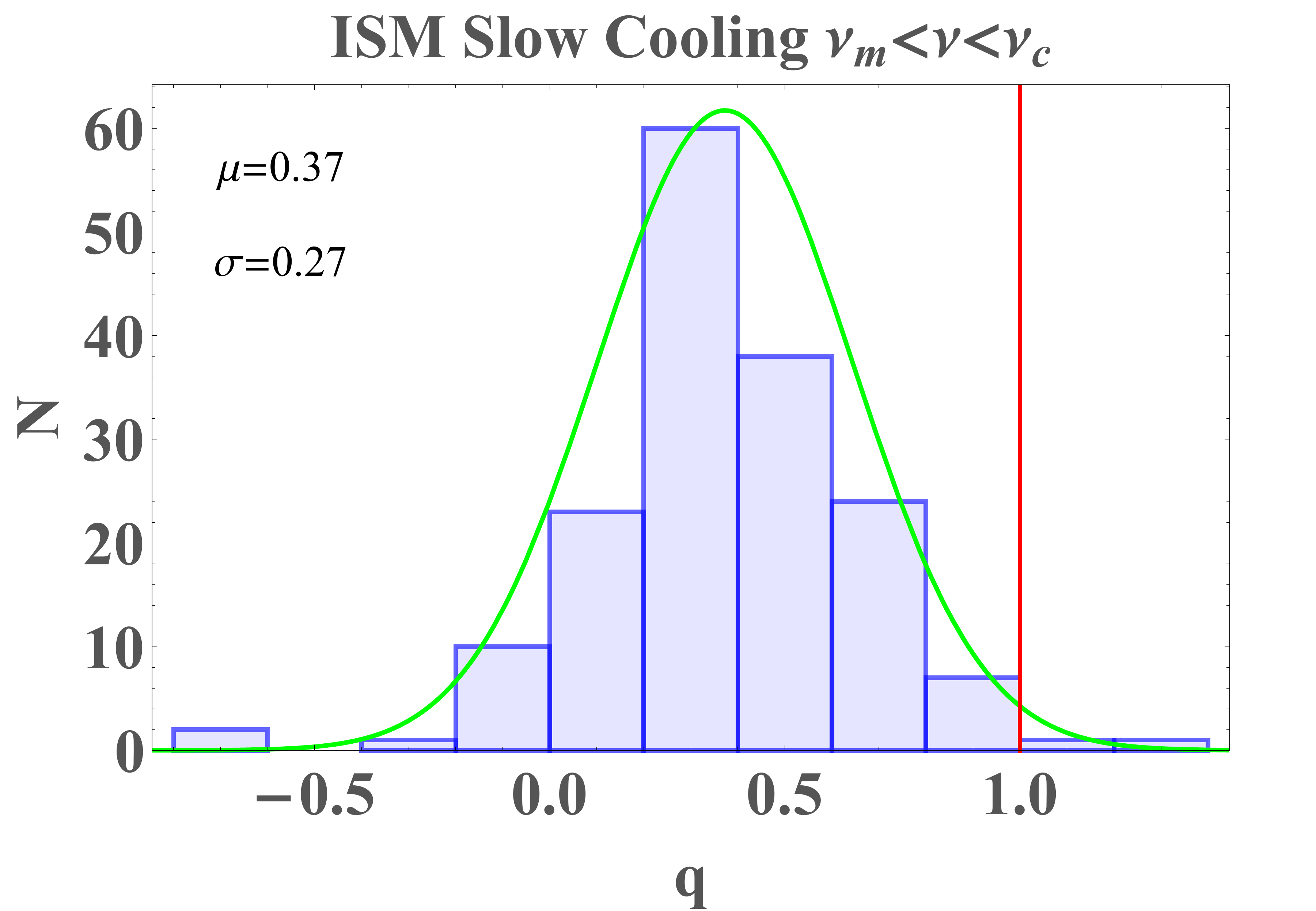}
    \includegraphics[scale=0.15]{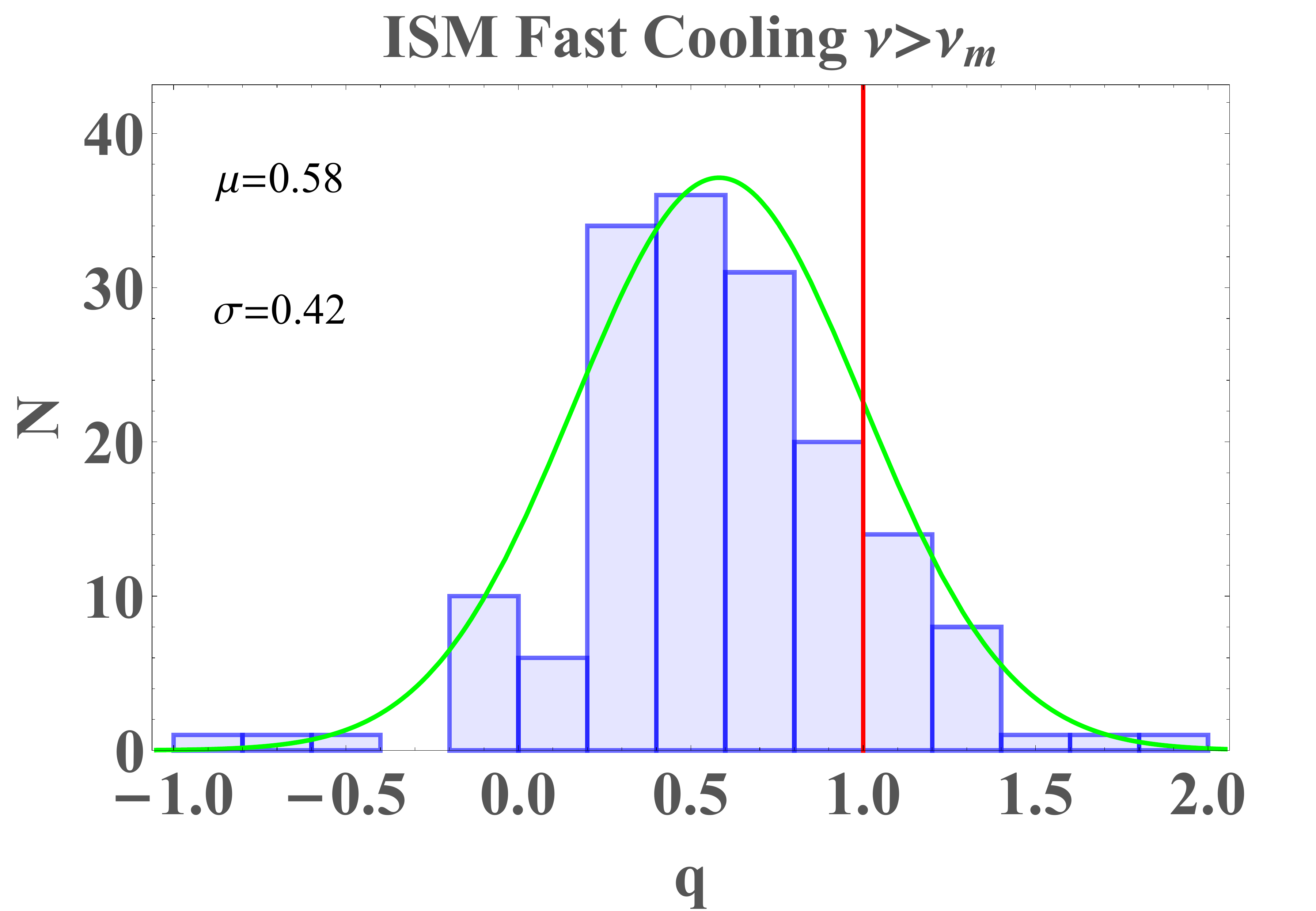}
    \includegraphics[scale=0.15]{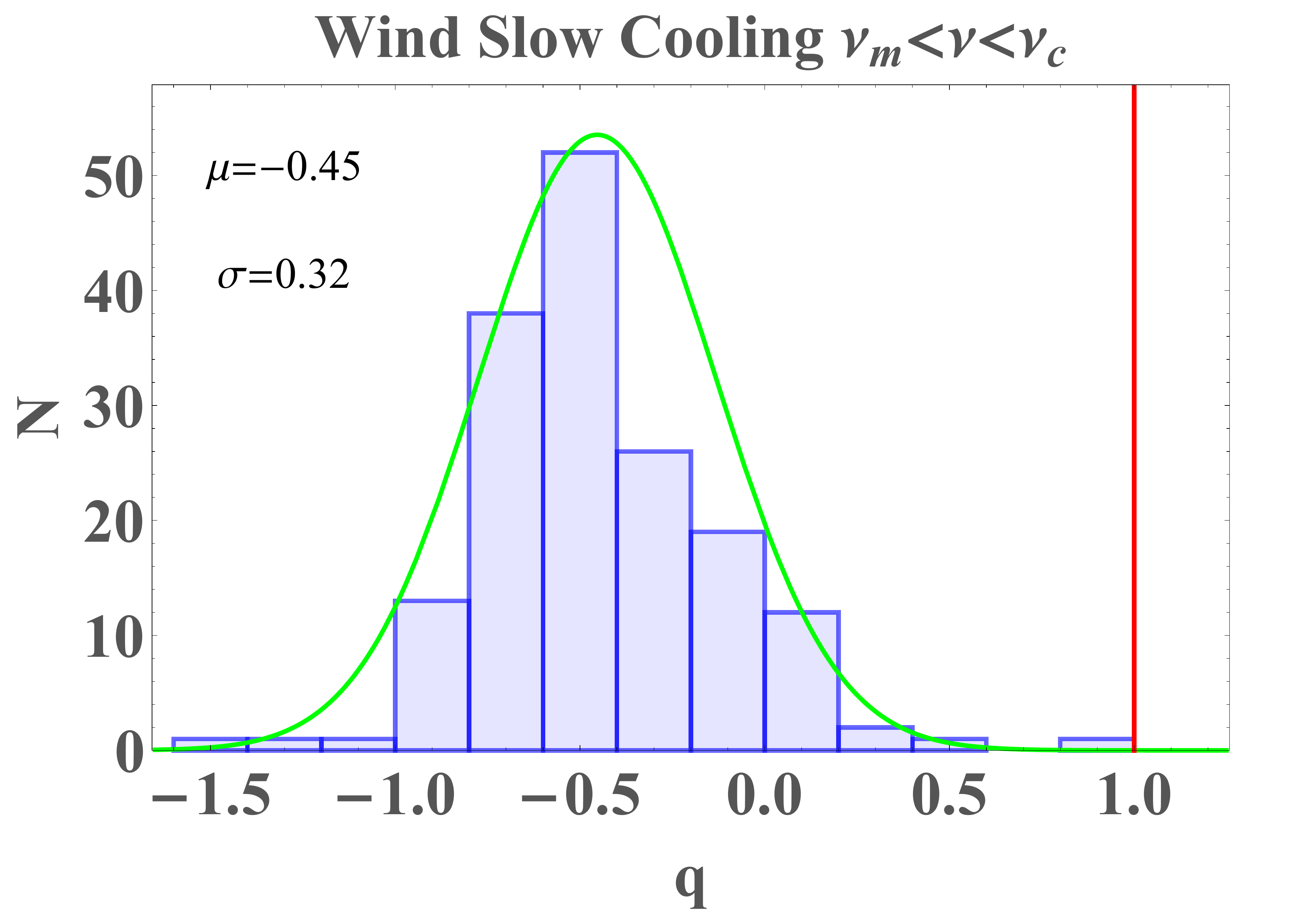}
    \includegraphics[scale=0.20]{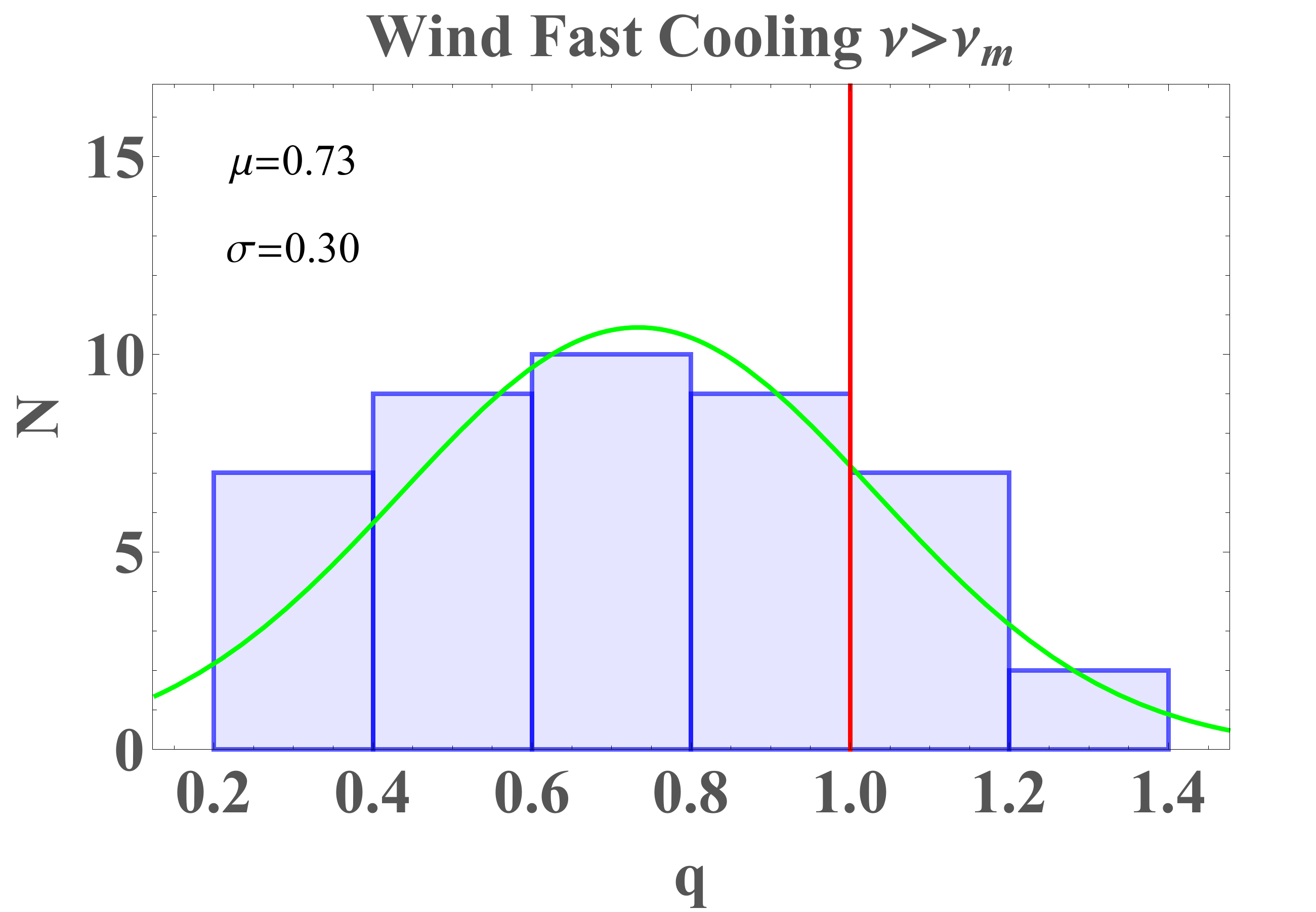}
    \includegraphics[scale=0.15]{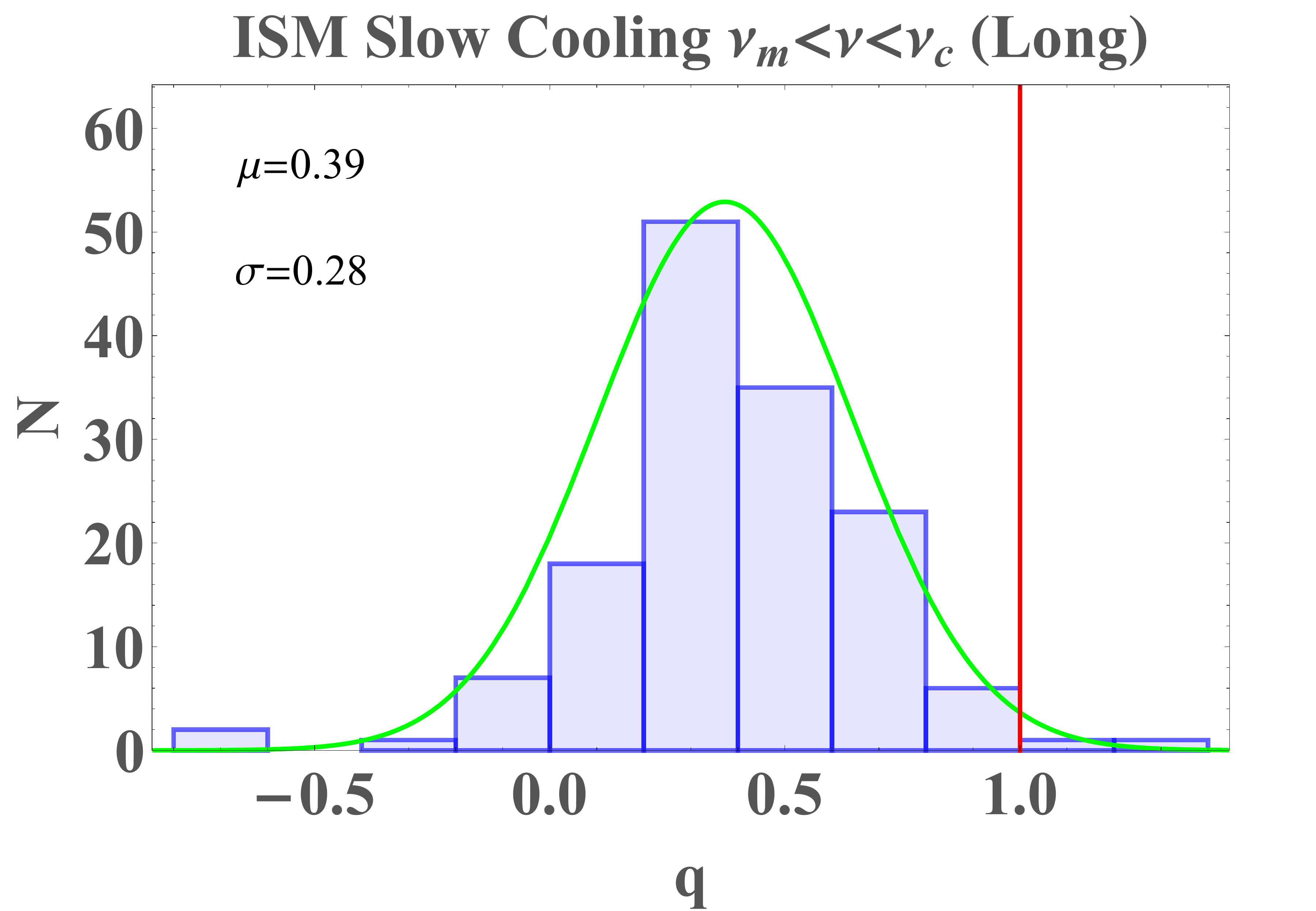}
    \includegraphics[scale=0.15]{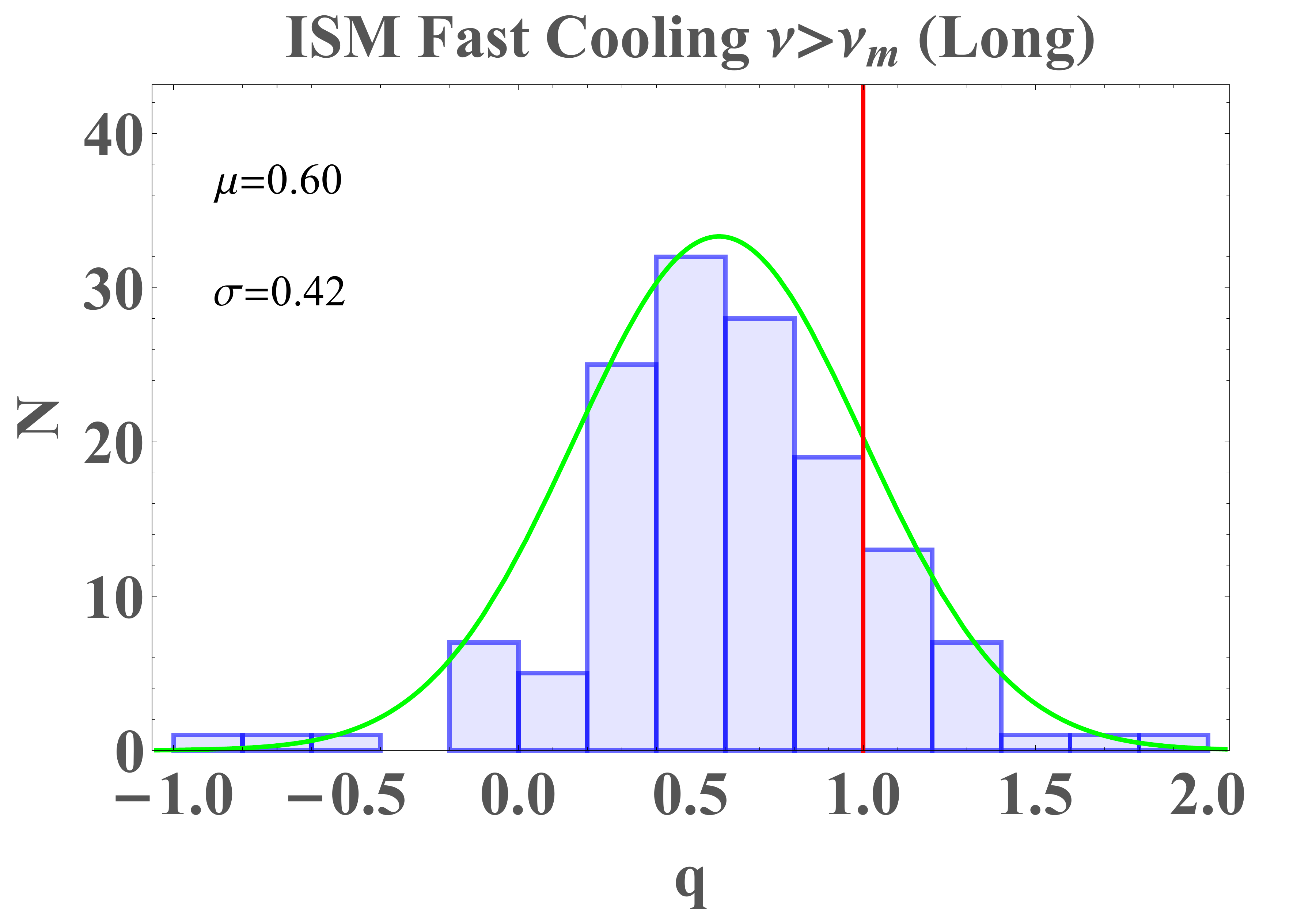}
    \includegraphics[scale=0.15]{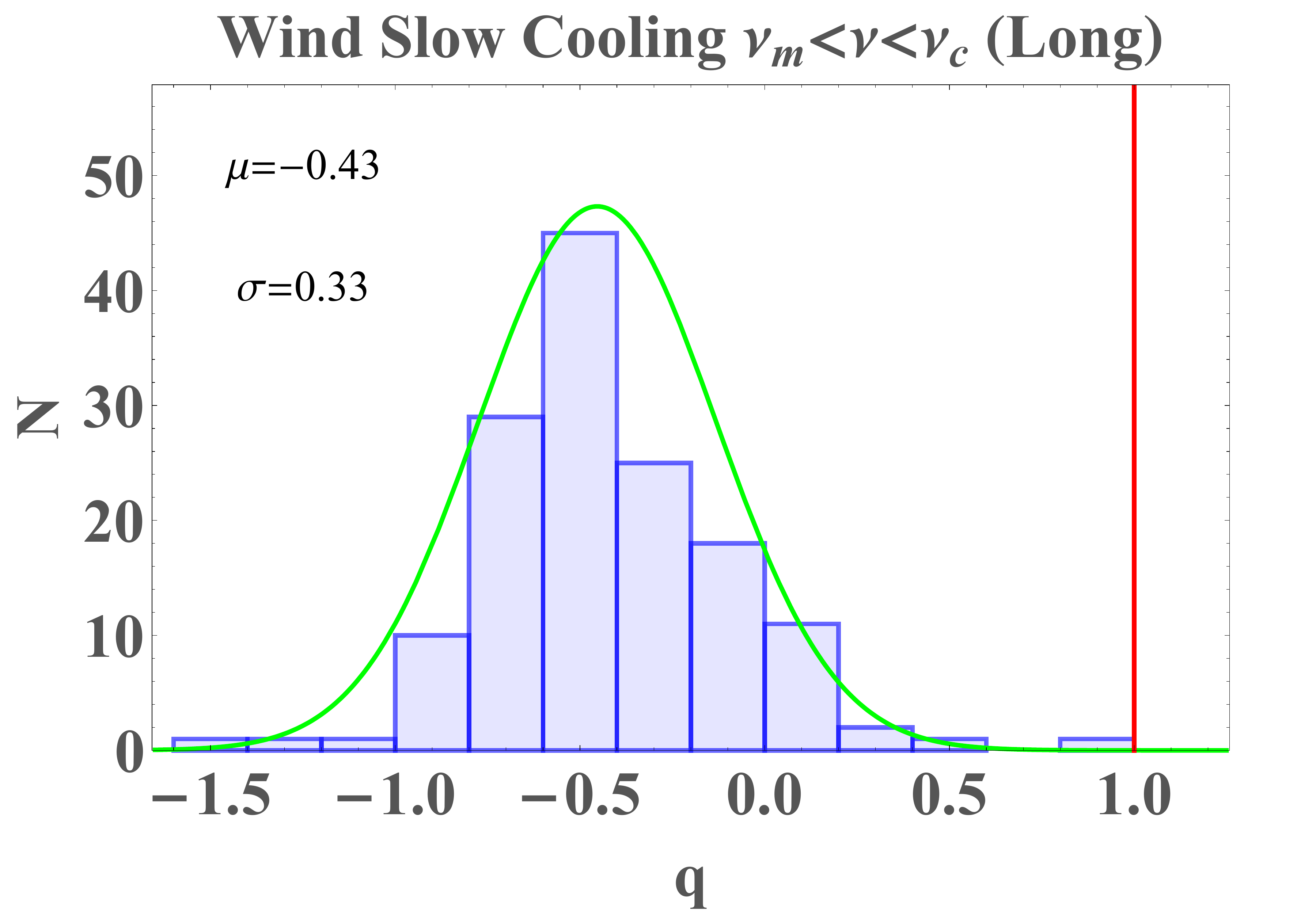}
    \includegraphics[scale=0.15]{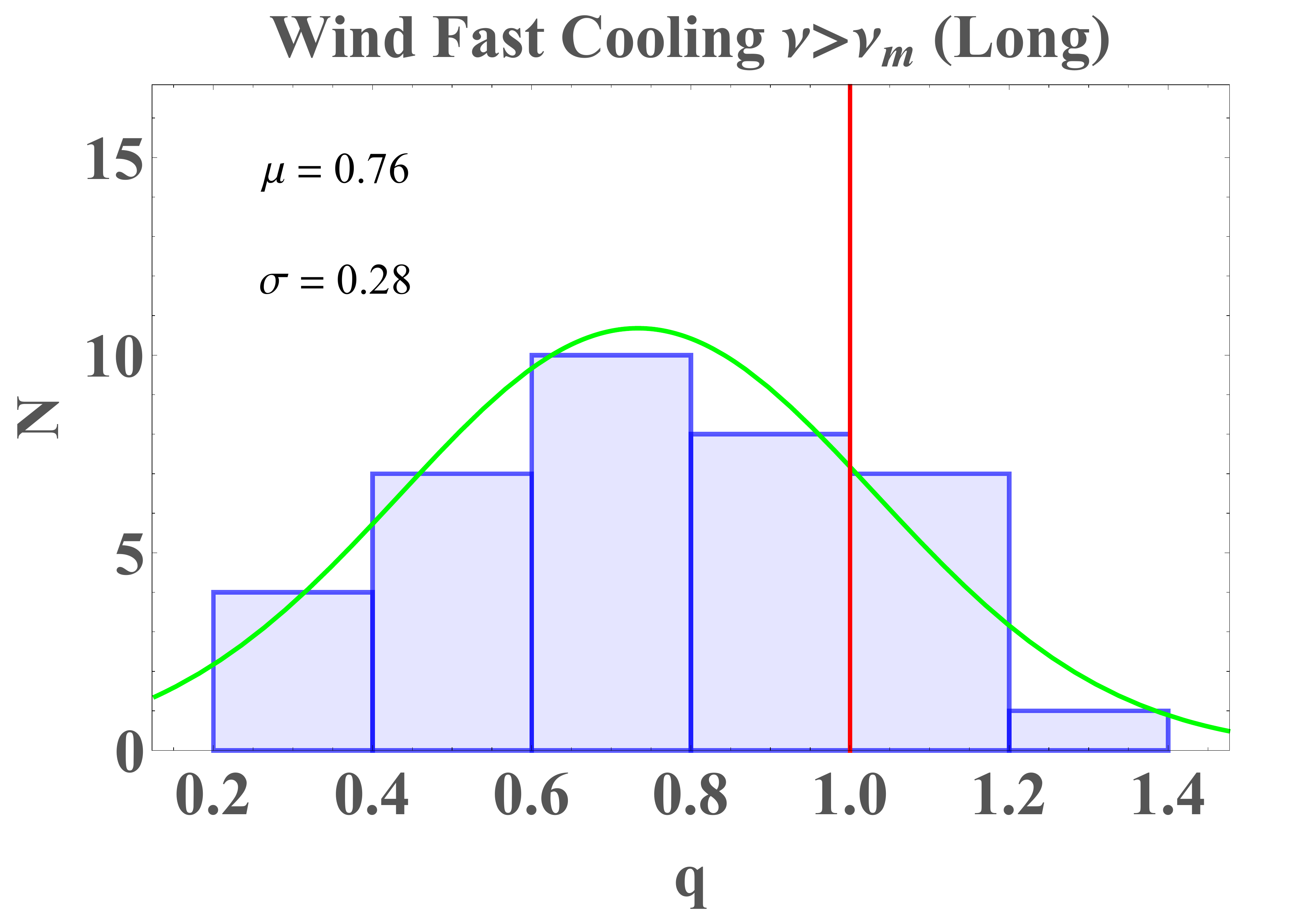}
    \includegraphics[scale=0.15]{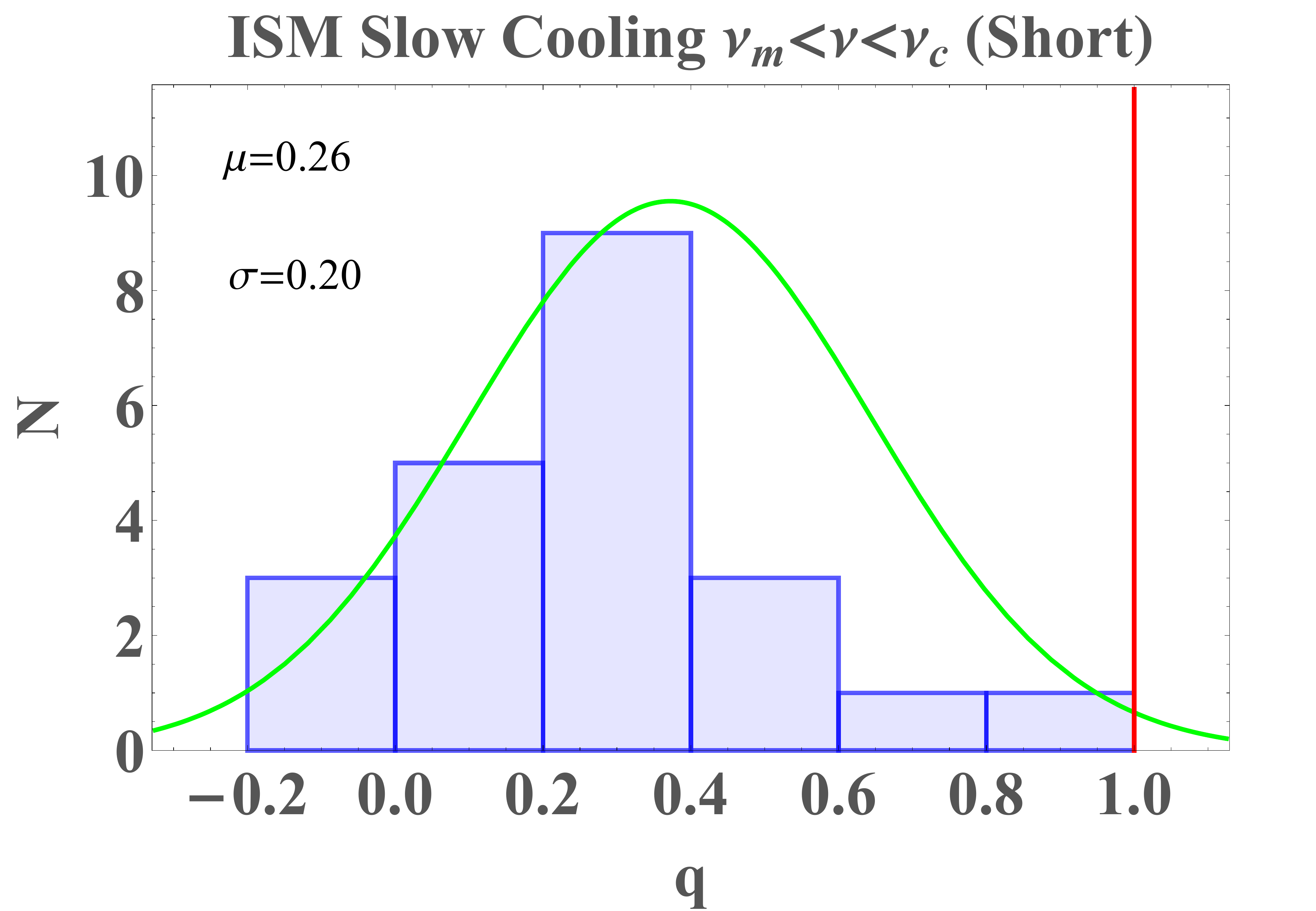}
    \includegraphics[scale=0.15]{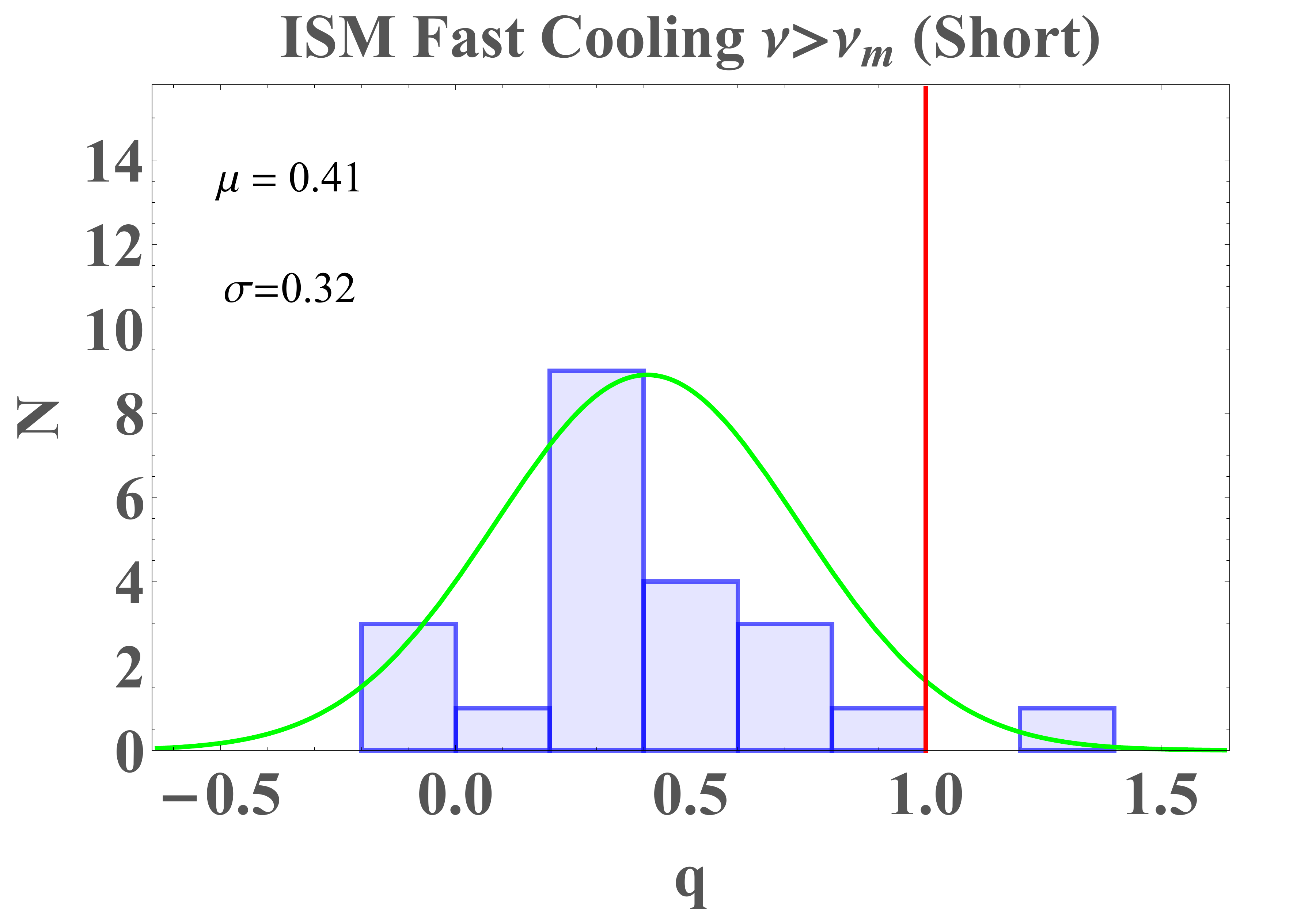}
    \includegraphics[scale=0.15]{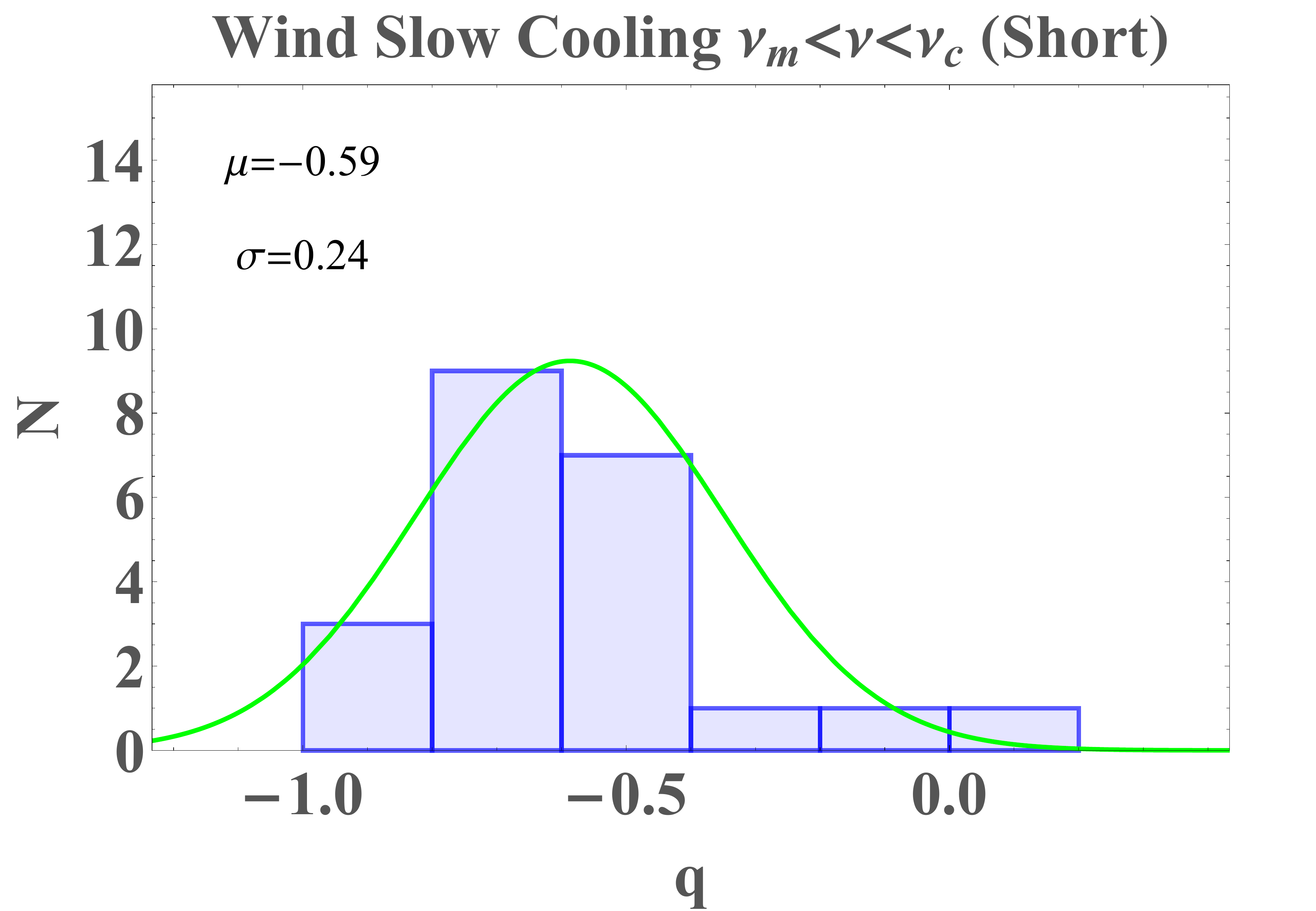}
    \includegraphics[scale=0.15]{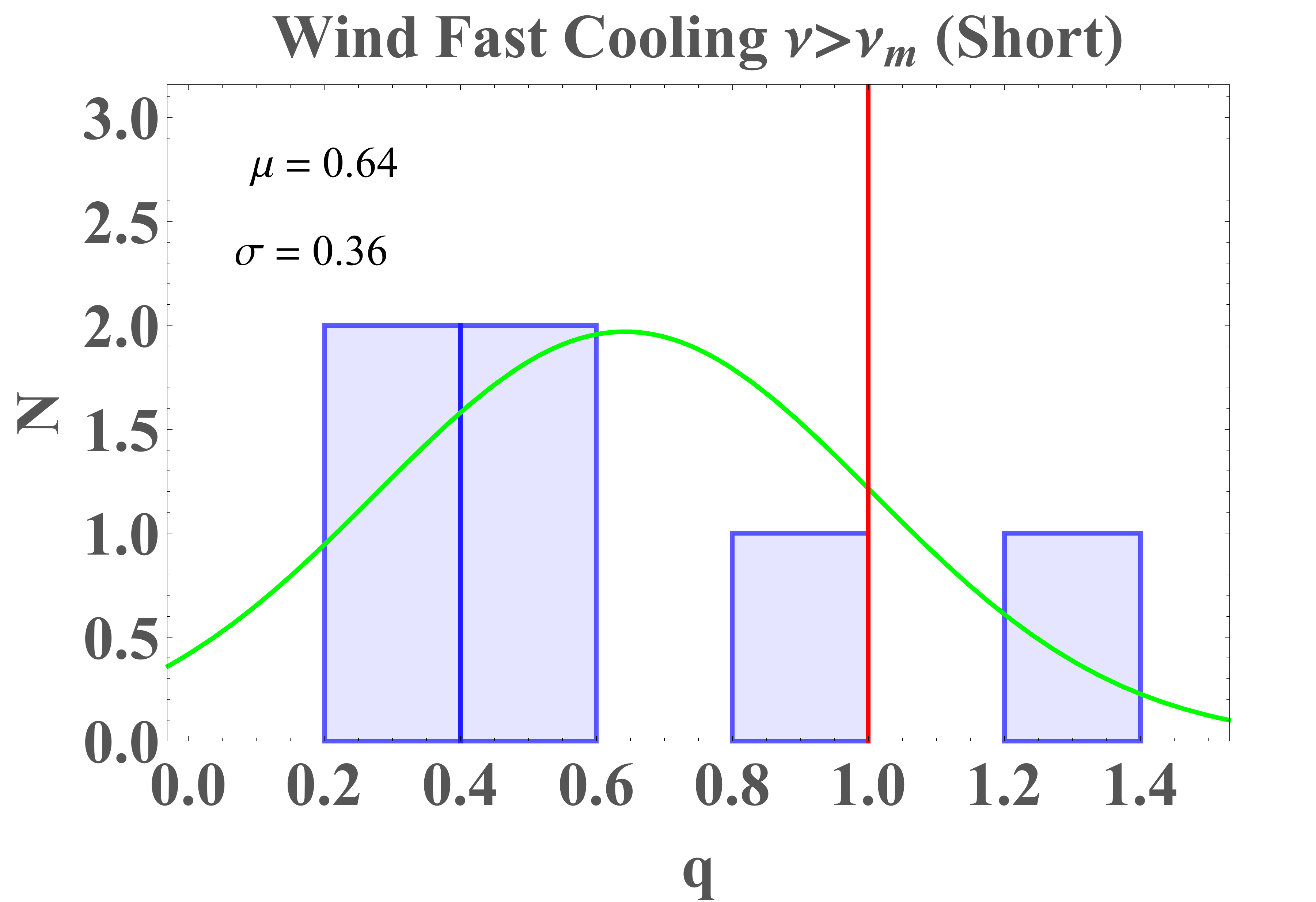}
    
    \caption{According to Table \ref{CR3}, the population of GRBs' $q$ value in each environment \bf{and category (All, lGRBs and sGRBs)}. The green curves show the Gaussian PDF for data with mean $\mu$ and standard deviation $\sigma$.
    The vertical lines refer to the limit $q=1$ since in this particular work we do not consider the impulsive energy injection for GRBs with $q>1$.}
    \label{fig_histogram_q0}
\end{figure}

\noindent {\bf In the current analysis we also consider with the given values of $\alpha$ and $\beta$ the distributions of the $q$ parameters derived analytically, shown in Table \ref{gaussian_parameters} and plotted in Fig. \ref{fig_histogram_q0}. Starting from the upper left panel we show the distribution of ISM slow cooling, in the middle-upper panel we present the ISM with Fast cooling, in the upper right panel we show the Wind Slow cooling, in the middle left upper panel we show the Wind Fast cooling for all GRBs. In the successive panels, from the middle-upper central panel to the lower middle central panel we present the same environments for the lGRBs, and from the right lower middle panel to the lower panels we present the same configurations for the sGRBs. As we can clearly see from Table 6 the error bars of the fit are sufficiently large to show that there is not a marked difference between the distributions of q for sGRBs and lGRBs, since the $\mu$ values all overlap in 1 $\sigma$.}

\begin{table*}
    \centering
    \begin{tabular}{|c|c|c|c|}
    \hline
    ISM SC (All) & ISM FC (All) & Wind SC (All) & Wind FC (All)
    \\
    $q=0.37\pm0.27$ & $q=0.58\pm0.42$ & $q=-0.45\pm0.32$ & $q=0.73\pm0.27$
    \\ \hline 
    ISM SC (lGRBs) & ISM FC (lGRBs) & Wind SC (lGRBs) & Wind FC (lGRBs)
    \\ 
    $q=0.39\pm0.28$ & $q=0.60\pm0.42$ & $q=-0.43\pm0.33$ & $q=0.76\pm0.28$
    \\ \hline
    ISM SC (sGRBs) & ISM FC (sGRBs) & Wind SC (sGRBs) & Wind FC (sGRBs)
    \\
    $q=0.26\pm0.20$ & $q=0.41\pm0.32$ & $q=-0.59\pm0.24$ & $q=0.41\pm0.43$
    \\ \hline
    \end{tabular}
    
    \caption{The Gaussian PDF parameter values referred to the fittings in Figure \ref{fig_histogram_q0}. The results are here reported in the form $q=\mu\pm\sigma$.}
    \label{gaussian_parameters}
\end{table*}

\section{The 3-D Fundamental Plane Relation}
\label{Results from SWIFT}

The CRs are crucial tools to pinpoint a sub-sample of GRBs potentially useful as standard candles. We aim to verify this property by using the 3-D Dainotti relation. We fit this relation with regards to the different astrophysical groups taken from Table \ref{CR2}, for our sample containing GRBs with known redshifts in the ($\log(T_{\rm a})$, $\log(L_\mathrm{{peak}})$, $\log(L_{\rm a})$) space applying a statistical method \citep{Dagostini2005}, a Bayesian approach which takes into account the error measurements in all the variables used.

\begin{figure}
    \centering
    \includegraphics[scale=0.20]{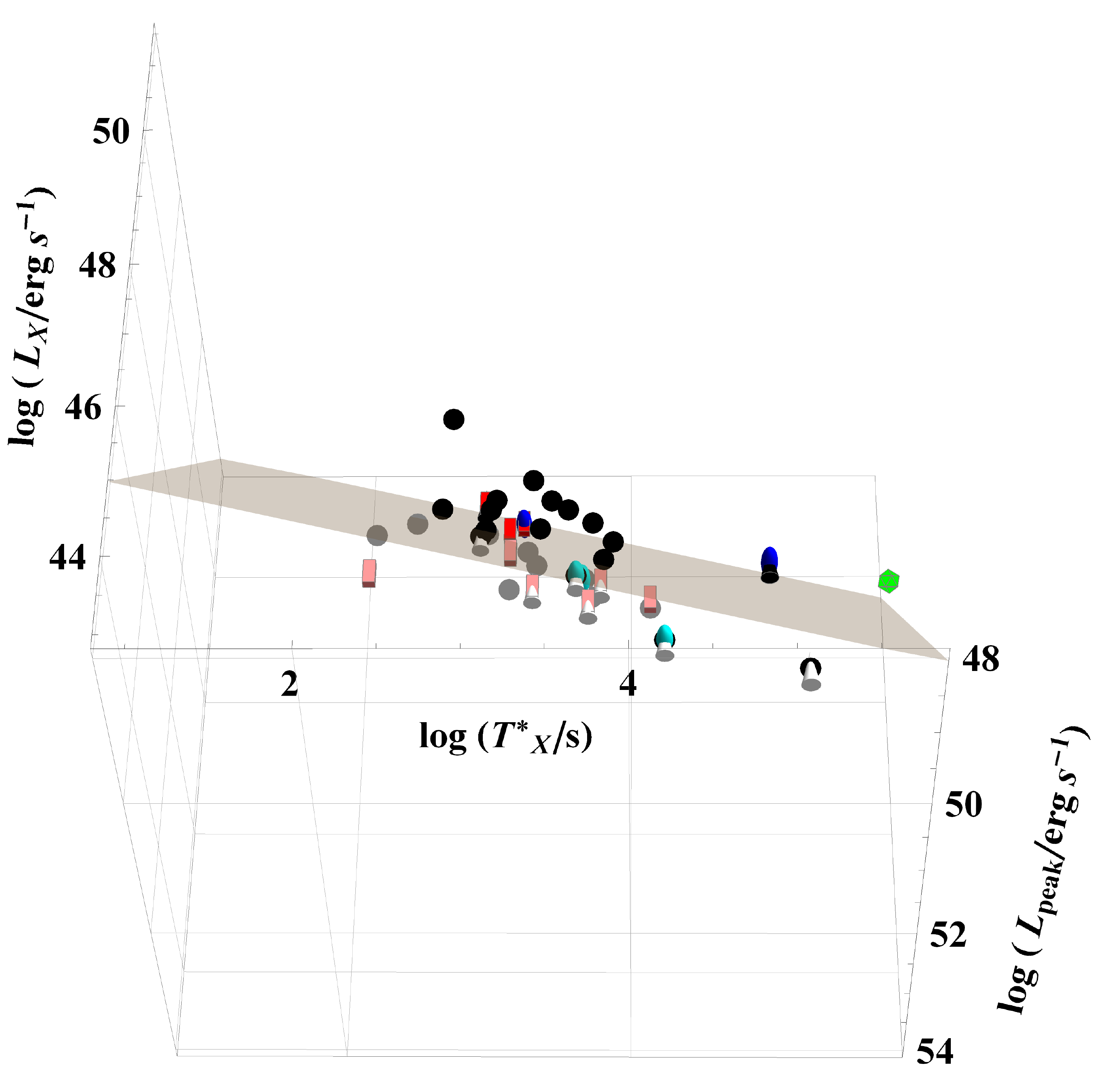}
    \includegraphics[scale=0.20]{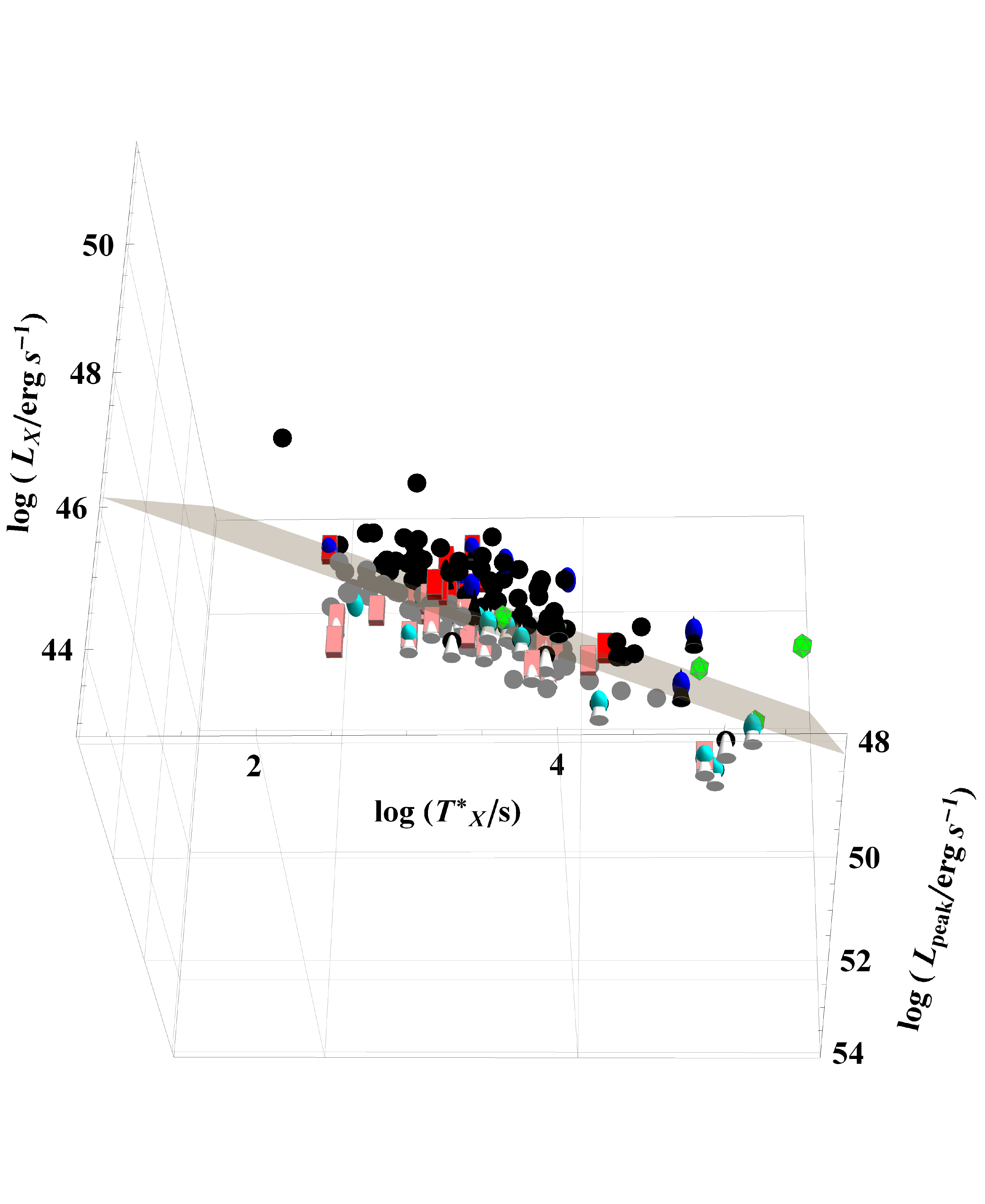}
    \includegraphics[scale=0.20]{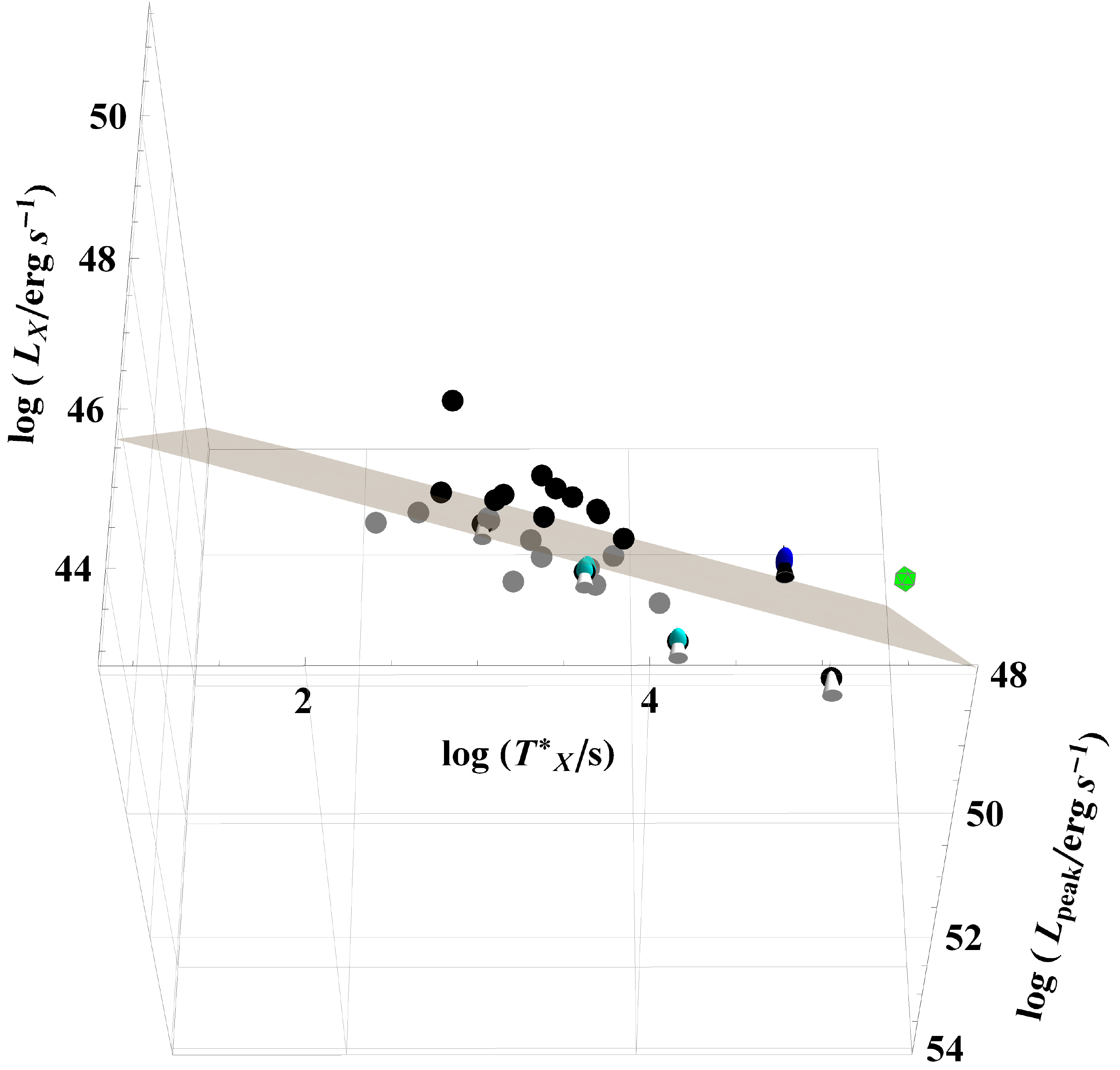}
    \includegraphics[scale=0.20]{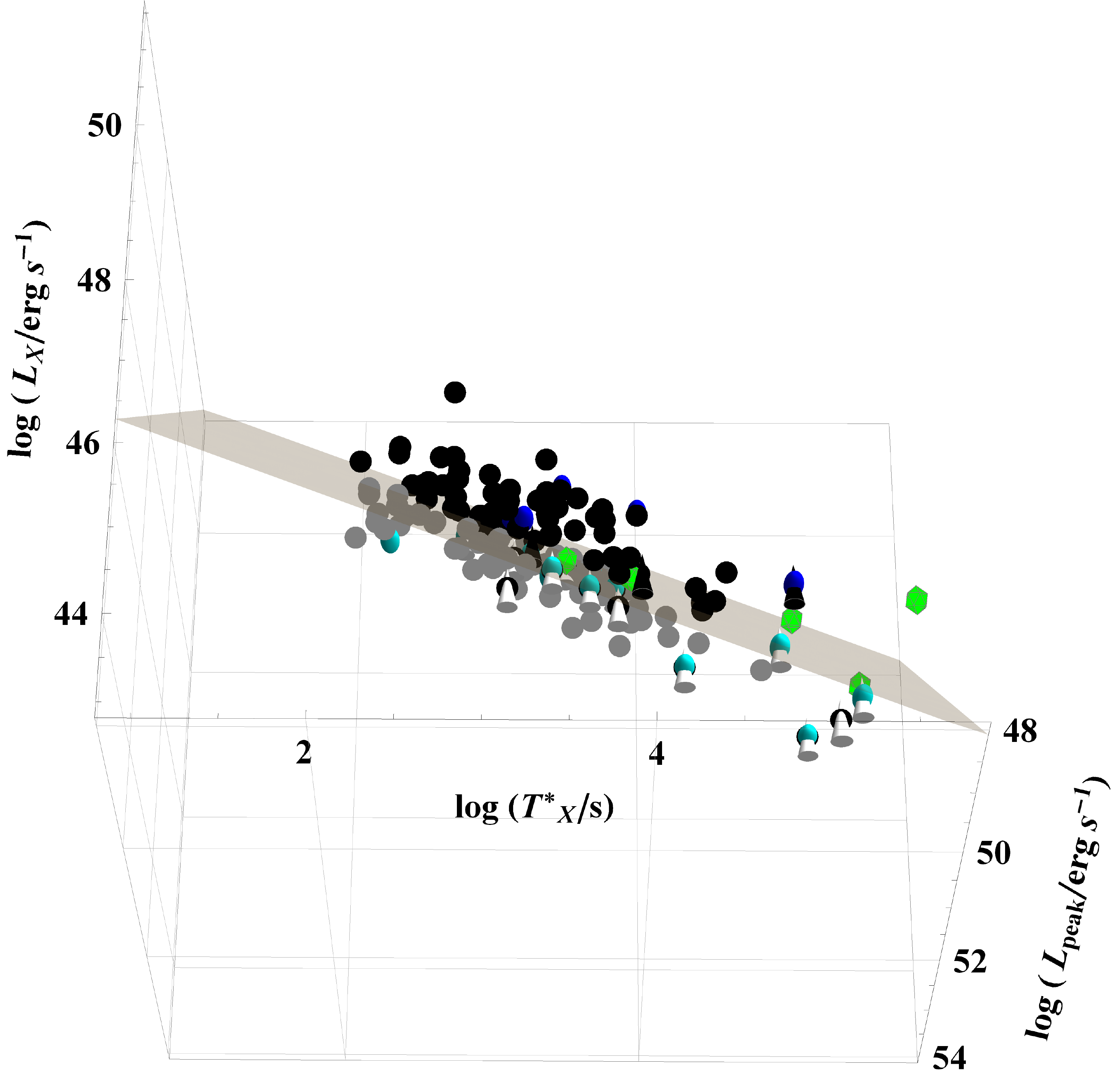}
    \includegraphics[scale=0.20]{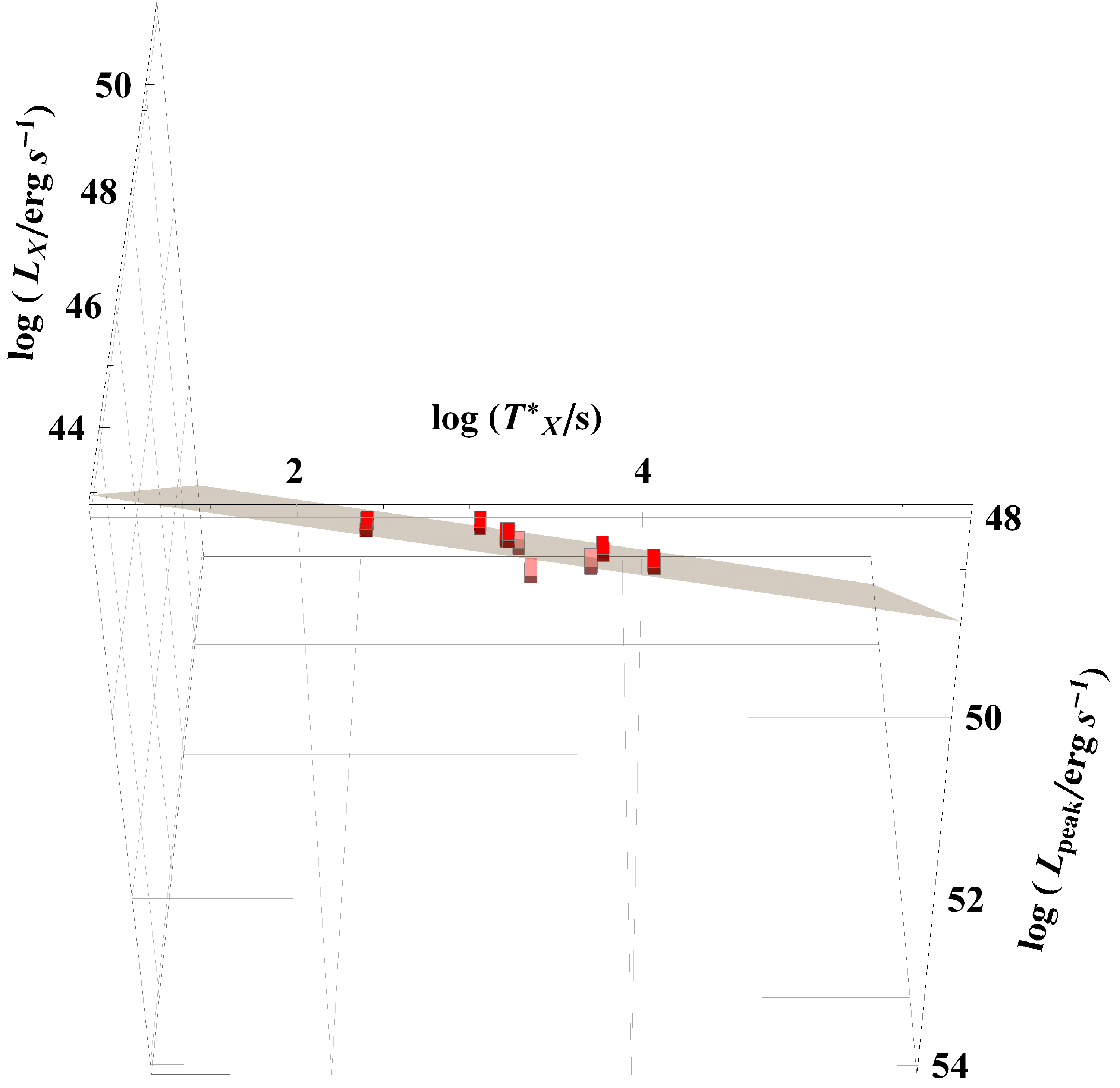}
    \includegraphics[scale=0.20]{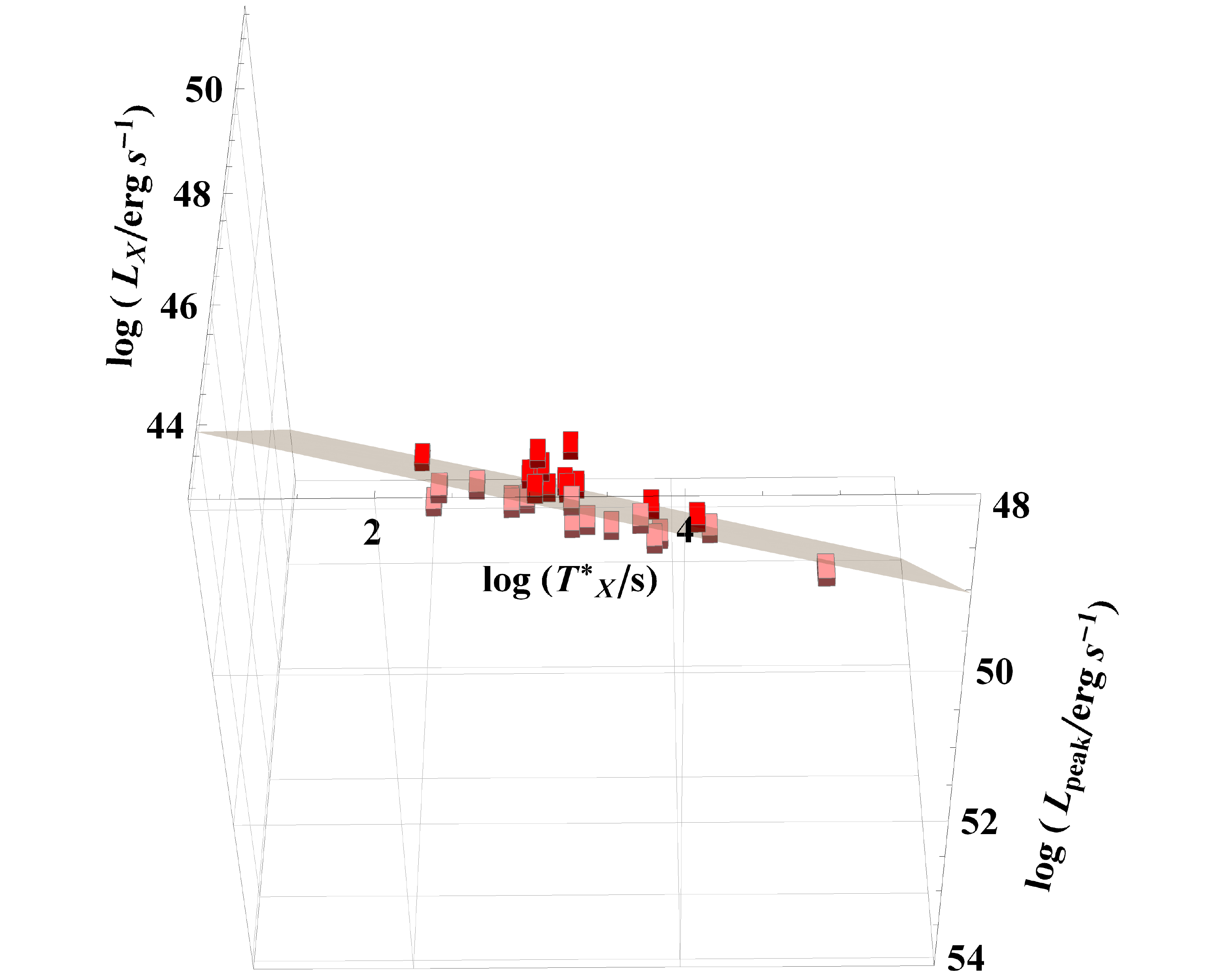}
    \caption{\bf{GRBs are divided into All ISM for all GRBs (upper left panel), All Wind for both lGRBs and sGRBs (upper right panel) groups, All ISM for lGRBs (middle left panel), All Wind for lGRBs (middle right panel), All ISM for sGRBs (lower left panel), All Wind for sGRBs (lower right panel) groups (see Table \ref{Dagostini_q0})}, and placed on the fundamental plane defined by Eq \ref{planeequation1}. Shapes and colors have been assigned as follows: GRB-SNe (black cones), XRFs (blue spheres), SEE (cuboids), lGRBs (black circles), UL GRBs (green truncated icosahedrons). The same color, but darker, label GRBs above the plane, while lighter colors label GRBs below the plane, except for UL GRBs which are all represented by bright green truncated icosahedrons. Here $q=0$ is considered.}
    \label{3D_AllISM_AllWIND_q0}
\end{figure}

\begin{figure}
    \centering
    \includegraphics[scale=0.22]{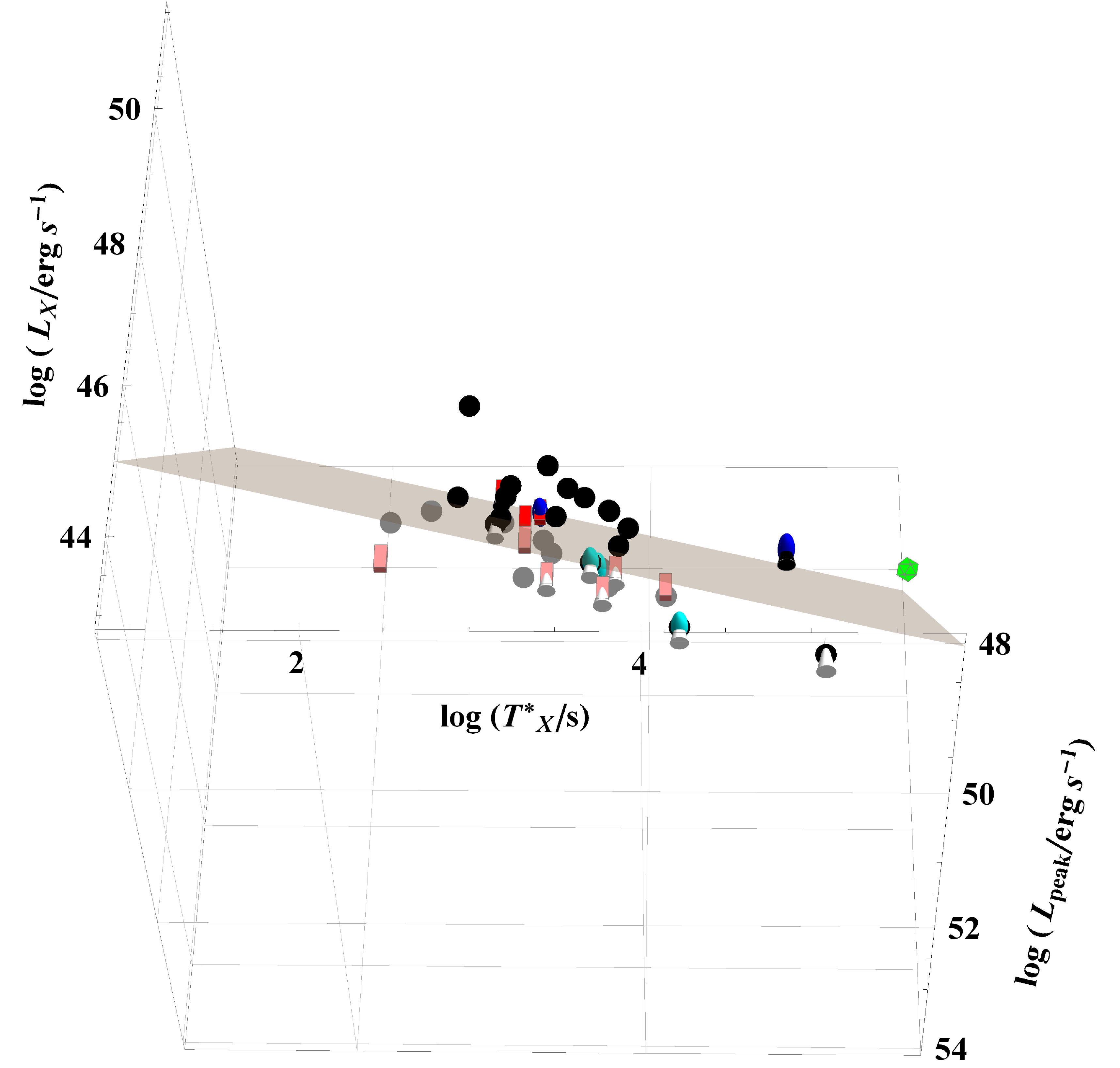}
    \includegraphics[scale=0.22]{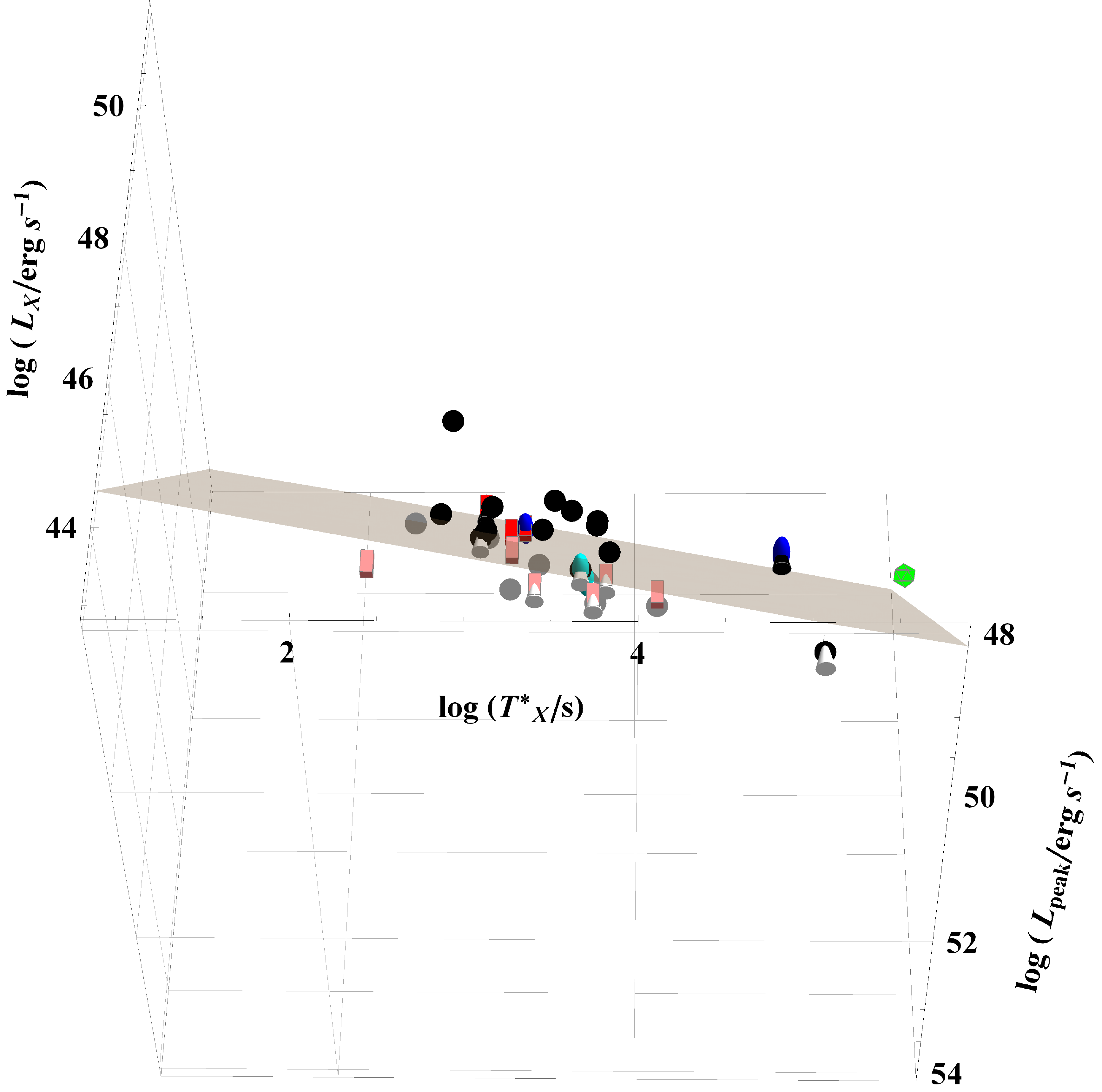}
    \includegraphics[scale=0.23]{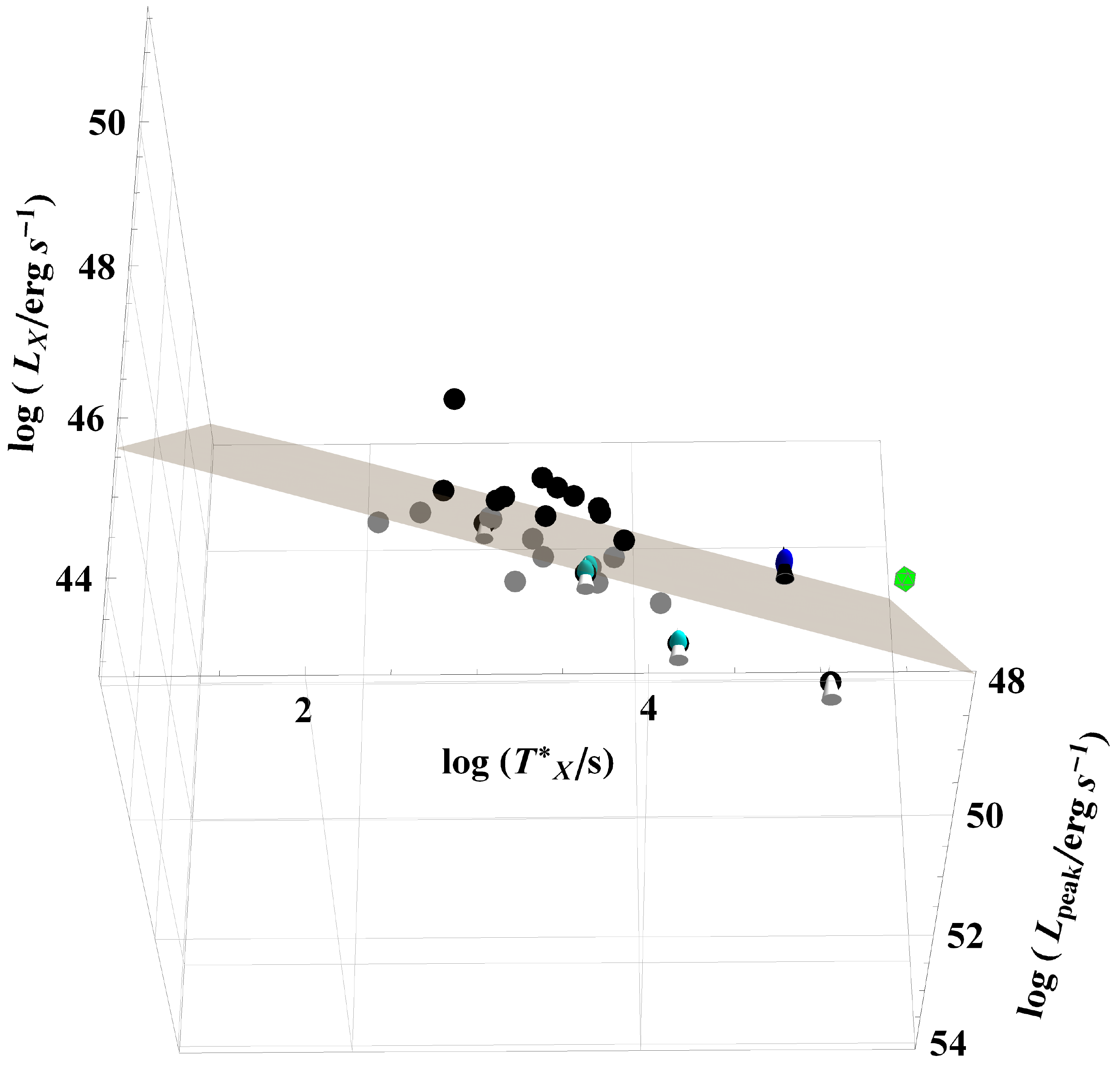}
    \includegraphics[scale=0.23]{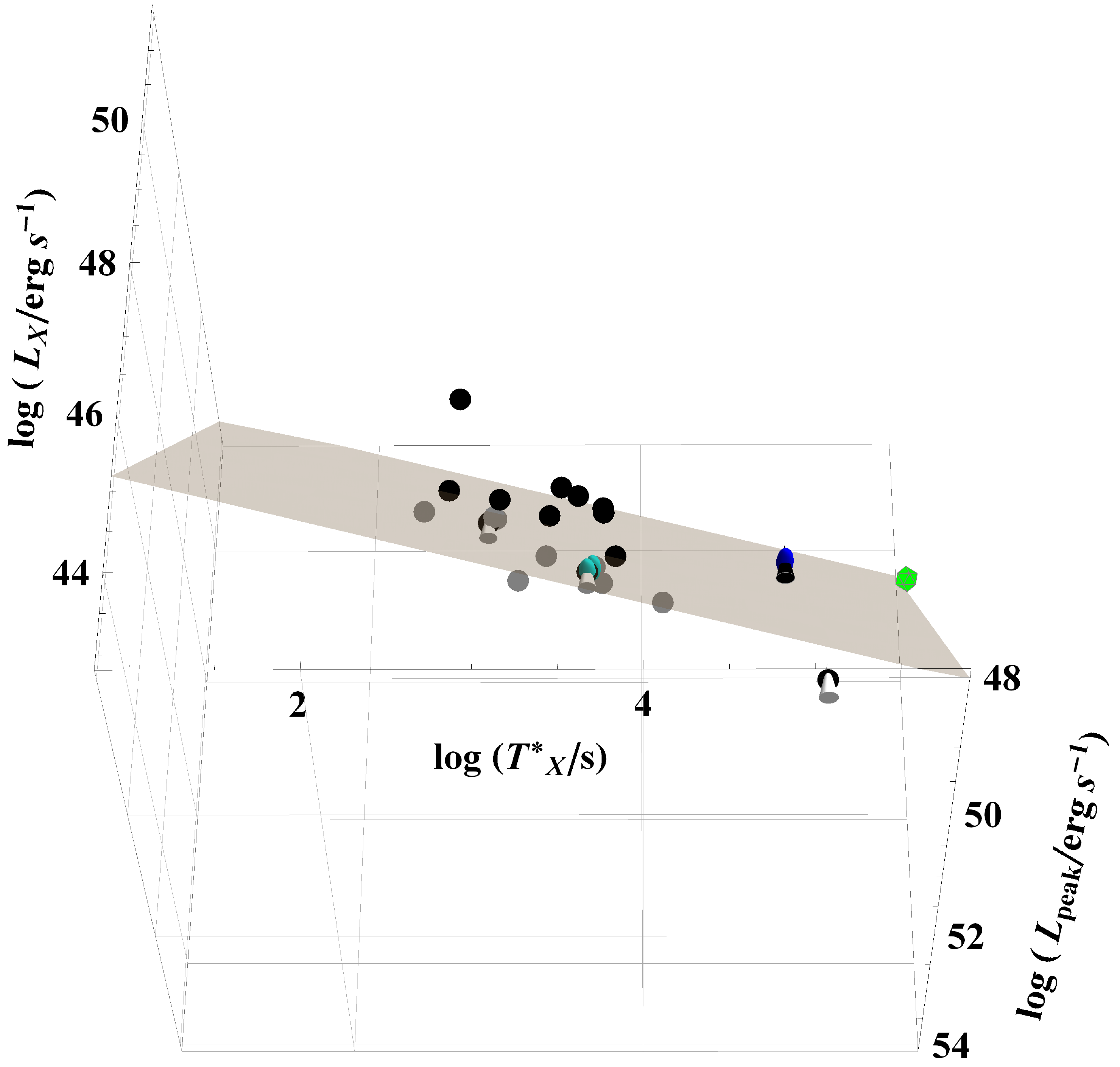}
    \includegraphics[scale=0.23]{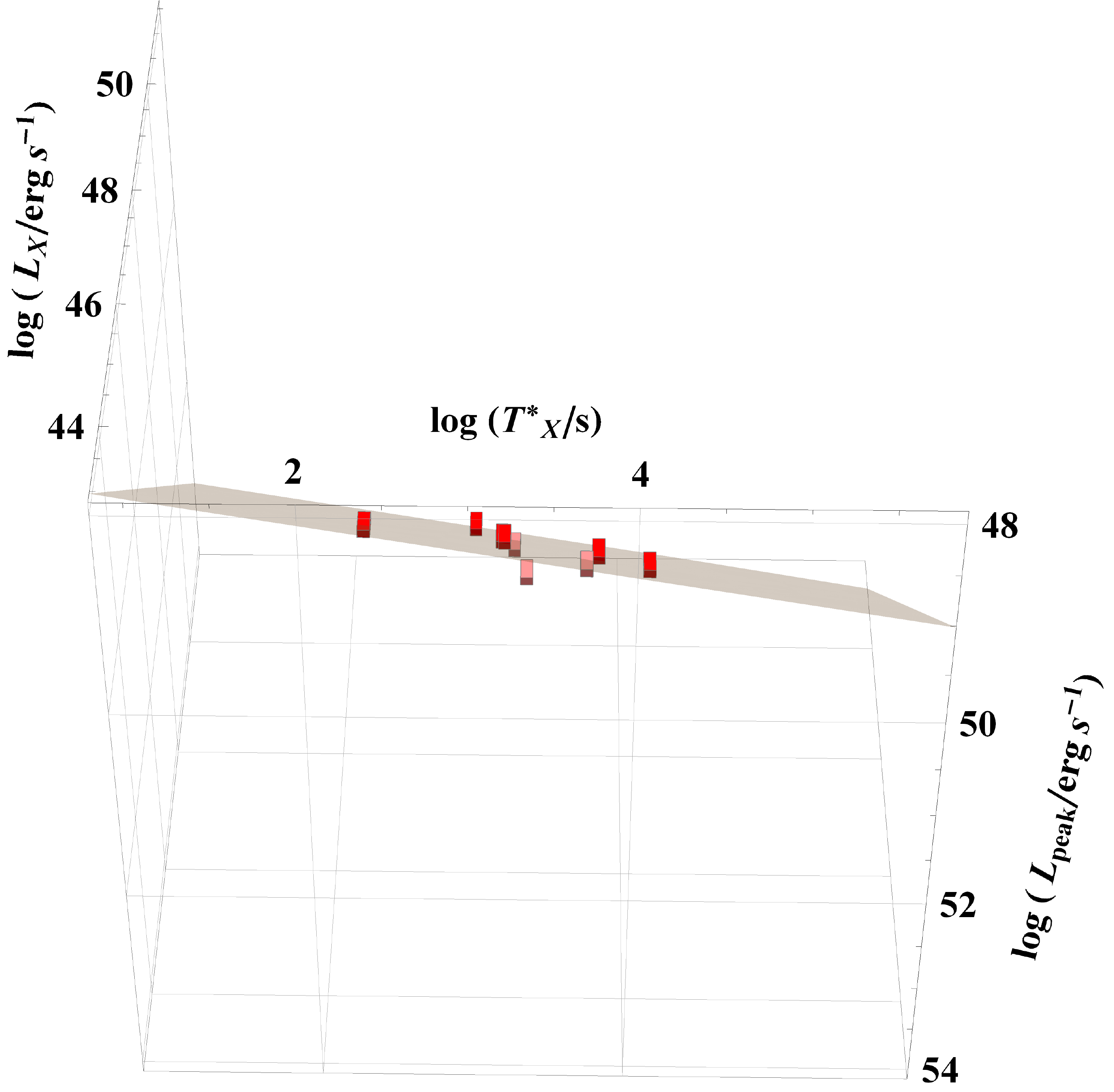}
    \includegraphics[scale=0.23]{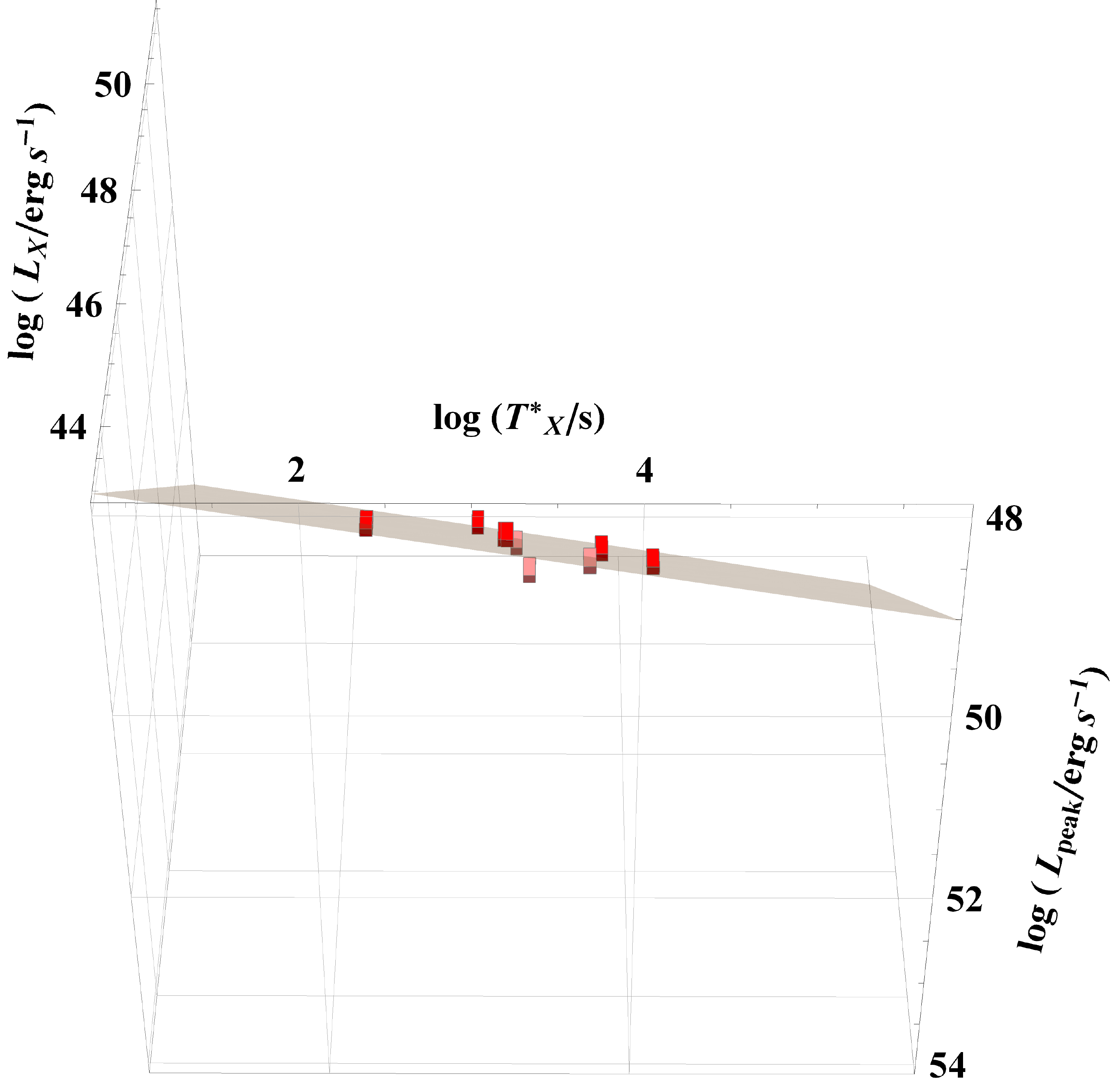}
    \caption{{\bf As in Figure \ref{3D_AllISM_AllWIND_q0} (for $q=0$): GRBs are divided into ISM SC for all GRBs (upper left) and ISM FC (upper right); GRBs divided into ISM SC lGRBs (middle left) and ISM FC lGRBs (middle right); GRBs divided into ISM SC sGRBs (lower left) and ISM FC sGRBs (lower right).}}
    \label{3D_ISMsc_ISMfc_q0}
\end{figure}

\begin{figure}
    \centering
    \includegraphics[scale=0.25]{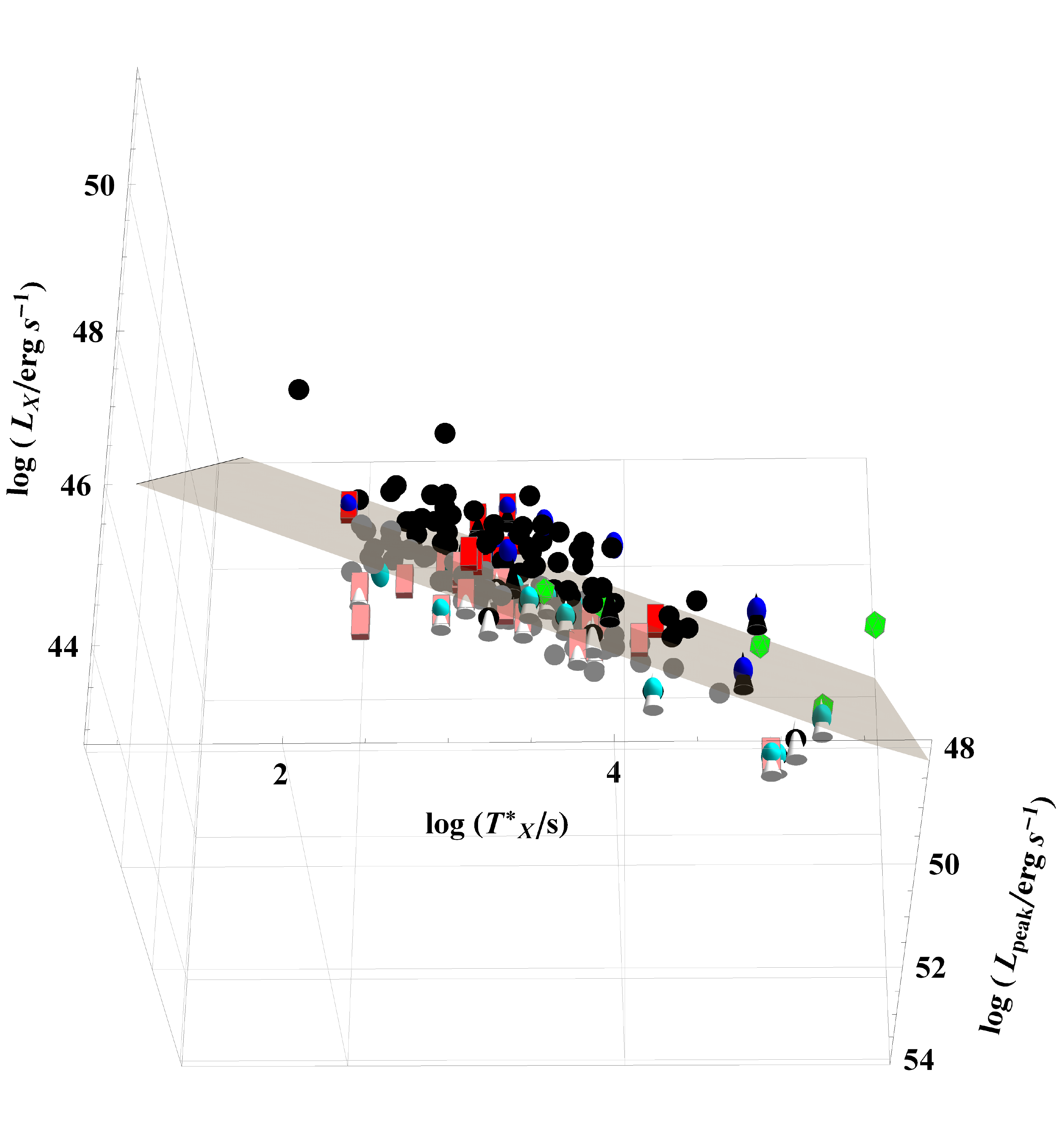}
    \includegraphics[scale=0.23]{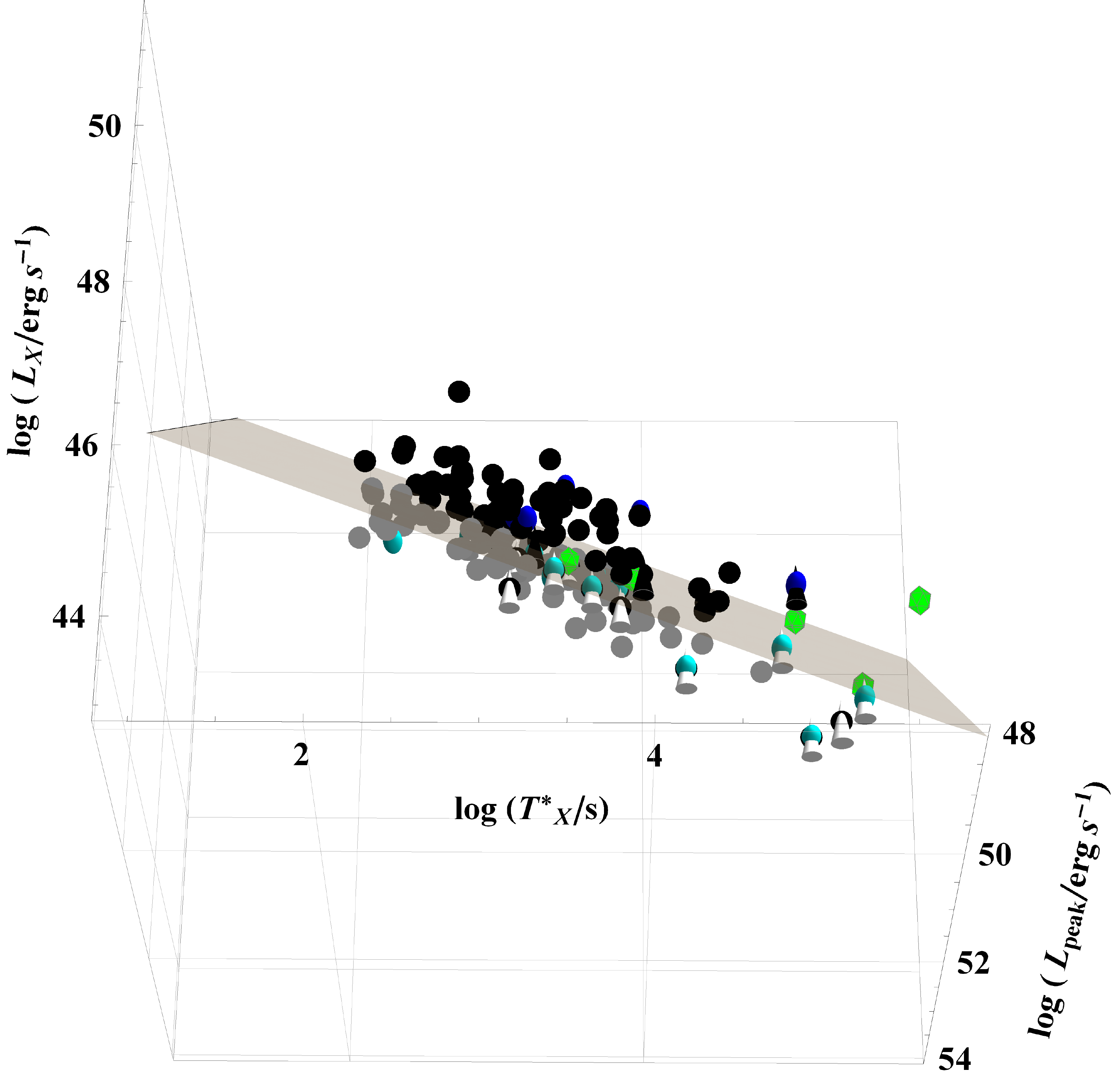}
    \includegraphics[scale=0.245]{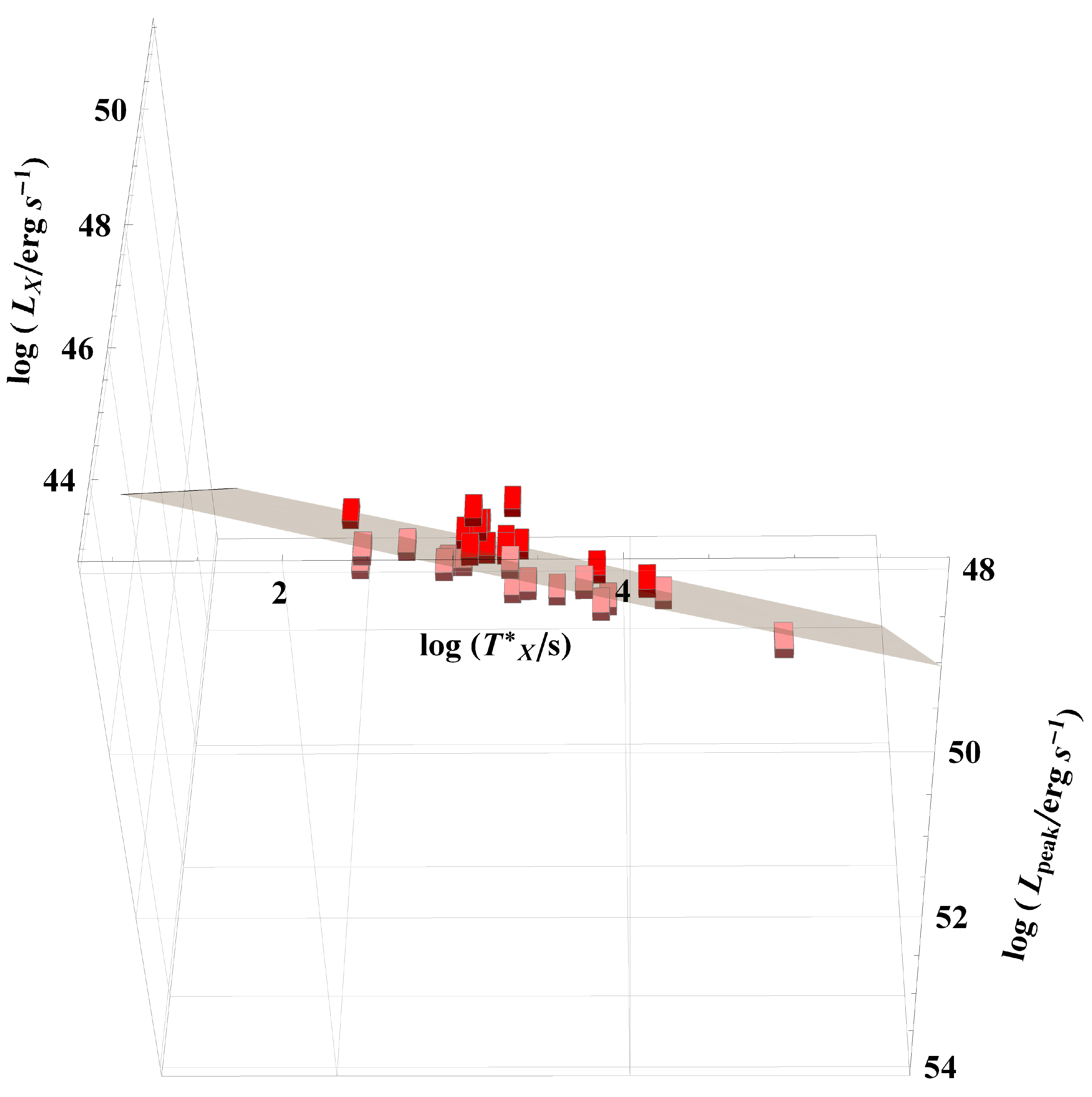}
    \caption{\bf{As in Figure \ref{3D_AllISM_AllWIND_q0} (for $q=0$), considering the Wind SC environment for (in the order): All, lGRBs and sGRBs.}} 
    \label{3D_WINDsc_WINDfc_q0}
\end{figure}

\begin{figure}
    \centering
    \includegraphics[scale=0.21]{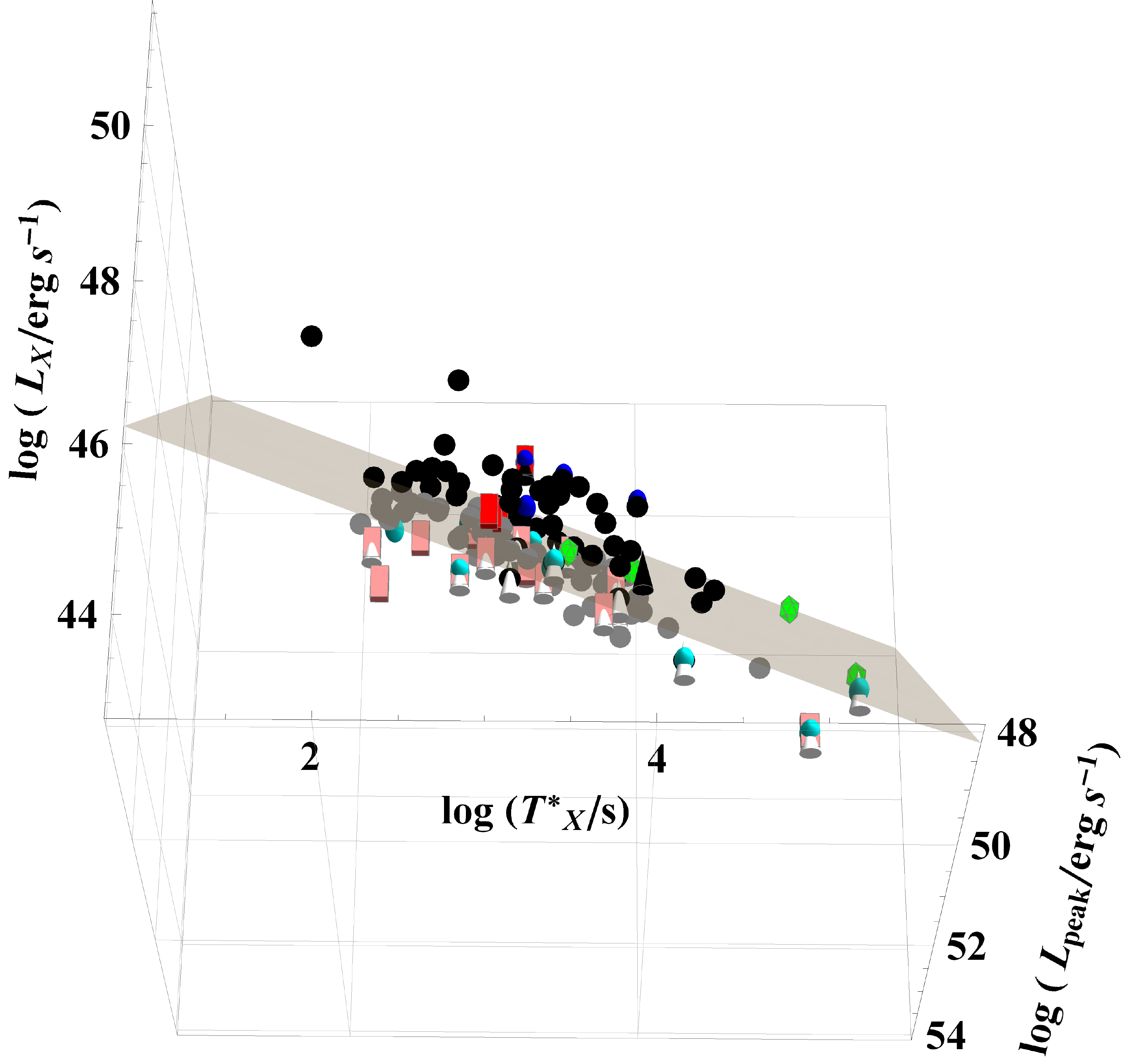}
    \includegraphics[scale=0.21]{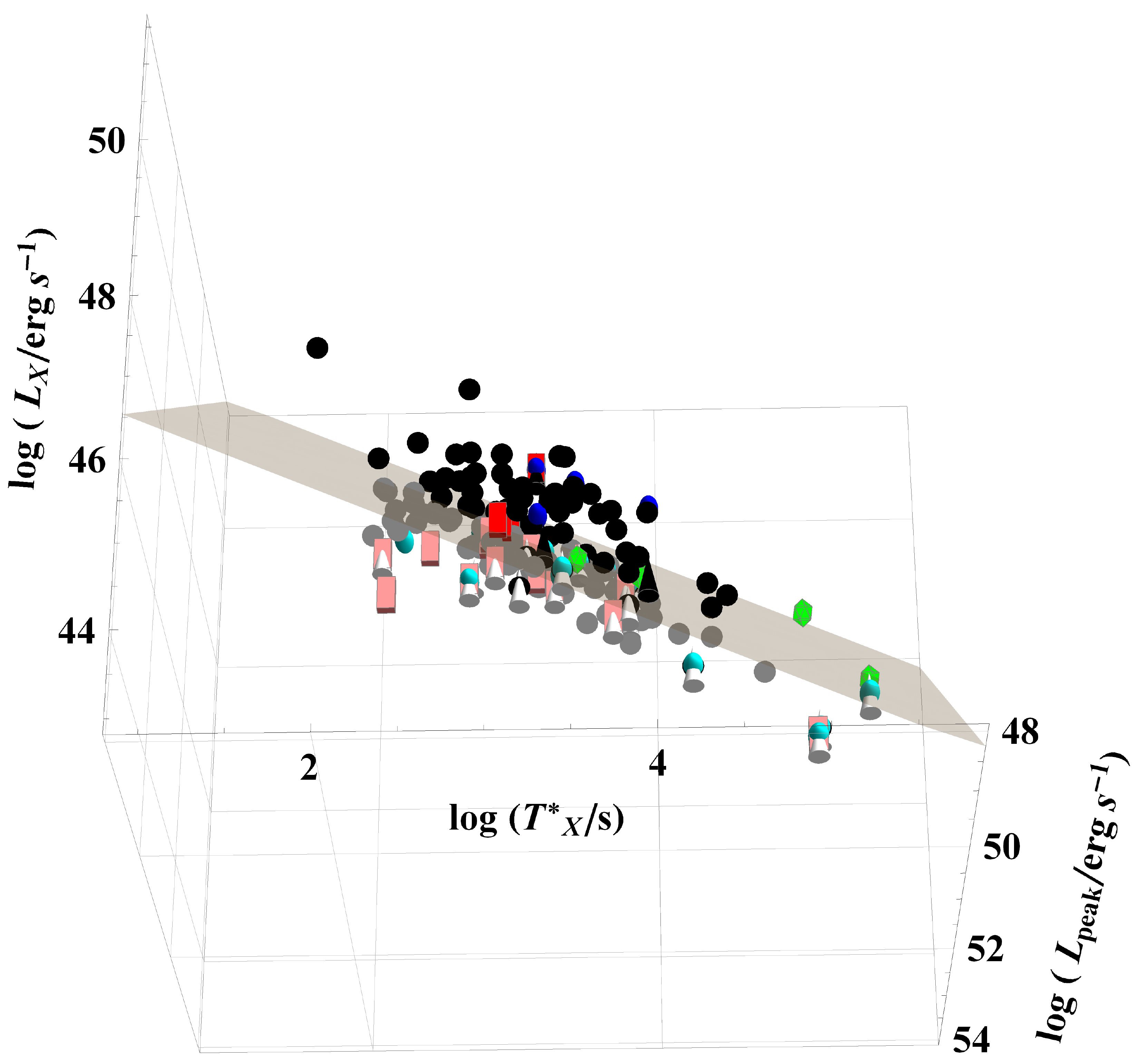}
    \includegraphics[scale=0.22]{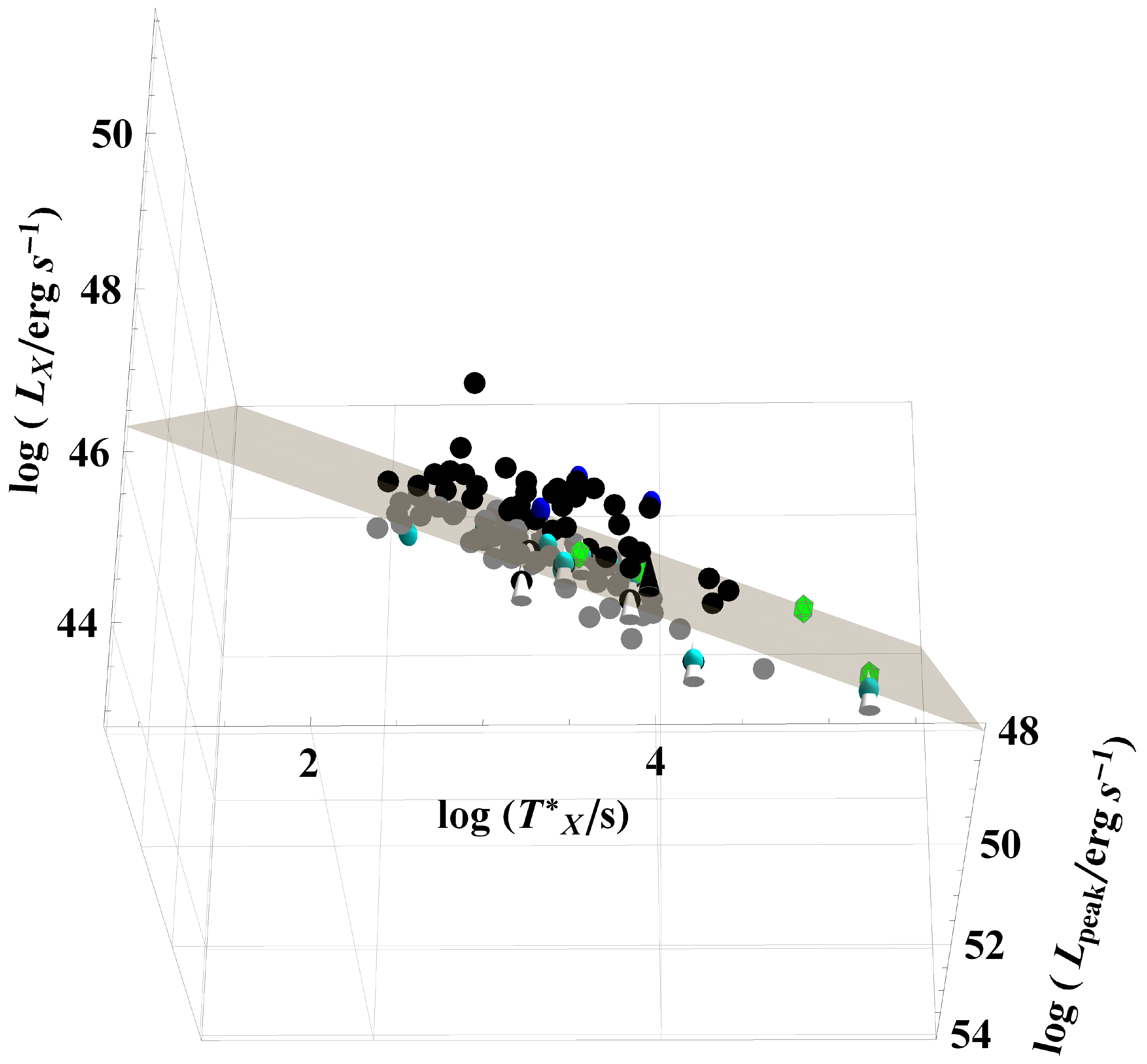}
    \includegraphics[scale=0.22]{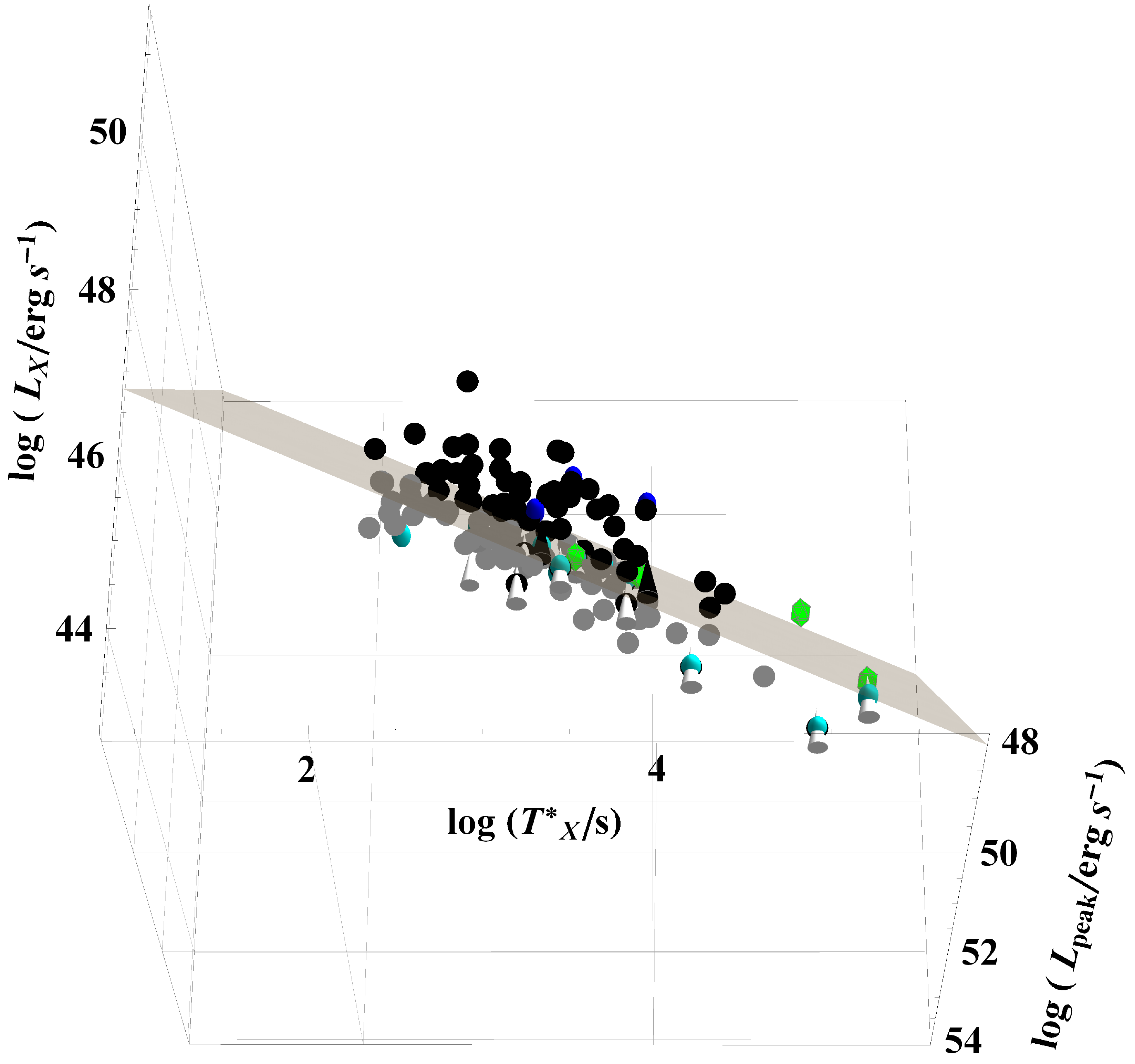}
    \includegraphics[scale=0.22]{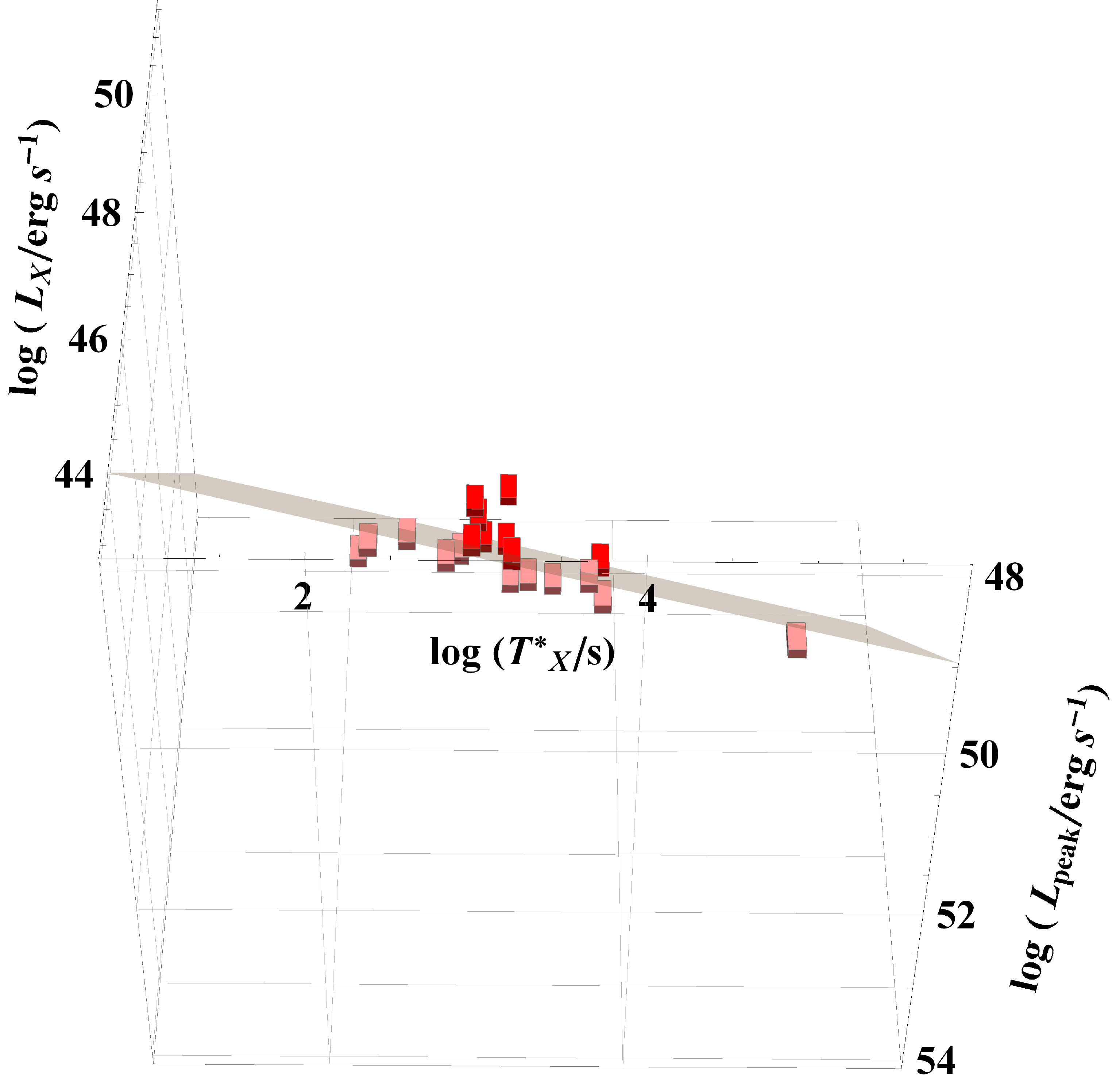}
    \includegraphics[scale=0.22]{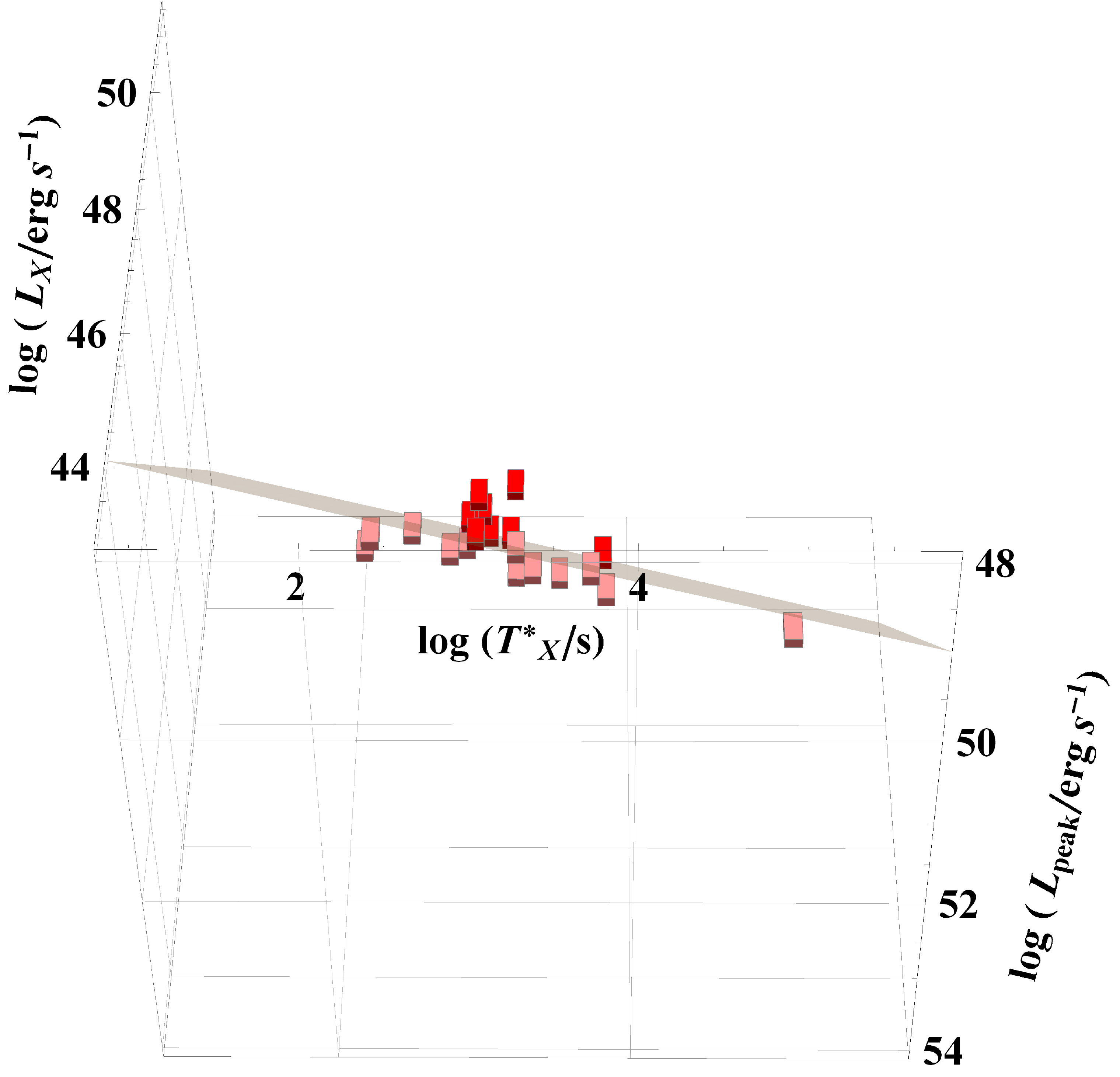}
    \caption{\bf{The GRBs divided into All ISM for all GRBs (upper left panel), All Wind for all GRBs (upper right panel), All ISM for lGRBs (middle left panel), All Wind for lGRBs (middle right panel), All ISM for sGRBs (lower left panel), and All Wind for sGRBs (lower right panel), see Table \ref{Dagostini_q0.5}}. These are placed on the fundamental plane defined by Eq. \ref{planeequation1}. The color-coded and symbol-coded are the same as Figure \ref{3D_AllISM_AllWIND_q0}. Here we consider $q=0.5$.}
    \label{3D_AllISM_AllWIND_q0.5}
\end{figure}

\begin{figure}
    \centering
    \includegraphics[scale=0.22]{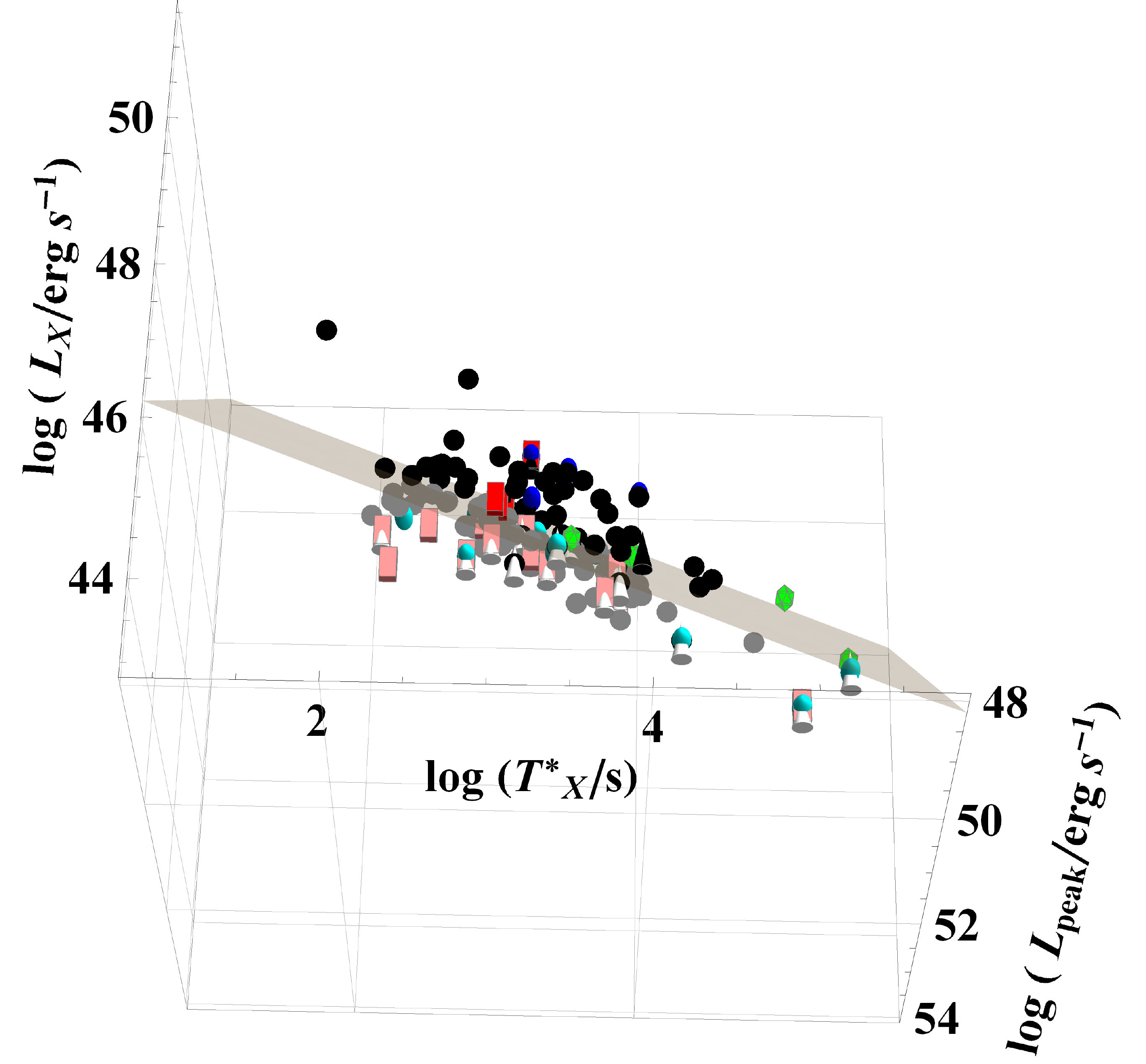}
    \includegraphics[scale=0.22]{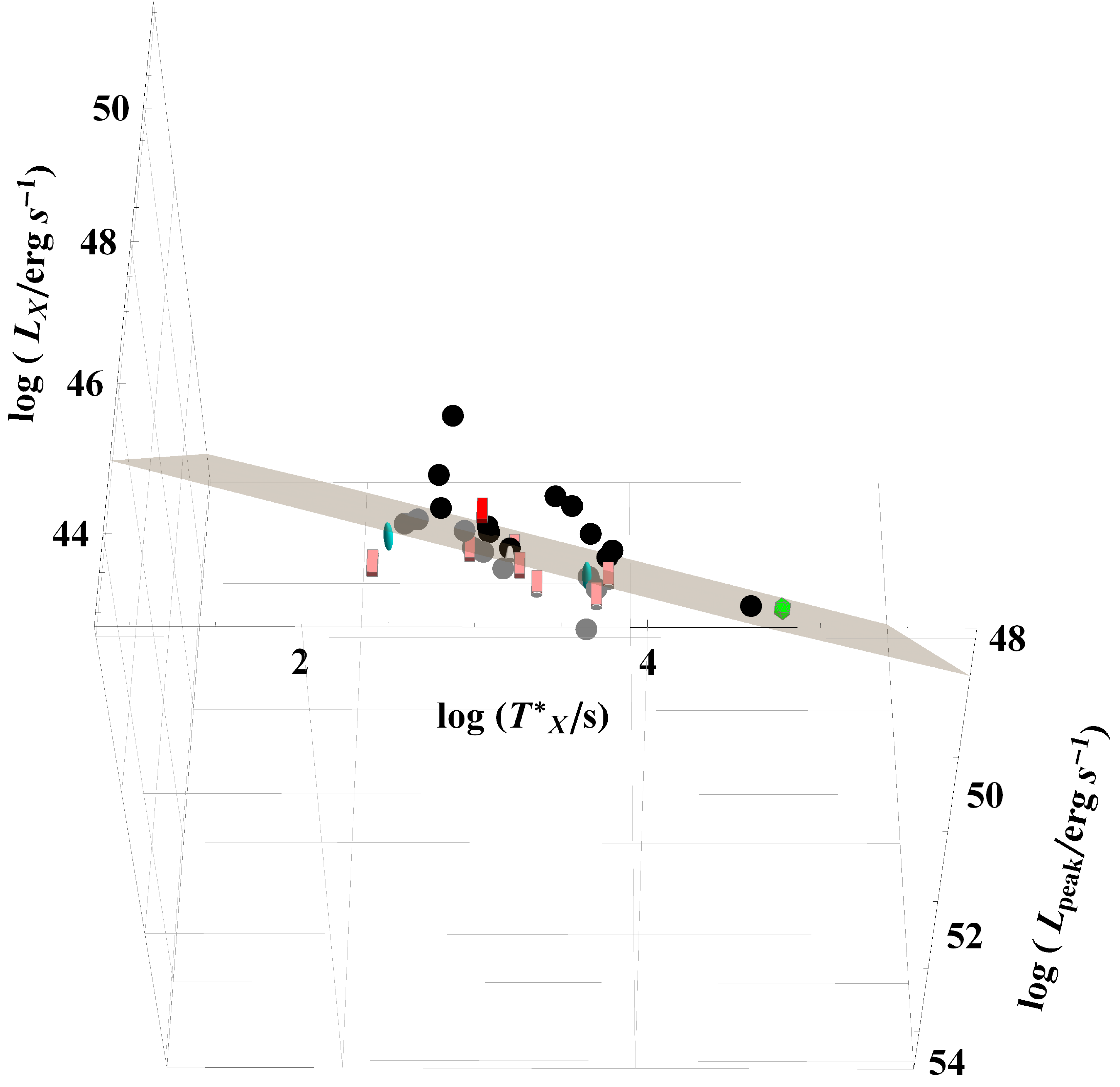}
    \includegraphics[scale=0.22]{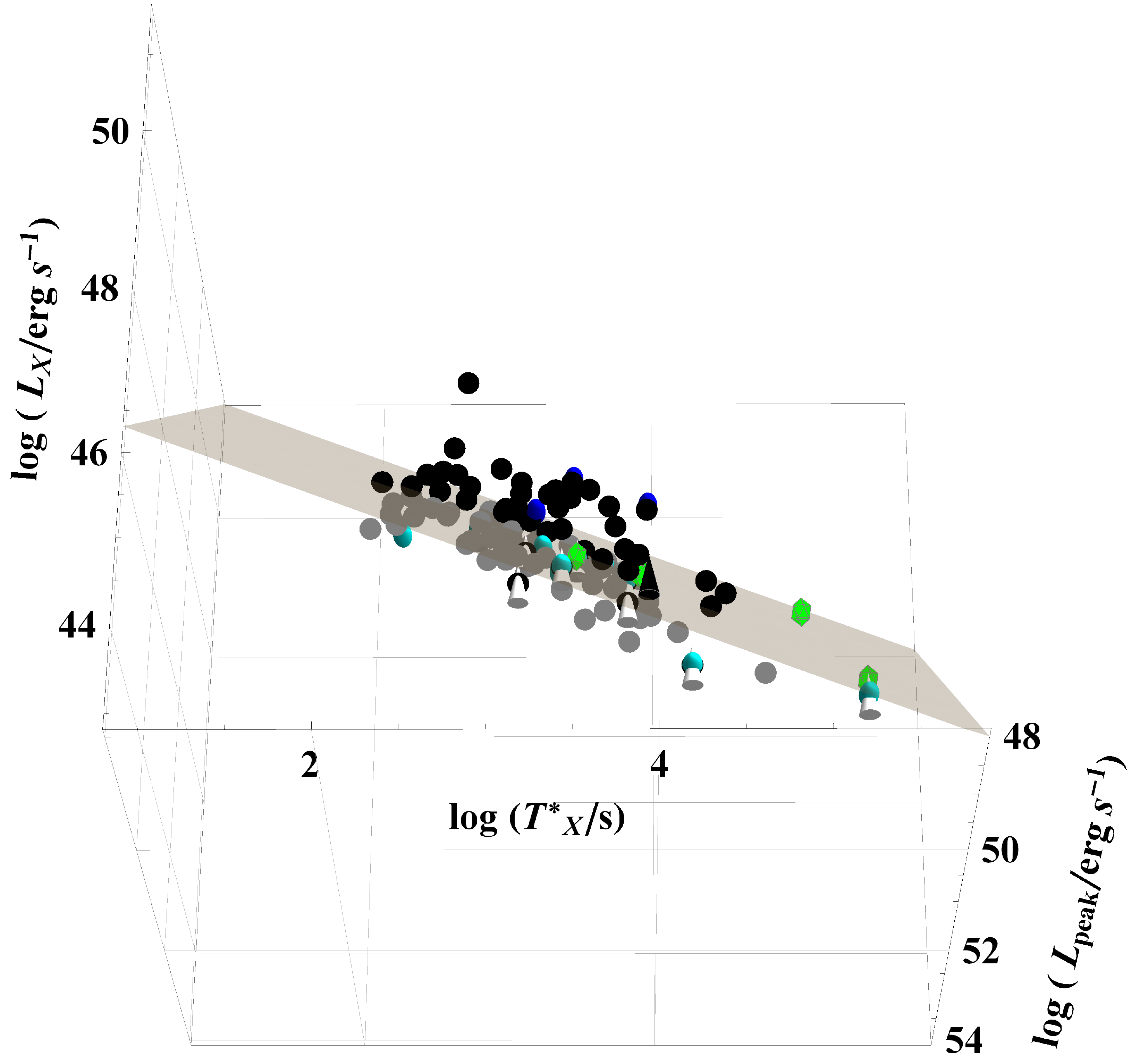}
    \includegraphics[scale=0.22]{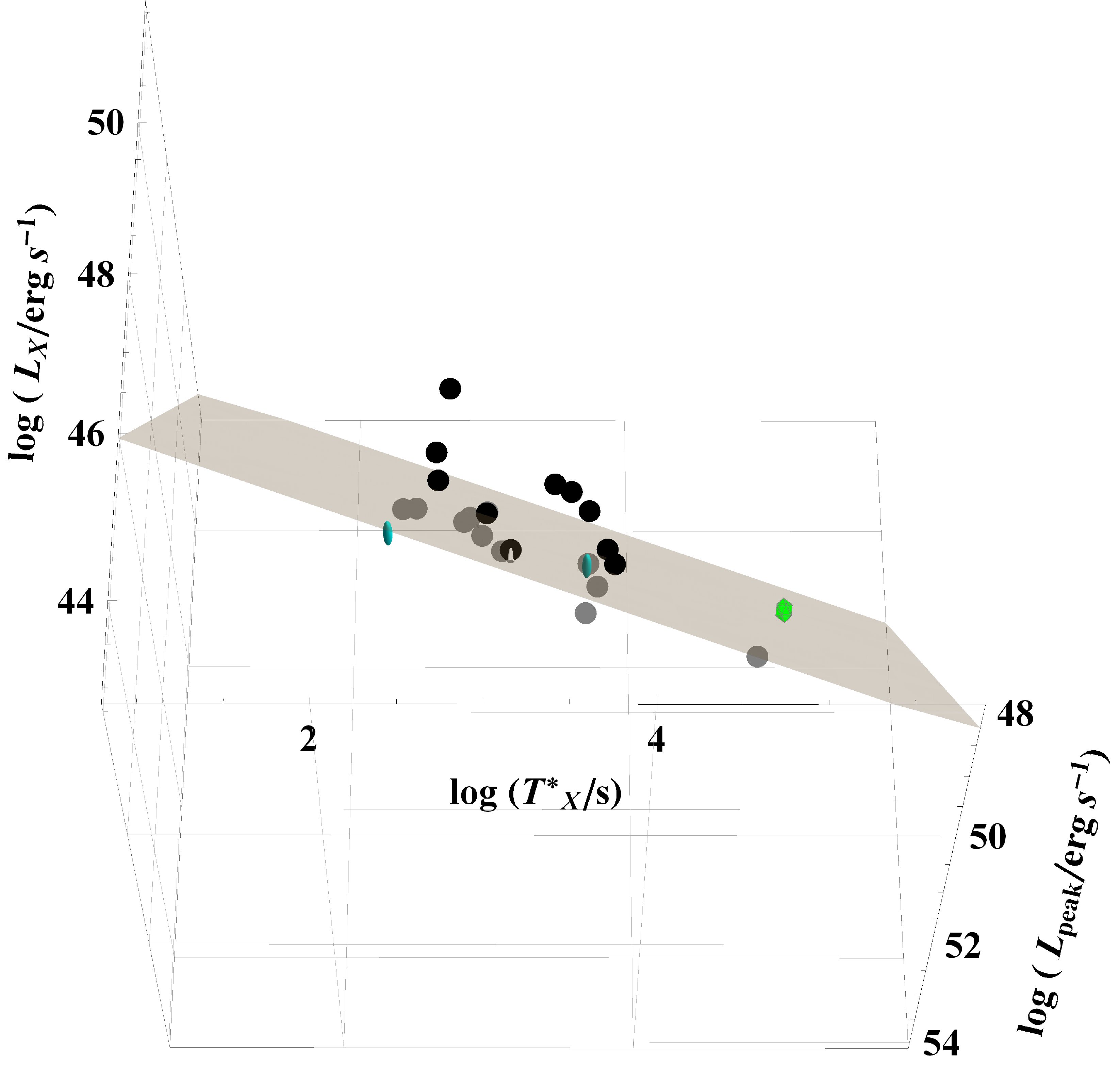}
    \includegraphics[scale=0.22]{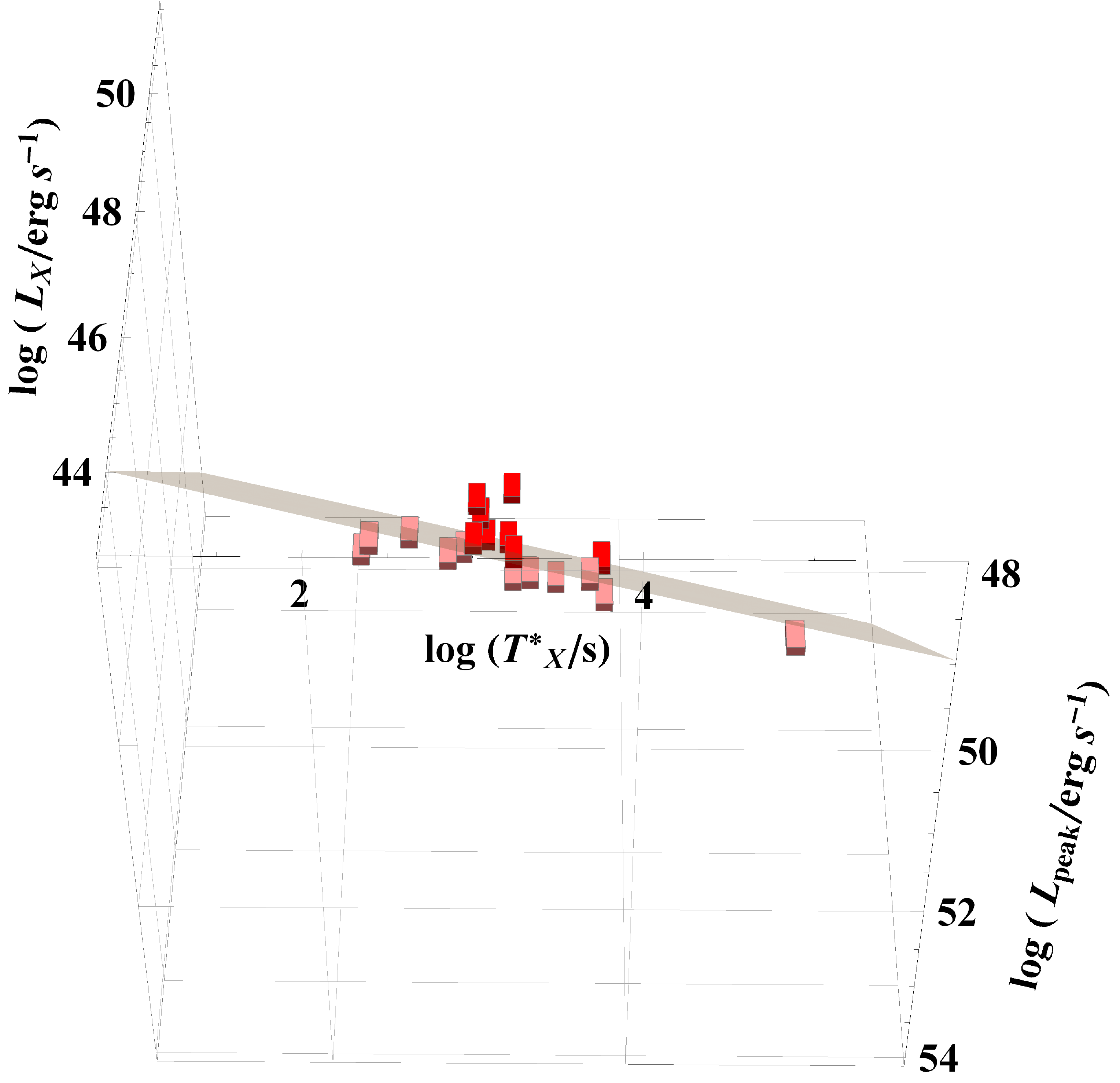}
    \includegraphics[scale=0.22]{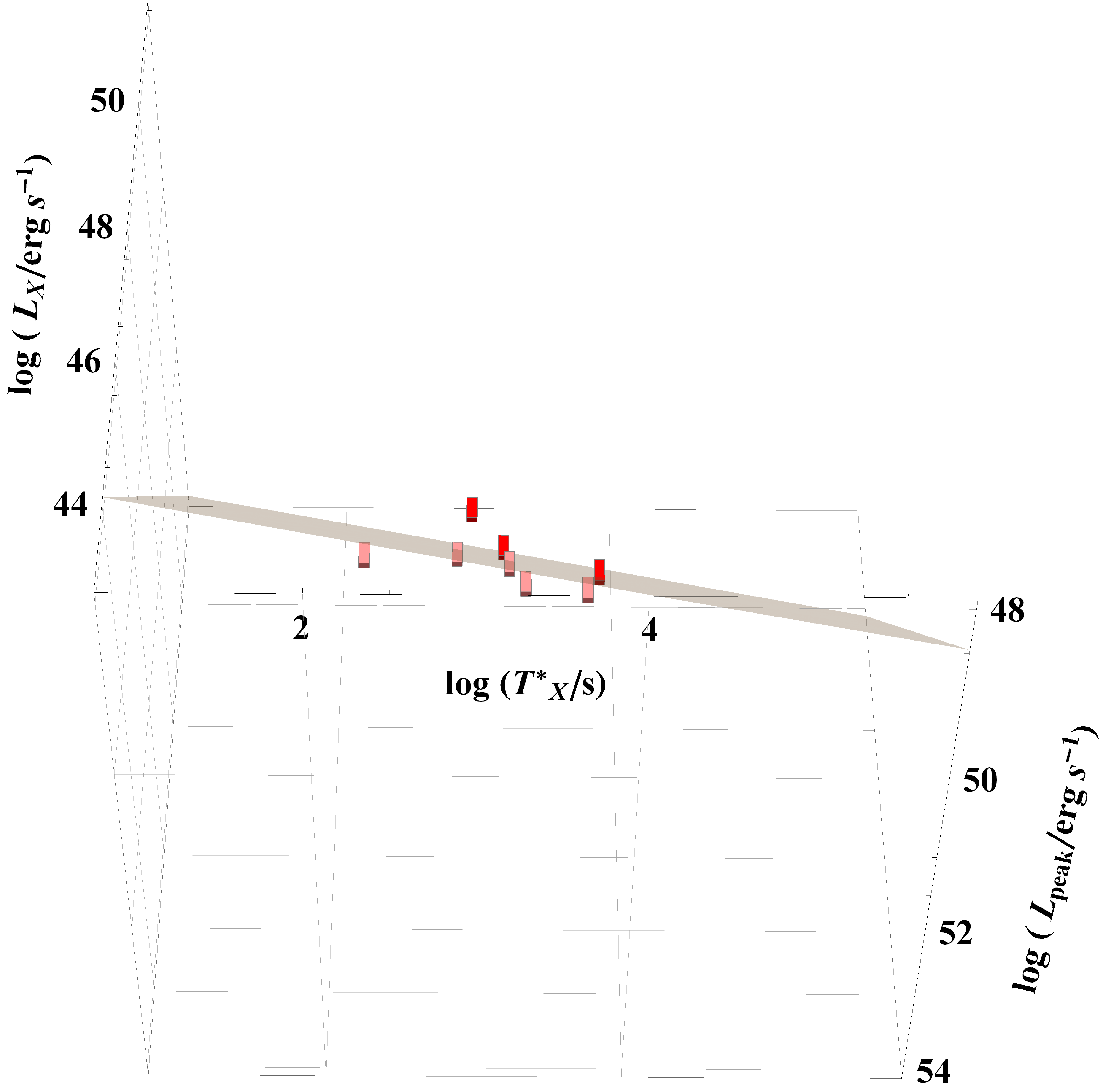}
    \caption{{\bf As in Figure \ref{3D_AllISM_AllWIND_q0.5} (for $q=0.5$) with GRBs divided into: ISM SC for all GRBs (upper left panel), ISM FC for all GRBs (lower panel), 
    ISM SC for lGRBs (middle left panel), ISM FC for lGRBs (middle right panel), 
    ISM SC for sGRBs (lower left panel), and ISM FC for sGRBs (lower right panel).}}
    \label{3D_ISMsc_ISMfc_q0.5}
\end{figure}

\begin{figure}
    \centering
    \includegraphics[scale=0.22]{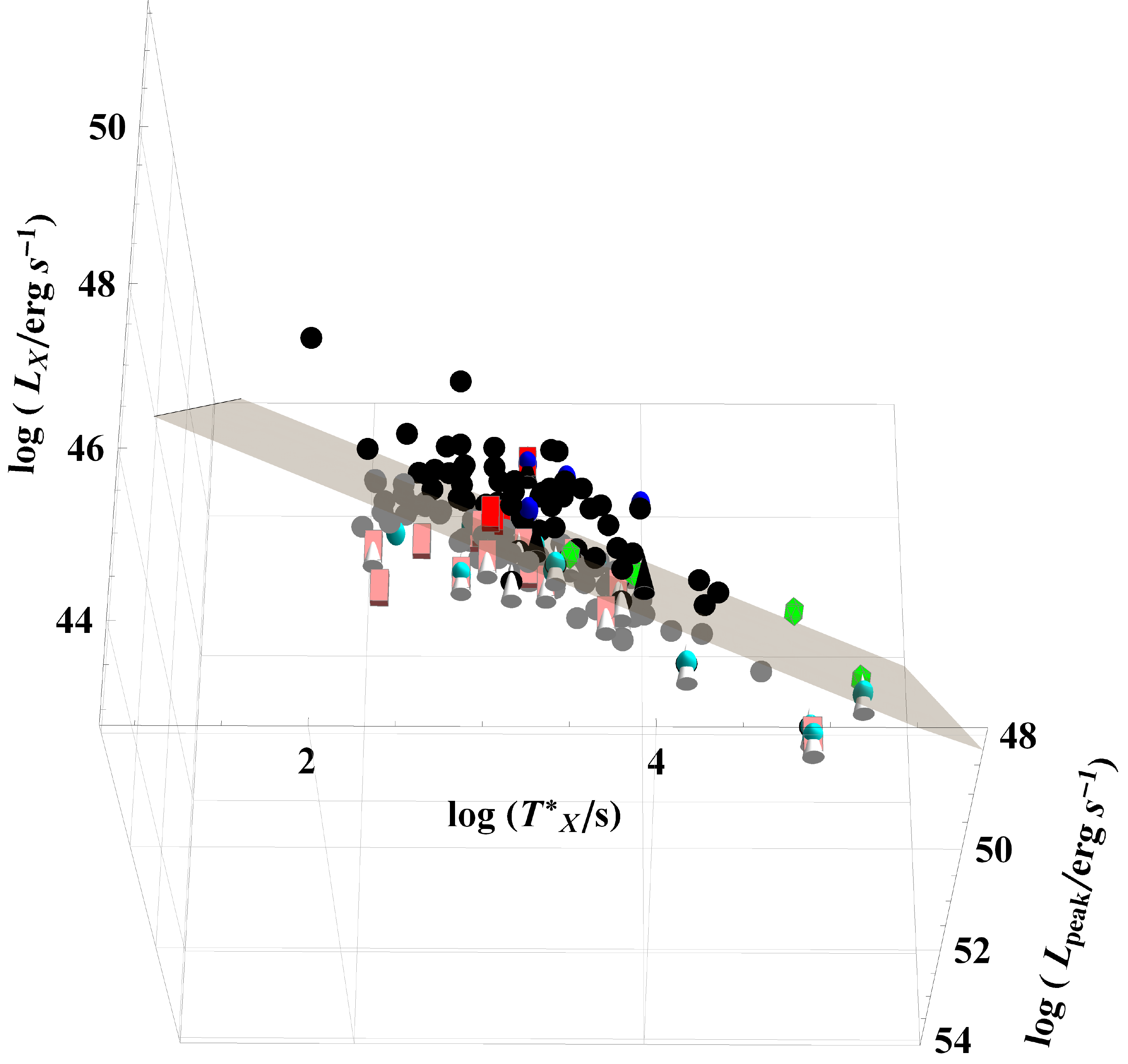}
    \includegraphics[scale=0.22]{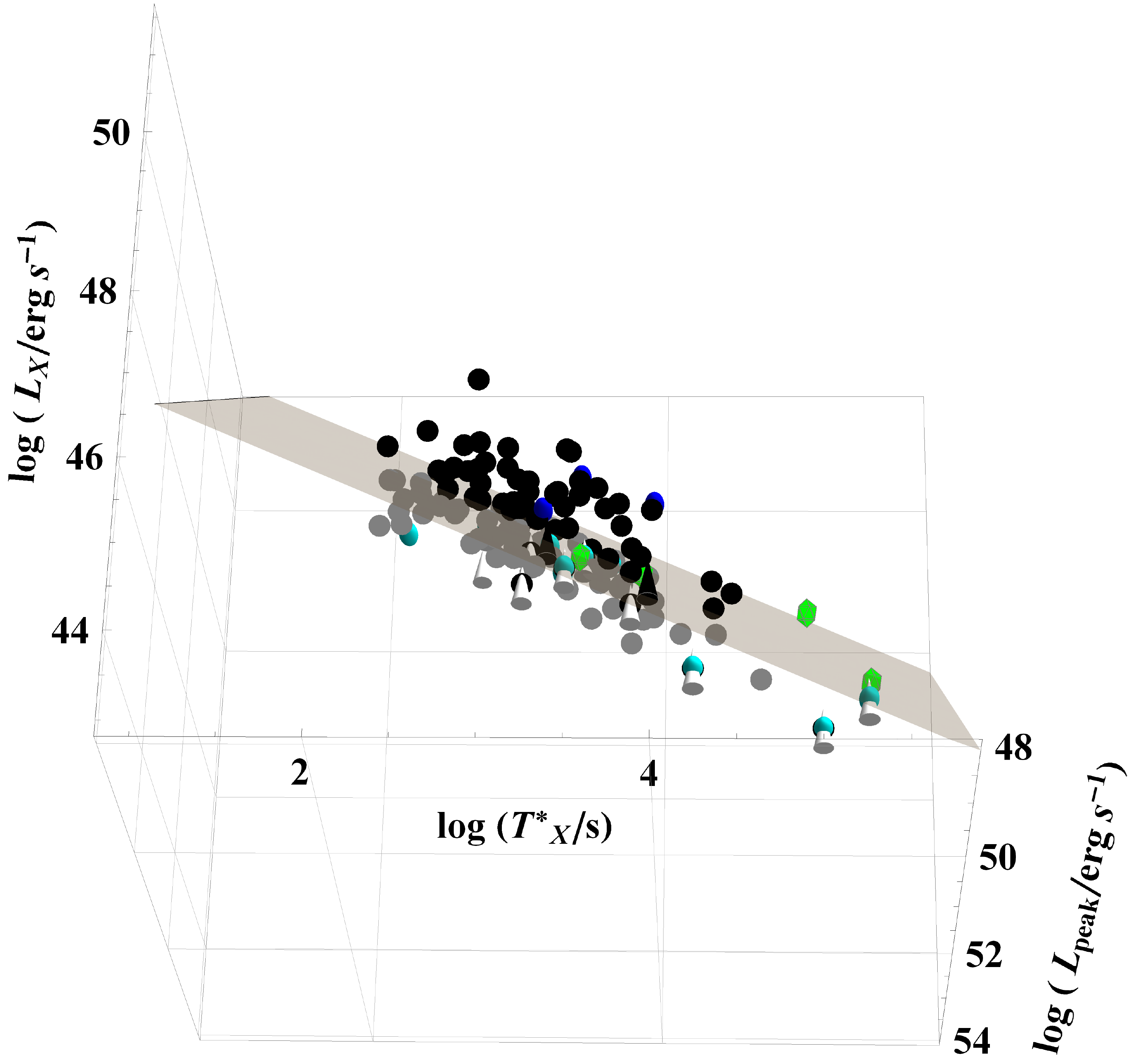}
    \includegraphics[scale=0.23]{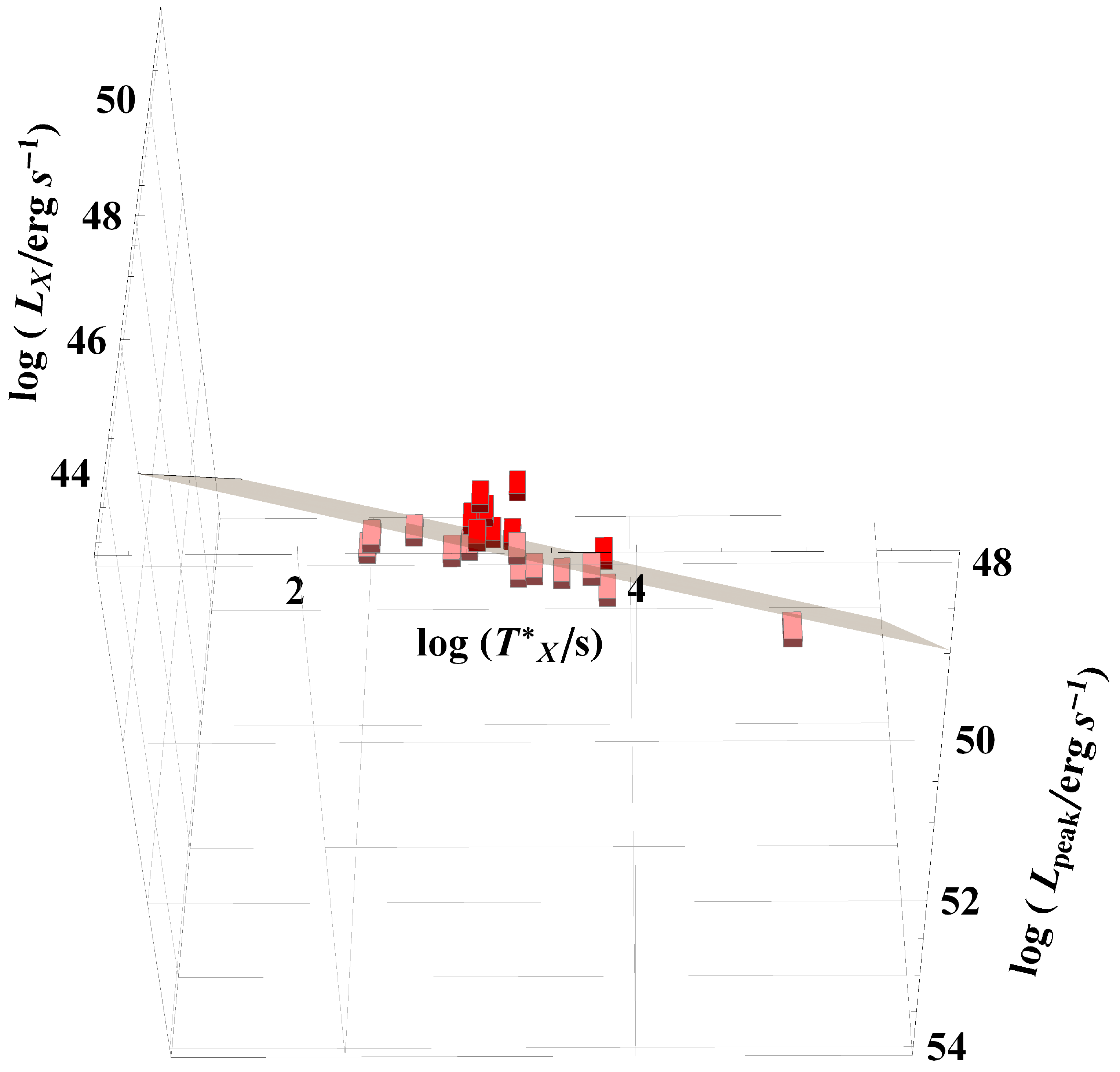}
    \caption{\bf{As in Figure \ref{3D_AllISM_AllWIND_q0.5} (for $q=0.5$), but with GRBs belonging to Wind SC environment considering the cases (in the order): All, lGRBs and sGRBs. The cases of Wind FC are based on few scattered data points thus their plotting will be omitted.}}
    \label{3D_WINDsc_WINDfc_q0.5}
\end{figure}

\begin{figure}
\centering
\includegraphics[width=0.4\columnwidth]{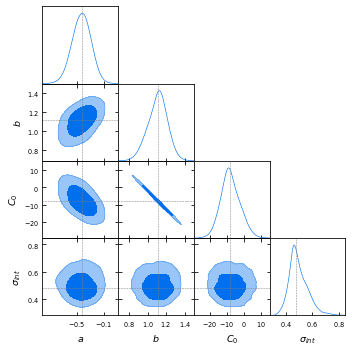}
\includegraphics[width=0.4\columnwidth]{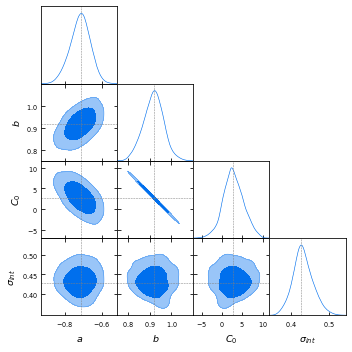}
\includegraphics[width=0.4\columnwidth]{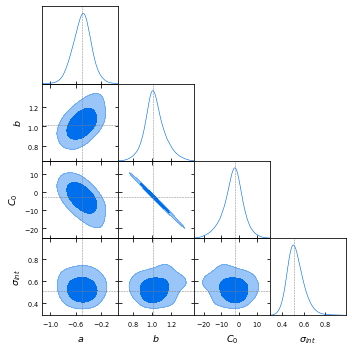}
\includegraphics[width=0.4\columnwidth]{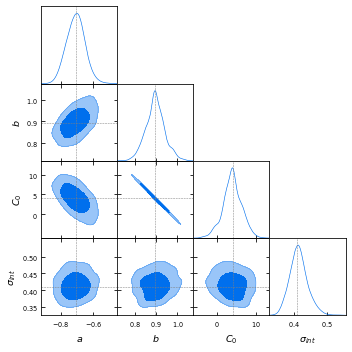}
\includegraphics[width=0.4\columnwidth]{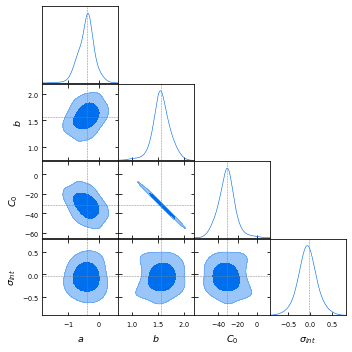}
\includegraphics[width=0.4\columnwidth]{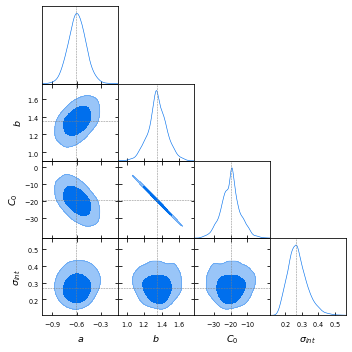}
\caption{\bf{Contour plots for GRB sub-classes taken from Table \ref{Dagostini_q0} $(q=0)$. In the order: All ISM (lGRBs and sGRBs), All Wind (lGRBs and sGRBs), All ISM (lGRBs), All Wind (lGRBs), All ISM (sGRBs), and All WIND (sGRBs). In these contour plots and in the following the dark blue contours indicate the $68\%$ confidence levels, the light blue ones correspond to $95\%$ confidence levels and the dashed lines mark the median values of the distributions.}} \label{contourgroups_q0}
\end{figure}

\begin{figure}
\centering
\includegraphics[scale=0.57]{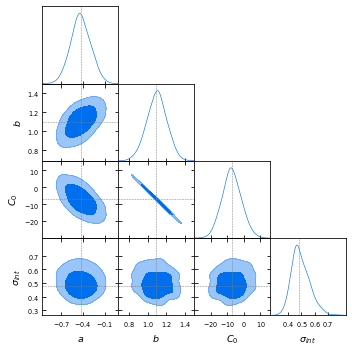}
\includegraphics[scale=0.57]{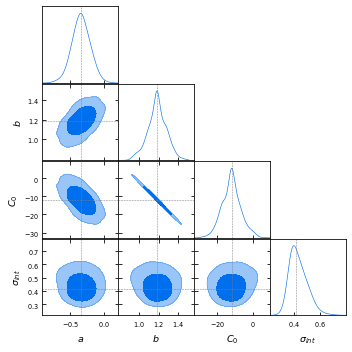}
\includegraphics[scale=0.57]{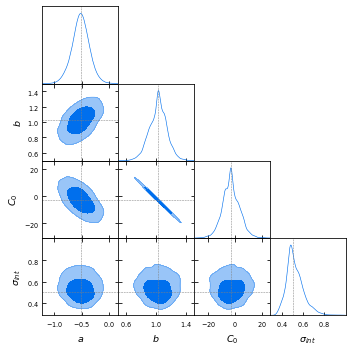}
\includegraphics[scale=0.57]{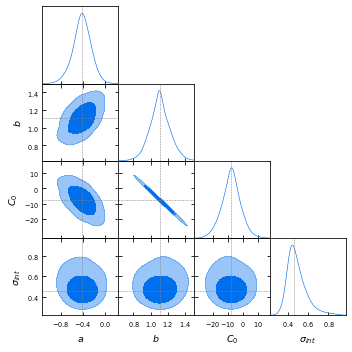}
\includegraphics[scale=0.57]{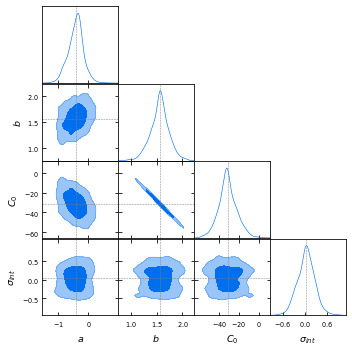}
\includegraphics[scale=0.57]{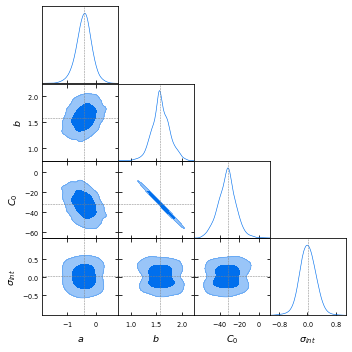}
\caption{\bf{Contour plots for groups taken from Table \ref{Dagostini_q0} $(q=0)$. In the order: ISM SC (all GRBs), ISM FC (all GRBs), ISM SC (lGRBs), ISM FC (lGRBs), ISM SC (sGRBs), and ISM FC (sGRBs).}} \label{contoursISM_q0}

\end{figure}

\begin{figure}

\centering

\includegraphics[width=0.45\columnwidth]{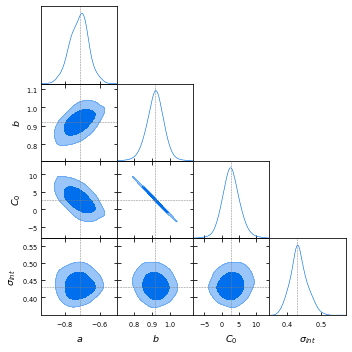}
\includegraphics[width=0.45\columnwidth]{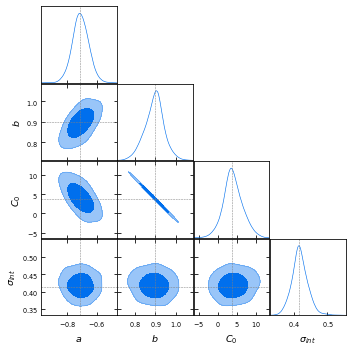}
\includegraphics[width=0.45\columnwidth]{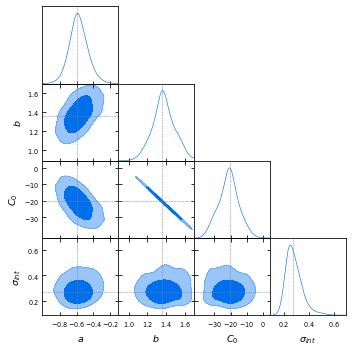}
\caption{\bf{Contour plots for sub-classes taken from Table \ref{Dagostini_q0} $(q=0)$, where the Wind SC is shown. In the order: All, lGRBs and sGRBs cases.}} \label{contoursWIND_q0}

\end{figure}

\begin{figure}

\centering

\includegraphics[width=0.42\columnwidth]{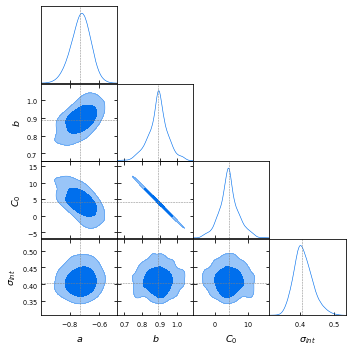}
\includegraphics[width=0.42\columnwidth]{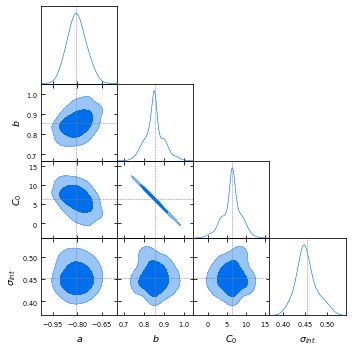}
\includegraphics[width=0.42\columnwidth]{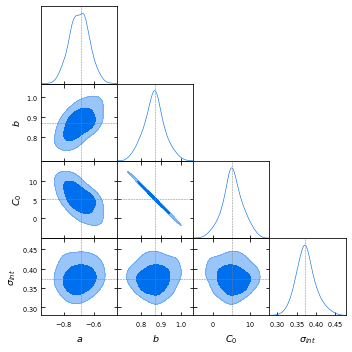}
\includegraphics[width=0.42\columnwidth]{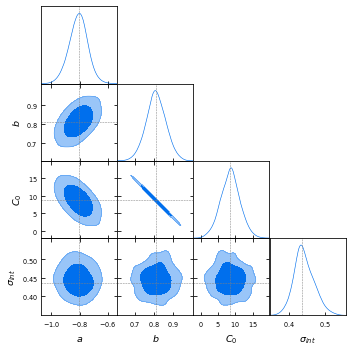}
\includegraphics[width=0.42\columnwidth]{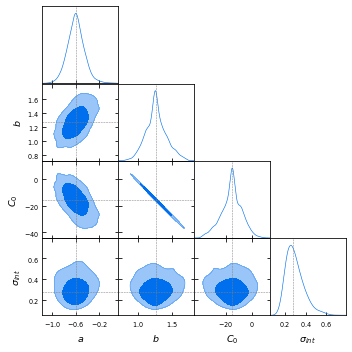}
\includegraphics[width=0.42\columnwidth]{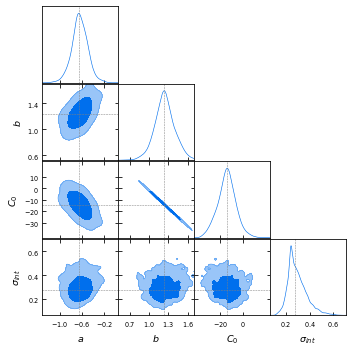}
\caption{\bf{Contour plots for groups taken from Table \ref{Dagostini_q0.5} $(q=0.5)$. In the order: All ISM (for all GRBs), All Wind for all GRBs, ISM for lGRBs, Wind for lGRBs, ISM for sGRBs, and WIND  for sGRBs.}} \label{contourgroups_q0.5}

\end{figure}

\begin{figure}

\centering

\includegraphics[scale=0.55]{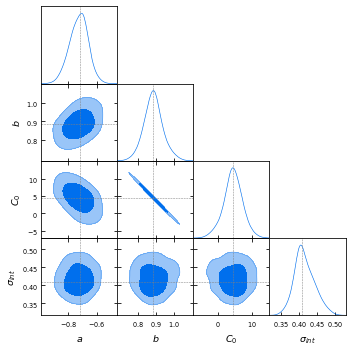}
\includegraphics[scale=0.55]{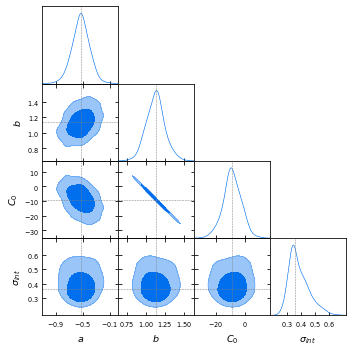}
\includegraphics[scale=0.55]{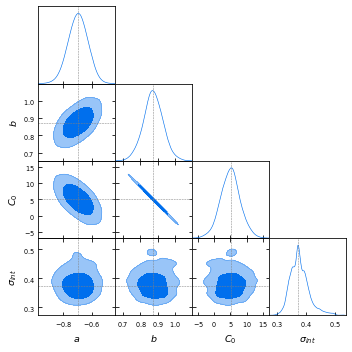}
\includegraphics[scale=0.55]{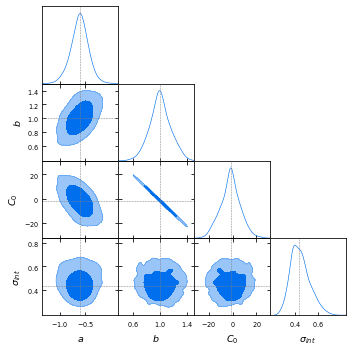}
\includegraphics[scale=0.55]{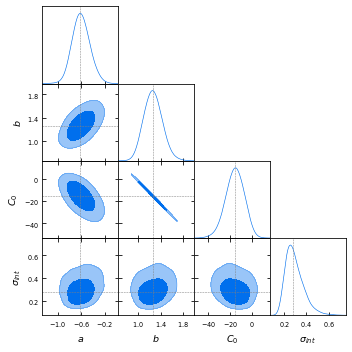}
\caption{{\bf Contour plots for groups taken from Table \ref{Dagostini_q0.5} $(q=0.5)$. In the order: ISM SC (lGRBs and sGRBs), ISM FC (lGRBs and sGRBs), ISM SC (lGRBs), ISM FC (lGRBs), and ISM SC (sGRBs).}} \label{contoursISM_q0.5}

\end{figure}

\begin{figure}

\centering

\includegraphics[width=0.43\columnwidth]{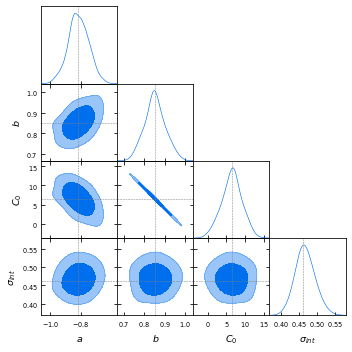}
\includegraphics[width=0.43\columnwidth]{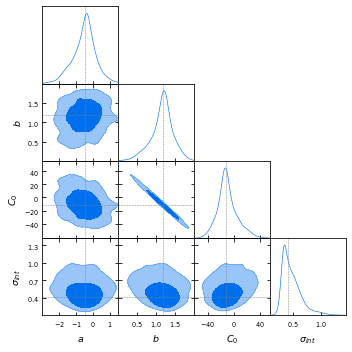}
\includegraphics[width=0.43\columnwidth]{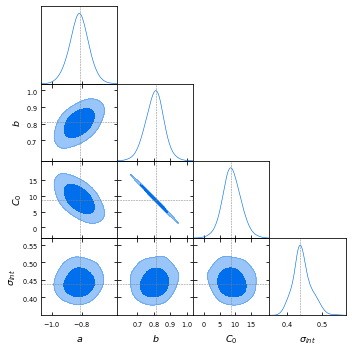}
\includegraphics[width=0.43\columnwidth]{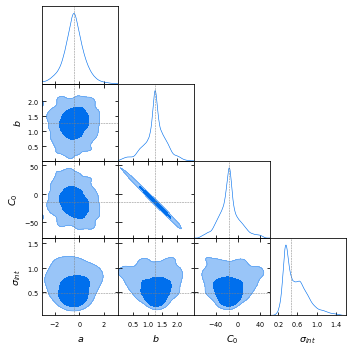}
\includegraphics[width=0.43\columnwidth]{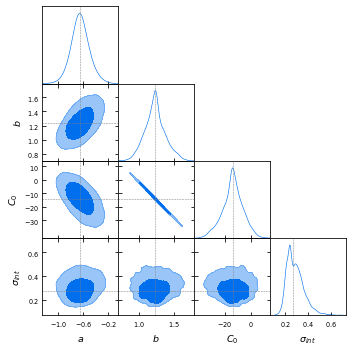}
\caption{{\bf Contour plots for sub-classes taken from Table \ref{Dagostini_q0.5} $(q=0.5)$ considering the Wind environment. In the order: WIND SC for all, WIND FC for all, WIND SC (for lGRBs), WIND FC (for lGRBs), and WIND SC (for sGRBs).}} \label{contoursWIND_q0.5}

\end{figure}

\begin{figure}

\centering

\includegraphics[scale=0.28]{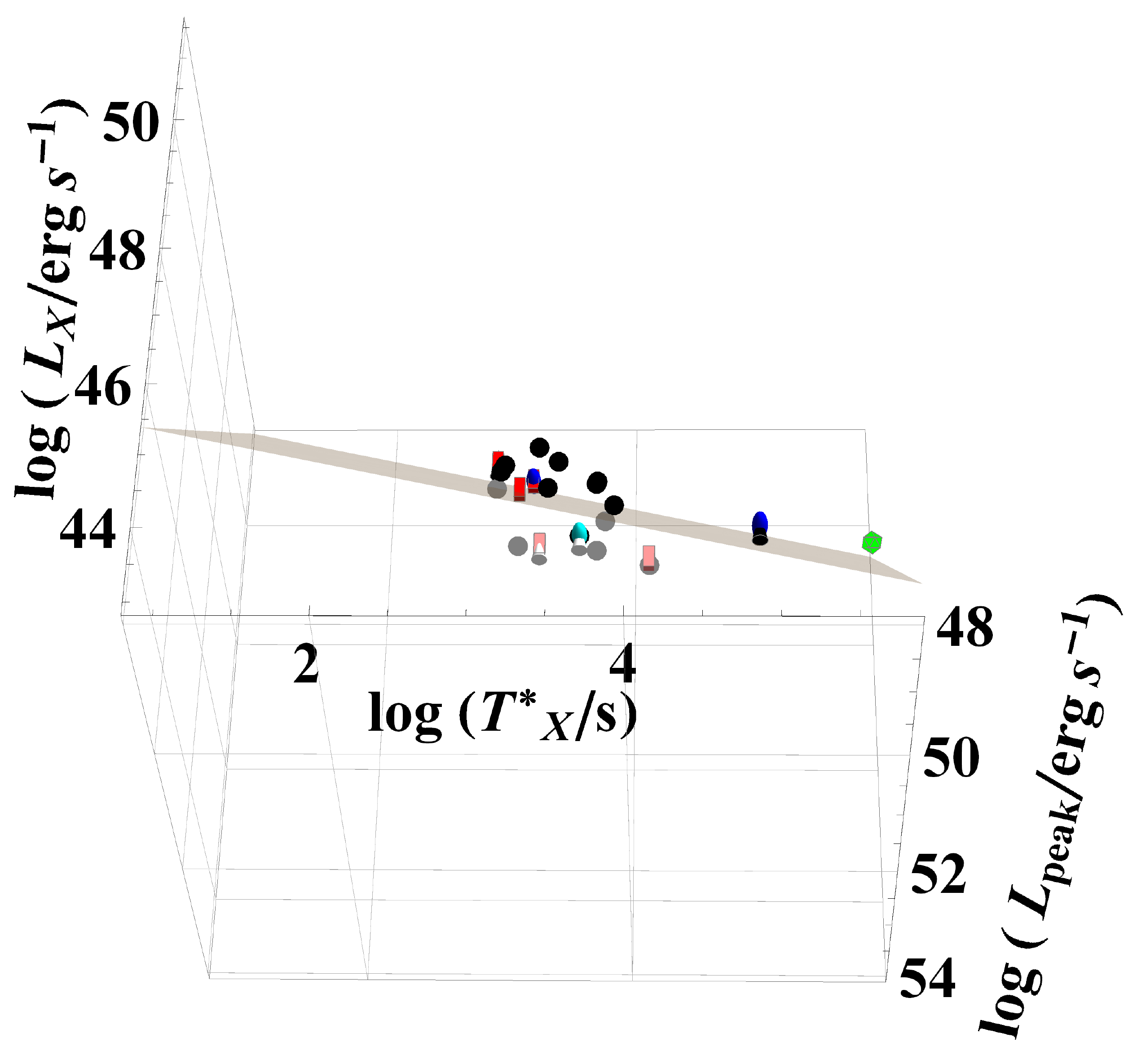}
\includegraphics[scale=0.21]{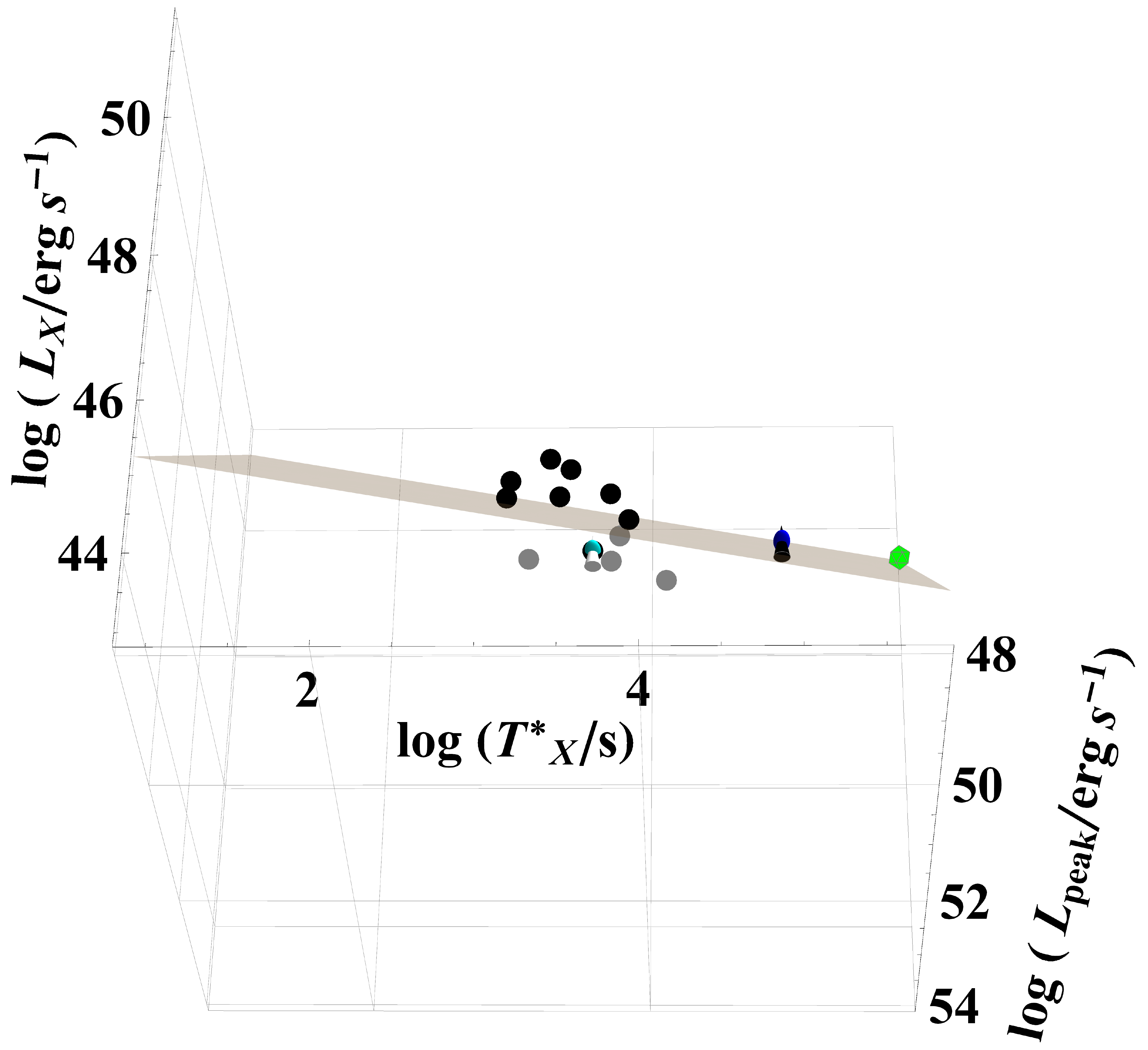}
\includegraphics[scale=0.21]{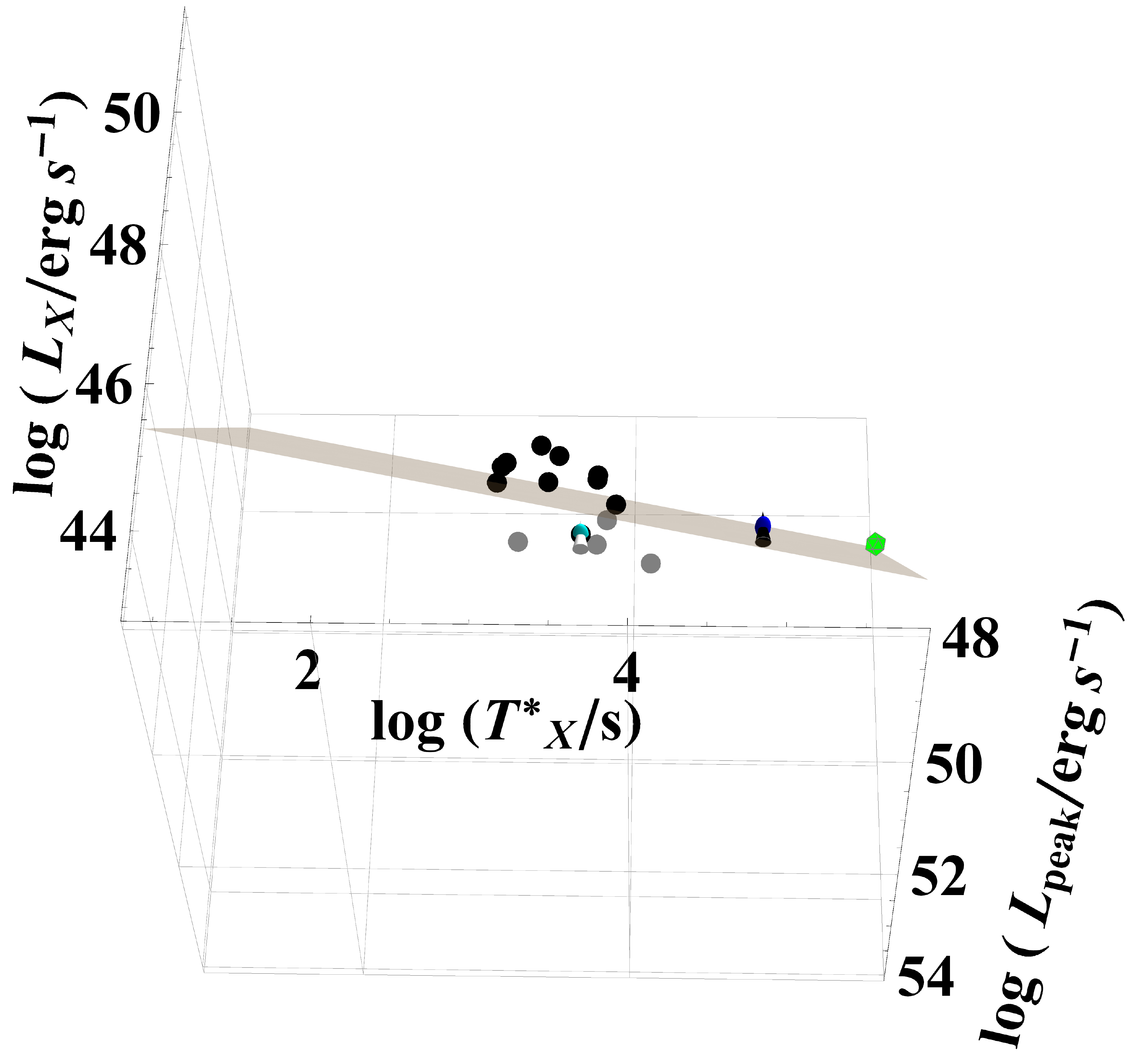}
\includegraphics[scale=0.21]{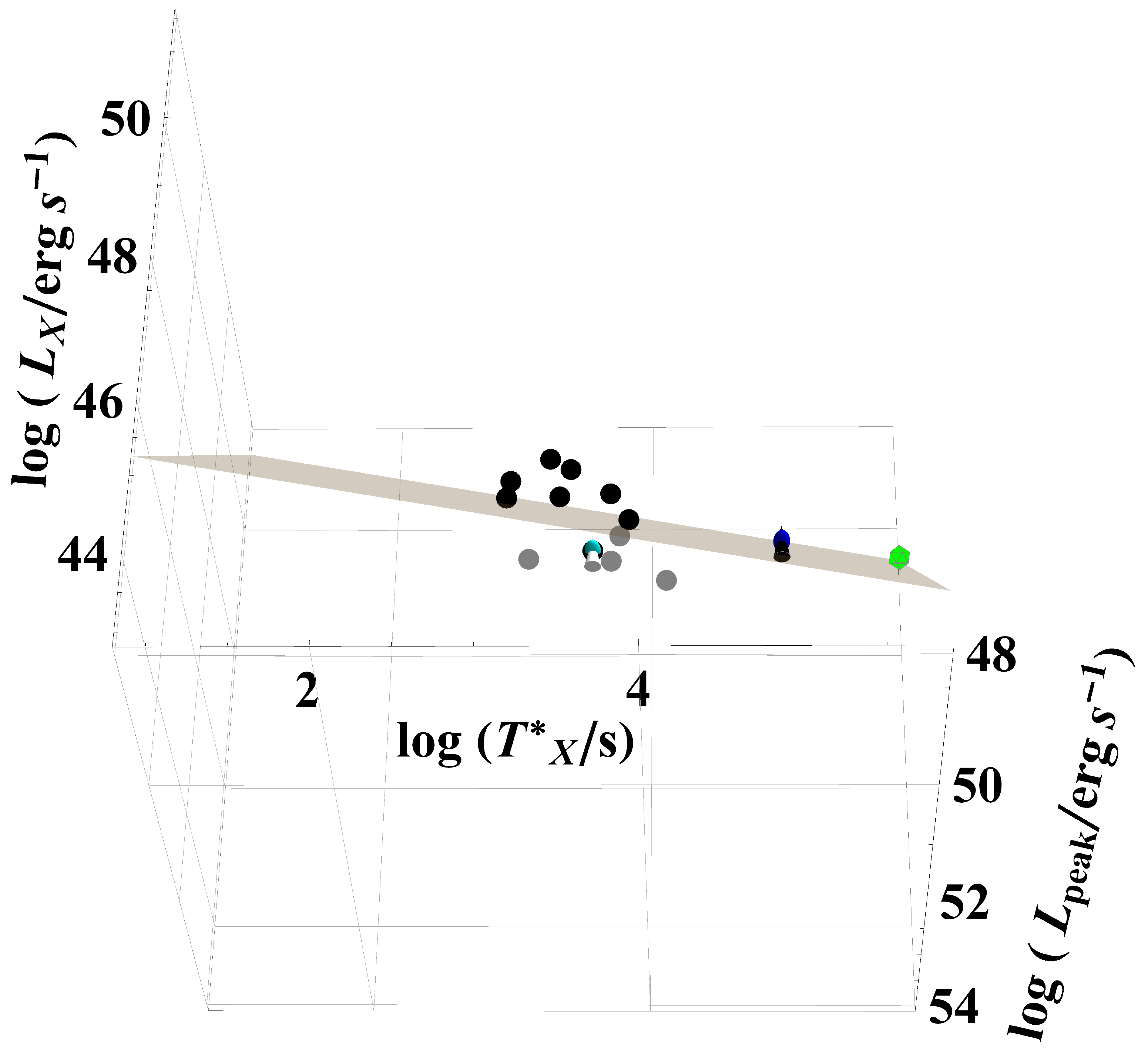}

\caption{\bf{The fundamental planes from Eq. \ref{planeequation1} plotting for the cases of $q$ compatible with $q=0$ reported in Table \ref{Dagostini_compatibleq0}. The GRBs are divided into ISM SC for all GRBs (upper left panel), ISM FC for all GRBs (upper right panel), ISM SC for lGRBs (lower left panel), ISM FC for lGRBs (lower right). The color-coded and symbol-coded are the same as Figure \ref{3D_AllISM_AllWIND_q0}. Here we consider the GRB categories for $q=0$ from Table \ref{CR3}.}} \label{contours_q0compatible_ISM}

\end{figure}

\begin{figure}

\centering

\includegraphics[scale=0.26]{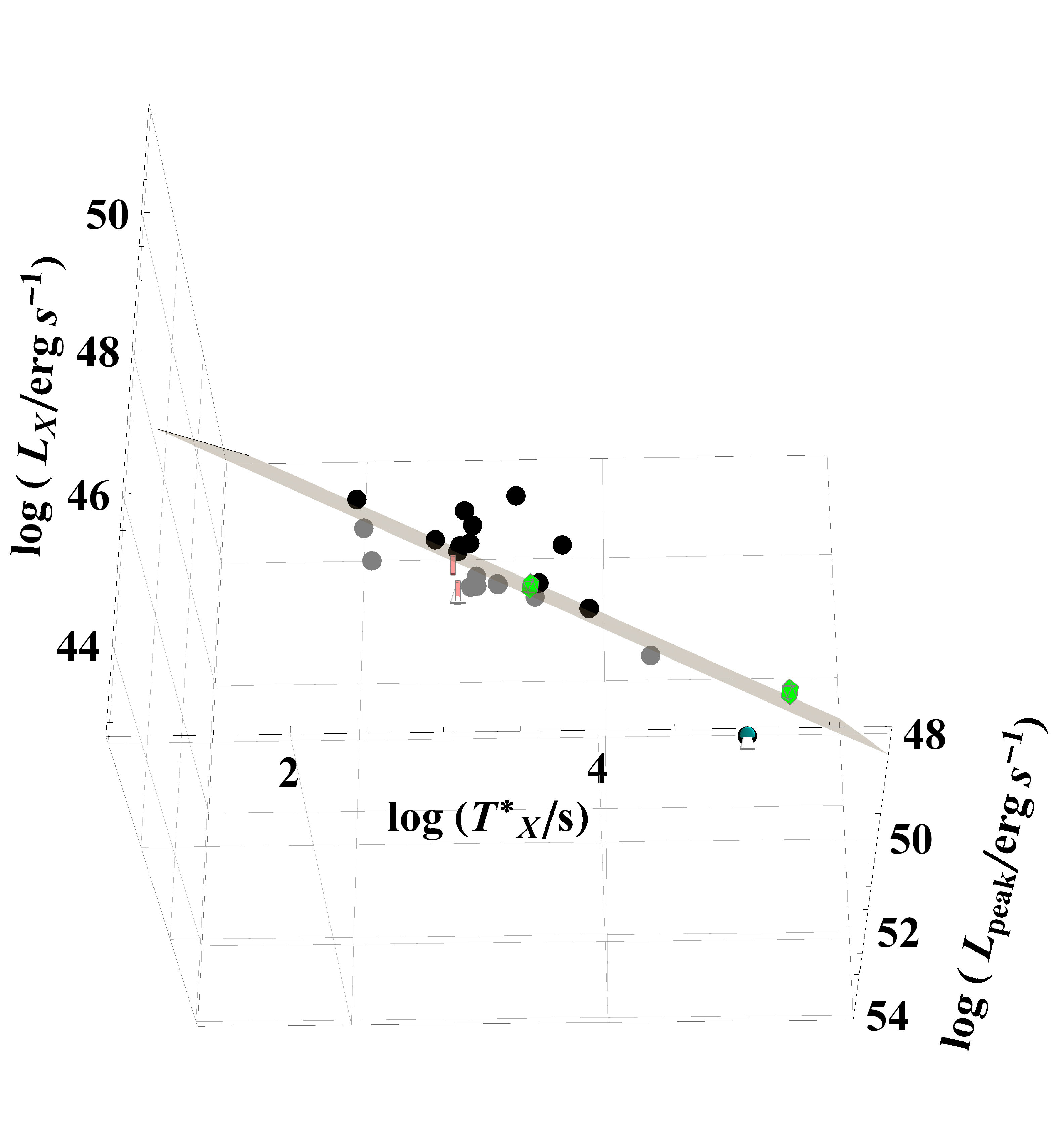}
\includegraphics[scale=0.21]{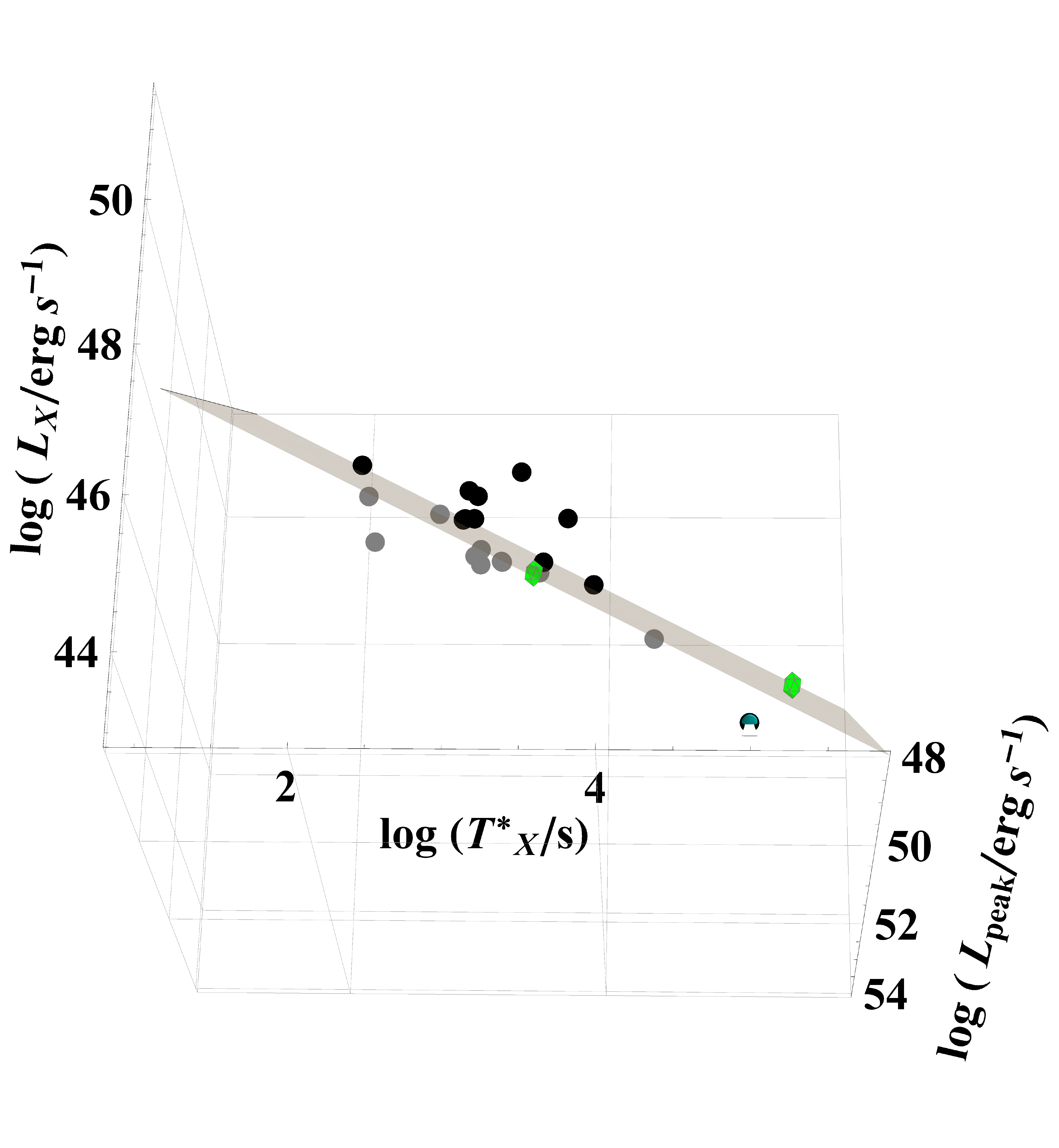}

\caption{\bf{GRBs from Table \ref{Dagostini_compatibleq0}, belonging to Wind SC, according to the same convention of Figure \ref{contours_q0compatible_ISM}. The cases reported here are: All (lGRBs and sGRBs) and lGRBs only. The FC is not reported due to the lack of a significant number of points.}} \label{contours_q0compatible_WIND}

\end{figure}

\begin{figure}

\centering

\includegraphics[scale=0.6]{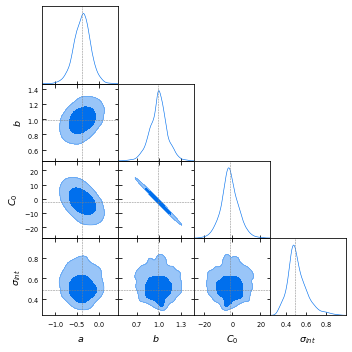}
\includegraphics[scale=0.6]{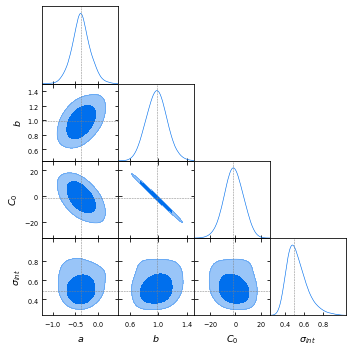}
\includegraphics[scale=0.6]{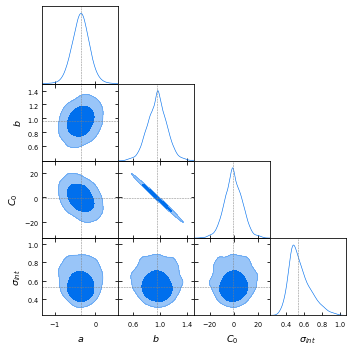}
\includegraphics[scale=0.6]{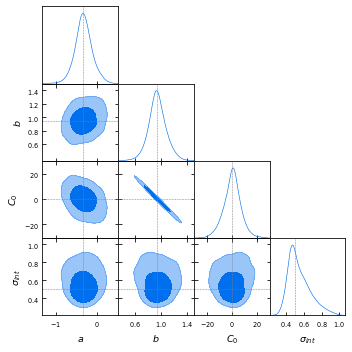}

\caption{\bf{Contour plots for groups taken from Table \ref{CR3} $(q=0)$. The upper left panel shows contours for the ISM SC for all GRBs, the right upper panel shows the ISM FC for all GRBs,
the lower-left panel shows the ISM Slow Cooling for lGRBs and the lower right panel shows the ISM Fast Cooling for lGRBs.}} \label{contour_qcomp0_ISM}

\end{figure}

\begin{figure}[H]

\centering

\includegraphics[scale=0.6]{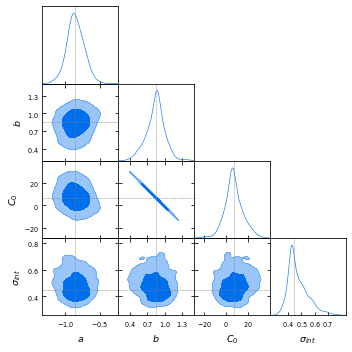}
\includegraphics[scale=0.6]{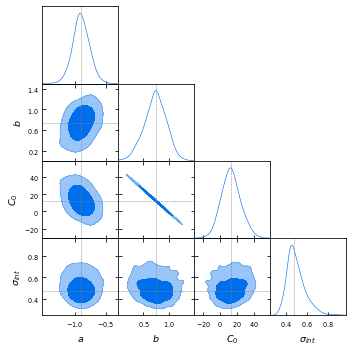}

\caption{\bf{Contour plots for groups taken from Table \ref{CR3} $(q=0)$. We consider the Wind SC for all GRBs (left panel) and the Wind SC for all lGRBs.}} \label{contour_qcomp0_WIND}

\end{figure}

\noindent A different fundamental plane is fit for each group of GRBs. The plane equation is:

\begin{equation}
\hspace{10ex}
\log L_a= C_{\rm o} + a \log T_{\rm a} + b \log L_\mathrm{{peak}}\ + \sigma_{\rm int},
\label{planeequation1}
\end{equation} 

\noindent where $C_{\rm o}$ is the plane's normalization, $\sigma_{\rm int}$, the intrinsic scatter of the sample, $a$ and $b$ are the parameters associated with $T_{\rm a}$ and $L_{\rm peak}$. We choose groups of GRBs that encompass: 1) all GRBs satisfying a constant density ISM environment; 2) all GRBs satisfying a wind environment (both points 1 and 2 are considered regardless of the cooling regime); 3) an ISM environment with SC; 4) an ISM environment with FC; 5) a wind environment with SC; 6) a wind medium with FC. Pictures of the fundamental plane, contour plots related to their best-fit parameters and their tabulated values are given in Figures from \ref{3D_AllISM_AllWIND_q0} to \ref{contour_qcomp0_WIND}, together with the Tables \ref{Dagostini_q0} and \ref{Dagostini_q0.5}. We investigate the parameters, in particular for the two groups used for the analysis, All ISM and All Wind and the All ISM {\bf for both sGRBs and lGRBs} and All Wind {\bf for both sGRBs and lGRBs}, and see how their best-fit parameters compare to the rest of our sample. In the following fitting results, we will express two quantities: the $R^2_{adj}$ and the $p$-value. The former is the square of the sample correlation coefficient between the dependent and the independent variables. The latter represents the probability that the correlation of the data is obtained by chance.
{\bf For $q=0$, see Table \ref{Dagostini_q0} for All ISM and All Wind the $a$, $b$, and $C_0$ parameters are compatible within 2 $\sigma$. If we now compare ISM FC and SC we find out that they are compatible in 1 $\sigma$ for the $a$, $b$, and for $C_0$ values. Instead, for the Wind SC, the $a$ parameter is compatible only within 3 $\sigma$ with the ISM FC and 2 $\sigma$ for ISM SC; the $b$ parameter is within 2 $\sigma$ with the ISM FC and SC, while the $C_0$ parameter is within 2 $\sigma$ for the ISM FC and for the ISM SC. As a note, the Wind FC has only 3 GRBs, thus preventing us from the possibility of reliably fitting the plane. 
For $q=0$ for lGRBs $a$, $b$ and $C_0$ are compatible with 1 $\sigma$ between All ISM and All Wind. In the ISM FC and SC, the $a$, $b$ all $C_0$ parameters are compatible in 1 $\sigma$, while the Wind SC $a$, $b$ and $C_0$ are compatible within 2 $\sigma$.
For $q=0$ for sGRBs, the $a$, $b$, $C_0$ parameters are compatible within 1 $\sigma$ for All Wind and All ISM. For ISM SC and ISM FC the $a$, $b$, $C_0$ parameters are the same since the sGRBs that satisfy these scenarios are the same. For Wind SC, ISM FC, and ISM SC all parameters are compatible within 1 $\sigma$.}
{\bf We now discuss the $q=0.5$ cases reported in Table \ref{Dagostini_q0.5}. When we consider All GRBs, the $a$, $b$, and $C_0$ parameters are compatible in 1 $\sigma$ between All ISM and All Wind categories. Considering ISM SC and ISM FC, $a$ values are compatible in 1 $\sigma$, while $b$ and $C_0$ values in 2 $\sigma$. Between ISM SC and Wind SC, all values are compatible in 1 $\sigma$. For Wind SC and ISM FC, the $a$, $b$ and $C_0$ values are compatible in 2 $\sigma$. 

Within the subclass of lGRBs, the parameter values of All ISM and All Wind are compatible in 1 $\sigma$ between themselves; this compatibility holds also for ISM SC and ISM FC. For Wind SC and ISM SC, $a$, $b$, and $C_0$ are compatible in 1 $\sigma$. Considering instead Wind SC and ISM FC, the same compatibility in 1 $\sigma$ is observed.
In the case of Wind SC and Wind FC, only the $a$ parameters are compatible in 1 $\sigma$, while for $b$ and $C_0$ the compatibility is observable in 2 $\sigma$. For the sGRBs case with $q=0.5$, all the values of $a$, $b$, and $C_0$ are all compatible in 1 $\sigma$.}
As a note for all cases in which the fitting of the fundamental plane is not leading to reliable results due to the small values of $R^2_{adj}$ and for the high $p$-value $>5\%$ which indicates that the correlation induced by this environment is drawn by chance, then we discard these fittings which are indicated by dashes. Thus, we have not plotted the corresponding fundamental plane and the fitted parameters are omitted. 



\noindent \textbf{If we consider the analytical computation regarding the $q$ values we can refer to Table \ref{CR3}, in which we also present the subclasses of lGRBs and sGRBs and the correspondent environments and energy regimes.
Taking the sub-classes of GRBs gathered according to the results presented in Table \ref{CR3} and correspondent to given astrophysical environments, we have fitted the correspondent fundamental planes and derived the best fit parameter values presented in Table \ref{Dagostini_compatibleq0}.}

\noindent \textbf{For all GRBs the $a$, $b$ and $C_0$ parameters are compatible in 1 $\sigma$ between ISM SC and ISM FC. For the Wind SC $a$ value is compatible with the others in 2 $\sigma$, while $b$, $C_0$ are compatible in 1 $\sigma$. Considering the lGRBs subclass only, we observe that $a$, $b$, and $C_0$ are compatible in 1 $\sigma$ for the ISM SC and ISM FC; the Wind SC, instead, presents an $a$ value compatible in 2 $\sigma$ with the one of ISM SC and in 3 $\sigma$ with the one of ISM FC, while the $b$ and $C_0$ values are compatible with the ISM environments in 1 $\sigma$.}   

\noindent As an additive analysis, considering the GRBs which have a computed value of $q$ compatible with zero (see Table \ref{CR3}), we perform the same fitting with the fundamental plane relation. The results are gathered in Table \ref{Dagostini_compatibleq0}. \textbf{In summary, for the cases of $q$ computed analytically we find again that the Wind SC environment arises interests since it is characterized by the highest number of fulfilling GRBs with a low central value of the $\sigma_{\rm int}$ parameter.}

\begin{table*}
\begin{center}
\begin{tabular}{|c|c|c|c|c|c|c|c|}

\hline
All GRBs $(q=0)$ & $a$ & $b$ & $C_o$ & $\sigma_{\rm int}$ & $R^{2}_{\rm adj}$& P-value & N \\ \hline

   All ISM & -0.43 $\pm$ 0.14 & 1.09 $\pm$ 0.10 & -7.02 $\pm$ 5.77 & 0.46 $\pm$ 0.07 & 0.82 & $2\times10^{-10}$ & 39 \\

   All Wind  & -0.72 $\pm$ 0.05 & 0.92 $\pm$ 0.05 & 2.80 $\pm$ 2.55 & 0.44 $\pm$ 0.02 & 0.83 & $4\times10^{-60}$ & 186 \\

   ISM SC &  -0.43 $\pm$ 0.14  & 1.09 $\pm$ 0.10 & -7.02 $\pm$ 5.38 &  0.46 $\pm$ 0.07 & 0.82 & $2\times10^{-10}$ & 39 \\

   ISM FC & -0.33 $\pm$ 0.14 &  1.19 $\pm$ 0.10 &  -12.35 $\pm$ 5.57 &  0.39 $\pm$ 0.08 & 0.85 & $2\times10^{-9}$ & 33 \\

   Wind SC & -0.72 $\pm$ 0.05 & 0.92 $\pm$ 0.05 & 2.80 $\pm$ 2.48 & 0.43 $\pm$ 0.02 & 0.83 & $4\times10^{-60}$ & 186 \\

   Wind FC  & - & - & - & - & - & - & 3 \\ \hline

lGRBs $(q=0)$ & $a$ & $b$ & $C_o$ & $\sigma_{int}$ & $R^{2}_{adj}$& P-value & N \\ \hline

   All ISM & $-0.50\pm0.16$ & $1.02\pm0.10$  &  $-2.94\pm5.54$  &  $0.48\pm0.08$ & 0.82 & $3\times10^{-9}$ & 30 \\

   All Wind  & $-0.71\pm0.06$  &  $0.89\pm0.05$ & $4.05\pm2.51$ & $0.41\pm0.02$ & 0.85 & $2\times10^{-55}$ & 157 \\

   ISM SC & $-0.50\pm0.16$ &  $1.02\pm0.12$  & $-2.94\pm6.25$  & $0.48\pm0.08$ & 0.82 & $3\times10^{-9}$ & 30 \\

   ISM FC & $-0.40\pm0.17$ & $1.11\pm0.13$ & $-8.16\pm7.02$ & $0.42\pm0.09$ & 0.85 & $6\times10^{-8}$ & 24 \\

   Wind SC & $-0.71\pm0.06$  & $0.89\pm0.05$ & $4.05\pm2.71$ & $0.41\pm0.03$ & 0.85 & $2\times10^{-55}$ & 157 \\

   Wind FC  & - & - & - & - & - & - & 3 \\
   \hline

sGRBs $(q=0)$ & $a$ & $b$ & $C_o$ & $\sigma_{int}$ & $R^{2}_{adj}$& P-value & N \\ \hline

   All ISM & $-0.41\pm0.26$ & $1.57\pm0.14$ & $-31.61\pm7.59$  &  $0.04\pm0.15$ & 0.96 & $0.005$ & 9 \\

   All Wind  & $-0.60\pm0.10$ &  $1.34\pm0.09$ & $-19.73\pm4.82$ & $0.24\pm0.06$ & 0.88 & $2\times10^{-9}$ & 28 \\

   ISM SC & $-0.41\pm0.27$ &  $1.57\pm0.14$  & $-31.61\pm7.31$  & $0.04\pm0.16$ & 0.96 & $0.005$ & 9 \\

   ISM FC & $-0.41\pm0.28$ & $1.57\pm0.16$ &  $-31.61\pm8.63$ & $0.04\pm0.15$ & 0.96 & $0.005$ & 9 \\

   Wind SC & $-0.60\pm0.11$ & $1.34\pm0.12$ & $-19.73\pm6.41$ & $0.24\pm0.06$ & 0.88 & $2\times10^{-9}$ & 28 \\

   Wind FC  & - & - & - & - & - & - & 0 \\
   \hline

\end{tabular}

\end{center}

\caption{The best-fit parameters for GRBs with redshift together with their number for each environment fulfilled from Table \ref{Redshifttable_p>2} in the case of $q=0$.}

\label{Dagostini_q0}
\end{table*}

\begin{table*}
\begin{center}
\begin{tabular}{|c|c|c|c|c|c|c|c|}

\hline
All GRBs $(q=0.5)$ & $a$ & $b$ & $C_o$ & $\sigma_{int}$ & $R^{2}_{adj}$& P-value & N \\ \hline

   All ISM & -0.73 $\pm$ 0.06 & 0.89 $\pm$ 0.05 & 4.38 $\pm$ 2.88 & 0.41 $\pm$ 0.03 & 0.81 & $4\times10^{-40}$ & 133 \\

   All Wind  & -0.85 $\pm$ 0.07 & 0.79 $\pm$ 0.06 & 9.72 $\pm$ 3.36 & 0.49 $\pm$ 0.03 & 0.77 & $7\times10^{-44}$ & 156 \\

   ISM SC & -0.73 $\pm$ 0.06 & 0.89 $\pm$ 0.05 & 4.38 $\pm$ 2.82 & 0.41 $\pm$ 0.03 & 0.81 & $4\times10^{-40}$ & 133 \\

   ISM FC & -0.53 $\pm$ 0.14 & 1.13 $\pm$ 0.12 & -8.60 $\pm$ 6.62 &  0.36 $\pm$ 0.08 & 0.78 & $4\times10^{-7}$ & 32 \\

   Wind SC & -0.81 $\pm$ 0.06 & 0.85 $\pm$ 0.05 & 6.47 $\pm$ 2.81 & 0.46 $\pm$ 0.03 & 0.80 & $2\times10^{-46}$ & 156 \\

   Wind FC & - & - & - & - & - & - & 11 \\

    \hline

lGRBs $(q=0.5)$ & $a$ & $b$ & $C_o$ & $\sigma_{int}$ & $R^{2}_{\rm adj}$& P-value & N \\ \hline

   All ISM & $-0.70\pm0.07$ & $0.87\pm0.06$  &  $5.13\pm2.95$ & $0.37\pm0.03$ & 0.84 & $8\times10^{-39}$ & 113 \\

   All Wind  & $-0.81\pm0.07$  & $0.81\pm0.06$ & $8.54\pm3.28$ & $0.43\pm0.03$ & 0.82 & $1\times10^{-44}$ & 135 \\

   ISM SC & $-0.70\pm0.07$ &  $0.87\pm0.06$  & $5.13\pm3.00$  & $0.37\pm0.03$ & 0.84 & $8\times10^{-39}$ & 113 \\

   ISM FC & $-0.61\pm0.17$ & $1.00\pm0.17$ & $-1.76\pm9.04$ & $0.38\pm0.09$ & 0.76 & $1\times10^{-6}$ & 24 \\

   Wind SC & $-0.81\pm0.07$ &  $0.81\pm0.06$ & $8.54\pm3.10$ & $0.43\pm0.03$ & 0.82 & $1\times10^{-44}$ & 135 \\

   Wind FC & - & - & - & - &- & - & 9 \\
   \hline

sGRBs $(q=0.5)$ & $a$ & $b$ & $C_o$ & $\sigma_{int}$ & $R^{2}_{\rm adj}$& P-value & N \\ \hline

   All ISM & $-0.61\pm0.14$ & $1.26\pm0.15$ & $-15.52\pm7.76$ & $0.23\pm0.08$ & 0.83 & $2\times10^{-5}$ & 19 \\

   All Wind  & $-0.64\pm0.15$  &  $1.24\pm0.14$ & $-14.49\pm7.23$ & $0.24\pm0.08$ & 0.82 & $8\times10^{-6}$ & 20 \\

   ISM SC & $-0.61\pm0.15$ & $1.26\pm0.16$ & $-15.52\pm8.25$ & $0.23\pm0.08$ & 0.83 & $2\times10^{-5}$ & 19 \\

   ISM FC & - & - & - & - & - & - & 8 \\

   Wind SC & $-0.64\pm0.15$ & $1.24\pm0.15$ & $-14.49\pm7.97$ & $0.24\pm0.08$ & 0.82 & $8\times10^{-6}$ & 20 \\

   Wind FC & - & - & - & - & - & - & 2 \\
   \hline
   
\end{tabular}

\end{center}

\caption{The best-fit parameters for GRBs with redshift together with their number for each environment fulfilled from Table  \ref{Redshifttable_p>2} in the case of $q=0.5$.}

\label{Dagostini_q0.5}
\end{table*}

\begin{table*}
\begin{center}
\begin{tabular}{|c|c|c|c|c|c|c|c|}

\hline
All GRBs $(q=0)$ & $a$ & $b$ & $C_o$ & $\sigma_{int}$ & $R^{2}_{adj}$& P-value & N \\ \hline

   ISM SC & -0.40 $\pm$ 0.19 & 0.98 $\pm$ 0.12 & -1.45 $\pm$ 6.38 & 0.45 $\pm$ 0.10 & 0.82 & $3\times10^{-5}$ & 22 \\

   ISM FC & -0.37 $\pm$ 0.20 & 0.99 $\pm$ 0.11 & -1.89 $\pm$ 6.07 & 0.46 $\pm$ 0.10 & 0.82 & $4\times10^{-5}$ & 20  \\

   Wind SC & -0.85 $\pm$ 0.13 & 0.85 $\pm$ 0.18 & 6.87 $\pm$ 9.52 & 0.43 $\pm$ 0.08 & 0.78 & $4\times10^{-8}$ & 26  \\

   Wind FC & - & - & - & - & - & - & 0 \\

\hline
lGRBs $(q=0)$ & $a$ & $b$ & $C_o$ & $\sigma_{int}$ & $R^{2}_{adj}$& P-value & N \\ \hline

   ISM SC & $-0.36\pm0.20$ & $0.96\pm0.13$ & $-0.28\pm6.81$ & $0.45\pm0.10$ & 0.80 & $3\times10^{-3}$ & 17 \\

   ISM FC & $-0.32\pm0.14$ & $0.95\pm0.10$ & $-0.23\pm5.12$ & $0.46\pm0.08$ & 0.80 & $4\times10^{-3}$ & 15 \\

   WIND SC & $-0.90\pm0.14$ & $0.72\pm0.20$ & $13.36\pm10.65$ & $0.44\pm0.10$ & 0.79 & $5\times10^{-8}$ & 24 \\

   WIND FC & - & - & - & - & - & - & 0 \\
   
   \hline




   
    \hline

\end{tabular}

\end{center}

\caption{The best-fit parameters for GRBs with redshift together with their number for each environment fulfilled from Table \ref{CR3} considering $q=0$. \bf{The cases of sGRBs are not reported due to the lack of a significant number of data points.}}

\label{Dagostini_compatibleq0}
\end{table*}

\section{Summary and Conclusions}

In summary, using GRB LCs detected by \emph{Swift} (more specifically two samples of 222 and 233 GRBs with known and unknown redshifts respectively), we consider phase II of the LCs during the time duration of the PE ($T_{\rm t}$ to $T_{\rm a}$), to test both the CRs and verify the 3--D Dainotti relation for our set of GRBs, for better understanding the emission mechanisms of GRBs as well as their cosmological implications as potential standard candles. The majority of GRBs do satisfy the CRs as a whole, and the most fulfilled CR sets in terms of percentage are the ones of the Wind SC environment ($q=0$ and $q=0.5$, with and without redshift). {\bf This trend is confirmed also in the sub-classes of lGRBs and sGRBs.} 

\noindent Furthermore, we also test the 3--D Dainotti relation for samples based on GRBs fulfilling CRs which lead to particular astrophysical environments. We find out that the sGRBs both for all ISM or ISM Fast Cooling with $q=0$ have the smallest $\sigma_{int}=0.04 \pm 0.15$ in terms of the fundamental plane relation with a probability of chance occurrence of $p=0.005$. This scatter is even smaller than the scatter found for phase III of the LCs ($\sigma=0.29 \pm 0.06$) when we consider the same ISM FC regime, but for all GRBs  \citep{Srinivasaragavan2020}. This $\sigma$ is compatible with 1 $\sigma$ with the platinum sample corrected for redshift evolution and selection effects, \citep{Dainotti2020a}, where $\sigma_{int}=0.22 \pm 0.10$. \textbf{Despite the estimated values of $\sigma_{int}$ for $q=0$ for the sGRBs are all compatible in 1 $\sigma$, it is worth noticing that the fundamental planes for sGRBs related to All ISM Environment with $q=0$ not only have the smallest $\sigma_{int}=0.04 \pm 0.15$ in terms of the fundamental plane relation (with a probability of chance occurring of $p=0.005$), but also the smallest error bar on the same parameter. This leads to the idea that this particular environment is able to put further constraints on the determination of the intrinsic scatter of GRBs from the fundamental plane relation.} 


\clearpage

\section*{Funding}
G. S. thank the United States Department of Energy funding scheme related to the Science Undergraduate Laboratory Internship (SULI) program from which also Bowden, Wynne, Wagner were supported. M.G.D. acknowledges the support from the NAOJ Division of Science and is particularly grateful to Dr. Cuellar for the support in managing the SULI students at SLAC. S.N. is partially supported by JSPS Grants-in-Aid for Scientific Research KAKENHI (A) 19H00693, Pioneering Program of RIKEN for Evolution of Matter in the Universe (r-EMU), and Interdisciplinary Theoretical and Mathematical Sciences Program (iTHEMS) of RIKEN. N.F. acknowledges the financial support from UNAM-DGAPA-PAPIIT through the grant IA102019.

\section*{Acknowledgements}
This work used the data supplied by the UK Swift Science Data Centre, University of Leicester. We thank S. Savastano and G. Sarracino for partially writing the Python codes for the D'Agostini method. We are grateful to L. Bowden, R. Wynne, R. Wagner, Z. Nuygen for fitting some of the GRB lightcurves. 

\clearpage

\bibliographystyle{aa}
\bibliography{bib_Swiftpaper_v1.bib}

\label{lastpage}
\end{document}